\documentclass[useAMS,usenatbib]{mn2e}
\pdfoutput=1
\usepackage{soul} 
\usepackage{dcolumn}
\usepackage{amsmath,amssymb,amstext}
\usepackage{epsfig}
\usepackage{changebar}
\usepackage{caption}

\usepackage{longtable,lscape}
\usepackage[version=3]{mhchem} 
\newcolumntype{d}[1]{D{.}{\cdot}{#1}}
\newcolumntype{.}{D{.}{.}{-1}}
\newcommand{\msun}{M$_\odot$}
\newcommand{\vlsr}{$v_{\rm{LSR}}$}

\newcommand{\kms}{km\,s$^{-1}$}
\newcommand{\Jm}{\emph{J}}
\newcommand{\Km}{\emph{K}}

\newcommand{\HII}{H{\sc ii}}
\newcommand{\UCHII}{UCH{\sc ii}}

\newcommand{\Trot}{$T_{\rm{rot}}$}
\newcommand{\Tkin}{$T_{\rm{kin}}$}

\newcommand{\water}{\ce{H2O}}
\newcommand{\methanol}{\ce{CH3OH}}
\newcommand{\ammonia}{\ce{NH3}}
\newcommand{\one}{(1,1)}
\newcommand{\two}{(2,2)}

\newcommand{\um}{${\rm \mu m}$}

\title[Ammonia towards dust emission in the G333~GMC]
  {Molecular line mapping of the giant molecular cloud associated with RCW~106 - IV. Ammonia towards dust emission}
\author[V.\,Lowe et al.]{
  \parbox{\textwidth}{
    V.~Lowe\thanks{E-mail:Vicki.Lowe@unsw.edu.au (UNSW)}$^{1,2}$,
    M.~R.~Cunningham$^{1}$,
    J.~S.~Urquhart$^{3,2}$,
    J.~P.~Marshall$^{4,1}$,
    S.~Horiuchi$^{5}$,
    N.~Lo$^{6}$,
    A.~J.~Walsh$^{7}$,
    C.~H.~Jordan$^{8,2}$,
    P.~A.~Jones$^{1}$
  }
  \vspace{0.4cm}\\
  \parbox{\textwidth}{
    $^{1}$Department of Astrophysics and Optics, School of Physics, University of New South Wales, Sydney, NSW 2052, Australia\\
    $^{2}$Australia Telescope National Facility, CSIRO Astronomy and Space Science, PO Box 76, Epping, NSW 1710, Australia \\
    $^{3}$Max-Planck-Institut f\"ur Radioastronomie, Auf dem H\"ugel 69, Bonn, Germany \\
    $^{4}$Departamento de F\'isica Te\'orica, Facultad de Ciencias, Universidad Aut\'onoma de Madrid, Cantoblanco, 28049, Madrid, Spain\\
    $^{5}$CSIRO Astronomy and Space Science, CDSCC, PO Box 1035, Tuggeranong, ACT 2901, Australia \\
    $^{6}$Departamento de Astronom\'ia,  Universidad de Chile,  Camino El Observatorio 1515, Las Condes, Santiago, Casilla 36-D, Chile\\
    $^{7}$International Centre for Radio Astronomy Research, Curtin University, Bentley, WA 6102, Australia\\
    $^{8}$School of Mathematics and Physics, University of Tasmania, Private Bag 21, Hobart, Tasmania 7001, Australia
  }
}

\begin{document}
\date{Accepted 2014 March 18.  Received 2014 March 18; in original form 2013 December 8}
\pagerange{\pageref{firstpage}--\pageref{lastpage}} \pubyear{2014}
\maketitle
\label{firstpage}
\begin{abstract}
Here we report observations of the two lowest inversion transitions of ammonia (\ammonia) with the 70-m Tidbinbilla radio telescope. The aim of the observations is to determine the kinetic temperatures in the dense clumps of the G333 giant molecular cloud associated with RCW~106 (hereafter known as the G333 GMC) and to examine the effect that accurate measures of temperature have on the calculation of derived quantities such as mass. This project is part of a larger investigation to understand the timescales and evolutionary sequence associated with high-mass star formation, particularly its earliest stages. Assuming that the initial chemical composition of a giant molecular cloud is uniform, any abundance variations within will be due to evolutionary state.

We have identified 63 clumps using SIMBA 1.2-mm dust continuum maps and have calculated gas temperatures for most (78 per cent) of these dense clumps. After using \textit{Spitzer} GLIMPSE 8.0\,\um\ emission to separate the sample into IR-bright and IR-faint clumps, we use statistical tests to examine whether our classification shows different populations in terms of mass and temperature. We find that in terms of log clump mass (2.44~--~4.12\,\msun) and log column density (15.3~--~16.6 cm$^{-2}$), that there is no significant population difference between IR-bright and IR-faint clumps, and that kinetic temperature is the best parameter to distinguish between the gravitationally bound state of each clump. The kinetic temperature was the only parameter found to have a significantly low probability of being drawn from the same population. This suggests that clump radii does not have a large effect on the temperature of a clump, so clumps of similar radii may have different internal heating mechanisms. We also find that while the IR-bright clumps have a higher median log virial mass than the IR-faint clumps (IR-bright:~2.88\,\msun; IR-faint:~2.73\,\msun), both samples have a similar range for both virial mass and FWHM (IR-bright:~log virial mass~=~2.03~--~3.68\,\msun, FWHM~=~1.17~--~4.50\,\kms; IR-faint:~log virial~mass~=~2.09~--~3.35\,\msun, FWHM~=~1.05~--~4.41\,\kms). There are 87~per~cent (40~of~46) of the clumps with masses larger than the virial mass, suggesting that they will form stars or are already undergoing star formation.
\end{abstract}




\section{INTRODUCTION}
\begin{figure*}
  \centering
  \includegraphics[trim=20 60 85 97,clip,width=0.99\textwidth]{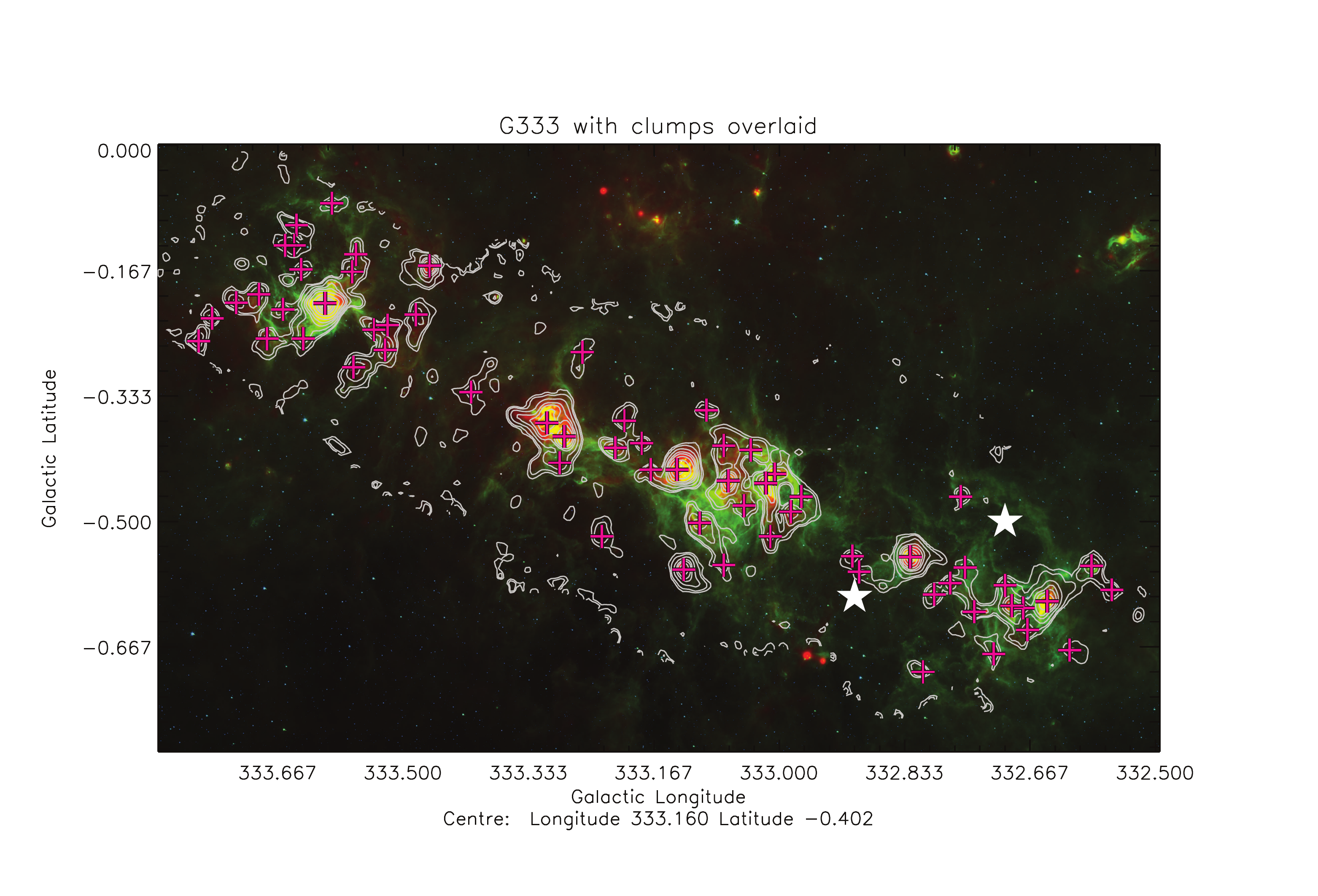}
  \caption{\label{fig:overview} The G333 giant molecular cloud as seen in the infrared. The shocked gas (\textit{Spitzer} 4.5\,\um), PAH$^+$ emission (\textit{Spitzer} 8.0\,\um) and cool dust (\textit{Herschel} 160\,\um) are shown in blue, green and red, respectively. The contours are from the SIMBA 1.2\,mm dust continuum emission at 0.1, 0.2, 0.5, 1, 1.5, 2.5, 5, 10 and 12\,Jy/beam (as per the contours of fig. 4 of \citealt{Mookerjea2004}). The position of the peak emission within SIMBA dust clumps identified with {\sc clumpfind} are overlaid with pink plus signs. The positions associated with RCW~106 have been overlaid with white stars \citep{Rodgers1960}.}
\end{figure*}

The life cycles of high-mass ($\ge$8\,\msun) stars have a major impact on the evolution of galaxies, while in turn, the position of a molecular cloud in the Galaxy has a major impact on the efficiency and type of star formation which occurs therein \citep{Luna2006}. However, exactly how these stars form, on what timescales and how they shape their environments during this active and energetic phase is poorly understood (see the review by \citealt{Zinnecker2007}).

The G333 giant molecular cloud (GMC), centred at~$l$~=~333.$^{\circ}$2, $b$~=~-0.$^{\circ}$4, is a 1.2\degr\ $\times$ 0.6\degr\ region of the southern Galactic Plane located at a distance of 3.6 kpc \citep{Lockman1979} and is the fourth most active star forming region in the Galaxy \citep{Urquhart2014}. The complex forms part of the Galactic Ring of molecular clouds at a Galactocentric radius of 3-5 kpc, it contains the bright \HII~regions RCW~106 (G332.9-0.6; Rodgers, Campbell \& Whiteoak 1960) and G333.6-0.2 \citep{Manchester1969}, as well as high-mass star forming clumps, MSX infrared sources and \textit{Spitzer} ``Extended Green Objects'' \citep{Cyganowski2008}. This cloud also contains the very young, high-mass star forming core G333.125$-$0.567 \citep{Becklin1973,Garay2004,Lo2007,Lo2011} which is an isolated core at an evolutionary stage without any radio continuum or near-infrared detection. We have used the Mopra radio telescope to study the G333~GMC extensively in 20 molecular tracers including $^{13}$CO \citep{Bains2006}, C$^{18}$O \citep{Wong2008}, HCN, HCO$^+$, and N$_2$H$^+$ \citep{Lo2009}.

In order to investigate a large sample of star forming clumps at different stages of evolution, we are conducting a multi-wavelength study of the G333~GMC. This region is within the area of the \textit{Spitzer} GLIMPSE (3.6, 4.5, 5.8, 8.0\,\um; \citealt{Benjamin2003,Churchwell2009}) and MIPSGAL (24, 70\,\um; \citealt{Carey2009}), and the \textit{Herschel} Hi-GAL (70, 160, 250, 350, 500\,\um; \citealt{Molinari2010}) surveys. \citet{Mookerjea2004} have defined multiple millimetre continuum clumps within this one molecular cloud, some associated with maser emission (e.g. \citealt{Breen2007}), as well as obvious polycyclic aromatic hydrocarbon (PAH) emission. We expect to have a range of evolutionary states within our clumps.

The ammonia (\ammonia) molecule can be used to probe the physical conditions within dense molecular gas. It has a large number of transitions that are sensitive to a variety of excitation conditions and can be detectable in both warm molecular gas and quiescent dark clouds. The rotational temperature and optical depth for each clump can be calculated from multiple (\Jm,\Km) inversion transitions and their hyperfine components, respectively \citep{Ungerechts1983}. Hence the kinetic temperature can be calculated \citep{Tafalla2004}; however the lower \one\ and \two\ transitions are only useful for constraining temperatures lower than 30\,K \citep{Danby1988,Hill2010}.

In this paper we report on the selection of dense clumps based on SIMBA dust emission and pointed \ammonia~\one\ and \two\ observations towards these clumps. Although finding the kinetic temperature is our main concern there are also several other physical parameters that can be determined from our measurements, including virial mass and column density. This is so we will have a sample of clumps within the G333~GMC, with clear selection criteria which will be used to compare the 3-mm molecular transitions and infrared \textit{Herschel} and \textit{Spitzer} dust surveys. We have chosen to study a single molecular cloud as the initial chemical composition of each clump should be very similar. Obtaining a reliable temperature measurement of each clump will allow us, in future papers, to calculate accurate molecular abundances for comparison between clumps.


\section{OBSERVATIONS AND DATA REDUCTION}
\label{sect:observations}


\subsection{Selection of clumps}
\label{sect:clumps}
We have utilised the 1.2-mm dust continuum emission from the SIMBA instrument on the Swedish-ESO Telescope (SEST; \citealt{Mookerjea2004}) as a tracer for dense molecular material. To complement the density tracers from \citet{Bains2006}, \citet{Wong2008} and \citet{Lo2009}, we have convolved the \citet{Mookerjea2004} beam size from 24\,arcsec to 36\,arcsec, to match the beam size of the Mopra radio telescope at 100\,GHz \citep{Ladd2005}. Using the {\sc starlink} tool {\sc cupid}, we have chosen the 2D {\sc clumpfind} algorithm to automatically, but robustly, detect physically realistic clumps. The {\sc clumpfind} algorithm works by identifying local maxima and includes all contiguous pixels down to the user-set contour level. If the contiguous pixels have already been identified in an earlier clump, then those pixels are included in that clump. For contiguous contours a ``friends-of-friends'' algorithm is used to assign the common pixels to a clump. These steps are repeated until the lowest level is reached.  For a complete description of the {\sc clumpfind} algorithm, see \citet{Williams1994}. Using a 3$\sigma$ detection limit we have identified 63 clumps. The catalogue of clumps found in this paper differ from those in \citet{Mookerjea2004} because of the differing resolutions (24 arcsec versus 36 arcsec). The global positions are shown in Fig.~\ref{fig:overview}, the local environment in Fig~\ref{app:3colour}, and the {\sc clumpfind} results can be found in \S\ref{sec:parameters} and Table \ref{tbl:cf_parameters_jsu}.


\subsection{Ammonia from the 70-m Tidbinbilla radio telescope}
\label{sec:Tidbinbilla}
Observations using the 70-m Tidbinbilla radio telescope were conducted in good weather conditions on 2011 July 11 and 12. Pointing errors were measured and corrected every few hours, by observing the quasar 1613-586, within 9$^\circ$ from the target sources, yielding residual pointing errors smaller than 2\,arcsec. System noise temperatures were in the range of $\sim$45-90\,K, but typically $\sim$50\,K. Sensitivity was $\sim$40-100\,mK, but typically $\sim$50\,mK. During these observations a left hand circularly polarized signal was recorded across two 64\,MHz bands centred at 23708 and 23870\,MHz, covering the \ammonia~\one\ and \two\ transitions in the first band and the (3,3) transition in the second band. With a bandwidth of 2048\,channels across 64\,MHz, each channel width was 31.25\,kHz (or a velocity channel width of 0.4\,\kms). Each source was observed for 2\,min on-source integration.

\begin{figure*}
  \centering
  \includegraphics[trim=100 20 190 40,clip,width=0.39\textwidth]{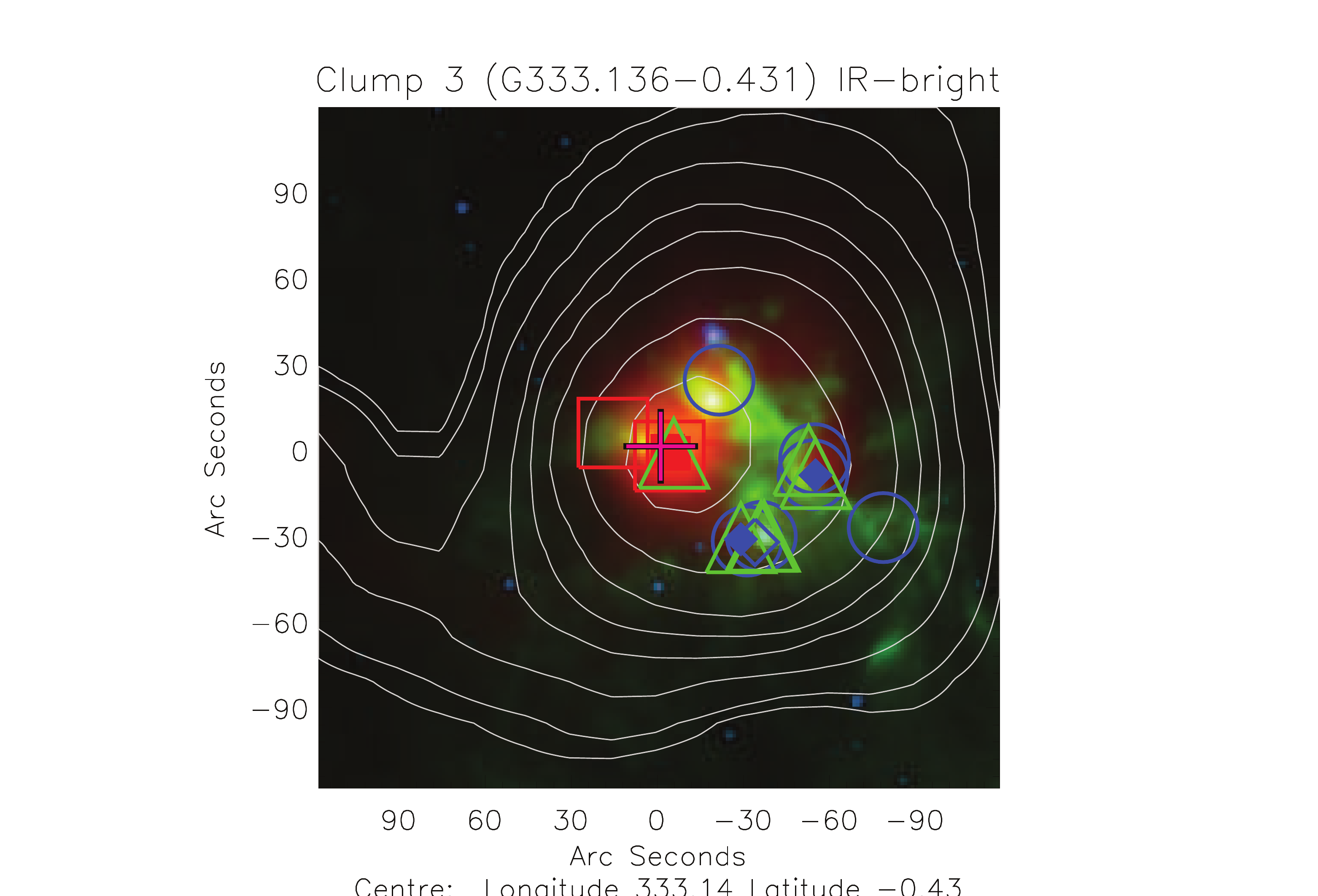}
  \includegraphics[width=0.49\textwidth, trim= 10 0 0 0]{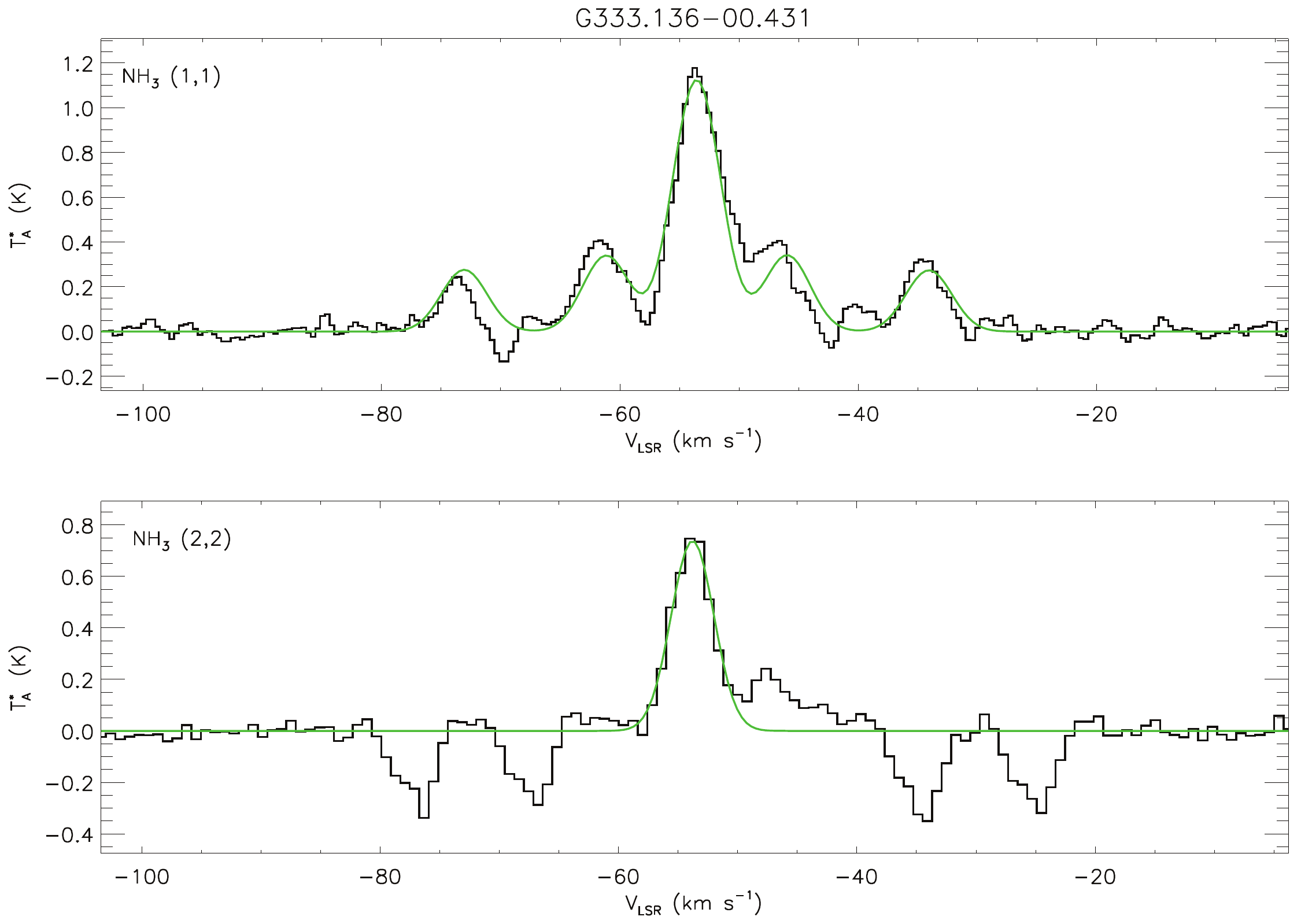}\\
  \vspace{1cm}
  \includegraphics[trim=100 20 190 40,clip,width=0.39\textwidth]{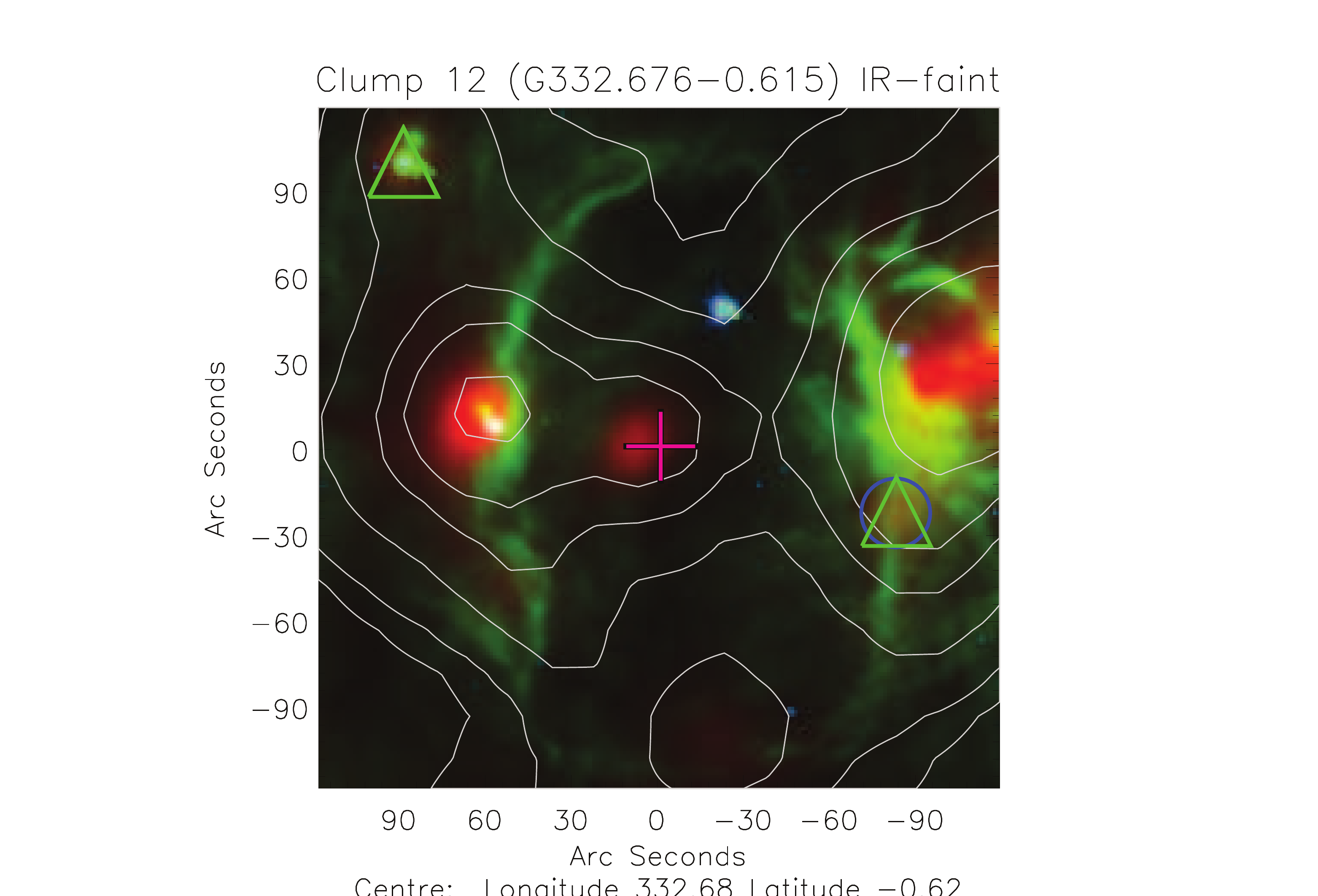}
  \includegraphics[width=0.49\textwidth, trim= 10 0 0 0]{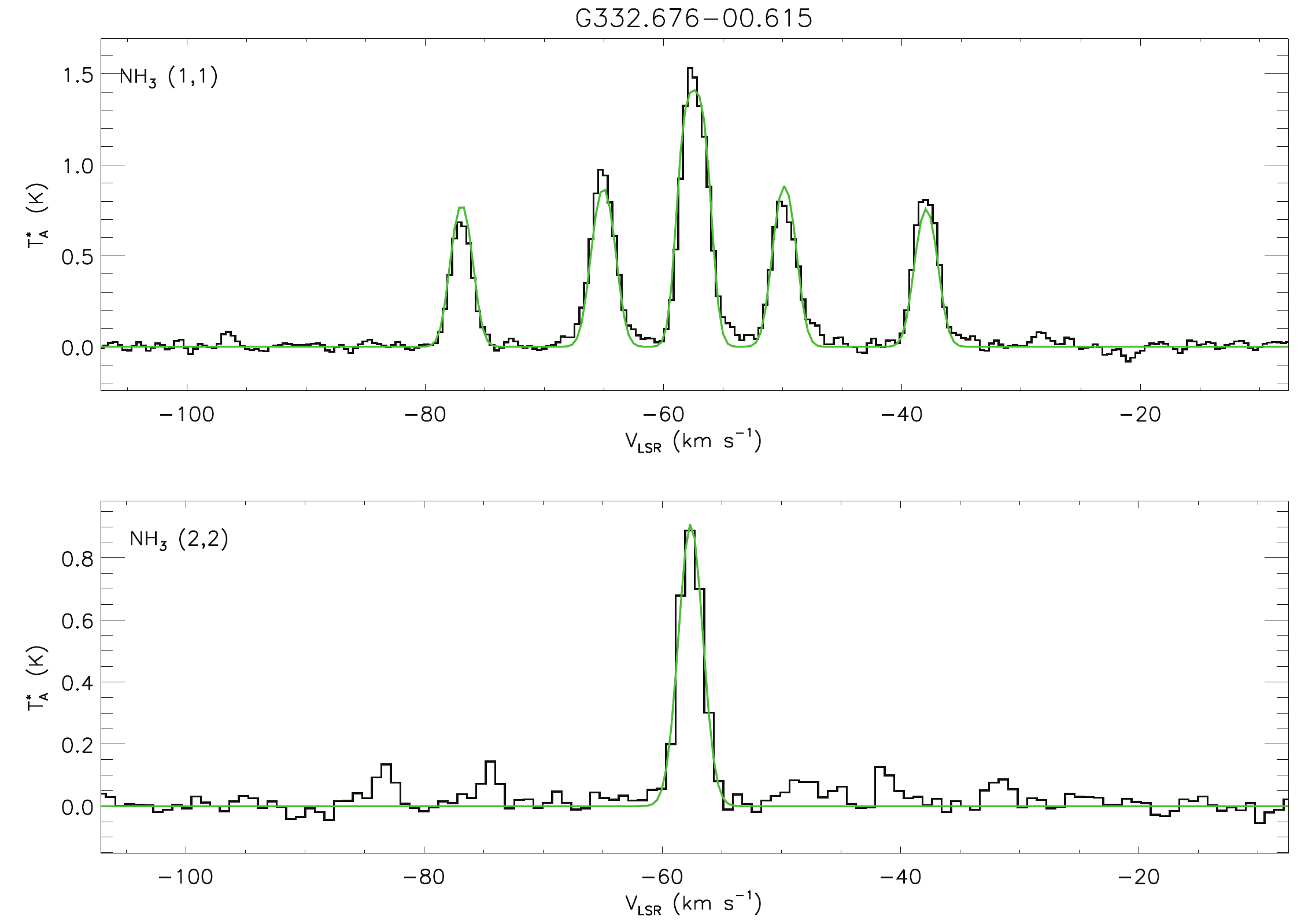}\\
  \vspace{1cm}
  \includegraphics[trim=100 20 190 40,clip,width=0.39\textwidth]{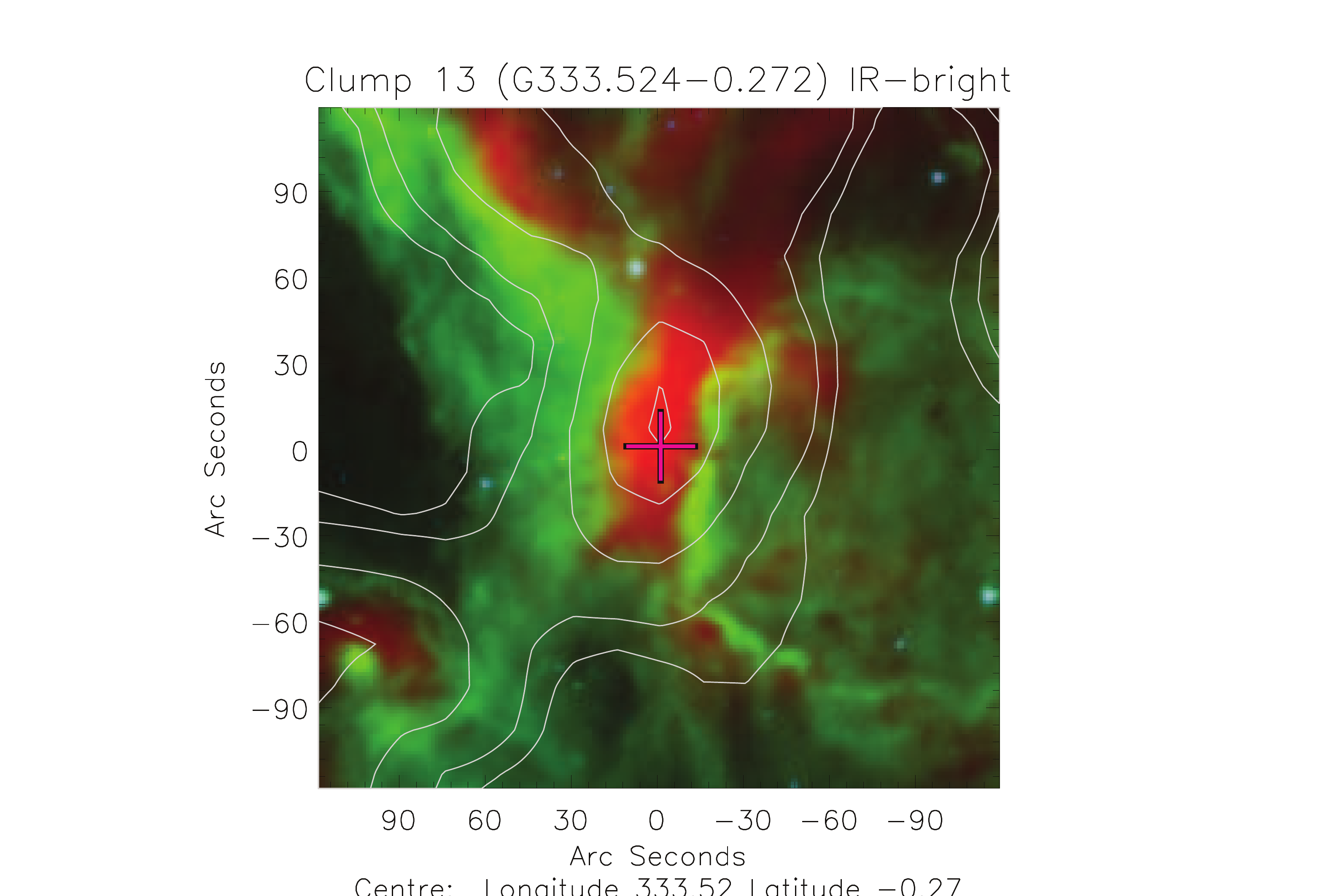}
  \includegraphics[width=0.49\textwidth, trim= 10 0 0 0]{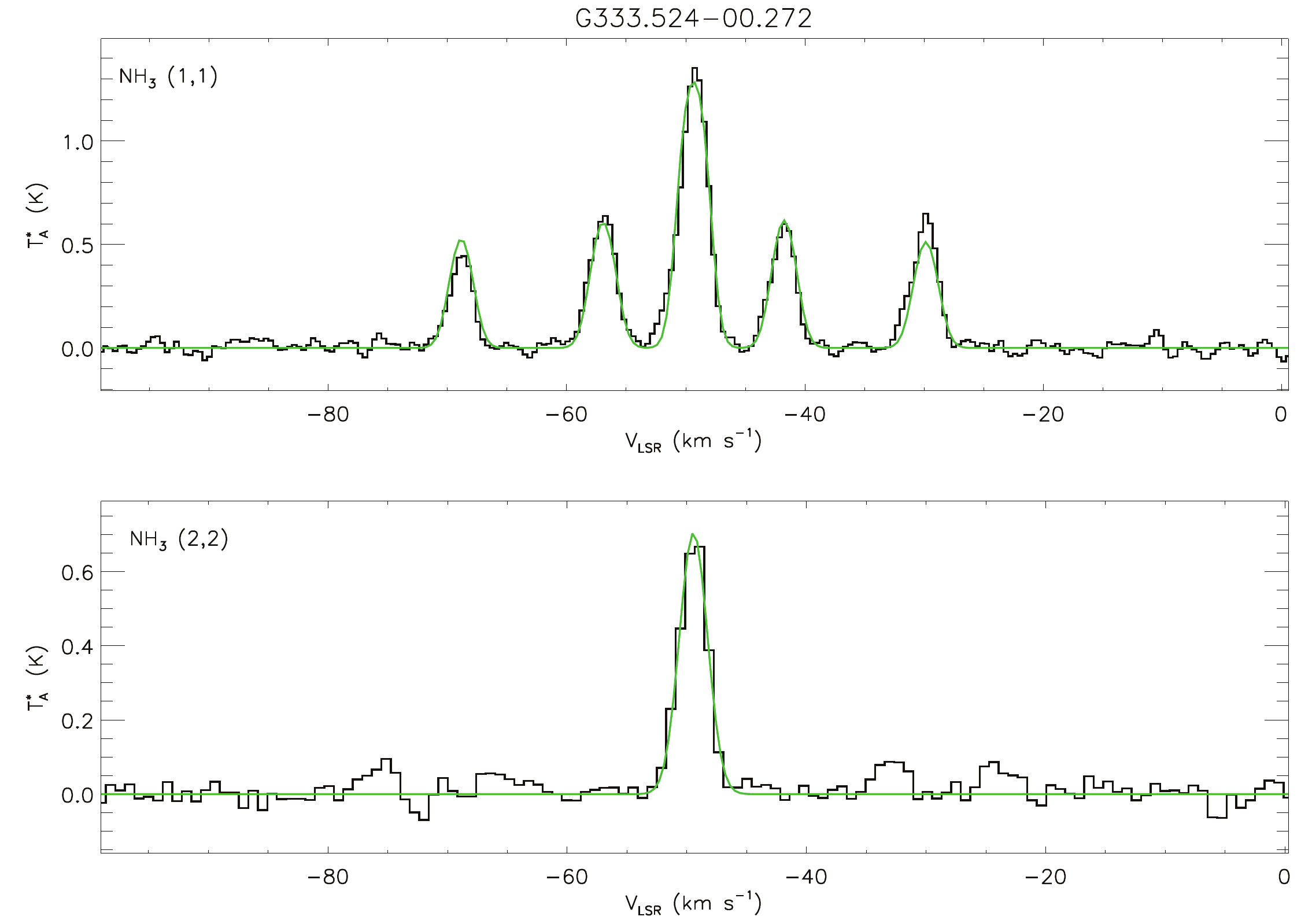}\\
  \vspace{0.2cm}
  \caption{\label{fig:3colour_spectra} Infrared composite images centred on the peak SIMBA dust clump (left) with accompanying NH$_3$ emission (right) detected towards a variety of environments within the G333 GMC. \textit{Spitzer} 4.5 and 8.0\,\um, and \textit{Herschel} 160\,\um\ are shown in blue, green and red, respectively. The SIMBA dust contours have been set to 0.1, 0.2, 0.5, 1, 1.5, 2.5, 5, 10, 15, 25 and 35\,mJy~beam$^{-1}$. \water~ masers are identified by blue circles. Class~I~\methanol~ masers are identified by unfilled (44.07\,GHz) and filled (95.1\,GHz) blue diamonds. Class~II~\methanol~ masers are identified by unfilled (6.7\,GHz) and filled (12.2\,GHz) green triangles. OH masers are identified by unfilled (1665/1667/1720\,MHz) and filled (6.035\,GHz) red squares. Each sub-image is centred on the peak dust emission found within that clump and emphasised by a pink plus sign. The plot showing the \ammonia~\one\ and \two\ spectra are presented on the upper and lower panels, respectively. The model fits to the data are shown in green. The full version of this figure is available in Fig.~\ref{app:3colour} and Fig.~\ref{app:spectra}, for the infrared composites and spectra, respectively.}
\end{figure*}

The \ammonia~\one\ and \two\ observations were reduced using the ATNF Spectral Analysis Package\footnote{http://svn.atnf.csiro.au/trac/asap} ({\sc asap}) and the graphical interface developed by Cormac Purcell\footnote{http://www.narrabri.atnf.csiro.au/mopra/SoftwareDepot/QuickData.py} (QuickData.py). The \ammonia~(3,3), if detected, will be used in future work to constrain the temperatures with large uncertainties. The individual on-off scans were subtracted to remove sky emission and visually inspected to identify and remove poor data. Where necessary a low-order polynomial was fitted to the channels free of emission and subtracted to remove baseline anomalies. The remaining scans were combined to produce a single spectrum for both transitions at each observed position. Hanning smoothing was used to improve the signal to noise of these spectra, which reduced the spectral resolution by a factor of 2. An example of the emission detected in a variety of environments within the G333~GMC are shown in Fig.~\ref{fig:3colour_spectra} with the full selection available in Fig.~\ref{app:spectra}.

\subsection{Cool dust from the \textit{Herschel} Space Observatory}
The G333~GMC was observed by \textit{Herschel} \citep{Pilbratt2010} with the PACS \citep{Poglitsch2010} and SPIRE \citep{Griffin2010} instruments in Parallel mode at 70, 160, 250, 350 and 500\,\um\ as part of the Open Time Key Programme ``Hi-Gal'' (PI: S. Molinari; \citealt{Molinari2010}). The observations comprised of a scan and cross-scan of two 2$^\circ$ fields (field 332.0 with obsid: 1342204046/47 and field 334.0 with  obsid: 1342204054/55) taken at a scan speed of 60\,arcsec\,s$^{-1}$. These observations were reduced in the Herschel Interactive Processing Environment (HIPE; \citealt{Ott2010}) to Level 1 (basic calibrated data product), then exported to the {\sc scanamorphos} map-making software to create a mosaic of the G333~GMC \citep{Roussel2013}. The pixel size of the final mosaic was 6.4\,arcsec per pixel at 160\,\um\ (equivalent to the instrument native pixel sizes in the red band) was chosen to decrease the low redundancy of the parallel mode scans (due to the high scan speed) and reduce the effect of correlated noise in the final image.


\section{RESULTS AND DERIVED PARAMETERS}
\label{sect:results}
We derive masses for the clumps from their integrated flux densities at 1.2-mm under the assumption that the clump is optically thin at this wavelength. The dust mass opacity is also quite uncertain, being based on emission models or laboratory dust measurements that may not reflect the actual material we observe. Each clump identified will be used as a mask and aided by a temperature derived from the \ammonia\ observations, accurate abundances will be derived within the clump region for comparing different molecular transitions at 3-mm. For this paper, the peak position within each clump was used to target the \ammonia~\one\ and \two\ observations. We report the results of the \ammonia\ observations and temperature calculations in this paper and will discuss the multi-molecular line abundances in a later paper.

\begin{table*}
  \setlength{\tabcolsep}{3pt}
  \caption{~The parameters of the 63 SIMBA 1.2-mm dust clumps identified with {\sc clumpfind}. The {\sc clumpfind} number is listed in Column 1, followed by the identifier for each source using the Galactic longitude and latitude designation for the peak pixel in Column 2. Columns 3 and 4 list the coordinates for the clump centroid. The peak flux, radii and total flux within the clumps is found in Columns 5, 6 and 7. Column 8 lists the nearby ($<$ 30 arcsec) infrared and maser associations for each clump, with the number denoting the quantity. The presence of a `n', `s', or `p' denotes \textit{Spitzer} 3.6, 4.5 or 8.0 \um\ emission \citep{Benjamin2003,Churchwell2009,SpitzerScience2009}. The presence of an `h' denotes the presence of a \textit{Herschel} 160\,\um\ source. The presence of an `o' denotes an OH maser, an `M$_1$' denotes a 44.07\,GHz Class~I~\methanol\ maser , an `M$_2$' denotes a 95.1\,GHz Class I methanol maser, an `m$_1$' denotes a 6.7\,GHz Class~II~\methanol\ maser, an `m$_2$' denotes a 12.2\,GHz Class~II~\methanol\ maser and a `w' denotes a \water~ maser. If an RMS object was within 30 arcsec of the SIMBA dust peak, then its type is listed in Column 9. Column 10 identifies whether the clump has been classified, via \textit{Spitzer} 8.0\,\um\ GLIMPSE emission, as IR-bright or IR-faint. Column 11 identifies which region the clump has been allocated to (see \S\ref{sec:variations} for more details).}\label{tbl:cf_parameters_jsu}
  \begin{tabular}{ccccccccccccc}
    \hline \hline
    ID &  Clump~name & \multicolumn{2}{c}{Clump~centroid} & Peak & \multicolumn{1}{c}{Radius} & Integrated & Associations$^\dagger$&RMS Type&IR Type&Region\\
    \cline{3-4}
    		&		&	RA	[J2000]	&	Dec	[J2000]	&	[Jy beam$^{-1}$]	&	[pc]	&	[Jy]	&	&	&	&	\\
    \hline
    \phantom{0}1	&	G333.604$-$0.210	&	16:22:08.5	&	-50:05:53	&	\phantom{}49.4	&	0.53	&	\phantom{}950.8	&	2wo$_{1,2}$M$_2$	&	\HII\ region	&	IR-bright	&	A	\\
    \phantom{0}2	&	G332.826$-$0.547	&	16:20:09.6	&	-50:53:09	&	\phantom{}15.5	&	0.46	&	\phantom{}263.7	&	2wo$_1$m$_1$b	&	\HII\ region/YSO	&	IR-bright	&	C	\\
    \phantom{0}3	&	G333.136$-$0.431	&	16:21:00.8	&	-50:35:13	&	\phantom{}14.5	&	0.55	&	\phantom{}409.5	&	nso$_{1,2}$M$_2$m$_1$	&	\HII\ region	&	IR-bright	&	B	\\
    \phantom{0}4	&	G333.286$-$0.387	&	16:21:30.3	&	-50:26:48	&	\phantom{0}9.6	&	0.49	&	\phantom{}198.1	&	nsw	&	\HII\ region	&	IR-bright	&	B	\\
    \phantom{0}5	&	G333.309$-$0.369	&	16:21:31.4	&	-50:24:42	&	\phantom{0}6.1	&	0.80	&	\phantom{}263.5	&	ns	&	Diffuse \HII\ region	&	IR-bright	&	B	\\
    \phantom{0}6	&	G333.465$-$0.160	&	16:21:19.4	&	-50:09:32	&	\phantom{0}5.6	&	0.43	&	\phantom{0}95.1	&	nspwo$_1$M$_1$m$_1$	&	$\cdots$	&	IR-bright	&	A	\\
    \phantom{0}7	&	G333.068$-$0.446	&	16:20:48.6	&	-50:38:37	&	\phantom{0}5.0	&	0.49	&	\phantom{0}99.3	&	nso$_{1,2}$	&	YSO	&	IR-bright	&	B	\\
    \phantom{0}8	&	G332.644$-$0.606	&	16:19:36.2	&	-51:03:10	&	\phantom{0}4.6	&	0.64	&	\phantom{}173.6	&	ns	&	Diffuse \HII\ region	&	IR-bright	&	C	\\
    \phantom{0}9	&	G333.127$-$0.564	&	16:21:35.2	&	-50:41:15	&	\phantom{0}3.9	&	0.48	&	\phantom{0}87.4	&	nsp2o$_{1}$M$_1$	&	$\cdots$	&	IR-faint	&	B	\\
    \phantom{}10	&	G332.691$-$0.612	&	16:19:51.6	&	-51:01:26	&	\phantom{0}2.9	&	0.43	&	\phantom{0}63.2	&	nsph	&	$\cdots$	&	IR-bright	&	C	\\
    \phantom{}11	&	G333.018$-$0.449	&	16:20:37.9	&	-50:41:22	&	\phantom{0}2.2	&	0.69	&	\phantom{0}67.1	&	nshw	&	Diffuse \HII\ region	&	IR-bright	&	B	\\
    \phantom{}12	&	G332.676$-$0.615	&	16:19:46.6	&	-51:02:15	&	\phantom{0}2.1	&	0.38	&	\phantom{0}38.6	&	ns	&	$\cdots$	&	IR-faint	&	C	\\
    \phantom{}13	&	G333.524$-$0.272	&	16:22:02.7	&	-50:11:42	&	\phantom{0}1.6	&	0.38	&	\phantom{0}34.0	&	nsp	&	$\cdots$	&	IR-faint	&	A	\\
    \phantom{}14	&	G332.985$-$0.487	&	16:20:36.6	&	-50:44:09	&	\phantom{0}1.5	&	0.45	&	\phantom{0}39.1	&	nsm$_1$	&	YSO	&	IR-bright	&	B	\\
    \phantom{}15	&	G333.218$-$0.402	&	16:21:17.0	&	-50:30:15	&	\phantom{0}1.5	&	0.33	&	\phantom{0}27.9	&	nspw	&	$\cdots$	&	IR-bright	&	B	\\
    \phantom{}16	&	G333.106$-$0.502	&	16:21:13.6	&	-50:39:42	&	\phantom{0}1.4	&	0.59	&	\phantom{0}60.1	&	nspm$_1$	&	YSO	&	IR-bright	&	B	\\
    \phantom{}17	&	G333.566$-$0.295	&	16:22:20.2	&	-50:11:22	&	\phantom{0}1.3	&	0.52	&	\phantom{0}51.6	&	nsp	&	$\cdots$	&	IR-faint	&	A	\\
    \phantom{}18	&	G333.074$-$0.399	&	16:20:37.1	&	-50:36:41	&	\phantom{0}1.3	&	0.61	&	\phantom{0}56.8	&	nsp	&	$\cdots$	&	IR-faint	&	B	\\
    \phantom{}19	&	G333.722$-$0.209	&	16:22:38.3	&	-50:00:43	&	\phantom{0}1.2	&	0.32	&	\phantom{0}23.4	&	nsph	&	$\cdots$	&	IR-bright	&	A	\\
    \phantom{}20	&	G332.585$-$0.559	&	16:19:08.3	&	-51:03:50	&	\phantom{0}1.2	&	0.43	&	\phantom{0}29.8	&	nsp	&	YSO	&	IR-bright	&	C	\\
    \phantom{}21	&	G333.006$-$0.437	&	16:20:27.7	&	-50:40:56	&	\phantom{0}1.2	&	0.41	&	\phantom{0}30.0	&	nspm$_1$	&	Diffuse \HII\ region	&	IR-bright	&	B	\\
    \phantom{}22	&	G332.700$-$0.585	&	16:19:45.7	&	-51:00:14	&	\phantom{0}1.0	&	0.30	&	\phantom{0}16.5	&	nspm$_1$	&	YSO and \HII\ region	&	IR-bright	&	C	\\
    \phantom{}23	&	G333.292$-$0.422	&	16:21:43.1	&	-50:27:34	&	\phantom{0}0.9	&	0.63	&	\phantom{0}41.8	&	nsp	&	$\cdots$	&	IR-faint	&	B	\\
    \phantom{}24	&	G333.692$-$0.198	&	16:22:30.1	&	-50:01:32	&	\phantom{0}0.9	&	0.40	&	\phantom{0}24.7	&	ns	&	$\cdots$	&	IR-faint	&	A	\\
    \phantom{}25	&	G332.773$-$0.582	&	16:20:04.3	&	-50:57:03	&	\phantom{0}0.9	&	0.36	&	\phantom{0}17.0	&	nsp	&	$\cdots$	&	IR-faint	&	C	\\
    \phantom{}26	&	G333.772$-$0.260	&	16:23:05.2	&	-50:00:40	&	\phantom{0}0.8	&	0.31	&	\phantom{0}15.8	&	nsp	&	$\cdots$	&	IR-faint	&	A	\\
    \phantom{}27	&	G333.539$-$0.245	&	16:21:59.6	&	-50:10:03	&	\phantom{0}0.8	&	0.26	&	\phantom{0}15.2	&	nsp	&	$\cdots$	&	IR-faint	&	A	\\
    \phantom{}28	&	G333.483$-$0.225	&	16:21:41.6	&	-50:11:59	&	\phantom{0}0.8	&	0.62	&	\phantom{0}29.7	&	nsp	&	$\cdots$	&	IR-faint	&	A	\\
    \phantom{}29	&	G333.633$-$0.257	&	16:22:25.9	&	-50:06:11	&	\phantom{0}0.8	&	0.50	&	\phantom{0}29.0	&	ns	&	YSO	&	IR-faint	&	A	\\
    \phantom{}30	&	G333.038$-$0.405	&	16:20:27.4	&	-50:38:16	&	\phantom{0}0.7	&	0.43	&	\phantom{0}23.3	&	ns	&	$\cdots$	&	IR-faint	&	B	\\
    \phantom{}31	&	G332.794$-$0.597	&	16:20:14.9	&	-50:56:28	&	\phantom{0}0.7	&	0.24	&	\phantom{0}10.8	&	nsp	&	$\cdots$	&	IR-bright	&	C	\\
    \phantom{}32	&	G332.670$-$0.644	&	16:19:54.6	&	-51:03:39	&	\phantom{0}0.7	&	0.44	&	\phantom{0}21.0	&	nsp	&	$\cdots$	&	IR-faint	&	C	\\
    \phantom{}33	&	G333.012$-$0.520	&	16:20:51.8	&	-50:44:01	&	\phantom{0}0.7	&	0.43	&	\phantom{0}17.3	&	nsp	&	$\cdots$	&	IR-faint	&	B	\\
    \phantom{}34	&	G333.568$-$0.168	&	16:21:47.8	&	-50:05:36	&	\phantom{0}0.7	&	0.43	&	\phantom{0}20.4	&	nsp	&	$\cdots$	&	IR-faint	&	A	\\
    \phantom{}35	&	G332.741$-$0.620	&	16:20:03.9	&	-50:59:52	&	\phantom{0}0.7	&	0.42	&	\phantom{0}16.0	&	nsphwM$_1$m$_{1,2}$o$_1$	&	$\cdots$	&	IR-faint	&	C	\\
    \phantom{}36	&	G333.754$-$0.230	&	16:22:53.0	&	-50:00:06	&	\phantom{0}0.7	&	0.16	&	\phantom{00}7.7	&	nspm$_1$h	&	YSO	&	IR-faint	&	A	\\
    \phantom{}37	&	G333.171$-$0.431	&	16:21:13.0	&	-50:33:44	&	\phantom{0}0.6	&	0.31	&	\phantom{0}10.9	&	nsp$\cdots$	&	$\cdots$	&	IR-faint	&	B	\\
    \phantom{}38	&	G333.681$-$0.257	&	16:22:41.5	&	-50:04:44	&	\phantom{0}0.6	&	0.47	&	\phantom{0}18.2	&	ns	&	$\cdots$	&	IR-faint	&	A	\\
    \phantom{}39	&	G333.642$-$0.106	&	16:21:51.3	&	-49:59:44	&	\phantom{0}0.6	&	0.40	&	\phantom{0}16.5	&	nsph	&	$\cdots$	&	IR-bright	&	A	\\
    \phantom{}40	&	G333.047$-$0.479	&	16:20:50.2	&	-50:40:40	&	\phantom{0}0.6	&	0.40	&	\phantom{0}16.6	&	nsp	&	$\cdots$	&	IR-faint	&	B	\\
    \phantom{}41	&	G332.971$-$0.467	&	16:20:27.0	&	-50:43:47	&	\phantom{0}0.6	&	0.41	&	\phantom{0}20.1	&	ns	&	$\cdots$	&	IR-bright	&	B	\\
    \phantom{}42	&	G333.206$-$0.366	&	16:21:01.9	&	-50:29:08	&	\phantom{0}0.5	&	0.35	&	\phantom{0}13.5	&	nsp	&	$\cdots$	&	IR-bright	&	B	\\
    \phantom{}43	&	G332.809$-$0.700	&	16:20:46.9	&	-50:59:57	&	\phantom{0}0.5	&	0.18	&	\phantom{00}6.8	&	nspM$_{1,2}$	&	$\cdots$	&	IR-bright	&	C	\\
    \phantom{}44	&	G332.558$-$0.591	&	16:19:08.9	&	-51:06:07	&	\phantom{0}0.5	&	0.20	&	\phantom{00}7.2	&	nsp	&	$\cdots$	&	IR-bright	&	C	\\
    \phantom{}45	&	G333.183$-$0.396	&	16:21:05.2	&	-50:31:34	&	\phantom{0}0.5	&	0.10	&	\phantom{00}5.0	&	nsp	&	$\cdots$	&	IR-faint	&	B	\\
    \phantom{}46	&	G333.521$-$0.239	&	16:21:53.0	&	-50:10:48	&	\phantom{0}0.5	&	0.21	&	\phantom{00}7.3	&	nsp	&	$\cdots$	&	IR-faint	&	A	\\
    \phantom{}47	&	G333.074$-$0.558	&	16:21:19.6	&	-50:42:48	&	\phantom{0}0.4	&	0.31	&	\phantom{00}8.1	&	nsph	&	$\cdots$	&	IR-bright	&	B	\\
    \phantom{}48	&	G332.759$-$0.467	&	16:19:29.4	&	-50:52:27	&	\phantom{0}0.4	&	0.12	&	\phantom{00}5.1	&	ns	&	$\cdots$	&	IR-faint	&	C	\\
    \phantom{}49	&	G333.236$-$0.520	&	16:21:51.7	&	-50:34:26	&	\phantom{0}0.4	&	0.45	&	\phantom{0}12.6	&	nsp	&	$\cdots$	&	IR-bright	&	B	\\
    \phantom{}50	&	G332.715$-$0.676	&	16:20:13.4	&	-51:02:50	&	\phantom{0}0.4	&	0.31	&	\phantom{00}8.7	&	ns	&	$\cdots$	&	IR-faint	&	C	\\
    \phantom{}51	&	G332.903$-$0.546	&	16:20:31.3	&	-50:49:47	&	\phantom{0}0.4	&	0.21	&	\phantom{00}5.8	&	nsp	&	$\cdots$	&	IR-faint	&	C	\\
    \phantom{}52	&	G333.097$-$0.352	&	16:20:30.7	&	-50:33:20	&	\phantom{0}0.4	&	0.23	&	\phantom{00}6.2	&	ns	&	$\cdots$	&	IR-bright	&	B	\\
    \phantom{}53	&	G333.563$-$0.145	&	16:21:38.7	&	-50:04:40	&	\phantom{0}0.3	&	0.31	&	\phantom{00}6.1	&	ns	&	$\cdots$	&	IR-faint	&	A	\\
    \phantom{}54	&	G333.636$-$0.165	&	16:22:04.0	&	-50:02:37	&	\phantom{0}0.3	&	0.10	&	\phantom{00}4.2	&	ns	&	$\cdots$	&	IR-faint	&	A	\\
    \phantom{}55	&	G333.645$-$0.133	&	16:21:59.0	&	-50:00:50	&	\phantom{0}0.3	&	0.30	&	\phantom{00}7.6	&	nsp	&	$\cdots$	&	IR-faint	&	A	\\
    \phantom{}56	&	G333.657$-$0.133	&	16:22:01.4	&	-50:00:01	&	\phantom{0}0.3	&	0.16	&	\phantom{00}3.8	&	nsp	&	$\cdots$	&	IR-faint	&	A	\\
  \end{tabular}
\end{table*}
\begin{table*}\addtocounter{table}{-1}
  \setlength{\tabcolsep}{3pt}
  \caption{-- {\emph {continued}}}
  \begin{tabular}{ccccccccccccc}
    \hline\hline
    ID &  Clump~name & \multicolumn{2}{c}{Clump~centroid} & Peak & \multicolumn{1}{c}{Radius} & Sum & Associations$^\dagger$&RMS Type&IR Type&Region\\
    \cline{3-4}
          &       &   RA  [J2000] &   Dec [J2000] &   [Jy beam$^{-1}$]  &   [pc]    &   [Jy]  &   &   &   \\
    \hline
    \phantom{}57	&	G333.660$-$0.218	&	16:22:25.0	&	-50:03:43	&	\phantom{0}0.3	&	0.19	&	\phantom{00}4.7	&	nsp	&	$\cdots$	&	IR-faint	&	A	\\
    \phantom{}58	&	G333.410$-$0.328	&	16:21:47.9	&	-50:19:02	&	\phantom{00}0.3	&	0.52	&	\phantom{0}10.4	&	nsph	&	$\cdots$	&	IR-bright	&	A	\\
    \phantom{}59	&	G333.595$-$0.077	&	16:21:30.0	&	-50:00:40	&	\phantom{00}0.3	&	0.32	&	\phantom{00}5.1	&	nsp	&	$\cdots$	&	IR-bright	&	A	\\
    \phantom{}60	&	G332.894$-$0.567	&	16:20:32.6	&	-50:51:16	&	\phantom{00}0.2	&	0.22	&	\phantom{00}4.4	&	nsp	&	$\cdots$	&	IR-faint	&	C	\\
    \phantom{}61	&	G333.262$-$0.275	&	16:20:54.0	&	-50:23:02	&	\phantom{00}0.2	&	0.28	&	\phantom{00}4.5	&	nsp	&	$\cdots$	&	IR-bright	&	B	\\
    \phantom{}62	&	G332.753$-$0.561	&	16:19:54.0	&	-50:56:47	&	\phantom{00}0.2	&	0.23	&	\phantom{00}4.6	&	nsp	&	$\cdots$	&	IR-bright	&	C	\\
    \phantom{}63	&	G332.614$-$0.671	&	16:19:43.5	&	-51:07:40	&	\phantom{00}0.2	&	0.33	&	\phantom{00}6.0	&	nsp	&	$\cdots$	&	IR-faint	&	C	\\
    \hline
  \end{tabular}\\
\vspace{0.1cm}
\footnotesize
\raggedright
$\dagger$ References -- Water (22\,GHz \water; \citealt{Braz1982,Breen2007,Breen2010,Walsh2011}), Class I (44.01\,GHz; \citealt{Slysh1994}, \citealt{Voronkov2014} and 95.1\,GHz; \citealt{Ellingsen2005}) and II (6.7\,GHz; \citealt{Caswell1996,Caswell1997,Caswell2009,Caswell2011,Ellingsen1996}; and 12.2\,GHz; \citealt{Breen2012}) methanol (\methanol) and hydroxyl (OH; 1665/1667/1720\,MHz; \citealt{Caswell1980,Caswell1995,Caswell1998} and 6.035 GHz; \citealt{Caswell1997}) masers.
\normalsize
\end{table*}
\setlength{\tabcolsep}{6pt}


\subsection{Clump~parameters}
\label{sec:parameters}
The parameters of the 63 SIMBA 1.2-mm dust clumps identified with {\sc clumpfind} can be found in Table \ref{tbl:cf_parameters_jsu}. The {\sc clumpfind} number is listed in Column 1, followed by the identifier for each source using the Galactic longitude and latitude designation for the peak pixel in Column 2. Columns 3 and 4 list the coordinates for the clump centroid. We find that the peak of the dust emission correlates poorly with the centroid of the emission, with a maximum deviation of 39.5\,arcsec ($\sim$1.3 convolved SIMBA beam). Since the size and shapes of clumps can differ when identified with different algorithms, we have performed our \ammonia\ observations towards the peak dust emission within each clump. The peak and integrated flux (in Jy~beam$^{-1}$ and Jy, respectively) within the clumps is found in Columns 5 and 7. {\footnotesize CUPID} outputs the pixel size of the clumps projected onto the equatorial plane, so the average projected values were then converted into parsecs by assuming a distance of 3.6\,kpc \citep{Lockman1979} and are displayed in Column 6.
Column 8 lists the nearby ($<$ 30\,arcsec) infrared and maser associations (see \S\ref{sec:separation} or Table \ref{tbl:cf_parameters_jsu} caption for further information). If there is a nearby ($<$ 30\,arcsec) RMS object, then its type is listed in Column 9. Column 10 identifies whether the clump has been classified, via \textit{Spitzer} GLIMPSE 8.0\,\um\ emission, as IR-bright or IR-faint. Column 11 identifies which region the clump has been allocated to (see \S\ref{sec:variations}).

The distribution of the clumps throughout the G333~GMC can be seen in Fig.~\ref{fig:overview}. A sample of the dust clumps and their accompanying \ammonia~\one\ and \two\ emission and Gaussian fits, where possible, can be seen in Fig.~\ref{fig:3colour_spectra}, with the full version available in Fig.~\ref{app:3colour} and Fig.~\ref{app:spectra}, for the infrared composites and spectra, respectively.


\subsection{Derivation of physical parameters}
\label{sect:formulae}
\begin{table}
  \centering
  \caption{Summary of calculated physical parameters.}
  \label{tbl:derived_parameters}
  \begin{minipage}{\linewidth}
    \small
    \begin{tabular}{lccccc}
      \hline\hline
      Parameter	&	Mean	&	1$\sigma$	&	Min	&	Max	\\
      \hline
      \Tkin [K]	&	18.1$\pm$0.5	&	3.6	&	12.5	&	35.3	\\
      \Trot [K]	&	22.0$\pm$1.0	&	7.4	&	13.6	&	\phantom{1}63.6$^\dagger$	\\
      Log[N(\ammonia)] [cm$^{-2}$]	&	15.9$\pm$2.3	&	0.24	&	15.3	&	16.6	\\
      Log[n(\ammonia)] [$\times 10^5$ cm$^{-3}$]	&	4.5$\pm$0.4	&	2.8	&	0.7	&	16.0	\\
      \hline
      \ammonia~\one~\vlsr~[\kms]	&	-52.9$\pm$1.1	&	8.1	&	-88.5$^\dagger$&	-42.57	\\
      \ammonia~\one~$\Delta$V [\kms]	&	2.2$\pm$0.9	&	0.87	&	1.1	&	4.5	\\
      $\tau_{(1,1)}$	&	1.9$\pm$0.1	&	0.86	&	0.16	&	3.76	\\
      \hline
      \ammonia~\two~\vlsr [\kms]	&	-53.1$\pm$1.2	&	8.2	&	-88.7$^\dagger$	&	-42.8	\\
      \ammonia~\two~$\Delta$V [\kms]	&	2.7$\pm$0.4	&	1.0	&	1.3	&	6.0	\\
      \hline
      Log(Clump mass) [\msun]	&	3.37$\pm$2.61	&	3.45	&	2.44	&	4.12	\\
      Log(Virial mass) [\msun]	&	2.99$\pm$2.18	&	3.03	&	2.03	&	3.68	\\
      \hline
    \end{tabular}\\
    \footnotesize \\
    $^\dagger$These values are outliers and their corresponding clumps and physical parameters have been excluded from \S\ref{sec:exclusions} onwards. The next highest \Tkin\ is Clump~21~(G333.006$-$0.437) with \Tkin=35.6\,K.
    \normalsize
  \end{minipage}
\end{table}
\begin{table*}
  \setlength{\tabcolsep}{3pt}
  \caption{Summary of physical parameters calculated following \S\ref{sect:formulae}. \ammonia~\one\ non-detections with a T$_{\rm{A}}^* < 0.1\,\mathrm{K}$ have been excluded.}
  \label{tbl:nh3_parameters}
  \begin{tabular}{clcccccccccccccc}
    \hline \hline
    && &&&&   \multicolumn{3}{c}{NH$_3$ (1,1)}&& \multicolumn{3}{c}{NH$_3$ (2,2)} & \multicolumn{2}{c}{Log(Mass)}  & Spectral\\
    \cline{7-9} \cline{11-13}
    ID&Clump~Name&	$T_{\rm{rot}}$ &$T_{\rm{kin}}$ &Log[$N({\rm{NH_3}})$] &Log[n(H$_2$)]&T$_{\rm{A}}^*$ & \vlsr & $\Delta$V & $\tau_{\rm{(m,1,1)}}$ &T$_{\rm{A}}^*$ & \vlsr & $\Delta$V & Clump~& Virial & features$^\dagger$\\
    &&  [K] & [K] &  [cm$^{-2}$] & [cm$^{-3}$] &[K]  &[\kms] & [\kms] &          &     [K] & [\kms] &[\kms] & [\msun] & [\msun] &\\
    \hline
    \phantom{0}1	&	G333.604$-$0.210	&	$\cdots$	&	$\cdots$	&	$\cdots$	&	$\cdots$	&	$<$0.2\phantom{0}	&	$\cdots$	&	$\cdots$	&	$\cdots$	&	$\cdots$	&	$\cdots$	&	$\cdots$	&	$\cdots$	&	$\cdots$	&	Aa	\\
    \phantom{0}2	&	G332.826$-$0.547	&	22.0	&	28.4	&	16.2	&	5.75	&	0.9	&	$-$57.62	&	4.50	&	1.16	&	0.6	&	$-$57.16	&	5.96	&	3.99	&	3.62	&	Hbhgb	\\
    \phantom{0}3	&	G333.136$-$0.431	&	23.8	&	31.9	&	15.5	&	4.97	&	1.1	&	$-$53.56	&	4.45	&	0.25	&	0.7	&	$-$53.79	&	4.23	&	4.12	&	3.68	&	HABhgab	\\
    \phantom{0}4	&	G333.286$-$0.387	&	22.5	&	29.2	&	16.1	&	5.62	&	0.9	&	$-$51.35	&	2.87	&	1.52	&	0.7	&	$-$51.46	&	3.50	&	3.85	&	3.25	&	Hbhgb	\\
    \phantom{0}5	&	G333.309$-$0.369	&	20.5	&	25.5	&	15.8	&	5.11	&	0.9	&	$-$50.29	&	2.62	&	0.85	&	0.5	&	$-$50.36	&	3.10	&	4.04	&	3.39	&	Hg	\\
    \phantom{0}6	&	G333.465$-$0.160	&	19.8	&	24.4	&	16.0	&	5.58	&	0.9	&	$-$42.57	&	2.62	&	1.38	&	0.5	&	$-$42.79	&	3.10	&	3.62	&	3.13	&	Hg	\\
    \phantom{0}7	&	G333.068$-$0.446	&	20.6	&	25.7	&	16.1	&	5.62	&	2.5	&	$-$53.01	&	2.02	&	2.25	&	1.8	&	$-$53.12	&	2.66	&	3.62	&	2.95	&	Hhg	\\
    \phantom{0}8	&	G332.644$-$0.606	&	20.1	&	24.9	&	16.1	&	5.50	&	0.9	&	$-$49.30	&	3.01	&	1.64	&	0.6	&	$-$49.43	&	3.93	&	3.87	&	3.41	&	Hbgb	\\
    \phantom{0}9	&	G333.127$-$0.564	&	17.3	&	20.3	&	16.5	&	6.03	&	1.6	&	$-$57.17	&	3.26	&	3.76	&	1.1	&	$-$57.49	&	4.31	&	3.69	&	3.35	&	Hbhgb	\\
    10	&	G332.691$-$0.612	&	18.3	&	22.0	&	16.2	&	5.78	&	0.6	&	$-$47.89	&	2.68	&	2.51	&	0.4	&	$-$47.98	&	3.23	&	3.51	&	3.13	&	Hbhgb	\\
    11	&	G333.018$-$0.449	&	25.6	&	35.6	&	16.1	&	5.47	&	0.4	&	$-$54.34	&	3.08	&	1.29	&	0.3	&	$-$54.52	&	3.25	&	3.27	&	3.46	&	Hbg	\\
    12	&	G332.676$-$0.615	&	16.3	&	18.8	&	16.2	&	5.83	&	1.4	&	$-$57.46	&	1.89	&	3.51	&	0.9	&	$-$57.62	&	2.40	&	3.38	&	2.78	&	Hhg	\\
    13	&	G333.524$-$0.272	&	17.3	&	20.3	&	16.0	&	5.63	&	1.3	&	$-$49.34	&	2.15	&	1.95	&	0.7	&	$-$49.45	&	2.76	&	3.28	&	2.90	&	Hhg	\\
    14	&	G332.985$-$0.487	&	18.4	&	22.1	&	15.9	&	5.46	&	1.6	&	$-$52.56	&	1.84	&	1.92	&	1.0	&	$-$52.61	&	2.31	&	3.29	&	2.83	&	Hhg	\\
    15	&	G333.218$-$0.402	&	20.1	&	24.8	&	15.9	&	5.59	&	1.1	&	$-$51.92	&	1.76	&	1.86	&	0.8	&	$-$52.06	&	2.29	&	3.08	&	2.65	&	Hhg	\\
    16	&	G333.106$-$0.502	&	18.1	&	21.5	&	16.0	&	5.44	&	1.0	&	$-$56.12	&	1.95	&	2.18	&	0.6	&	$-$56.16	&	2.69	&	3.50	&	3.00	&	Hhg	\\
    17	&	G333.566$-$0.295	&	15.1	&	17.1	&	15.9	&	5.39	&	1.1	&	$-$46.09	&	1.05	&	3.27	&	0.6	&	$-$46.12	&	1.34	&	3.57	&	2.41	&	Hahga	\\
    18	&	G333.074$-$0.558	&	15.9	&	18.2	&	16.0	&	5.42	&	1.0	&	$-$55.77	&	1.74	&	2.75	&	0.6	&	$-$55.95	&	2.30	&	3.52	&	3.31	&	Hg	\\
    19	&	G333.722$-$0.209	&	15.6	&	17.8	&	16.0	&	5.70	&	1.5	&	$-$46.58	&	1.41	&	2.85	&	0.8	&	$-$46.69	&	1.92	&	3.20	&	2.46	&	Hhg	\\
    20	&	G332.585$-$0.559	&	15.6	&	17.8	&	16.0	&	5.58	&	0.8	&	$-$50.99	&	1.47	&	2.80	&	0.4	&	$-$51.03	&	2.07	&	3.30	&	2.61	&	Hhg	\\
    21	&	G333.006$-$0.437	&	35.3	&	63.6	&	16.6	&	6.20	&	0.2	&	$-$56.20	&	4.46	&	1.80	&	0.2	&	$-$55.85	&	5.17	&	2.64	&	3.56	&	Hbgb	\\
    22	&	G332.700$-$0.585	&	19.1	&	23.2	&	15.9	&	5.63	&	0.5	&	$-$59.15	&	2.40	&	1.46	&	0.3	&	$-$59.76	&	2.46	&	2.89	&	2.88	&	Hbg	\\
    23	&	G333.292$-$0.422	&	18.0	&	21.4	&	15.9	&	5.31	&	0.9	&	$-$49.94	&	1.96	&	1.56	&	0.5	&	$-$49.95	&	2.50	&	3.34	&	3.03	&	Hg*	\\
    24	&	G333.692$-$0.198	&	12.5	&	13.6	&	16.0	&	5.61	&	0.6	&	$-$50.37	&	1.25	&	3.53	&	0.2	&	$-$50.36	&	1.95	&	3.39	&	2.44	&	Hg	\\
    25	&	G332.773$-$0.582	&	18.7	&	22.6	&	15.6	&	5.25	&	0.8	&	$-$55.86	&	1.72	&	0.95	&	0.4	&	$-$55.92	&	2.12	&	2.92	&	2.68	&	Hg	\\
    26	&	G333.772$-$0.260	&	16.6	&	19.3	&	15.9	&	5.62	&	0.7	&	$-$48.95	&	2.43	&	1.51	&	0.3	&	$-$49.02	&	2.74	&	2.98	&	2.91	&	Hg	\\
    27	&	G333.539$-$0.245	&	16.5	&	19.2	&	15.7	&	5.49	&	0.6	&	$-$48.20	&	1.39	&	1.51	&	0.3	&	$-$48.32	&	1.64	&	2.97	&	2.35	&	Hbgb	\\
    28	&	G333.483$-$0.225	&	14.3	&	16.0	&	16.0	&	5.42	&	0.8	&	$-$48.58	&	1.47	&	2.89	&	0.4	&	$-$48.62	&	1.99	&	3.36	&	2.77	&	Hg	\\
    30	&	G333.038$-$0.405	&	17.5	&	20.6	&	16.3	&	5.88	&	0.3	&	$-$51.75	&	4.41	&	1.85	&	0.2	&	$-$52.96	&	5.49	&	3.03	&	2.74	&	Hbgb	\\
    31	&	G332.794$-$0.597	&	20.7	&	26.0	&	15.9	&	5.73	&	0.3	&	$-$54.89	&	2.41	&	1.19	&	0.2	&	$-$55.10	&	2.71	&	2.65	&	2.78	&	Hbgb	\\
    32	&	G332.670$-$0.644	&	18.0	&	21.5	&	15.7	&	5.27	&	0.4	&	$-$49.58	&	1.96	&	1.01	&	0.2	&	$-$49.58	&	2.15	&	3.04	&	2.89	&	Hbb	\\
    33	&	G333.012$-$0.520	&	16.9	&	19.7	&	15.8	&	5.38	&	0.8	&	$-$52.94	&	2.25	&	1.18	&	0.4	&	$-$52.94	&	2.41	&	3.01	&	2.99	&	Hg	\\
    34	&	G333.568$-$0.168	&	13.1	&	14.3	&	16.0	&	5.58	&	0.9	&	$-$88.53	&	1.29	&	3.46	&	0.4	&	$-$88.69	&	1.81	&	3.27	&	2.50	&	Hg	\\
    35	&	G332.741$-$0.620	&	15.4	&	17.6	&	15.8	&	5.39	&	0.8	&	$-$49.81	&	1.48	&	2.05	&	0.4	&	$-$49.88	&	2.10	&	3.04	&	2.60	&	Hg	\\
    36	&	G333.754$-$0.230	&	16.0	&	18.4	&	16.0	&	6.01	&	0.6	&	$-$49.70	&	1.69	&	2.70	&	0.3	&	$-$49.96	&	2.35	&	2.69	&	2.30	&	Hg	\\
    37	&	G333.171$-$0.431	&	19.3	&	23.5	&	16.0	&	5.72	&	1.1	&	$-$51.07	&	2.51	&	1.53	&	0.7	&	$-$51.07	&	3.14	&	2.70	&	2.93	&	Hg	\\
    38	&	G333.681$-$0.257	&	19.4	&	23.7	&	15.3	&	4.84	&	0.2	&	$-$47.04	&	1.11	&	0.77	&	0.1	&	$-$47.69	&	2.34	&	2.93	&	2.39	&	Hagb	\\
    39	&	G333.642$-$0.106	&	13.3	&	14.6	&	16.1	&	5.71	&	0.8	&	$-$87.95	&	1.71	&	3.06	&	0.3	&	$-$87.98	&	2.11	&	3.17	&	2.72	&	Hg	\\
    40	&	G333.047$-$0.479	&	16.8	&	19.5	&	15.8	&	5.41	&	0.6	&	$-$52.55	&	1.99	&	1.40	&	0.3	&	$-$52.79	&	2.76	&	2.99	&	2.85	&	Hg	\\
    42	&	G333.206$-$0.366	&	$<$9.3	&	$<$9.5	&	$<$14.0	&	$<$3.67	&	0.2	&	$-$48.49	&	2.21	&	0.16	&	$\cdots$	&	$\cdots$	&	$\cdots$	&	$\cdots$	&	2.87	&	H	\\
    43	&	G332.809$-$0.700	&	17.7	&	21.0	&	15.8	&	5.75	&	0.5	&	$-$52.98	&	1.17	&	2.44	&	0.3	&	$-$53.17	&	1.87	&	2.56	&	2.03	&	Hagb	\\
    44	&	G332.558$-$0.591	&	19.0	&	23.0	&	16.0	&	5.91	&	0.7	&	$-$49.64	&	2.28	&	1.93	&	0.4	&	$-$49.61	&	2.49	&	2.54	&	2.66	&	Hg	\\
    45	&	G333.183$-$0.396	&	17.3	&	20.4	&	15.8	&	6.01	&	0.7	&	$-$49.95	&	1.73	&	1.70	&	0.4	&	$-$49.96	&	2.57	&	2.44	&	2.11	&	Hg	\\
    46	&	G333.521$-$0.239	&	16.1	&	18.5	&	15.8	&	5.69	&	0.9	&	$-$48.16	&	1.54	&	1.79	&	0.4	&	$-$48.34	&	2.04	&	2.67	&	2.35	&	Hg	\\
    47	&	G333.074$-$0.399	&	17.0	&	19.8	&	16.1	&	5.82	&	1.2	&	$-$53.69	&	2.75	&	1.80	&	0.6	&	$-$53.78	&	3.44	&	2.72	&	2.62	&	Hhg	\\
    48	&	G332.759$-$0.467	&	17.0	&	19.9	&	15.6	&	5.73	&	0.6	&	$-$52.87	&	1.82	&	0.93	&	0.2	&	$-$52.80	&	2.44	&	2.47	&	2.25	&	Hg*	\\
    50	&	G332.715$-$0.676	&	17.0	&	19.8	&	16.0	&	5.72	&	0.2	&	$-$45.64	&	1.92	&	2.37	&	0.1	&	$-$45.38	&	1.52	&	2.68	&	2.79	&	Hbgb	\\
    51	&	G332.903$-$0.546	&	15.0	&	16.9	&	15.7	&	5.59	&	0.6	&	$-$55.37	&	1.15	&	1.91	&	0.3	&	$-$55.61	&	1.37	&	2.62	&	2.09	&	Hbg	\\
    52	&	G333.097$-$0.352	&	17.1	&	20.0	&	15.6	&	5.45	&	0.5	&	$-$51.60	&	1.71	&	1.05	&	0.2	&	$-$51.59	&	1.85	&	2.55	&	2.48	&	Hg	\\
    53	&	G333.563$-$0.145	&	15.9	&	18.2	&	15.4	&	5.12	&	0.2	&	$-$44.76	&	1.90	&	0.52	&	0.1	&	$-$44.60	&	4.10	&	2.58	&	2.72	&	Hbgb	\\
    60	&	G332.894$-$0.567	&	15.2	&	17.2	&	15.9	&	5.77	&	0.5	&	$-$56.42	&	1.67	&	1.96	&	0.2	&	$-$56.40	&	2.07	&	2.49	&	2.44	&	Hg	\\
    \hline
  \end{tabular}\\
  \vspace{0.1cm}
  \footnotesize
  \raggedright
  $\dagger$ The presence of a `H' denotes a \ammonia~\one\ hyperfine, `A/a' denotes an absorption feature, `B/b' denotes a blended spectra, `g' denotes a Gaussian was fit to the \ammonia~\two, `hg' denotes that a \ammonia~\two\ hyperfine structure could be seen but since the signal-to-noise was below 3$\sigma$, a Gaussian was used to model the fit, and `g*' denotes a \ammonia~\two\ characteristic not included in any of the other cases (see \S\ref{sect:formulae} for further details).
  \normalsize
\end{table*}
\setlength{\tabcolsep}{6pt}
The spectra presented here are calibrated to the corrected antenna temperature scale  ($T_{\rm A}^*$) and, since the main beam efficiency has not been well determined at this frequency for the 70-m Tidbinbilla radio telescope, we do not convert these to the telescope-independent main-beam temperature scale ($T_{\rm mb}$). However, all our analysis relies on the ratio of the two ammonia transitions, which are essentially independent of calibration as they probe the same volume of gas, so the choice of temperature scale is not important.

We have followed the same fitting routine as described in \citet{Urquhart2011} and include an overview as follows. For the sources with a detectable \ammonia~\one~hyperfine structure ($\sim$75 per cent), we simultaneously fit all 18 hyperfine components and derive the optical depth and FWHM. For the remaining \ammonia~\one~sources and all the \ammonia~\two~sources, a single Gaussian profile was fitted to the main line to obtain a corrected antenna temperature. The \ammonia~\one~and \two~FWHM were obtained for all sources by fitting the hyperfine components of their spectra with their respective main line emission in order to remove the effect of line broadening due to the optical depth. An example of the resulting fits are shown in Fig.~\ref{fig:3colour_spectra} and the full selection of reduced \ammonia~\one~and \two~inversion spectra and the resulting fits can be seen in Fig.~\ref{app:spectra}. Emission from the ammonia inversion transition is detected towards 50~(79~per~cent) and 49~(78~per~cent) of the
63 SIMBA 1.2-mm dust continuum sources, for \ammonia~\one\ and \two, respectively. There were 13 (21 per cent) clumps without any detectable \ammonia. We present a summary of the detection rates in Table~\ref{tbl:derived_parameters}. The standard error of the mean ($s/\sqrt{n}$) is included with the mean. The results of the calculations and spectral fits are given in Table~\ref{tbl:nh3_parameters}. The presence of a `H' denotes a \ammonia~\one\ hyperfine, `g' denotes a Gaussian was fit to the \ammonia~\two, and `hg' denotes that a \ammonia~\two\ hyperfine structure could be seen but since the signal-to-noise was below 3$\sigma$, a Gaussian was used to model the fit. For two sources, Clumps~23 (G333.292$-$0.422) and 48 (G332.759$-$0.467), a `g*' is used to indicate that the clump has \ammonia~\two\ emission at the G333 \vlsr\ as well as at -76\,\kms\ and -69\,\kms, respectively. Features associated with the \ammonia~\one\ and \two\ were shown with an upper and lower case, respectively: `A/a' denotes an
absorption feature and `B/b' denotes a blended
spectra.

In this section we will describe the method used to determine the physical properties of the millimetre clumps identified within the G333~GMC. The fitted (\vlsr, FWHM) and calculated (optical depth and corrected antenna temperatures) parameters are presented in Table \ref{tbl:nh3_parameters}. The distribution of various derived parameters are shown in Fig.~\ref{fig:histo_SFNSF}. The uncertainties in the detection rates have been calculated using binomial statistics.


\subsubsection{Optical depth, $\tau$}
The total optical depth of the \ammonia\ transition, $\tau_{\rm{(1,1)}}$, was obtained numerically from the ratio of the satellite to main line intensity of the hyperfine fits described in \S \ref{sec:Tidbinbilla} \citep{Ho1983}:
\begin{equation}\label{eq:tau}
\frac{\Delta T^{*}_{a}\left(J,K,m\right)}{\Delta T^{*}_{a}\left(J,K,s\right)} = \frac{1-e^{-\tau\left(J,K,m\right)}}{1-e^{-\alpha\tau\left(J,K,m\right)}}
\end{equation}
where \textit{J} and \textit{K} are the principal quantum numbers corresponding to the total angular momentum and its projection along the molecular axis, $\Delta T^{*}_{a}$ is the observed brightness temperature, m and s refer to the main and satellite hyperfine components, $\tau$ is the optical depth, and $\alpha$ is the line intensity ratio of the satellite-to-main line [$\alpha$ = 0.28 and 0.22 for the (1,1) satellites]. In using this equation, we have assumed that the hyperfine components have the same beam-filling factor and excitation temperature. This is an acceptable assumption as the energy separations and probability of special excitation mechanisms that differentiate between the hyperfine components are small.


\subsubsection{Rotation temperature, \Trot}
The rotational temperature, \Trot, for the \ammonia~\one\ and \two\ transitions can be calculated using the main line intensities and the \ammonia~\one\ main quadrupole transition optical depth \citep{Ho1983}:
\begin{equation}
T_\mathrm{rot} = \frac{-T_0}{\ln \left\{\frac{-0.282}{\tau _{(1,1,m)}} \ln \left[1-\frac{\Delta T_{\mathrm{a}}^{*}\left(2,2\right)}{\Delta T_{\mathrm{a}}^{*}\left(1,1,m\right)}\left(1-\mathrm{e}^{-\tau _(1,1,m)} \right) \right] \right\}}~[{\rm{K}}]
\end{equation}
where $T_0=\frac{E_{(2,2)}-E_{(1,1)}}{k_B}\approx 41.5$\,K is the temperature of the energy difference between the \ammonia~\one\ and \two\ levels, $T_{\mathrm{a}}^{*}$ is the observed brightness temperature and $\tau$ is the optical depth. Fourteen clumps did not have a detectable \ammonia~\two\ emission, whilst thirteen clumps did not have a detectable \ammonia~\one\ emission. Clump~42 (G333.206$-$0.366), has a detectable \ammonia~\one, using the root mean square of the \two\ band, we calculate an upper limit of 9.3\,K and 9.5\,K for \Trot\ and \Tkin, respectively. For all other clumps (1, 29, 41, 49, 54-59, 61-63), fitting the \ammonia~\one\ and/or \two\ spectra was not possible so no physical parameters have been calculated for these clumps.


\subsubsection{Kinetic temperature, \Tkin}
For \Tkin $<$ $T_0 \approx 41.5$\,K, an empirical relationship between the rotational and kinetic temperature can be calculated \citep{Swift2005}:
\begin{equation}
T_\mathrm{rot} = \frac{T_\mathrm{kin}} {1+\frac{T_\mathrm{kin}}{T_0}\ln\left[ 1+0.6\times \exp\left(\frac{-15.7}{T_\mathrm{kin}}\right)\right]}~[{\rm{K}}]
\end{equation}
however, at high kinetic temperatures this expression underestimates the \Trot. From the original sample of 63 SIMBA clumps, 49 had detectable \ammonia~\two\ emission. However, one clump displays a \Trot\ which has large uncertainties [Clump~21 (G333.006$-$0.437)].


\subsubsection{Column density, N}
By assuming that all hyperfine lines have the same excitation temperature and that the excitation conditions are homogeneous along the beam, then the column density at a given transition (e.g. N$_{(1,1)}$) can be written \citep{Mangum1992}:
\begin{equation}\label{eq:coldensity}
N_{\rm{(1,1)}} = 6.60 \times 10^{14} \Delta v_{\rm{(1,1)}}\tau_{\rm{(1,1)}}\frac{T_\mathrm{rot}}{\nu_{\rm{(1,1)}}} ~~[{\rm{cm}}^{-2}]
\end{equation}
where $\Delta v_{\rm{(1,1)}}$ is the FWHM of the \ammonia\ (1,1) transition in \kms\ and $\nu_{\rm{(1,1)}}$ is the transition frequency in GHz.
We estimate the total ammonia column density following \citep{Li2003}:
\begin{align}\label{eq:coldensity_total}
N_{\rm{NH_3}} &= N_{\rm{(1,1)}} \left[ 1+\frac{1}{3}{\rm{exp}}\left(\frac{23.1}{T_\mathrm{rot}}\right)+\frac{5}{3}{\rm{exp}}\left(\frac{-41.2}{T_\mathrm{rot}}\right)\right.\nonumber \\
&\left.+\frac{14}{3}{\rm{exp}}\left(\frac{-99.4}{T_\mathrm{rot}}\right)  \right] ~~[{\rm{cm}}^{-2}]
\end{align}


\subsubsection{Volume density, n}
By assuming a fixed \ammonia/H$_2$ abundance of 10$^{-8}$ \citep{Johnstone2010}, the \ammonia\ volume density, $n$, can be calculated:
\begin{equation}\label{eq:voldensity}
n = \frac{N}{2R} ~~[{\rm{cm}}^{-3}]
\end{equation}
where $N$ is the column density in~cm$^{-2}$ and $R$ is the clump radius in~cm. The mean volume density within the G333~GMC is 4.5~$\times~10^5$\,cm$^{-3}$, with a standard deviation of 2.8~$\times~10^5$\,cm$^{-3}$.


\subsubsection{Clump (M$_{clump}$) and virial (M$_{virial}$) masses}
Assuming the dust is optically thin and using the $T_{kin}$ derived from \ammonia, we calculate the clump mass, $M_{clump}$, as \citep{Hildebrand1983}:
\begin{equation}
  \label{eqn:mclump}
  M_{clump}=\frac{F_\nu D^2}{\kappa_\nu B_\nu\left(T_{kin}\right)} \left[ M_{\sun}\right]
\end{equation}
where $F_{\nu}$ is the 1.2-mm flux density, $D$ is the distance, $\kappa_\nu$ is the dust mass opacity coefficient \citep{Ossenkopf1994} and $B_{\nu}\left(T_{kin}\right)$ is the Planck function at the dust temperature, $T_{kin}$.

Assuming that we have a Gaussian volume density distribution throughout the clump and that the gravitational energy within a clump is balanced by the turbulent energy, we calculate the virial mass, $M_{virial}$, following \citep{Protheroe2008}:
\begin{equation}
  M_{virial}= 444 \left(\frac{R}{1\, \mathrm{pc}}\right)\left(\frac{v_{\mathrm{FWHM}}}{1\,\mathrm{km s}^{-1}}\right)^2 \left[ M_{\sun}\right]
\end{equation}
where $R$ is the radius of the SIMBA 1.2-mm dust clump and $v_{\mathrm{FWHM}}$ is the FWHM.
\begin{figure*}
  \centering
  \includegraphics[angle=90,width=0.32\textwidth, trim= 10 0 45 65,clip]{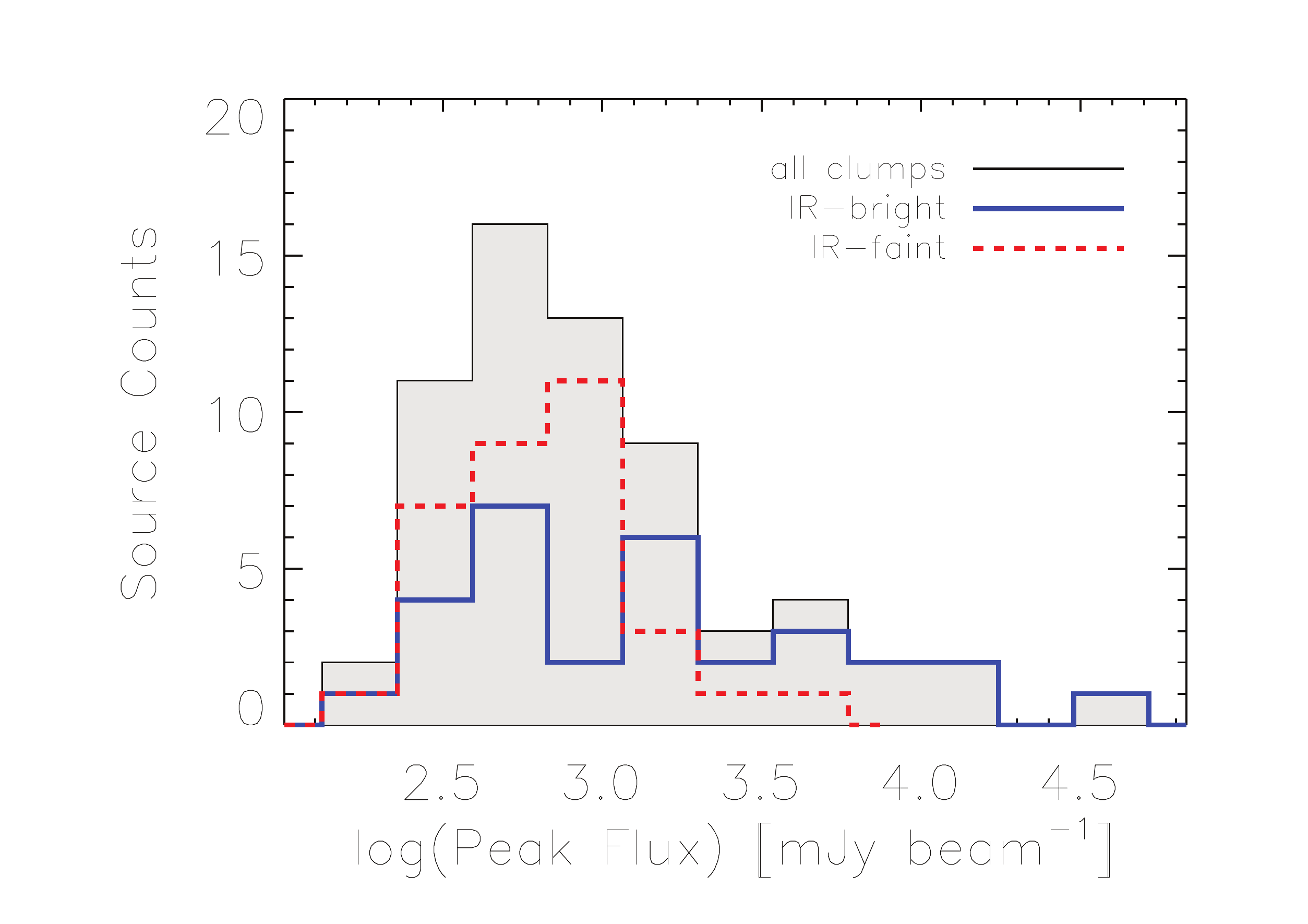}
  \includegraphics[angle=90,width=0.32\textwidth, trim= 10 0 45 65,clip]{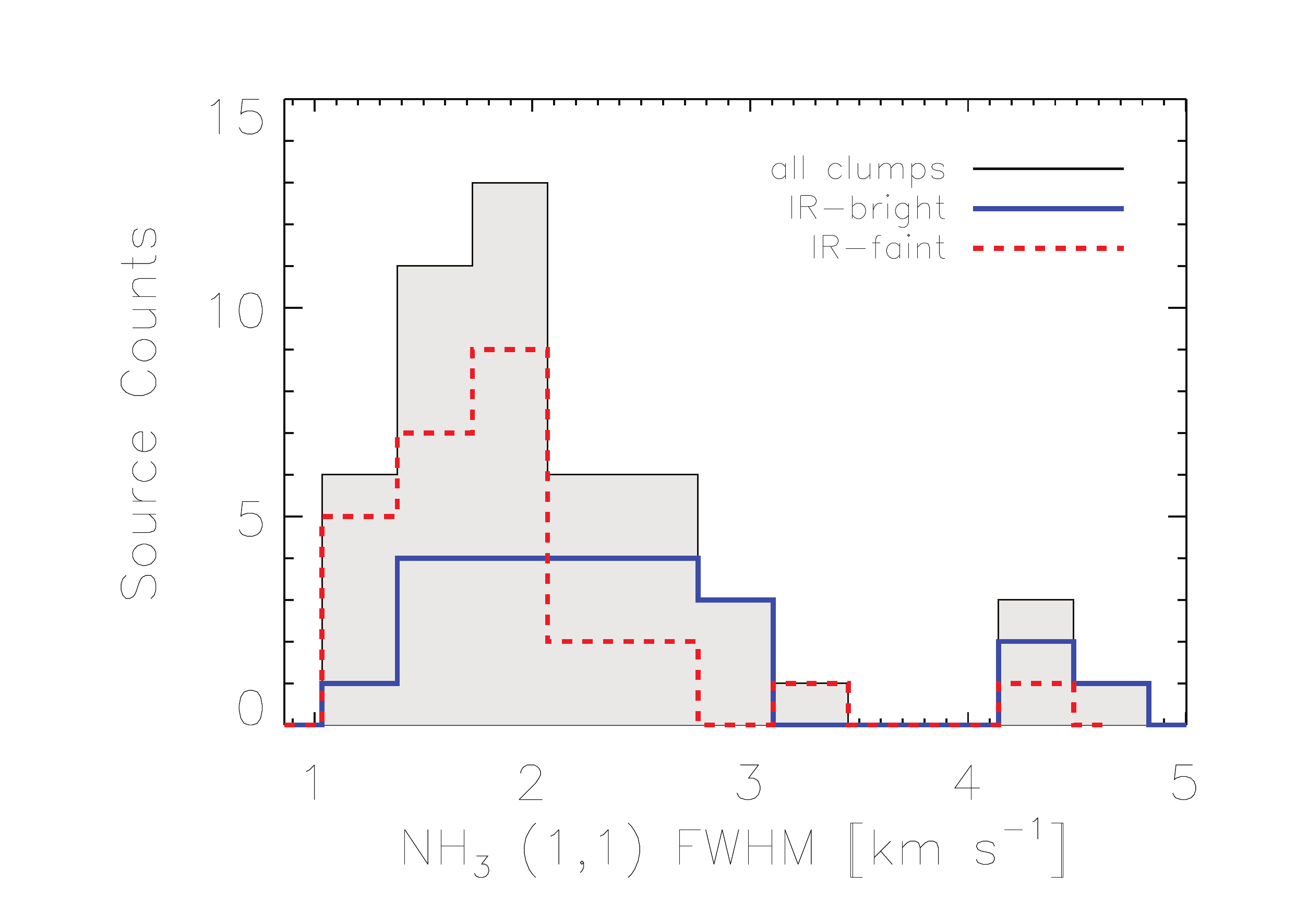}
  \includegraphics[angle=90,width=0.32\textwidth, trim= 10 0 45 65,clip]{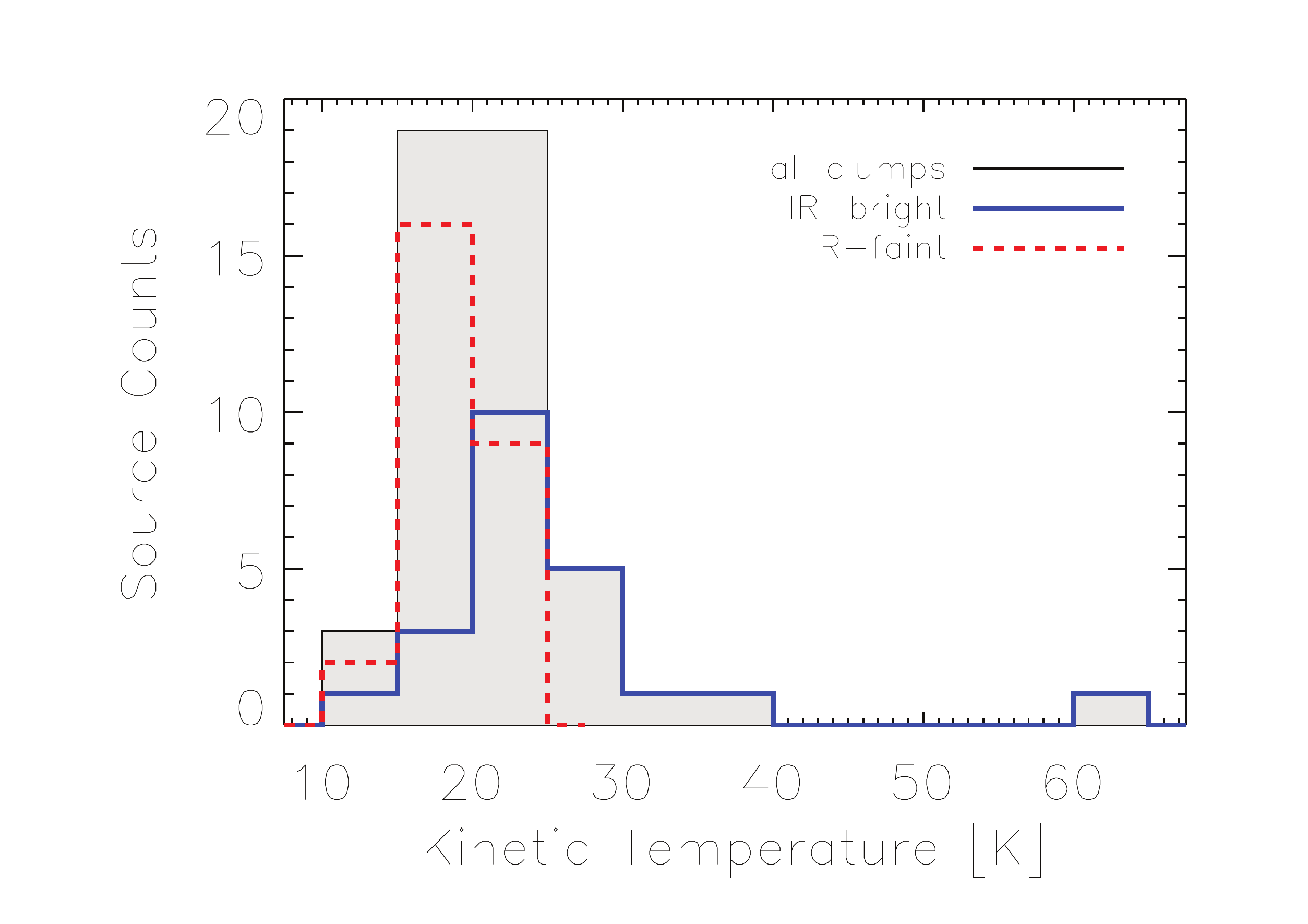}
  \includegraphics[angle=90,width=0.32\textwidth, trim= 10 0 45 65,clip]{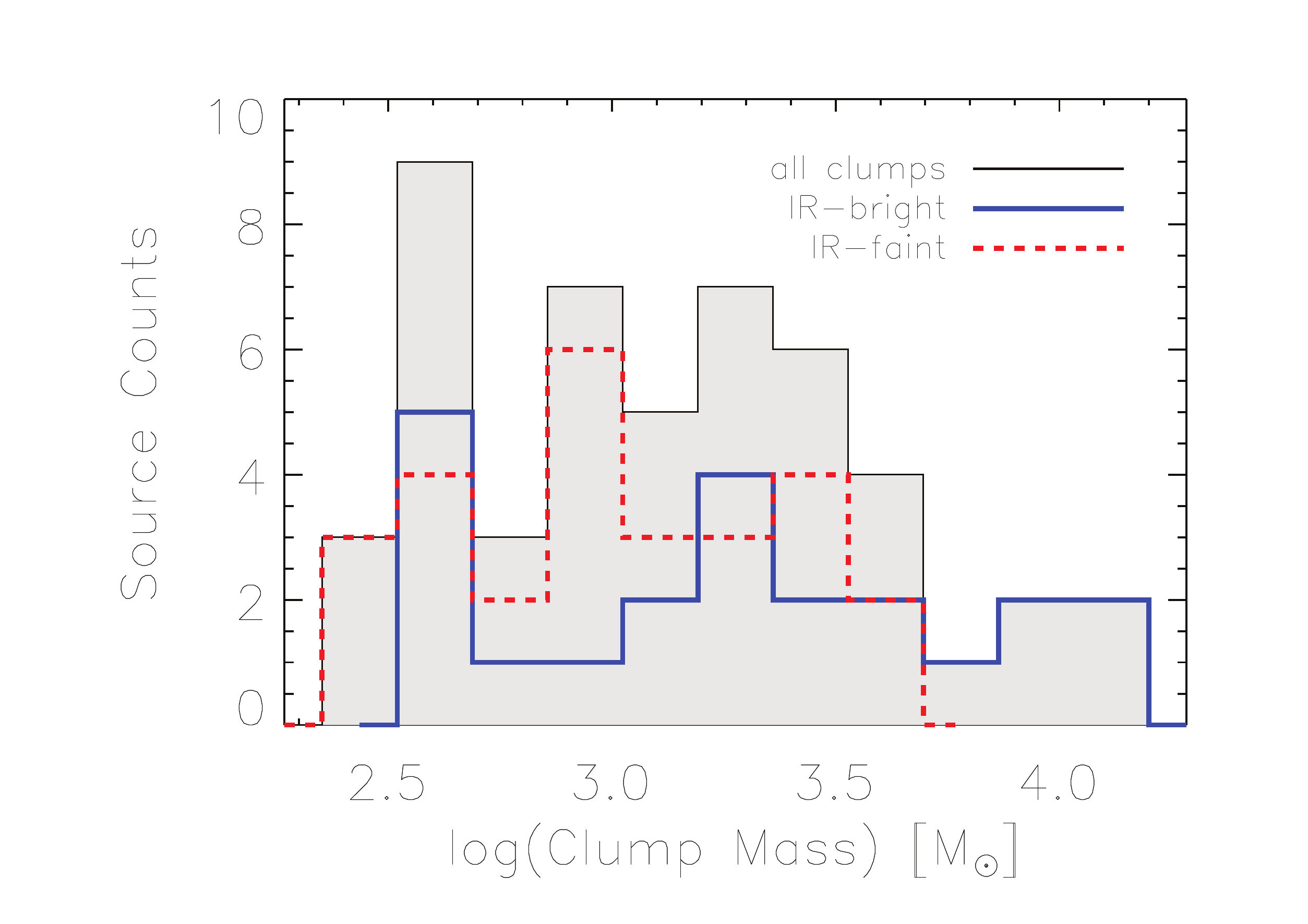}
  \includegraphics[angle=90,width=0.32\textwidth, trim= 10 0 45 65,clip]{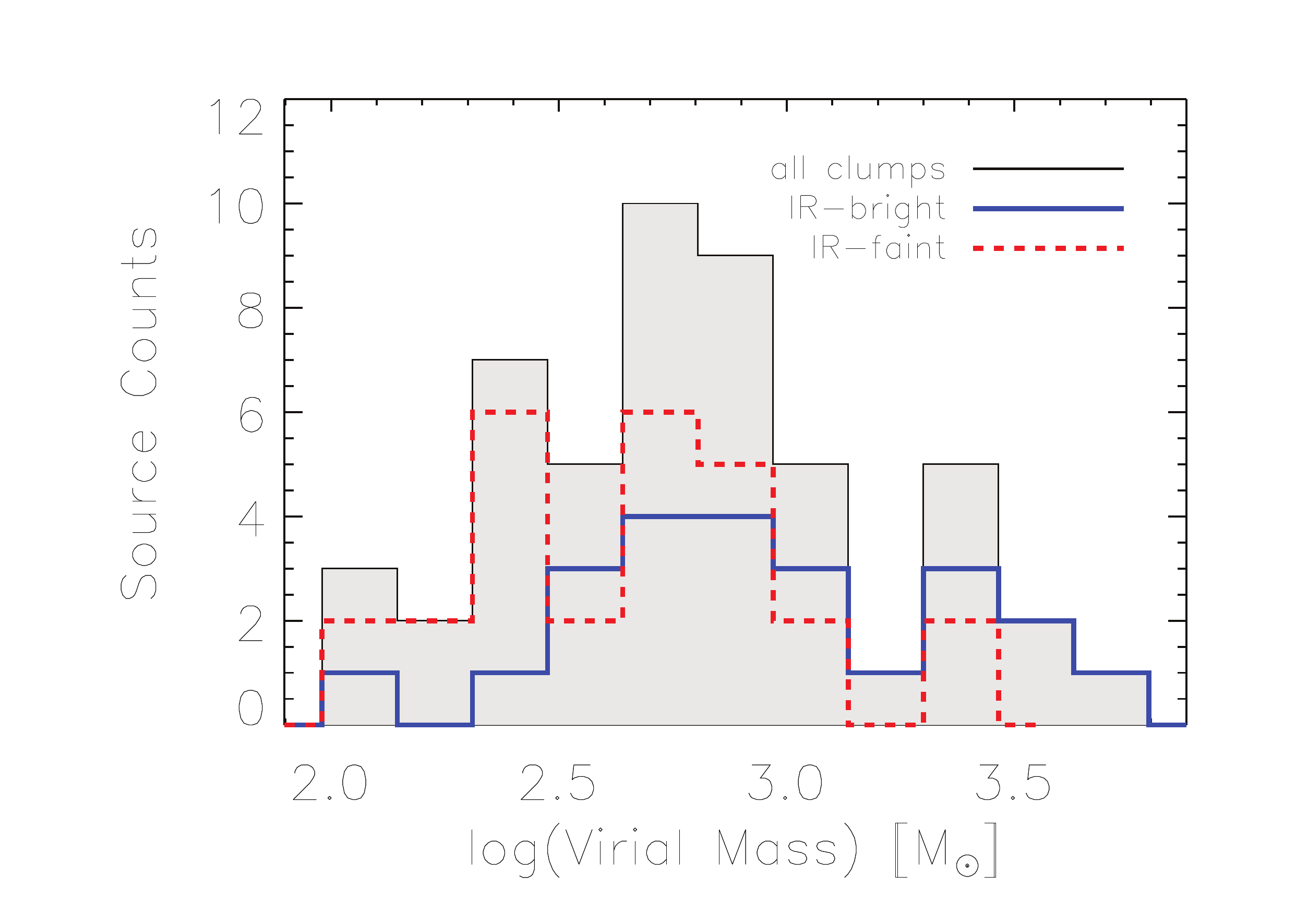}
  \includegraphics[angle=90,width=0.32\textwidth, trim= 10 0 45 65,clip]{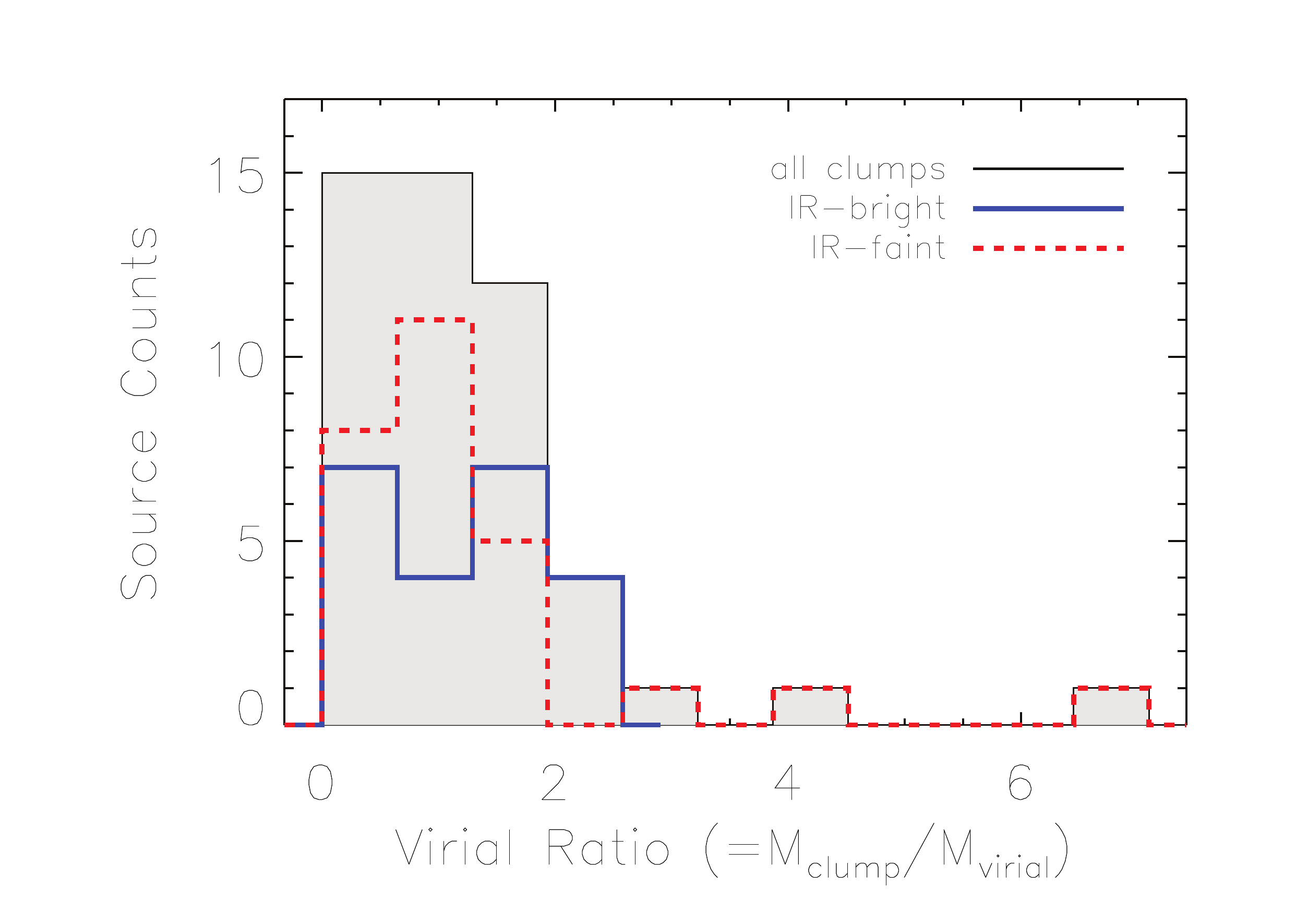}
  \caption{Distribution of the parameters from IR-bright (solid blue) and IR-faint (dashed red) clumps. The solid black line indicates the distribution for clumps with detectable \ammonia~\one\ and \two.}
  \label{fig:histo_SFNSF}
\end{figure*}
\begin{table*}
  \setlength{\tabcolsep}{3pt}
  \caption{Summary of derived parameters for IR-bright and IR-faint clumps.}
  \label{tbl:SFNSF_parameters}
  \begin{minipage}{\linewidth}
    \centering
    \begin{tabular}{lcccccccccccc}
      \hline\hline
      &\multicolumn{6}{c}{IR-bright clumps}&\multicolumn{6}{c}{IR-faint clumps} \\
      \cline{2-7}\cline{8-13}
      Parameter   &   Number  &   Mean    &   1$\sigma$   &   Median   &   Min &   Max &   Number  &   Mean    &   1$\sigma$  &   Median   &   Min &   Max \\
      \hline
      \Trot~[K]  &   20  &   19.6$\pm$0.6    &   2.6  &   19.5    &   15.6    &   25.6    &   26  &   16.6$\pm$0.3    &   1.5 &   16.7    &   12.5    &   19.4    \\
      \Tkin~[K]  &   20  &   24.2$\pm$1.0    &f   4.5  &   23.8    &   17.8    &   35.6    &   26  &   19.3$\pm$0.4    &   2.3 &   19.4    &   13.6    &   23.7    \\
      Log[N(\ammonia)] [cm$^{-2}$]    &   20  &   16.0$\pm$0.04    &   0.2 &   16.0 &   15.5    &   16.2    &   26  &   15.9$\pm$0.05    &   0.3    &   15.9 &   15.3    &   16.5    \\
      Log[n(\ammonia)] [$\times 10^2$ cm$^{-3}$]    &   20  &   5.57$\pm$0.05    &   0.23 &   5.61 &   4.97    &   5.91    &   26  &   5.56$\pm$0.06    &   0.29    &   5.60 &   4.84    &   6.03    \\
      \hline
      \ammonia~\one~\vlsr~[\kms]   &   21  &   -51.84$\pm$0.82  &   3.78   &   -51.92  &   -59.15  &   -42.57  &   26  &   -50.97$\pm$0.71  &   3.63    &   -49.95  &   -57.46  &   -44.76  \\
      \ammonia~\one~FWHM [\kms]   &   21  &   2.44$\pm$0.19    &   0.86    &   2.40    &   1.17    &   4.50    &   26  &   1.90$\pm$0.14    &   0.70    &   1.78    &   1.05    &   4.41    \\
      $\tau_{(1,1)}$  &   21  &   1.64$\pm$0.16    &   0.74    &   1.64    &   0.16    &   2.85    &   26  &   1.96$\pm$0.18    &   0.90    &   1.82    &   0.52    &   3.76    \\
      \hline
      \ammonia~\two~\vlsr~[\kms]  &   20  &   -52.11$\pm$0.85  &   3.80    &   -52.34  &   -59.76  &   -42.79  &   26  &   -51.11$\pm$0.72  &   3.68    &   -49.96  &   -57.62  &   -44.60  \\
      \ammonia~\two~FWHM [\kms]   &   20  &   2.95$\pm$0.22    &   0.98    &   2.70    &   1.85    &   5.96    &   26  &   2.50$\pm$0.18    &   0.92    &   2.35    &   1.34    &   5.49    \\
      \hline
      Log($M_{clump}$) [\msun]    &   20  &   3.31$\pm$0.12    &   0.53    &   3.30    &   2.54    &   4.12    &   26  &   2.99$\pm$0.07    &   0.36    &   2.99    &   2.44    &   3.69    \\
      Log($M_{virial}$) [\msun]   &   21  &   2.95$\pm$0.09    &   0.42    &   2.88    &   2.03    &   3.68    &   26  &   2.67$\pm$0.07    &   0.34    &   2.73    &   2.09    &   3.35    \\
      \hline
    \end{tabular}\\
  \end{minipage}
\end{table*}


\section{ANALYSIS AND DISCUSSION}
\label{sect:discussion}
\begin{figure*}
\centering
\includegraphics[trim=20 60 85 98, clip,width=0.98\textwidth]{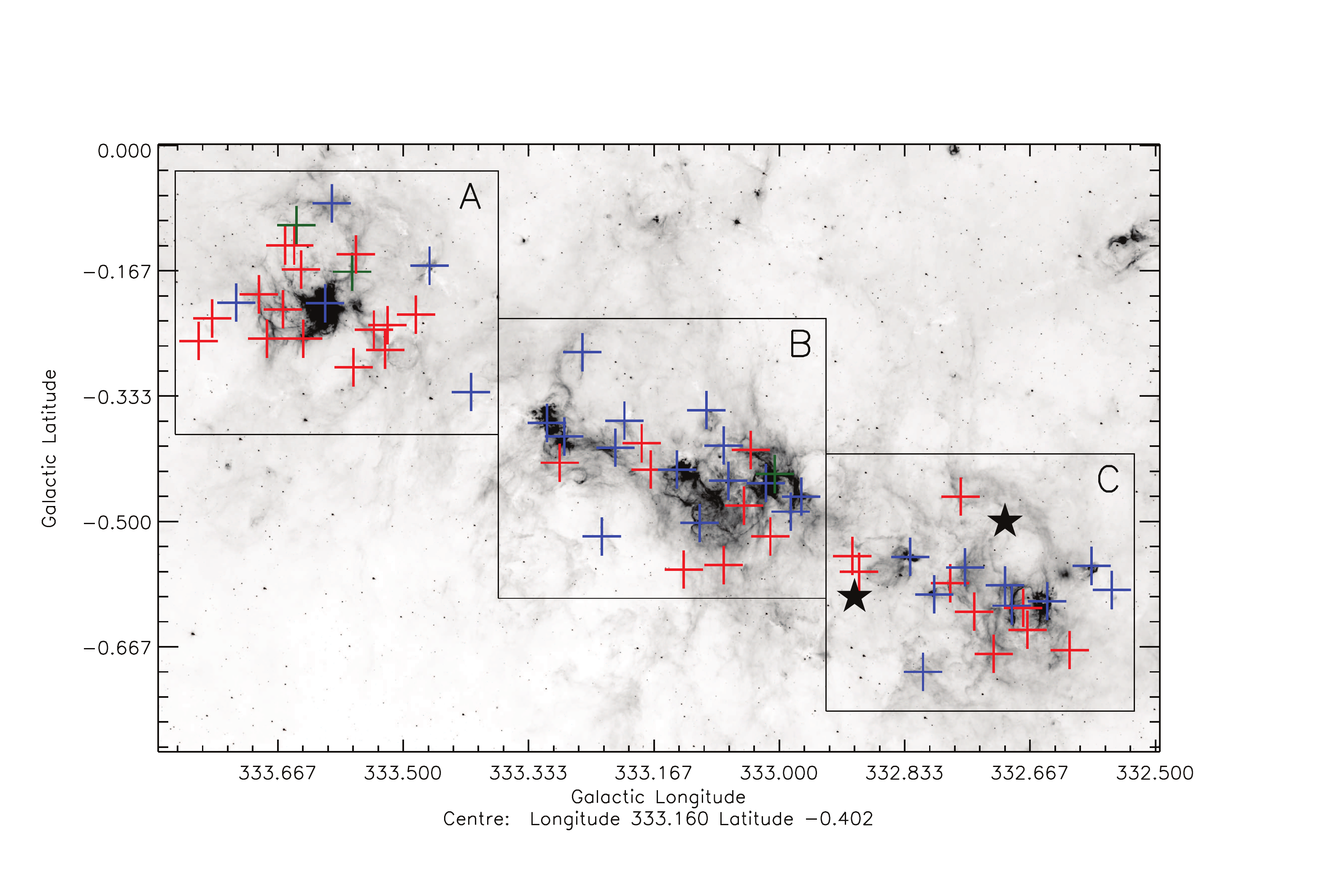}
\caption{\label{fig:distro_SFNSF} The distribution of IR-bright (blue pluses) and IR-faint (red pluses) clumps throughout the G333 giant molecular cloud. The clumps excluded as per \S \ref{sec:exclusions} are shown as green pluses. The clumps have also been separated into three regions (A, B, C). The positions associated with RCW~106 have been overlaid with black stars \citep{Rodgers1960}.}
\end{figure*}


\subsection{Separation into clumps with and without 8\,\um\ emission}
\label{sec:separation}
Through visual inspection of \textit{Spitzer} GLIMPSE 8\,\um\ emission we separated our sample of 63 SIMBA 1.2-mm clumps into two categories: clumps with (IR-bright) or without (IR-faint) significant 8\,\um\ emission (Fig.~\ref{fig:distro_SFNSF}), since the 8\,\um\ band is sensitive to ionised PAH (PAH$^+$) emission which is excited by the ultraviolet radiation from high-mass stars in photon dominated regions (PDRs). MIPSGAL 24\,\um\ emission, which represents heated very small dust grains was not used due to saturation in the 24\,\um\ band within the G333~GMC.

We identified 30 IR-bright and 33 IR-faint clumps. Column 8 of Table~\ref{tbl:cf_parameters_jsu} lists the nearby ($<$ 30\,arcsec) star forming signatures (see table caption for details). The Red MSX Source (RMS) Survey uses a multi-wavelength follow-up to disentangle their colour-selected MSX and 2MASS point sources into stages of star formation \citep{Lumsden2013}. There are RMS objects associated within 30 arcsec of the SIMBA dust peak for 15 of 63 clumps (24 per cent). The type of RMS object is listed in Column 9 of Table~\ref{tbl:cf_parameters_jsu}. Possible types include YSO, \HII\ region/YSO, \HII\ region and diffuse \HII\ region. Masers are known to be signposts for star formation so the water (\water), Class I and II (6.7\,GHz and 12.2\,GHz) methanol (\methanol) and hydroxyl (OH; 1665/1667/1720\,MHz and 6.035\,GHz) masers have been overlaid to show potential sites of star formation. The different classes of \methanol\ masers indicate different pumping mechanisms. Class~I~\methanol\ masers are collisionally pumped and associated with regions with outflows and shocks. Class~II~\methanol\ masers are radiatively pumped and exclusively associated with high-mass star forming regions \citep{Minier2003,Breen2013}. The infrared (4.5, 8.0, 160\,\um) dust emission has also been included to show regions of shocked gas, PDRs, and cool dust, respectively.

We find that our SIMBA dust clumps are associated with a variety of environments within the G333~GMC. Each of our clumps are associated with varying amounts of infrared emission, as seen in Fig.~\ref{fig:3colour_spectra} and \ref{app:3colour}. Emission at 8.0\,\um\ is common, which indicates that the G333~GMC is undergoing star formation, but there are also pockets of 160\,\um\ emission at sites of cool dust which may show regions where the embedded protostar has not significantly warmed its surroundings. It is known that modelling the spectral energy distribution of the dust can be used to identify clumps at different stages of star formation and that \ammonia\ is a good tracer of the dense cold gas which traces the molecular gas involved in star formation (e.g.~\citealt{Dunham2011,Urquhart2011,Wienen2012}). By combining the information from both the \ammonia\ and dust emission for the G333~GMC, we will develop a better understanding of the star formation occurring therein.


\subsubsection{Main cloud clumps}
\label{sec:exclusions}
From our original sample of 63 SIMBA dust clumps, we have detected \ammonia~\one\ towards 50 clumps (79 per cent) and \two\ towards 49 clumps (78 per cent). No \ammonia\ was detected towards Clump~1 (G333.604-0.210) due to self-absorption against a strong continuum emission generated by the \HII\ region, RCW~106 \citep*{Rodgers1960}. The majority of clumps with no detectable amounts of \ammonia\ are associated with the weakest millimetre continuum sources within the G333~GMC. The \ammonia~\one\ and \two\ detection rates for the IR-bright and IR-faint clumps are not obviously different. Of the 30 IR-bright clumps, 23 (77 per cent) and 22 (73 per cent) of the clumps had detectable \ammonia~\one\ and \two\ emission, respectively. Of the 33 IR-faint clumps, 27 clumps (82 per cent) had both \ammonia~\one\ and \two\ emission.

Two clumps (one clump from each category) was identified via \ammonia~\one~velocity information to be outside of the \vlsr\ range of the main G333~GMC and are excluded from the analysis in the later portions of this paper unless explicitly included. These clumps are Clump~34 (G333.586$-$0.168) with a  \vlsr\ of -88.53\,\kms\ (a clump with trace amounts of diffuse 8\,\um\ emission), and  Clump~39 (G333.642$-$0.106) with a \vlsr\ of -87.95\,\kms\ (a YSO which has cleared its surrounding \HII~region). The \vlsr\ for the excluded clumps corresponds with the Norma-Cygnus arm and the included clumps corresponds with the Norma-Cygnus and/or Scutum-Crux arms (fig.~3 of \citealt{Vallee2008}). At temperatures much greater than 30\,K, \ammonia~\one\ and \two\ cannot accurately be used as a temperature probe and any physical parameters using this kinetic temperature calculation would also be inaccurate \citep{Danby1988,Hill2010}. As Clump~21 (G333.006$-$0.437) had a derived kinetic temperature of 63.6\,K, it has been
removed from the following analysis. We note that Clump~3 (G333.136$-$0.431) has a \Tkin~of 31.9\,K and Clump~11 (G333.018$-$0.449) has a \Tkin~of 35.6\,K, but has been retained for the following analysis.

From the original sample of 63 SIMBA 1.2\,mm dust clumps, we are left with 29 IR-bright and 31 IR-faint clumps (Table~\ref{tbl:SFNSF_parameters}). Their distribution (with the excluded three clumps) for the derived parameters is shown in Fig.~\ref{fig:histo_SFNSF}. We note that derived parameters could not be calculated for all clumps due to the non-detection of \ammonia~\one\ and/or \two. The black shaded histogram shows the total number of clumps with that derived parameter. IR-bright and IR-faint clumps are shown by the solid blue and dashed red line, respectively. The distributions in these histograms generally have skewed Gaussian-like profiles. The kinetic temperature, the FWHM of the \ammonia~\one\ transition and the dust and virial masses show that the distribution is skewed in favour of higher values for all these parameters for the IR-bright clumps, while the IR-faint clumps cluster at the lower end.  These are not wholly independent parameters, as we have used the same spectrum to measure the FWHM and temperature for the calculations of the virial and clump mass, respectively. However, these histograms show a qualitative difference between the two populations. We investigate these variations further with statistical methods in the next section.


\subsection{Importance of accurate kinetic temperatures}
\label{sec:tkin}
\begin{figure}
\centering
\includegraphics[angle=90,width=0.49\textwidth, trim=0 0 20 40,clip]{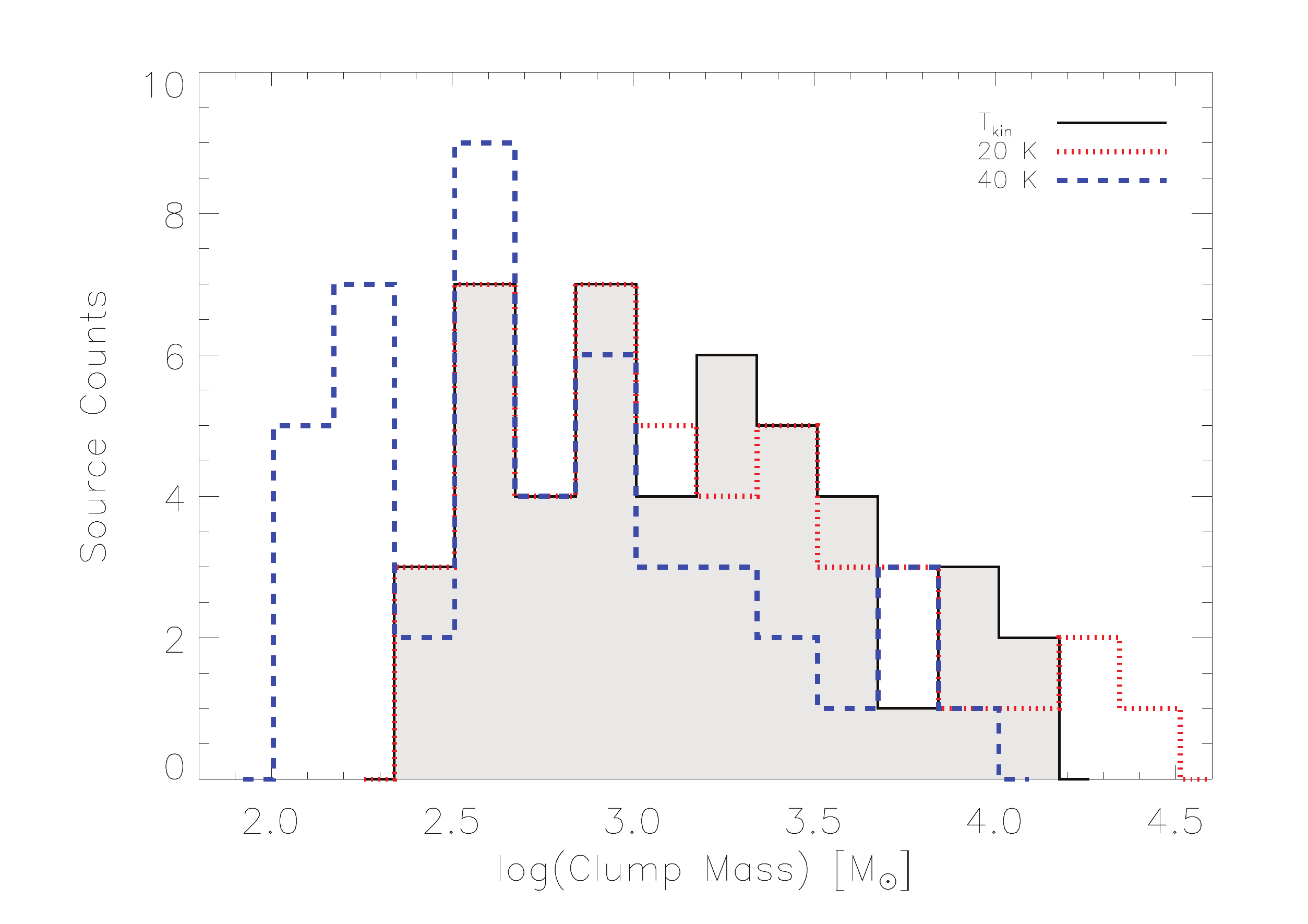}
\caption{\label{fig:tkin} The distribution of clump mass as a function of temperature. The solid black line was calculated from the temperature found from \ammonia, whilst the dotted red and dashed blue were calculated from the assumption of 20 and 40\,K, respectively.}
\end{figure}

It is common to assume 20\,K for isolated clumps and 40\,K for \HII~regions (e.g. \citealt{Mookerjea2004} for the G333~GMC). Since \ammonia~\one\ and \two\ are closely separated ($\Delta\nu\approx 28\,\mathrm{MHz}$), we are able to obtain the \Trot\ and \Tkin\ with the same observation per clump. In Fig.~\ref{fig:tkin}, we display the effect that kinetic temperature has on the calculation of clump mass [Eqn. (\ref{eqn:mclump})]. We find that clumps with \Tkin=20\,K are more massive, whereas clumps with \Tkin= 40\,K are less massive. From the dust mass for clumps derived using \Tkin, we found a mean of 2390 \msun, a median of 1089\,\msun, a standard deviation of 2920\,\msun\ and a range between 278--13036\,\msun. For the masses calculated assuming 40\,K, we found a mean of 1285 \msun, a median of 483\,\msun, a standard deviation of 1992\,\msun\ and a range between 105--9951\,\msun. For the masses calculated using 20\,K, we found a mean of 3064 \msun, a median of 1151\,\msun, a standard deviation of 4750\,\msun\ and
a range between 251--23723\,\msun. We find that T=20\,K overestimates the mass whereas T=40\,K underestimates it. Therefore, an accurate temperature is important for determining if a clump is bound and will collapse onto itself, or if it is unbound and remains a starless core.

\subsection{Variations in physical parameters throughout the cloud}
\label{sec:variations}
\begin{figure*}
  \centering
  \includegraphics[trim=20 60 85 98, clip,width=0.99\textwidth]{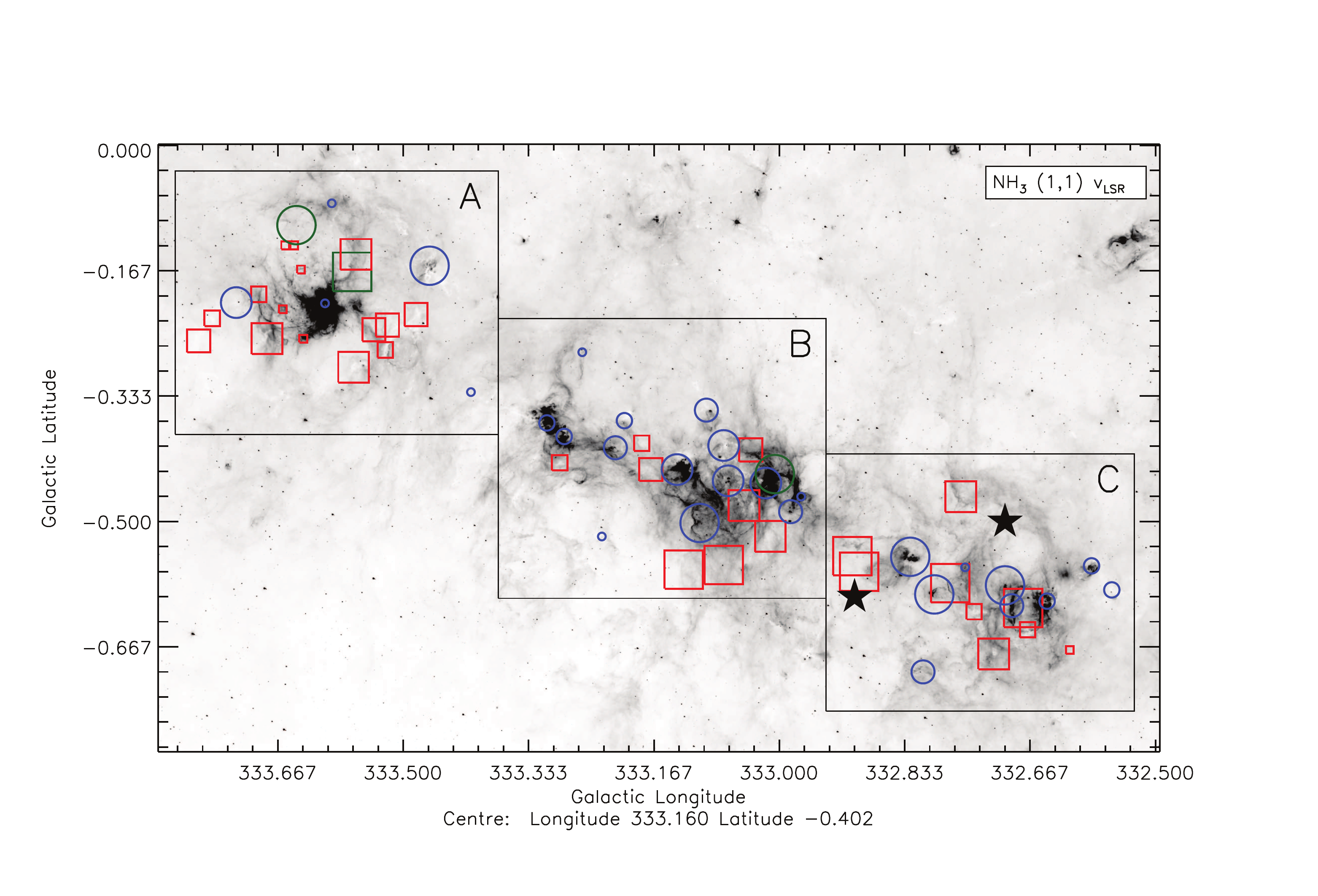}
  \caption{\label{fig:distro_vlsr11} The distribution of the \vlsr\ of \ammonia~\one\ throughout the G333 giant molecular cloud. Each clump was classified as IR-bright (blue circles) or IR-faint (red squares) clumps using the presence of \textit{Spitzer} GLIMPSE 8.0\,$\mu m$ emission (grey-scale image). The samples were separated into non-detections and detections, with the detections further split into four sub-samples at the 25th, 50th and 75th percentile. The IR-bright and IR-faint clumps which were excluded, as per \S\ref{sec:exclusions}, are shown as green circles and squares, respectively, and the symbol size for these clumps corresponds with the percentile calculations for the included clumps. The clumps have also been separated into three regions (A, B, C). The positions associated with RCW~106 have been overlaid with black stars \citep{Rodgers1960}.}
\end{figure*}
\begin{figure*}
  \centering
  \includegraphics[trim=20 60 85 98, clip,width=0.99\textwidth]{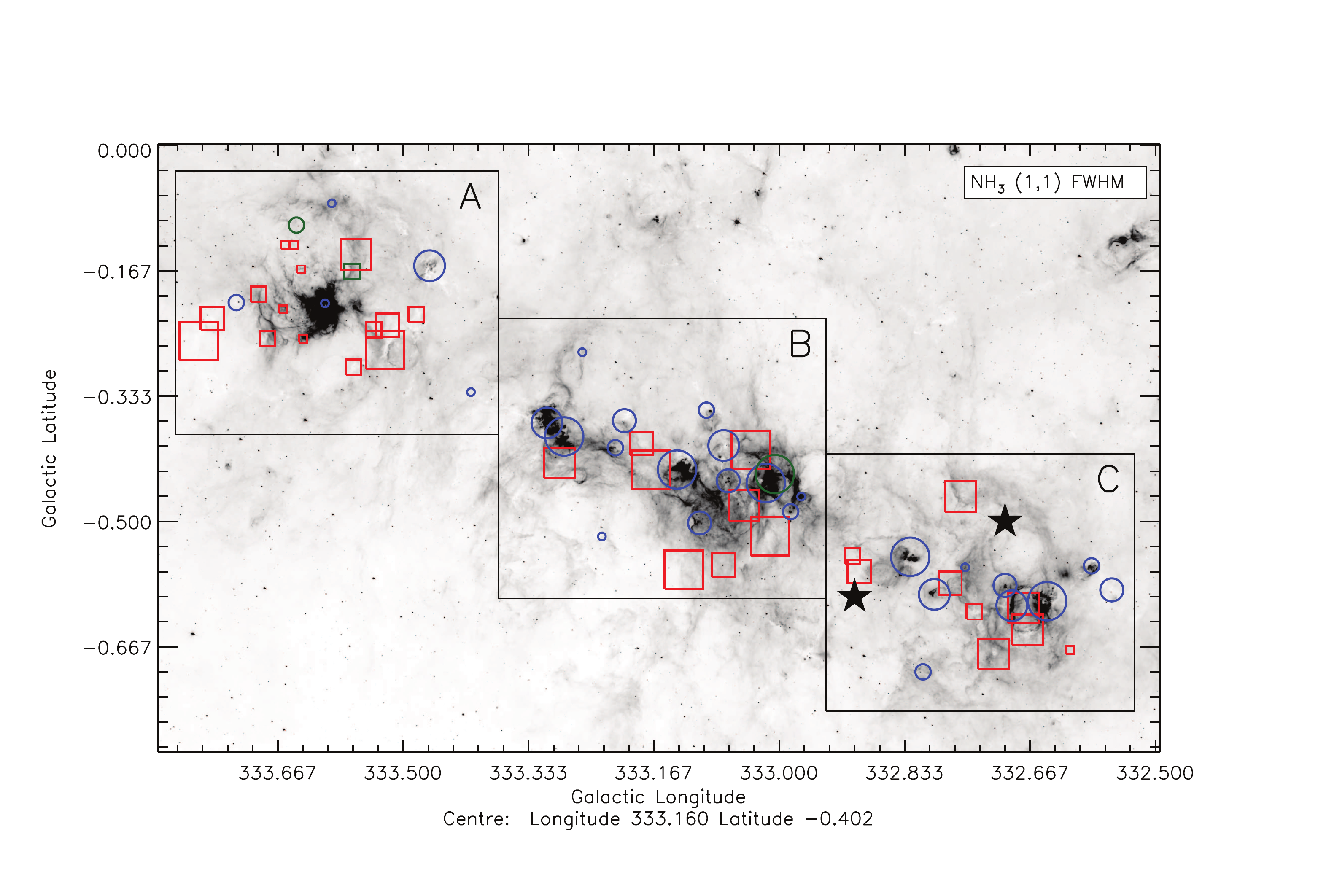}
  \caption{\label{fig:distro_fwhm11} The distribution of the FWHM of \ammonia~\one\ throughout the G333 giant molecular cloud. Each clump was classified as IR-bright (blue circles) or IR-faint (red squares) using the presence of \textit{Spitzer} GLIMPSE 8.0\,$\mu m$ emission (grey-scale image). The samples were separated into non-detections and detections, with the detections further split into four sub-samples at the 25th, 50th and 75th percentile. The IR-bright and IR-faint clumps which were excluded, as per \S\ref{sec:exclusions}, are shown as green circles and squares, respectively, and the symbol size for these clumps corresponds with the percentile calculations for the included clumps. The clumps have also been separated into three regions (A, B, C). The positions associated with RCW~106 have been overlaid with black stars \citep{Rodgers1960}.}
\end{figure*}
\begin{figure*}
  \centering
  \includegraphics[trim=20 60 85 98, clip,width=0.99\textwidth]{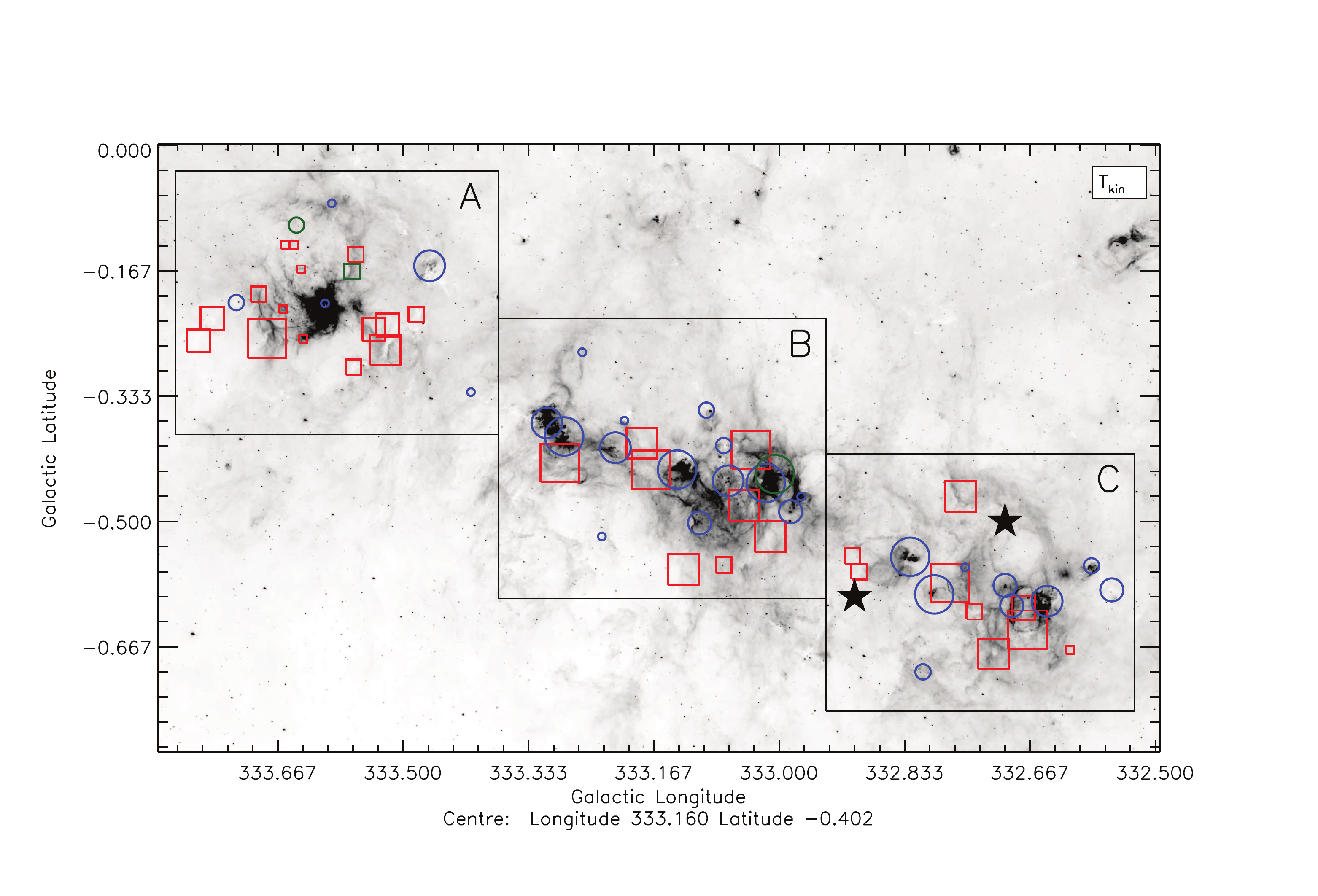}
  \caption{\label{fig:distro_tkin} The distribution of the \Tkin\ throughout the G333 giant molecular cloud. Each clump was classified as IR-bright (blue circles) or IR-faint (red squares) clumps using the presence of \textit{Spitzer} GLIMPSE 8.0\,$\mu m$ emission (grey-scale image). The samples were separated into non-detections and detections, with the detections further split into four sub-samples at the 25th, 50th and 75th percentile. The IR-bright and IR-faint clumps which were excluded, as per \S\ref{sec:exclusions}, are shown as green circles and squares, respectively, and the symbol size for these clumps corresponds with the percentile calculations for the included clumps. The clumps have also been separated into three regions (A, B, C). The positions associated with RCW~106 have been overlaid with black stars \citep{Rodgers1960}.}
\end{figure*}

The majority of clumps can be grouped into three regions (A, B, C; Fig.~\ref{fig:distro_SFNSF}) and the distribution of physical parameters throughout the G333~GMC is shown in Table~\ref{tbl:grouping} and Fig.~\ref{fig:distro_vlsr11}, \ref{fig:distro_fwhm11} and \ref{fig:distro_tkin}. For Fig.~\ref{fig:distro_SFNSF}, each clump was classified as IR-bright (blue pluses) or IR-faint (red pluses) clumps using the presence of \textit{Spitzer} GLIMPSE 8.0\,$\mu m$ emission (grey-scale image). The excluded clumps are shown with green pluses. For the \ammonia~\one~FWHM (Fig.~\ref{fig:distro_fwhm11}) and \Tkin\ (Fig.~\ref{fig:distro_tkin}), the samples were separated into non-detections and detections, with the detections further split into four sub-samples (Q1, Q2, Q3, Q4) at the 25th, 50th and 75th percentile by taking the medians of the \ammonia~\one~FWHM and \Tkin, respectively. The \ammonia~\one~\vlsr\ was separated in a similar fashion but the absolute difference between the \vlsr\ of the G333~GMC ($\sim$50\,\kms; \citealt{Bains2006}) and the \ammonia~\one~\vlsr\ was used. The values of these subdivisions can be seen in Table~\ref{tbl:grouping}. The symbol size correlates with the percentile sub-sample, and IR-bright and IR-faint clumps are shown as blue circles and red squares, respectively. The IR-bright and IR-faint clumps which were excluded, as per \S\ref{sec:exclusions}, are shown as green circles and squares, respectively, and the symbol size for these clumps corresponds with the percentile calculations for the included clumps.

Region~A contains one of the brightest infrared sources in the southern Galactic Plane (G333.6$-$0.2; \citealt{Becklin1973}), however \ammonia\ for the associated Clump~1 (G333.604$-$0.210) was not fit due to \ammonia\ self-absorption, so we were unable to extract any physical parameters from the largest and most active clump within the G333~GMC. It also appears to be surrounded by IR-faint clumps. Region~A contains 5 and 15 IR-bright and IR-faint clumps, respectively. This asymmetry could be due to the destructive interaction of the \HII\ region with the GMC. There are two other \HII\ regions [G333.6$-$0.1 \citep{Goss1970} and IRAS~16175-5002 \citep{Ellingsen1996}] within this region and both are associated with IR-bright clumps. However, only IRAS~16175-5002 has detectable \ammonia\ emission. The \vlsr\ of the clumps within Region~A does not significantly differ from the \vlsr\ of the G333~GMC. However, the two clumps which were excluded, as per \S\ref{sec:exclusions}, due to having a \vlsr\ different to the main G333 cloud, were found in this region.

Region~A shows relatively small \vlsr\ deviations ($<$ 1.5 \kms\ from the \vlsr\ of the G333~GMC) for IR-faint clumps. However, with over half the included IR-bright clumps showing non-detections (with, at least, one due to \ammonia\ self-absorption), we cannot reliably comment further on the \vlsr\ of the IR-bright clumps in Region~A. Clump~6 (G333.465$-$0.160) contains the largest \ammonia~\one~FWHM for IR-bright clumps in this region and is associated with the \UCHII\ region, IRAS~16175$-$5002 (e.g. \citealt{Walsh1998}). However, there are two IR-faint clumps [Clumps~13 (G333.524$-$0.272) and 26 (G333.772$-$0.260)] in the highest percentile. Both of these clumps have an excess of 160\,\um\ emission with Clump~13 (G333.524$-$0.272) showing possible compression between two expanding 8\,\um\ bubbles. Most (60\,\%) of the \Tkin\ values for IR-bright clumps are non-detections. Half (53\,\%) of the IR-faint clumps are below the 50th percentile, and a third are non-detections. Clump~13~(G333.524$-$0.272) shows the second highest \Tkin\ (20.3\,K) for the IR-faint clumps and Clump~38~(G333.681$-$0.106) shows the highest (measured) \Tkin\ (23.7\,K) of the region.

Region~B appears to be the opposite and contains 14 and 8 IR-bright and IR-faint clumps, respectively. There are three clusters of \HII\ regions and they correspond with IR-bright clumps; however there are also IR-bright clumps that are not \HII\ regions. The IR-faint clumps appear to be on the outskirts of \HII\ regions. The IR-bright clumps in Region~B show smaller deviations from the \vlsr\ towards the north of the GMC and the clumps with larger deviations clustered towards the south. The IR-faint clumps have larger \vlsr\ deviations and are closer to the south. This region shows the largest \ammonia~\one~FWHM for IR-faint clumps, whereas the IR-bright clumps are somewhat equally spaced throughout all quartiles. The majority of high \Tkin\ appear to be associated with, or adjacent to, the \HII\ regions, however Clump~9 (G333.127$-$0.564) has a \Tkin\ of 20.3\,K and appears to have excess 160\,\um\ emission coincident with the 1.2-mm flux peak, and Class I and II \methanol\ masers.

Region~C also contains multiple \HII~regions (including RCW~106) but the number of IR-bright and IR-faint clumps are equal (with 9 clumps in each category). \citet{Rodgers1960} identified two bright areas in RCW~106: one at G332.9$-$0.6 and the other at G333.7$-$0.5, with their extent covering 20\,arcmin~$\times$~7\,arcmin and 12\,arcmin~$\times$~12\,arcmin, respectively. This covers the majority of SIMBA clumps within Region~C. For Region~B and C, we note that the IR-faint clumps are often towards the edge of each region. This suggests that star formation on these scales are propagating outwards from the central and more evolved parts of the regions, which suggests sequentially triggered star formation. Region~C has a larger proportion of IR-faint clumps showing larger \vlsr\ deviations, with the largest deviations seen towards the boundary of Regions~B with~C. The largest values for the \ammonia~\one~FWHM are seen towards the IR-bright clumps that are associated with \HII~regions. For the IR-faint clumps, the \Tkin\ tend towards the larger values. Clump~25 (G332.773$-$0.582) and Clump~32 (G332.670$-$0.644) are IR-faint clumps however they show high \Tkin\ (22.6 and 21.5\,K, respectively) but have low values for \ammonia~\one~FWHM (1.72 and 1.96\,\kms). These clumps are also adjacent to IR-bright clumps so appear to be sites of future or currently unseen star formation.

The 63 clumps within the G333~GMC can be grouped into three regions. The physical parameters throughout the GMC are influenced by the numerous \HII~regions. Region~A contains the most number of IR-faint clumps (15), whilst Region~B contains the most number of IR-bright clumps (14), and Region~C contains the same number (nine) of IR-bright and IR-faint clumps. Region~A and~B contain the most number of \ammonia\ non-detections, with eight and three, respectively. This may be because of the effect that the \HII~regions, G333.6$-$0.2 and RCW~106, have on the surrounding molecular gas. Region~B contains seven clumps with large (Q4; 2.15$-$4.41\,\kms) \ammonia~\one\ FWHM, whilst both Regions~A and C have two each. This could indicate that the \HII~regions in Region~A and C are sweeping molecular material into Region~B. Region~A and C have the most clumps with low (Q1; 13.6$-$18.2\,K) \Tkin, whereas Region~B has the warmest [Q3-Q4; 13 clumps with \Tkin\ above the median (19.4\,K)]. Region~C also contains the most numbers of clumps with the largest (Q4; 5.37$-$7.46\,\kms) deviation from the \vlsr\ of the G333~GMC.
\begin{table}
  \setlength{\tabcolsep}{3pt}
  \centering
  \caption{Variations in physical parameters within the regions in the G333~GMC. The samples were separated into non-detections and detections, with the detections further split into four sub-samples (Q1, Q2, Q3, Q4) at the 25th, 50th and 75th percentile. The \ammonia~\one~\vlsr\ was separated in a similar fashion but the absolute difference between the \vlsr\ of the G333~GMC ($\sim$50 \kms; \citealt{Bains2006}) and the \ammonia~\one~\vlsr\ was used. Excluded clumps as per \S\ref{sec:exclusions} were not used to calculate the quartiles but a~+~indicates the quartile it would belong to.}
  \label{tbl:grouping}
  \begin{minipage}{\linewidth}
    \centering
    \small
    \begin{tabular}{ccccccc}
      \hline\hline
       & \multicolumn{6}{c}{Region}\\
      FWHM & \multicolumn{2}{c}{A} & \multicolumn{2}{c}{B} & \multicolumn{2}{c}{C} \\
      \cline{2-7}
       $[$\kms] & IR-bright & IR-faint & IR-bright & IR-faint & IR-bright & IR-faint \\
      \hline
      2.15--4.41 & 0 & 2 & \phantom{0}3+ & 4 & 2 & 0 \\
      1.82--1.99 & 1 & 1 & 2 & 2 & 2 & 4 \\
      1.54--1.74 & 0 & 2 & 3 & 2 & 2 & 2 \\
      1.11--1.48 & \phantom{0}1+ & \phantom{0}5+ &3 & 0 & 2 & 2 \\
      non-detection & 3 & 5 & 3 & 0 & 1 & 1 \\
      \hline\hline
       & \multicolumn{6}{c}{Region}\\
      \Tkin & \multicolumn{2}{c}{A} & \multicolumn{2}{c}{B} & \multicolumn{2}{c}{C} \\
      \cline{2-7}
       $[$K] & IR-bright & IR-faint & IR-bright & IR-faint & IR-bright & IR-faint \\
      \hline
      20.6--23.7 & 0 & 1 & \phantom{0}3+ & 3 & 2 & 2 \\
      19.5--20.4 & 1 & 1 & 3 & 4 & 1 & 2 \\
      18.4--19.3 & 0 & 4 & 2 & 0 & 3 & 1 \\
      13.6--18.2 & \phantom{0}1+ & \phantom{0}4+ &2 & 1 & 2 & 3 \\
      non-detection & 3 & 5 & 4 & 0 & 1 & 1 \\
      \hline\hline
       & \multicolumn{6}{c}{Region}\\
      \ammonia~\one~\vlsr & \multicolumn{2}{c}{A} & \multicolumn{2}{c}{B} & \multicolumn{2}{c}{C} \\
      \cline{2-7}
       $[$\kms] & IR-bright & IR-faint & IR-bright & IR-faint & IR-bright & IR-faint \\
      \hline
      5.37--7.46 & \phantom{0}1+ & \phantom{0}0+ & \phantom{0}1+ & 2 & 3 & 4 \\
      2.55--5.24 & 1 & 3 & 4 & 2 & 0 & 2 \\
      1.05--1.84 & 0 & 4 & 3 & 2 & 2 & 0 \\
      0.05--0.66 & 0 & 3 & 3 & 2 & 3 & 2 \\
      non-detection & 3 & 5 & 3 & 0 & 1 & 1 \\
      \hline
      Total & \phantom{0}5+ & 15+ & 14+ & 8 & 9 & 9 \\
      \hline
    \end{tabular}
  \end{minipage}
\end{table}


\subsection{Correlations of physical parameters}
\label{sec:correlations}
We have calculated the linear Pearson correlation coefficient and the linear fit to test the level of correlation between physical parameters for the IR-bright and IR-faint clumps.

The linear Pearson correlation coefficient, $r$, for different pairs of physical parameters,  were found with the {\sc correlate} function found in IDL. For each $\left(x_i,y_i\right)$ pair, with $i=1,....,N$, $r$ can be written following \citet[chap. 14]{Press1992} as:
\begin{equation}\label{eq:correlate}
r = \frac{\sum_{i}\left(x_i-\bar{x}\right)\left(y_i-\bar{y}\right)}{{\sqrt{\sum_{i}\left(x_i-\bar{x}\right)^2}}{\sqrt{\sum_i\left(y_i-\bar{y}\right)^2}}}
\end{equation}

The linear fits between different physical parameters were found with the {\sc linfit} function found in IDL. It fits a linear model ($y = a+bx$), while minimising the chi-square error statistic. The chi-square error statistic, $\chi^2$, can be written following \citet[chap. 15]{Press1992} as:
\begin{equation}\label{eq:chisq}
\chi^2(a,b) = \displaystyle\sum_{i=1}^{N} \left(y_i - a - bx_i\right)^2
\end{equation}

\begin{figure}
  \centering
  \includegraphics[trim=0 0 31 0,clip,angle=90,width=0.49\textwidth]{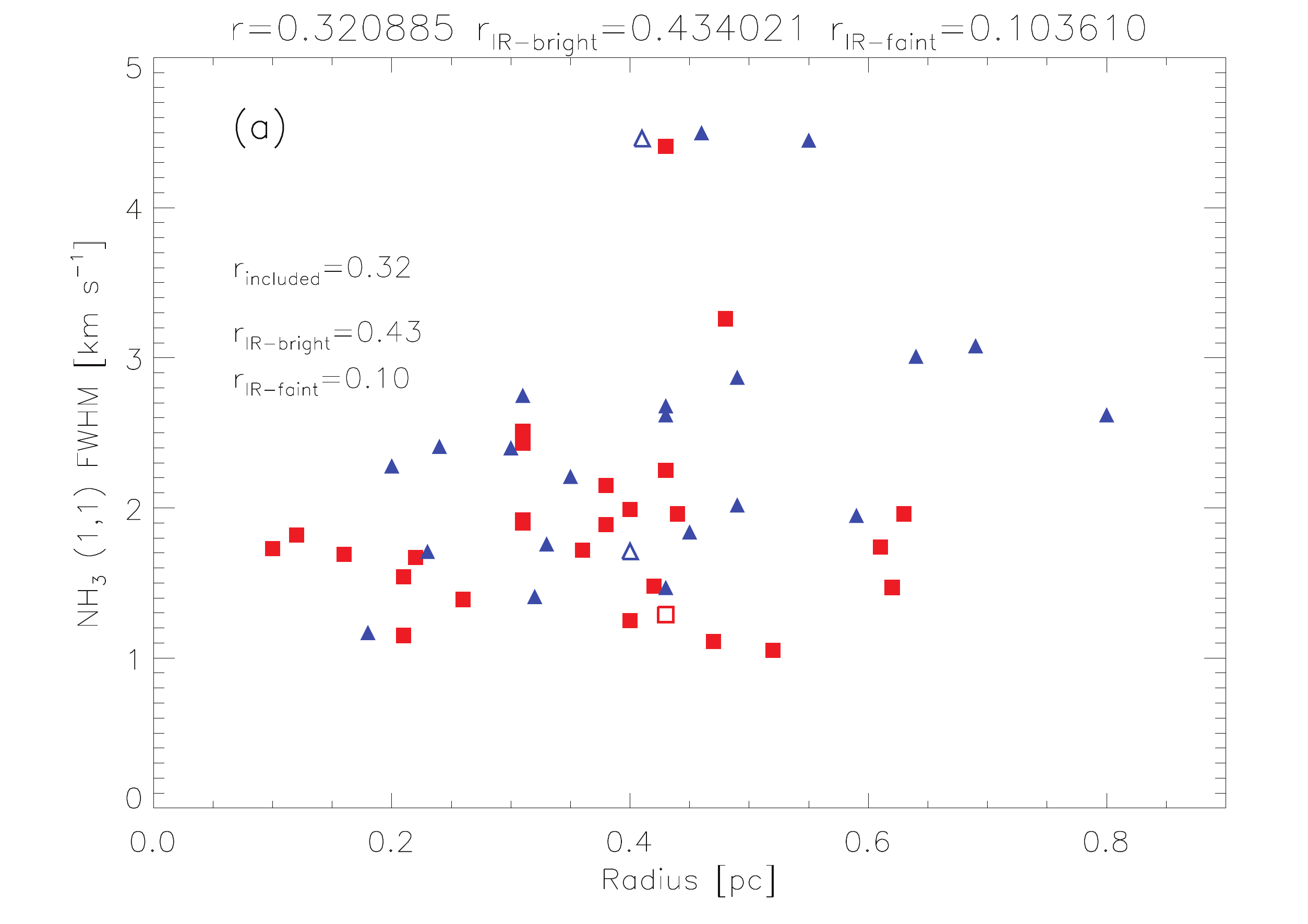}
  \includegraphics[trim=0 0 35 0,clip,angle=90,width=0.49\textwidth]{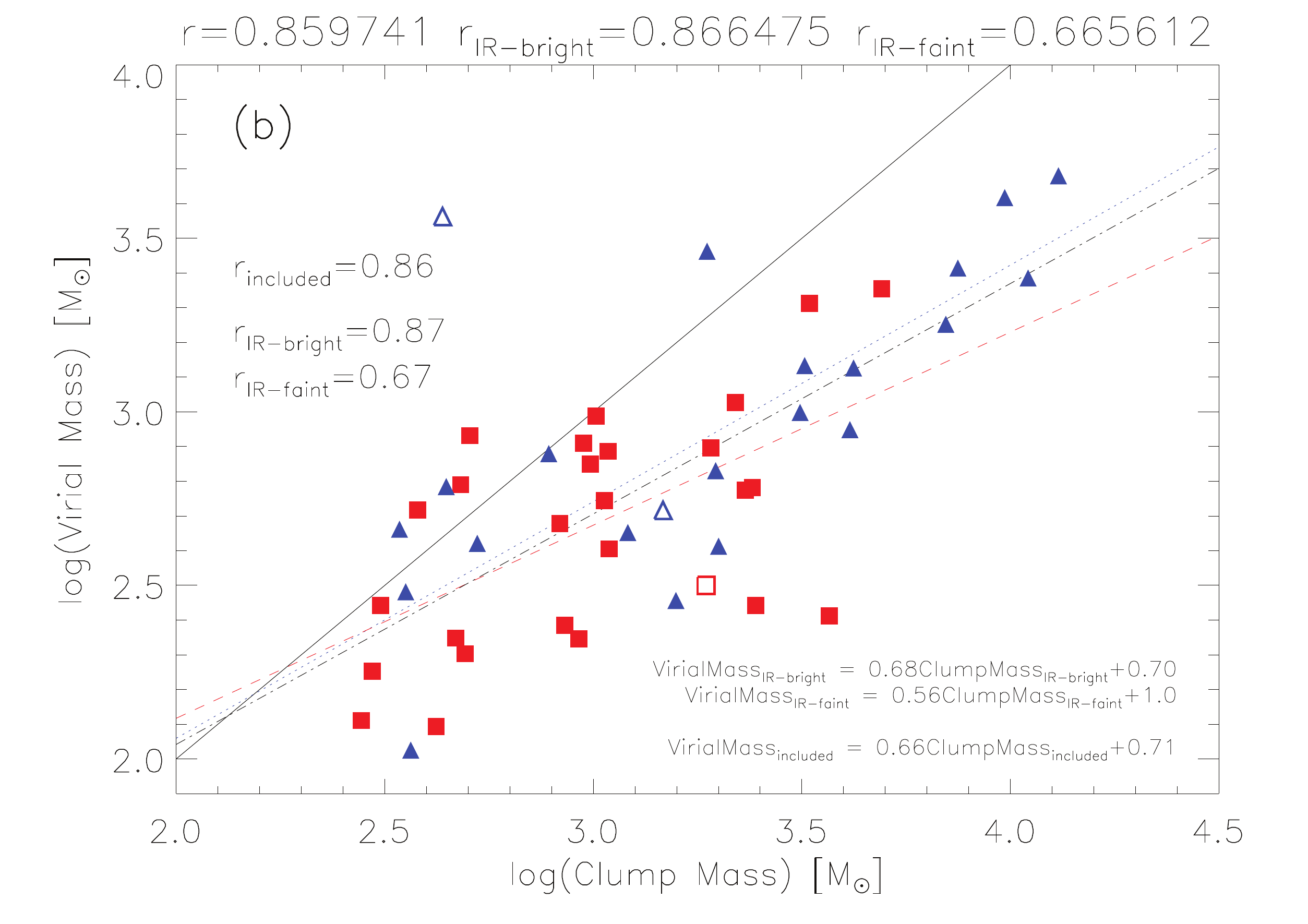}
  \includegraphics[trim=0 0 31 0,clip,angle=90,width=0.49\textwidth]{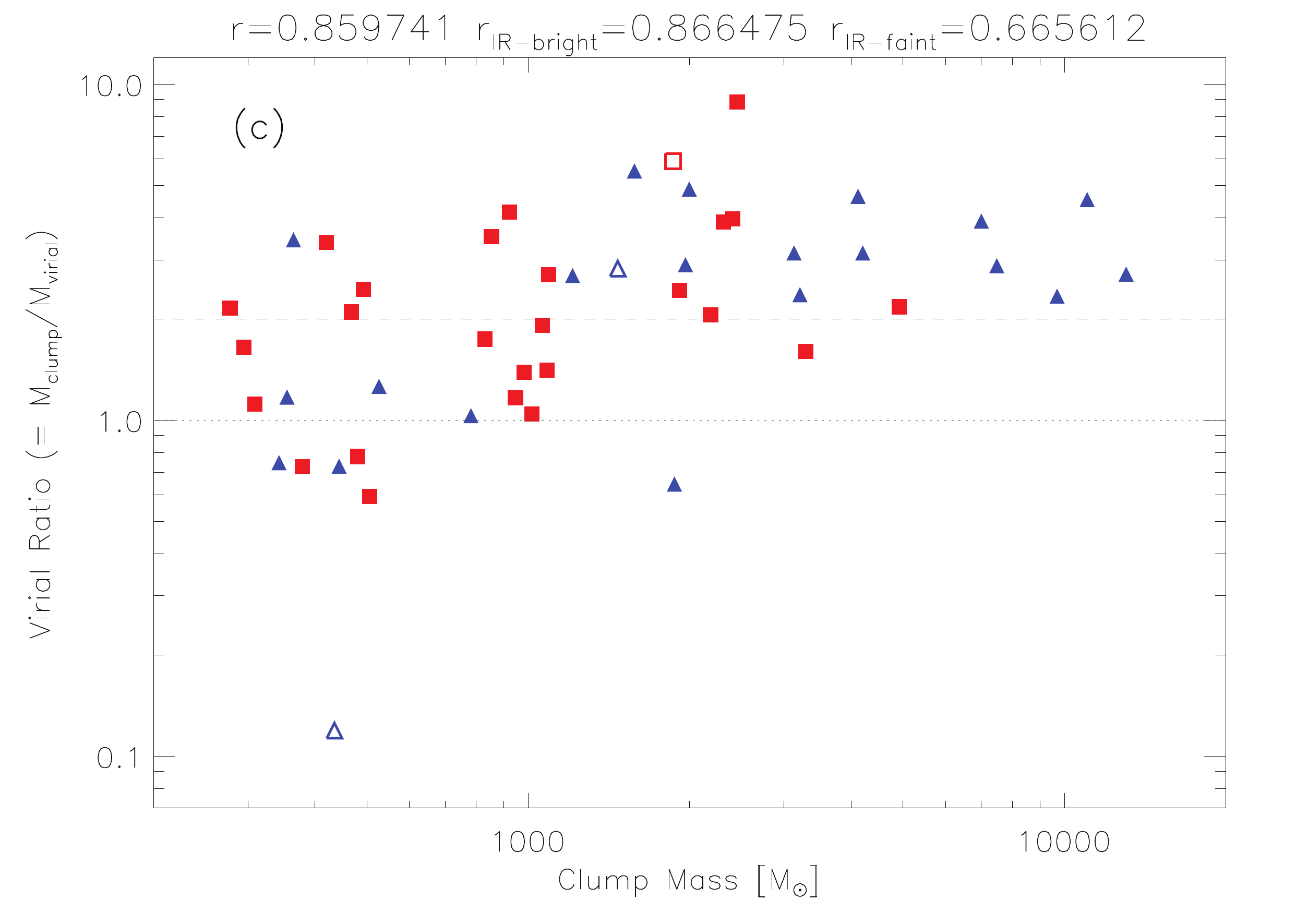}
  \caption{Correlations between physical parameters. IR-bright clumps are represented by blue filled triangles. IR-faint clumps are represented by red filled squares. The excluded IR-bright and IR-faint clumps are represented by blue hollow triangles and red hollow squares, respectively. For (b): The blue dotted line displays the fit to the IR-bright clumps. The red dashed line displays the fits to the IR-faint clumps. The black dashed dot line displays the fit to both samples. These fits do not include the clumps that have been excluded, as per \S\ref{sec:exclusions}. The black solid line displays the 1:1 line. For (c): The green dotted line displays the equilibrium between clump and virial masses, while the green dashed line shows the equilibrium between kinetic and magnetic energy.}
  \label{fig:correlation}
\end{figure}

The relationship between clump radius and FWHM can be seen in Fig.~\ref{fig:correlation}(a). There is significant scatter in the data, as seen by the medium positive correlation for IR-bright clumps ($r_\mathrm{IR-bright}~=~0.43$) and low positive correlation for IR-faint clumps ($r_\mathrm{IR-faint}~=~0.10$), which shows a poor linear fit to the data. This implies that before the onset of 8\,\um\ emission, \ammonia\ is not correlated with the physical parameters of the clump but once they become IR-bright, the FWHM has some correlation with 8\,\um\ emission. There are four clumps with \ammonia~\one~FWHM $\sim$4.5\,\kms\ that deviate from the other clumps. Three of these clumps are IR-bright, however one (Clump~21 G333.006$-$0.437) has been excluded from our analysis as per \S\ref{sec:exclusions}. Two of the included clumps (Clump~3 G333.136$-$0.431 and Clump~30 G333.038$-$0.405) belong to Region~B but one is IR-bright and the other is IR-faint. The third included clump belongs to Region~C.

The relationship between the virial and clump mass can be seen in Fig.~\ref{fig:correlation}(b). However, the G333~GMC contains numerous \HII~regions which have free-free emission which can contribute to the 1.2\,mm emission. This may contaminate surrounding clumps resulting in smaller clump masses [Eqn. (\ref{eqn:mclump})]. Of these 32~clumps, 72~per~cent (23~of~32) were classified as IR-faint clumps. The masses for 14~clumps could not be calculated because no \ammonia\ emission was detected in 12~clumps; the emission profile was too complex, due to self-absorption, for Clump~1 (G333.604$-$0.210); and \ammonia~\two\ was not detected in Clump~42 (G333.206$-$0.366). There are high positive correlations for both IR-bright and IR-faint clumps, with $r_\mathrm{IR-bright}=0.87$ and $r_\mathrm{IR-faint}=0.67$. The three clumps which have been previously excluded in \S\ref{sec:exclusions}, were included in Fig.~\ref{fig:correlation} as unfilled blue triangles and unfilled red squares to represent IR-bright and IR-faint clumps, respectively; however, they were not included in the fits. We can see that the Clump~21 (G333.006$-$0.437) with a \Tkin\ of 63.6\,K contains a significantly larger virial mass and would skew further analysis. However, the clumps with a \vlsr\ different to the G333~GMC do not deviate much from the scatter of the other clumps and would not greatly influence our later results. The clumps that are gravitationally bound are shown below the black solid 1:1 line. Three clumps, with a clump mass smaller than the virial mass, have been classified as clumps with signs of star formation. For the six clumps with a smaller clump mass, four clumps have masses within 30~per~cent of their corresponding virial mass.

The virial ratio ($= M_{clump}/M_{virial}$) versus the clump mass is shown in Fig.~\ref{fig:correlation}(c). The green dotted horizontal line marks the line of gravitational instability (i.e. $M_{clump} = M_{virial}$). Clumps below this line are unbound and will not collapse under gravity. There are 87~per~cent (40~of~46) of the clumps with masses larger than the virial mass, suggesting that they will form stars or are already undergoing star formation. The IR-bright and IR-faint clumps have a mean virial ratio of 2.7 and 2.8, respectively, and a median virial ratio of 2.8 and 2.1, respectively. However, magnetic fields can prevent the collapse of a clump and by assuming the equipartition of kinetic and magnetic energy, the relationship between clump and virial masses can be found to be $M_{clump}=2M_{virial}$ (e.g.~\citealt{Bertoldi1992}) and is shown by the dashed green horizontal line.
By including magnetic fields, the number of clumps which are gravitationally unstable becomes 70 per cent (14~of~20) and 54 per cent (14~of~26), for IR-bright and IR-faint clumps, respectively.


\subsection{Kolmogorov--Smirnov test}
\label{sec:kstest}
The Kolmogorov--Smirnov (K-S) test is used to test whether our IR-bright and IR-faint clumps are likely to be drawn from the same population. We perform a K-S test on the physical parameters derived from the \ammonia~\one~and \two\ spectra. To compare the cumulative distribution functions of the physical parameters for the IR-bright and IR-faint clumps, $S_\mathrm{IR-bright}(x)$ and $S_\mathrm{IR-faint}(x)$, the K-S statistic, can be written following \citet[chap. 14]{Press1992} as:
\begin{equation}
  \mathrm{D} = \max_{-\infty < x < \infty}|S_\mathrm{IR-bright}(x) - S_\mathrm{IR-faint}(x)|
\end{equation}
To calculate the significance of the KS test:
\begin{equation}\label{eq:Q_KS}
  \mathrm{Q}_{KS} (\lambda) =  2 \sum_{j=1}^\infty (-1)^{j-1}\mathrm{e}^{-2j^2\lambda^2}
\end{equation}
where
\begin{equation}
  \lambda = \left[{\sqrt{N_{eff}} + \dfrac{0.11}{\sqrt{N_{eff}}}}\right] D
\end{equation}
and $N_{eff}$ is the effective number of data points, with
\begin{equation}
  N_{eff} = \dfrac{N_1N_2}{N_1+N_2}
\end{equation}
where $N_1$ and $N_2$ are the two distributions to be tested.

The significance and K-S statistic for the clump and derived parameters can be seen in Table~\ref{tbl:ks}. The significances were calculated to be 0.0078, 0.3747, 0.0162, 0.0002, 0.0446 and 0.0056 for peak flux, clump radius, \ammonia~\one\ FWHM, kinetic temperature, clump mass, and virial mass, respectively. We note that the significance of the cumulative distribution functions between the two populations for the kinetic temperature are over the 0.003 significance level (3$\sigma$) than if they were drawn from the same population \citep{Wall2012}. Through accurate kinetic temperatures, we confirm that 8\,\um\ emission traces the warm dust and gas. In Fig.~\ref{fig:histo_SFNSF}, the \ammonia~\one\ FWHM displays a difference between the two populations, however the significance in the K-S test is below the 3$\sigma$ level. The same outcome is not apparent through the clump radius or either the clump or virial masses. The effect of turbulence is important to high-mass star formation as it can encourage the development of stars above the mass threshold. However, if the amount of turbulence is too high, it can discourage star formation \citep{Padoan1995}. As the NH3 (1,1) linewidths (1.1--4.5\,\kms) are larger than the thermal component ($\sim$0.28\,\kms\ for \Tkin=30\,K), all our clumps show varying amounts of turbulence. However, the K-S test was unable to find a significant difference between the IR-bright and IR-faint clumps, so we are unable to find a link between the amount of turbulence and 8\,\um\ emission. The K-S test was unable to distinguish a significant difference for the clump radius. This is important because a significant difference would indicate that different algorithms and input settings would drastically affect the parameters derived from the clumps.
\begin{table}
  \setlength{\tabcolsep}{3pt}
  \centering
  \caption{Results from the Kolmogorov--Smirnov (K-S) test of IR-bright and IR-faint clumps. Column 2 shows the significance (Q$_{KS}$) of the K-S test and Column 3 shows the maximum deviation (D) between the cumulative distribution function of the IR-bright and IR-faint clumps.}
  \label{tbl:ks}
  \begin{minipage}{\linewidth}
    \centering
    \small
    \begin{tabular}{lcc}
      \hline\hline
      \multicolumn{3}{c}{Clumps with detectable \ammonia~\one\ and \two}\\
      \hline
      Parameter	& Q$_{KS}$	& D\\
      \hline
      Peak flux	& 0.0059	& 0.42 \\
      Clump radius		& 0.4694	& 0.21 \\
      Total flux	& 0.0436	& 0.34 \\
      \ammonia~\one~FWHM		& 0.0136	& 0.43 \\
      Column density	& 0.1917	& 0.30 \\
      Kinetic temperature		& 0.0002	& 0.59 \\
      Clump mass	& 0.0595	& 0.35 \\
      Virial mass	& 0.0041	& 0.48 \\
      \hline
      \multicolumn{3}{c}{Clumps in the main cloud$^\dagger$}\\
      \hline
      Parameter	& Q$_{KS}$	& D \\
      \hline
      Peak flux 	& 0.0078	& 0.42 \\
      Clump radius		& 0.3742	& 0.23 \\
      Total flux 	& 0.0297	& 0.36 \\
      \ammonia~\one~FWHM		& 0.0162	& 0.44 \\
      Column density	& 0.3667	& 0.26 \\
      Kinetic temperature		& 0.0002	& 0.61 \\
      Clump mass	& 0.0446	& 0.39 \\
      Virial mass	& 0.0056	& 0.48 \\
      \hline
    \end{tabular}\\
    \vspace{0.1cm}
    \footnotesize
    \raggedright
    $\dagger$The clumps designated as ``main cloud'' are those with calculated parameters from the \ammonia\ spectra but exclude those as per \S\ref{sec:exclusions}.
    \normalsize
  \end{minipage}
\end{table}


\subsection{Environments around interesting sources}
\label{sec:sources}
Here we discuss sources that deserve further detail than those provided by Columns 8, 9 and 10 of Table \ref{tbl:cf_parameters_jsu}. Infrared composite images centered on the peak emission of the dust clumps identified by {\sc clumpfind} can be found in Fig.~\ref{app:3colour}. \ammonia~\one\ and \two\ emission detected towards the peak emission of each clump can be found in Fig.~\ref{app:spectra}.

\textit{Clump~1 (G333.604$-$0.210).} This is the brightest IR source in the southern Galactic Plane \citep{Becklin1973}. \citet{Fujiyoshi2005,Fujiyoshi2006} found that this \HII~region is excited by a cluster of O and B stars. \citet{Fujiyoshi2006} also found a complex H90$\alpha$ radio recombination line spectrum which they suggest may be caused by champagne outflows. \ammonia~emission is weak and shows absorption but both \one\ and \two\ show similar intensities.

\textit{Clump~2 (G332.826$-$0.547) and Clump~60 (G332.894$-$0.567).} These clumps appear to be related to a 8\,\um\ bubble. Clump~2 (G332.894$-$0.567) coincides with an \HII\ region and YSO \citep{Urquhart2008} which is within the extent of RCW~106 \citep{Rodgers1960}. There are \water, Class~II \methanol\ and OH masers. There is bifurcated 8\,\um\ emission which may explain the broad \ammonia~\one\ and \two\ spectrum. The \ammonia~\one\ spectrum appears to be blended with no clear baseline between the hyperfine transitions. The \HII\ region and YSO also coincides with a 160 \um\ diffuse source. Clump~60 (G332.894$-$0.567) is not associated with any known maser activity and there is no detectable \ammonia\ emission.

\textit{Clump~3 (G333.136$-$0.431).} This clump contains an assortment of masers however the peak of the SIMBA dust clump coincides with a 160\,\um\ diffuse source as well as OH and 6.7\,GHz Class~I~\methanol\ masers. The peak of this SIMBA dust clump is surrounded by 8\,\um\ emission with an assortment of maser activity. There are four clusters of maser activity around the peak with two clusters containing 6.7\,GHz Class~I~\methanol\ and \water\ masers as well as 44.07 and 95.1\,GHz class~II \methanol\ masers. It is rare to see the two classes of \methanol\ masers in the same cluster. Class~II \methanol\ masers are closely associated with YSOs, while Class~I \methanol\ masers are associated with shocks and cloud-cloud collisions. There may be triggered star formation occurring in this 8\,\um\ arc which has allowed for the two classes of masers to be seen at the same time. A collision between clouds would create the Class~I \methanol\ masers but also trigger star formation which leads to outflows, YSOs and the Class~II \methanol\ masers.

\textit{Clump~4~(G333.286$-$0.387) and Clump~5~(G333.309$-$0.369).} These clumps appear to be local peaks in a larger clump. Clump~4 contains multiple point sources with a water maser and was classified by the RMS survey as a \HII~region. Clump~5 was classified by the RMS survey as a diffuse \HII~region.

\textit{Clump~6~(G333.465$-$0.160).} This clump contains a mixture of Class~I and II \methanol\ masers as well as a water maser. It has a lot of 8\,\um\ emission around a diffuse 160\,\um\ source. This looks like Clump~2~(G332.826$-$0.547) but with significantly more 8\,\um\ emission. This may be a later evolutionary stage but the \ammonia~\one\ and \two\ do not show blended spectra, which could indicate turbulence, like Clump~2~(G332.826$-$0.547).

\textit{Clump~7~(G333.068$-$0.446).} Satellite lines can be seen in both the \ammonia~\one\ and \two. There are Class~II \methanol\ masers associated with the SIMBA dust peak and a Class~II \methanol\ and water maser offset by $\sim$1\,arcmin on either side of the SIMBA dust peak.

\textit{Clump~8~(G332.644$-$0.606), Clump~10~(G332.691$-$0.612), Clump~12~(G332.676$-$0.615) and Clump~22~(G332.700$-$0.585).} There are two 160\,\um\ sources within 8\,\um\ arcs on either side of another 160\,\um\ source. There is a 6.7\,GHz Class~I~\methanol\ and water maser along the west 8\,\um\ arc, however neither is coincident with the SIMBA dust peak. The ionising source of this bubble is unknown; however, further analysis is required to show if it is expanding outwards from Clump~12~(G332.676$-$0.615). These clumps are also within the region classified by \citet{Rodgers1960} as RCW~106.

\textit{Clump~9 (G333.127$-$0.564).} This is a relatively isolated, cold dust clump which coincides with the second brightest SiO source within this GMC \citep{Lo2007}. There are low levels of the \ammonia~\two\ satellite lines.

\textit{Clump~10 (G332.691$-$0.612).} This is an elongated \HII\ region and could be star formation triggered by a bow shock of a nearby protostellar object. This clump could be associated with nearby Clump~12 (G332.676$-$0.615). The maser at the top corresponds to a  YSO and \HII\ region.

\textit{Clump~11 (G333.018$-$0.449) and Clump~21 (G333.006$-$0.437).} These two clumps are on the edge of the same 8\,\um\ arc. The SIMBA dust clumps appear to coincide with 160\,\um\ emission. Both clumps have \water\ masers on the far side of the 8\,\um\ arc. Both clumps have high \Tkin. Clump~11 (G333.018$-$0.449) has a \Tkin\ of 35.6\,K and Clump~21 (G333.006$-$0.437) has a \Tkin\ of 63.6\,K.

\textit{Clump~13 (G333.524$-$0.272), Clump~27 (G333.539$-$0.245) and Clump~46 (G333.521$-$0.239).} Clump~13 (G333.524$-$0.272) appears to follow the 160\,\um\ emission which appears to be compressed between two 8\,\um\ arcs. There is no known maser emission in this region. Clump~27 (G333.539$-$0.245) also appear to follow the compressed 160\,\um\ emission however the intensity is lower. Clump~46 (G333.521$-$0.239) is also associated but the 160\,\um\ emission is the weakest of the three.

\textit{Clump~14 (G332.985$-$0.487), Clump~18 (G333.074$-$0.558), Clump~19 (G333.722$-$0.209), Clump~39 (G333.642$-$0.106), Clump~47 (G333.074$-$0.399), Clump~51 (G332.903$-$0.546) and Clump~58 (G333.410$-$0.328).} These clumps contain a compact 160\,\um\ object towards the position of the peak SIMBA position. There is no 4.5 or 8\,\um\ emission nearby. It is likely to be the earliest stage of star formation.

From the environment around select clumps, there are a variety of star formation stages within the G333 GMC. This ranges from isolated, compact 160\,\um\ objects, to areas with complex infrared emission and multiple masers. There also appears to be possible triggered star formation [e.g. Clump~3~(G333.136$-$0.431)]. In future papers, we will be combining the molecular and dust information for all 63 clumps identified in this paper, to characterise the clumps in relation to the evolution of star formation within the G333~GMC.


\section{Summary and Conclusions}
\label{sect:conclusion}
In this paper we have presented data from the two lowest inversion transitions of \ammonia\ in the G333 giant molecular cloud with the 70-m Tidbinbilla radio telescope. Our main conclusions are as follows:

(1) There were 63 SIMBA 1.2-mm dust clumps identified with {\sc clumpfind} and pointed \ammonia\ observations were conducted towards the peak of each clump. The peak of each clump was offset from the clump centroid by between 1.6 and 39.5 arcsec.

(2) A variety of environments were identified within the G333~GMC and the sample was separated into 30 IR-bright and 33 IR-faint clumps. One clump from each category was identified via \ammonia~\two\ emission, to not be associated with the \vlsr\ of the G333~GMC, and one clump was identified to have an inaccurate \Tkin; resulting in 29 IR-bright and 31 IR-faint clumps. The \ammonia~\one\ and \two\ were fitted for 20 IR-bright and 26 IR-faint clumps, with mean \Tkin\ values of 24.2\,K and 19.3\,K, respectively.

(3) We found that the clumps appeared to cluster into three regions. The IR-bright clumps were found predominately towards the centre of each region, especially for Regions B and C. This may indicate sequentially triggered star formation in the vicinity around IR-faint clumps.

(4) Using \ammonia~\one\ and \two, the clump masses were found to be between 278 and 13036 \msun. The dust temperature is often assumed: 20\,K for isolated clumps and 40\,K for \HII~regions; however this assumption can significantly shift the masses of the clumps (e.g. for 20\,K the masses of our clumps were 251--23723\,\msun, and for 40\,K the masses of our clumps were 105--9951\,\msun) and can affect whether the clump is determined to be bound and will collapse onto itself, or if it is unbound and remains a starless core. Hence accurate measurements of kinetic temperatures are needed for studies of the gravitationally bound state of clumps.

(5) The largest clumps are also the most unstable. We find that the clumps with signs of star formation are predominately associated with the least stable clumps. This may be due to feedback from an internal heating source. We found that 70~per~cent of the clumps had masses larger than the virial mass (187--6848\,\msun), suggesting that they will form stars or are already undergoing star formation.

(6) A K-S test showed that the \Tkin\ for IR-bright and IR-faint clumps have a much smaller probability of being drawn from the same population. A tentative difference between the two populations in terms of FWHM could indicate a contribution due to feedback. No difference was seen for the other physical parameters. We conclude that IR-bright and IR-faint clumps can have the same masses but that the warmer regions are more evolved.

This is the fourth paper in a series of papers utilising molecular line observations of the G333 giant molecular cloud associated with RCW~106 (G333~GMC). In this paper we reported on the selection of dense clumps based on SIMBA dust emission and pointed \ammonia~\one\ and \two\ observations towards these clumps. Future papers in the series will analyse the individual clumps and combine the information presented in this paper with the molecular 3-mm tracers presented in \citet{Bains2006}, \citet{Wong2008} and \citet{Lo2009}, and dust from the \textit{Spitzer} Space Telescope and the \textit{Herschel} Space Observatory.


\section*{Acknowledgments}
We would like to thank the anonymous referee for their comments that have greatly improved this paper. Partial financial support for this work was provided by the Australian Research Council. JPM is supported by Spanish grant AYA 2011-26202. JSU was supported by a CSIRO OCE postdoctoral grant. NL is supported by a CONICYT/FONDECYT postdoctorado, under project no. 3130540. NL acknowledges partial support from the ALMA-CONICYT Fund for the Development of Chilean Astronomy Project 31090013, Center of Excellence in Astrophysics and Associated Technologies (PFB 06) and Centro de Astrof\'{i}sica FONDAP\,15010003. The Tidbinbilla 70-m Radio Telescope is part of the NASA Deep Space Network and is operated by CSIRO Astronomy and Space Science. \textit{Herschel} is an ESA space observatory with science instruments provided by European-led Principal Investigator consortia and with important participation from NASA. This research has made use of NASA's Astrophysics Data System. This research has made use of the SIMBAD database, operated at CDS, Strasbourg, France. This research has made use of the NASA/ IPAC Infrared Science Archive, which is operated by the Jet Propulsion Laboratory, California Institute of Technology, under contract with the National Aeronautics and Space Administration. This work is based in part on archival data obtained with the Spitzer Space Telescope, which is operated by the Jet Propulsion Laboratory, California Institute of Technology under a contract with NASA. Support for this work was provided by an award issued by JPL/Caltech. This paper made use of information from the Red MSX Source survey database at www.ast.leeds.ac.uk/RMS which was constructed with support from the Science and Technology Facilities Council of the UK.

\vspace{-0.2cm}
\section*{SUPPORTING INFORMATION}
Additional Supporting Information may be found in the online version of this article.\\\\
{\bf Figure \ref{app:3colour}.} Infrared composite images centred on the peak of the 63 SIMBA dust clumps.\\
{\bf Figure \ref{app:spectra}.} \ammonia~\one~ and \two~spectra towards the peak of the 63 SIMBA dust clumps.\\
\clearpage


\renewcommand\thefigure{A}
\setcounter{figure}{0}

\begin{figure*}
\centering
\includegraphics[trim=100 20 190 40,clip,width=0.32\textwidth]{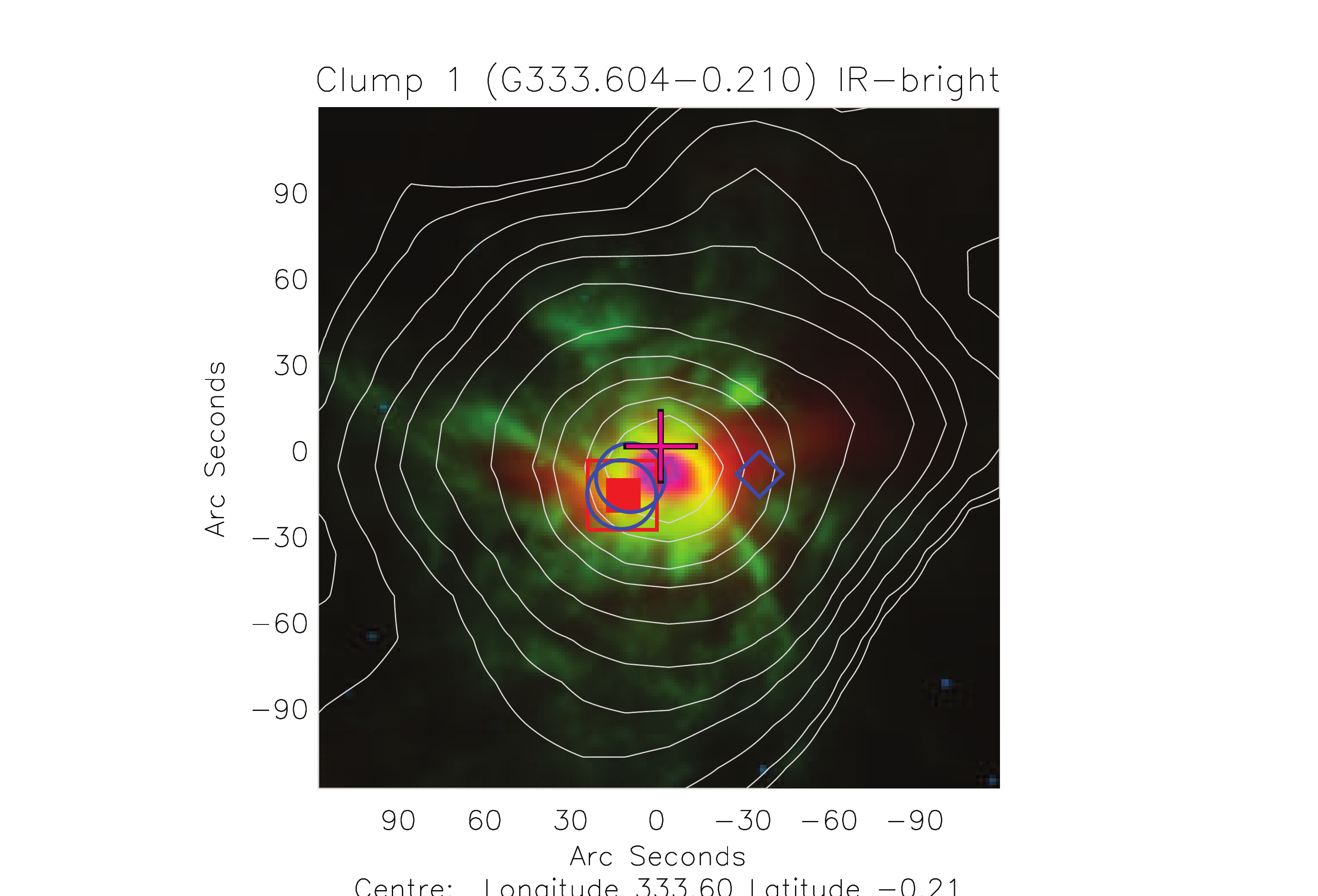}
\includegraphics[trim=100 20 190 40,clip,width=0.32\textwidth]{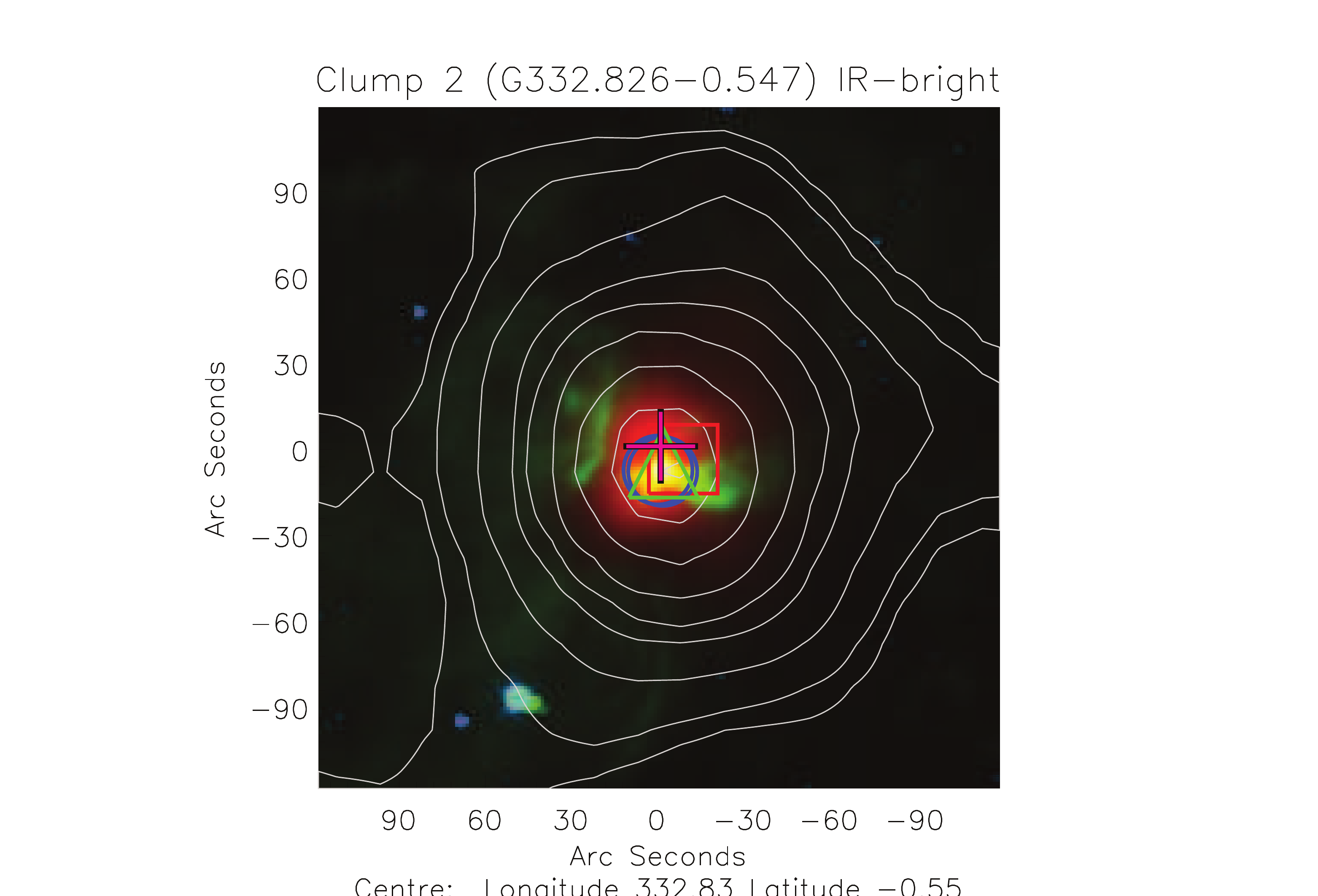}
\includegraphics[trim=100 20 190 40,clip,width=0.32\textwidth]{FIGURES/FigA/clump3_3colour}\\
\vspace*{0.5cm}
\includegraphics[trim=100 20 210 40,clip,width=0.32\textwidth]{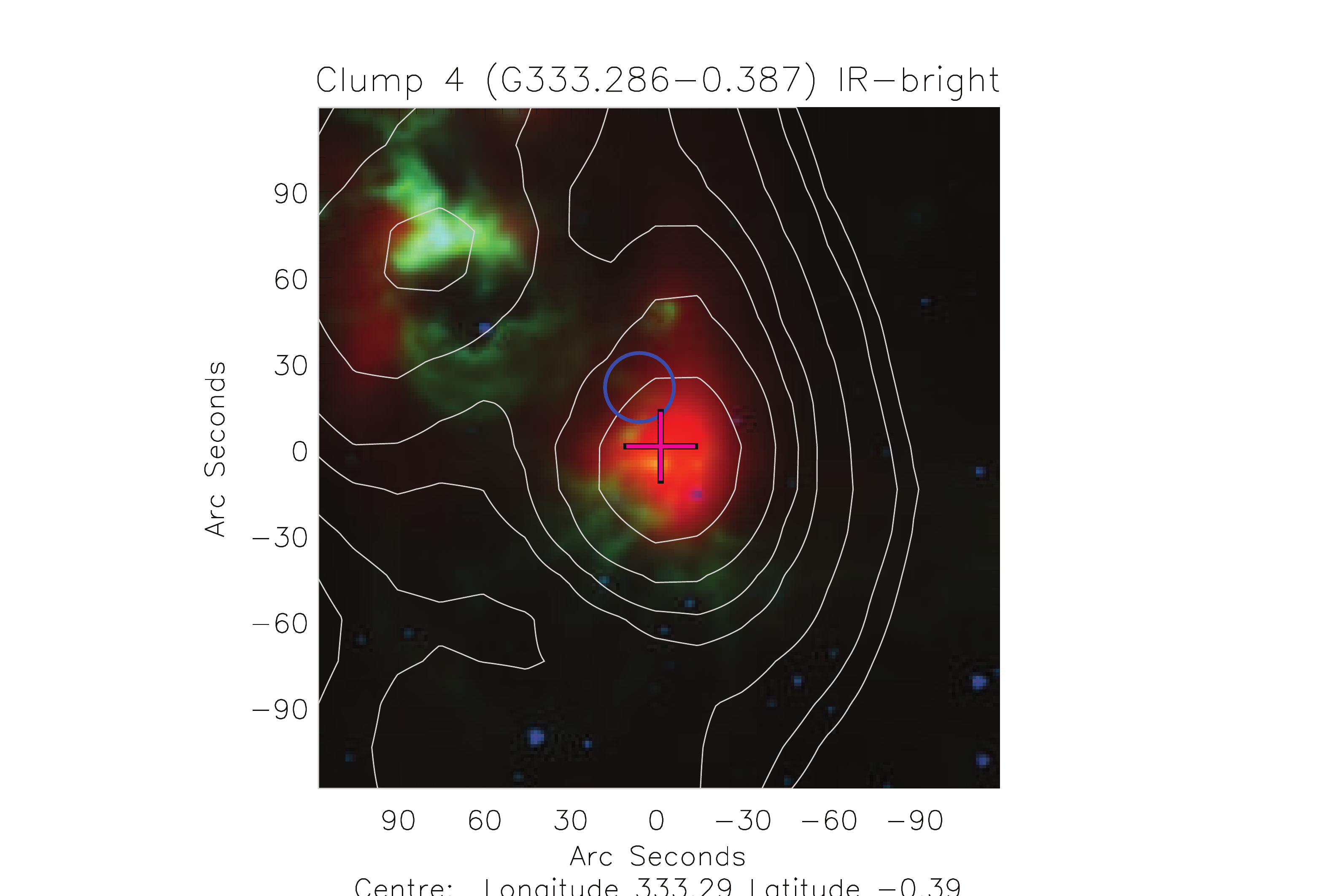}
\includegraphics[trim=100 20 210 40,clip,width=0.32\textwidth]{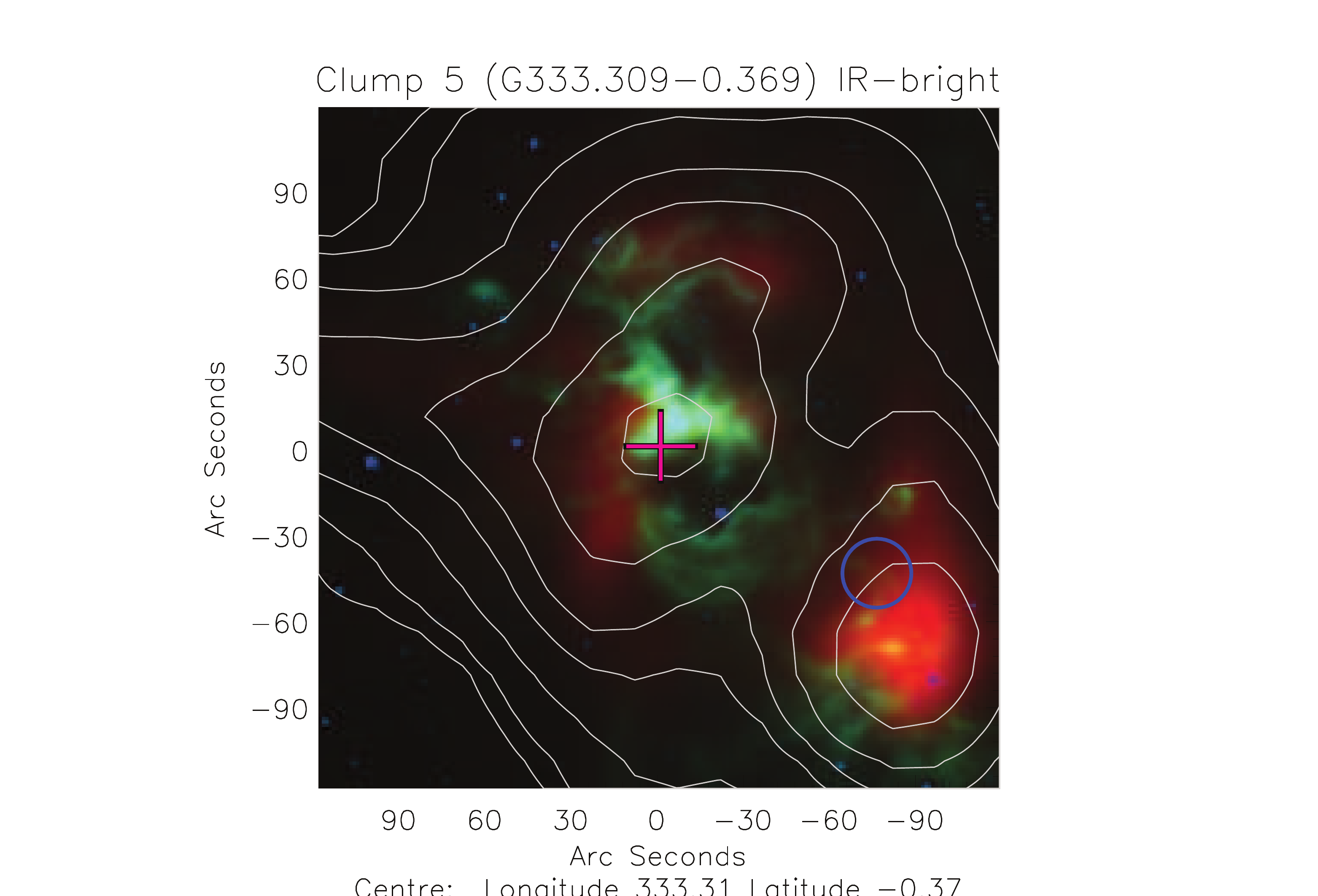}
\includegraphics[trim=100 20 210 40,clip,width=0.32\textwidth]{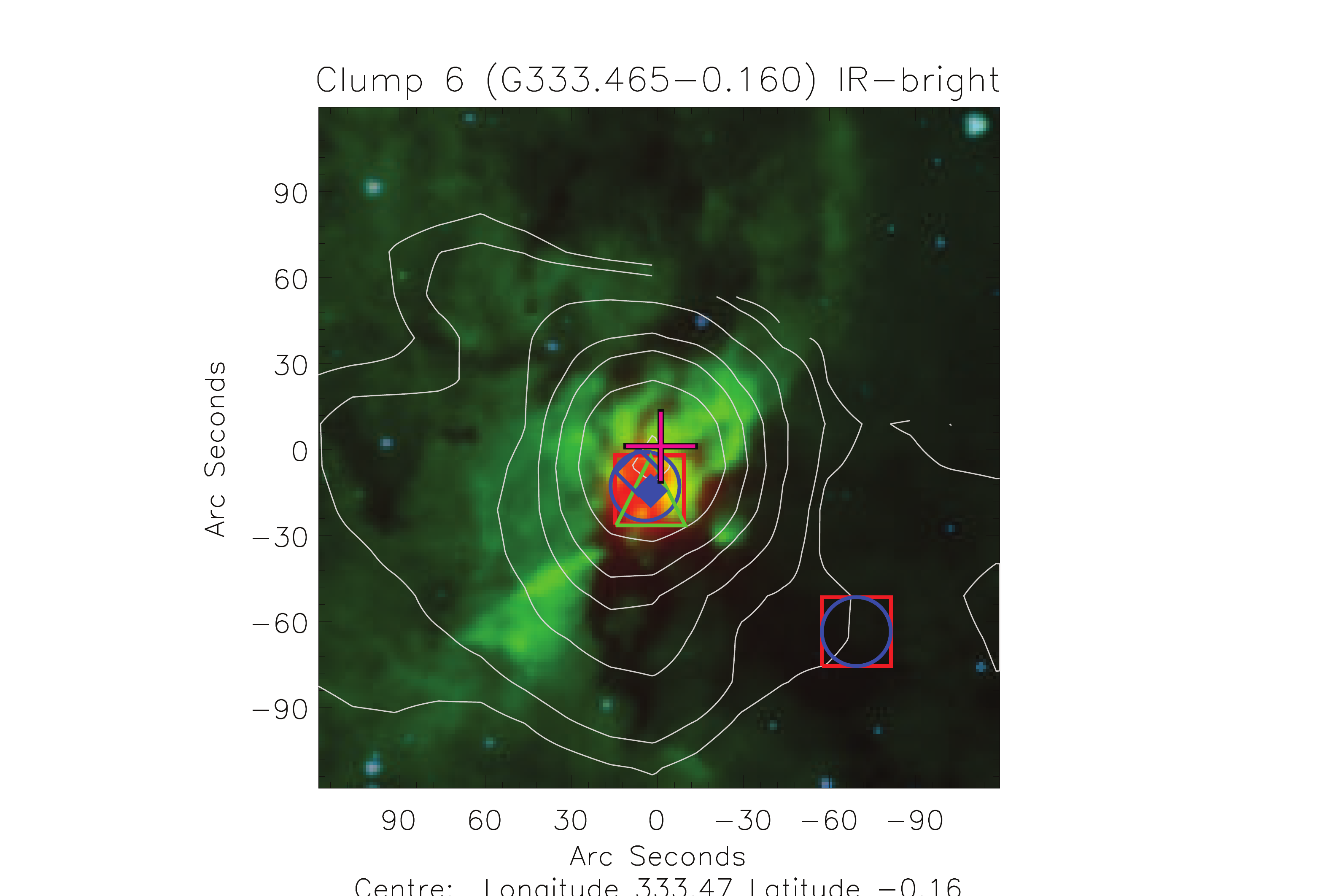}\\
\vspace*{0.5cm}
\includegraphics[trim=100 20 190 40,clip,width=0.32\textwidth]{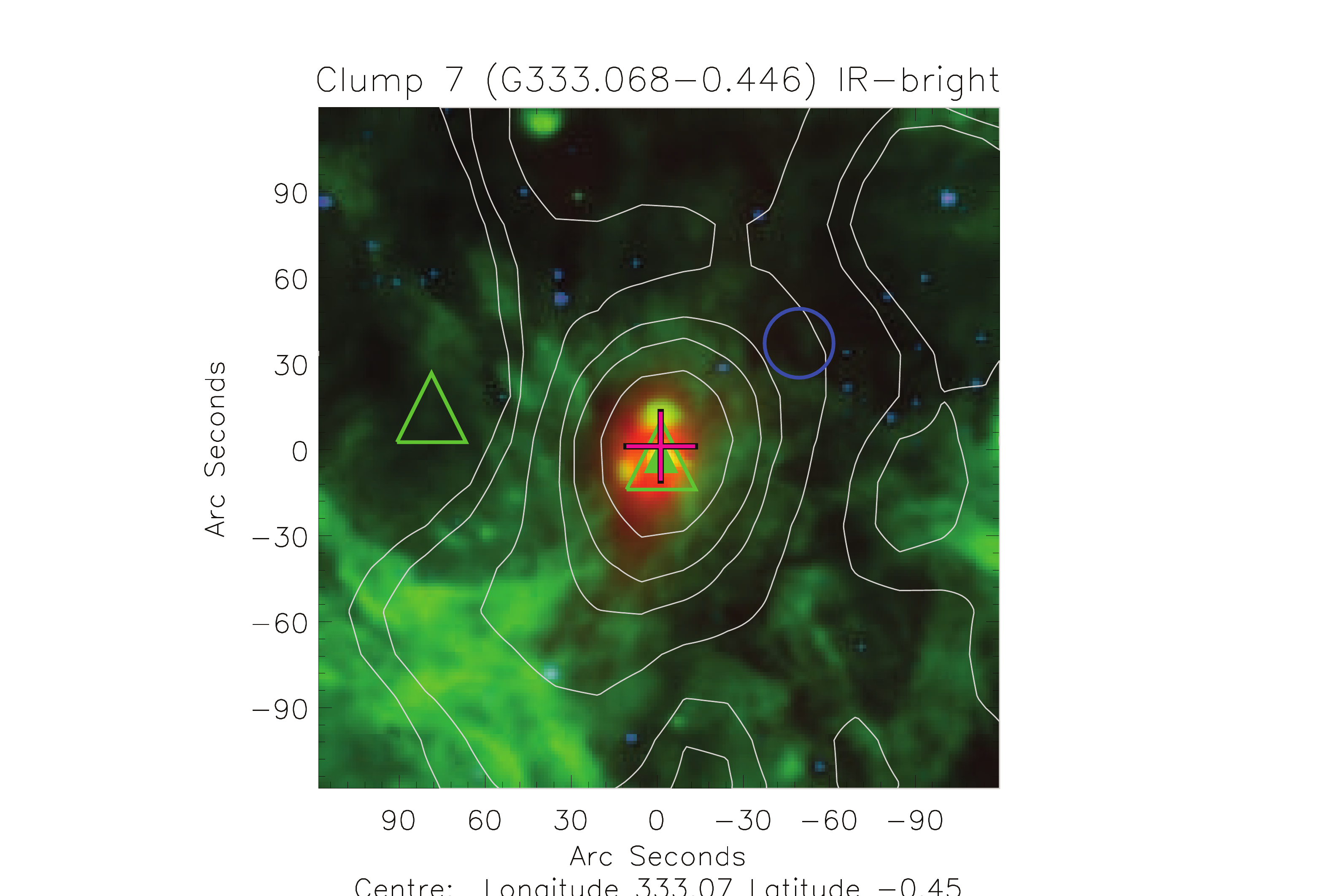}
\includegraphics[trim=100 20 190 40,clip,width=0.32\textwidth]{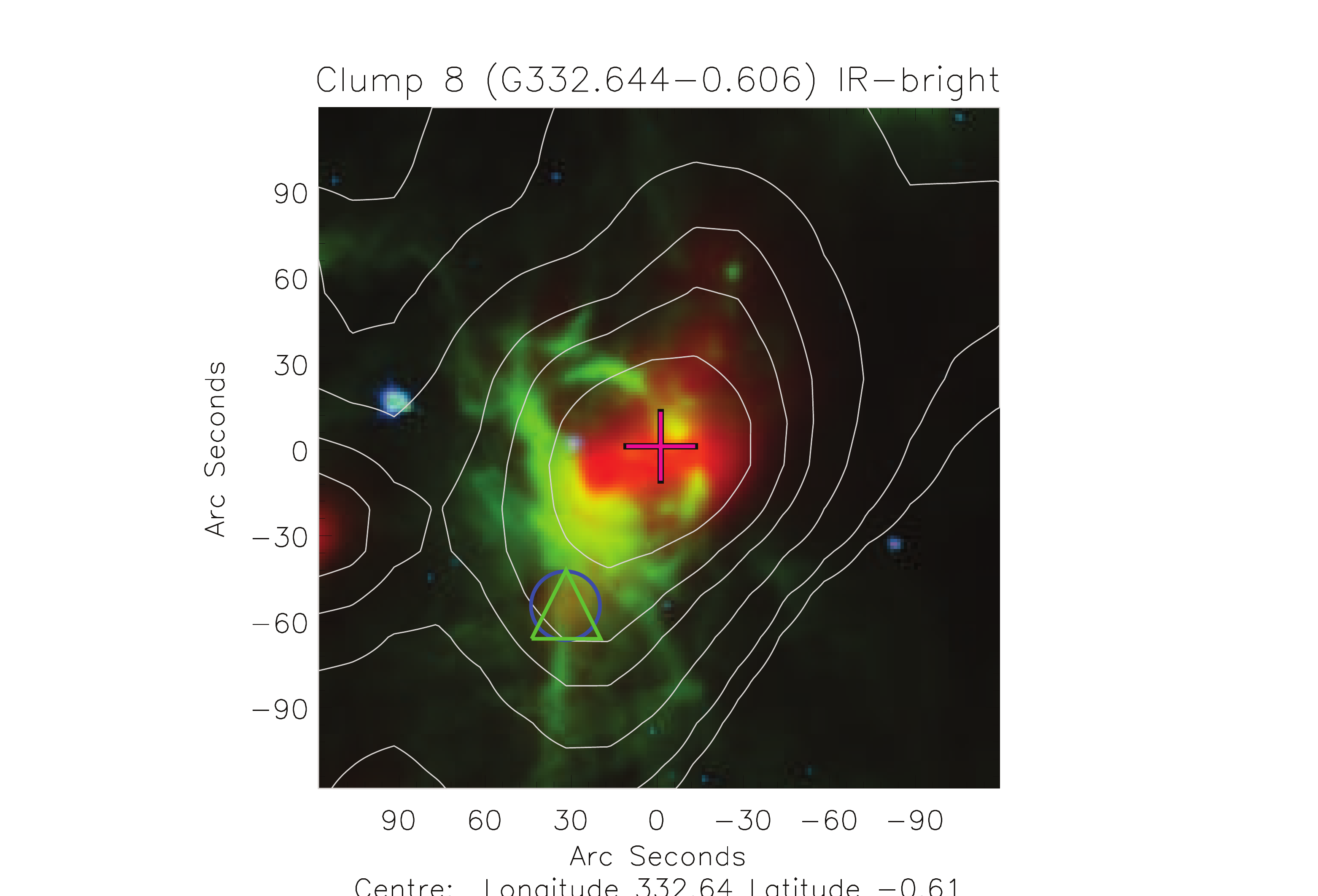}
\includegraphics[trim=100 20 190 40,clip,width=0.32\textwidth]{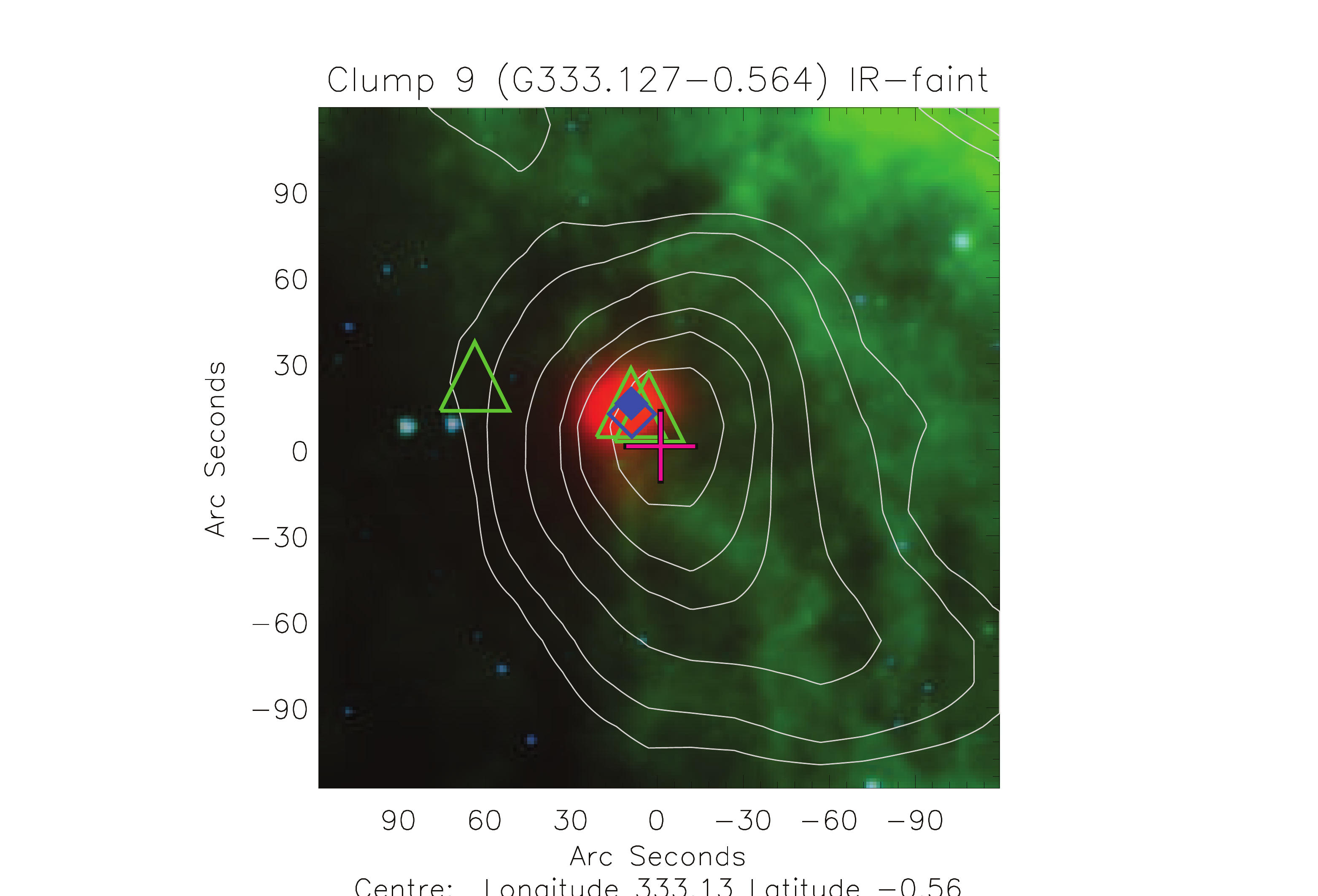}\\
\vspace*{0.5cm}
\includegraphics[trim=100 20 210 40,clip,width=0.32\textwidth]{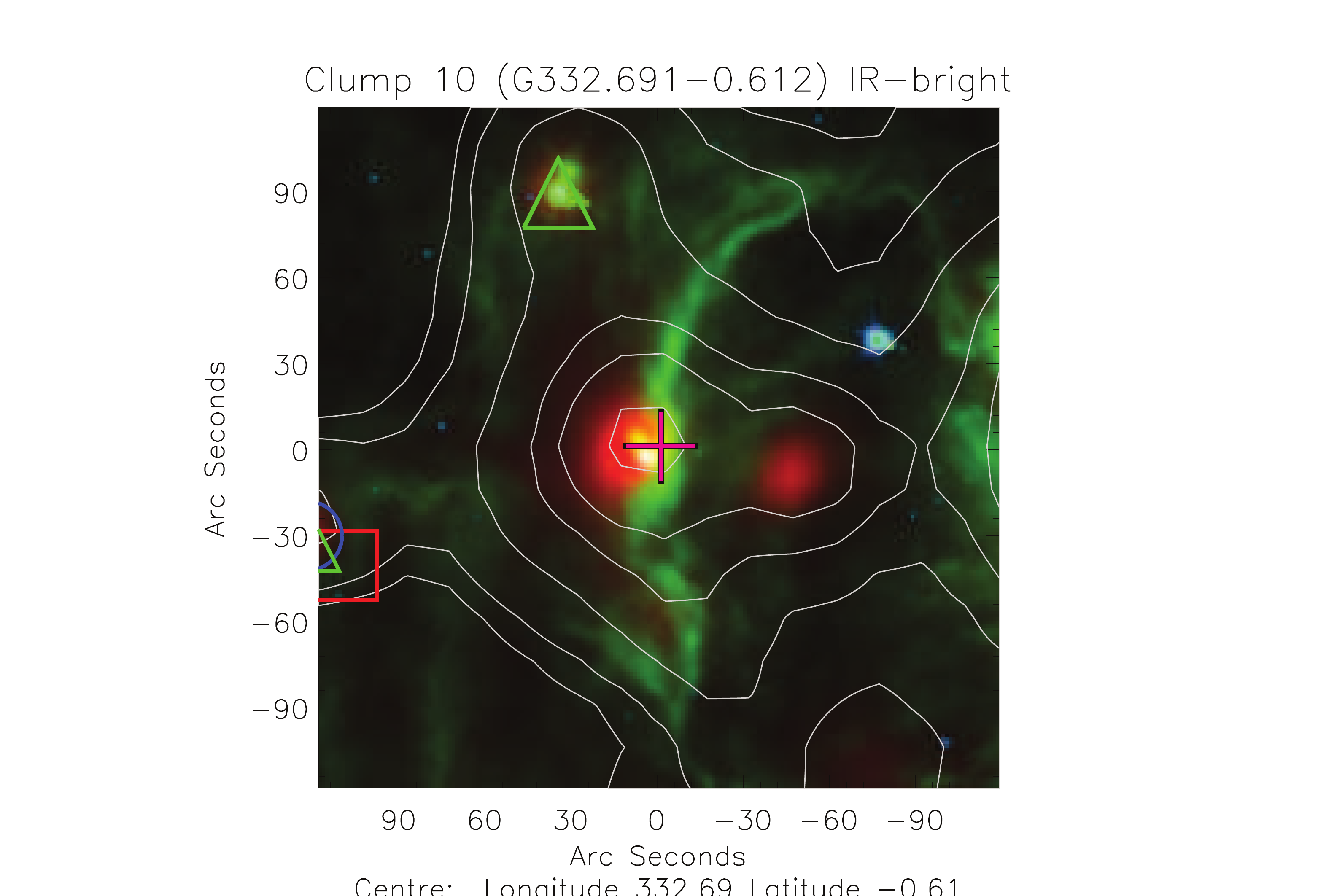}
\includegraphics[trim=100 20 210 40,clip,width=0.32\textwidth]{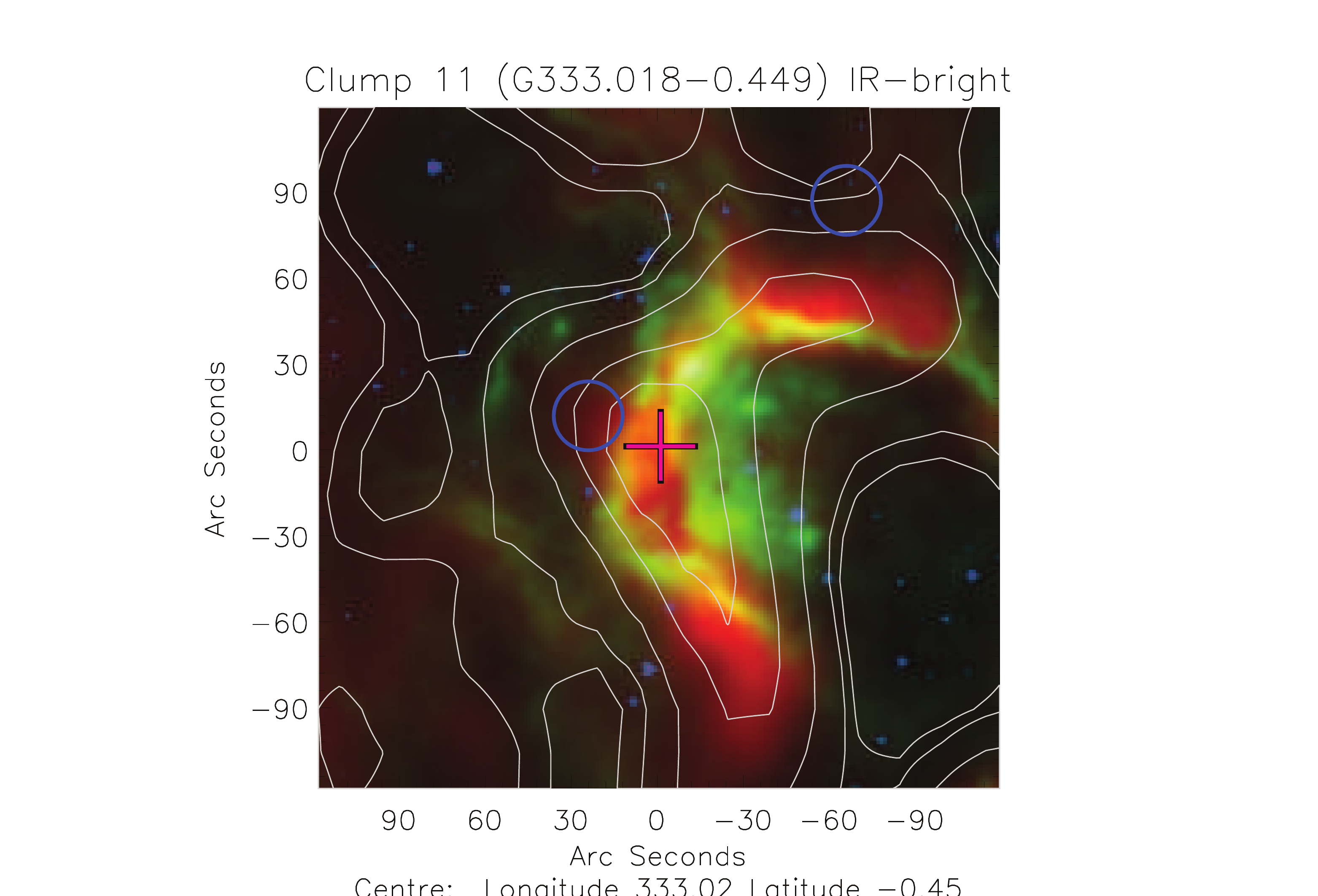}
\includegraphics[trim=100 20 210 40,clip,width=0.32\textwidth]{FIGURES/FigA/clump12_3colour}\\
\vspace*{-0.05cm}
\caption{Infrared composite images centered on the peak dust emission of each clump identified with {\sc clumpfind}. \textit{Spitzer} 4.5 (shocked gas) and 8.0 \um\ (PAH$^+$ emission), and \textit{Herschel} 160 \um\ (cool dust) are shown in blue, green and red, respectively. The SIMBA dust contours have been set to 0.1, 0.2, 0.5, 1, 1.5, 2.5, 5, 10, 15, 25 and 35 mJy beam$^{-1}$. \water~ masers are identified by blue circles. Class~I~\methanol~ masers are identified by unfilled (44.07 GHz) and filled (95.1 GHz) blue diamonds. Class~II~\methanol~ masers are identified by unfilled (6.7 GHz) and filled (12.2 GHz) green triangles. OH masers are identified by unfilled (1665/1667/1720 MHz) and filled (6.035 GHz) red squares.}
\label{app:3colour}
\end{figure*}

\begin{figure*}
\centering
\includegraphics[trim=100 20 190 40,clip,width=0.32\textwidth]{FIGURES/FigA/clump13_3colour}
\includegraphics[trim=100 20 190 40,clip,width=0.32\textwidth]{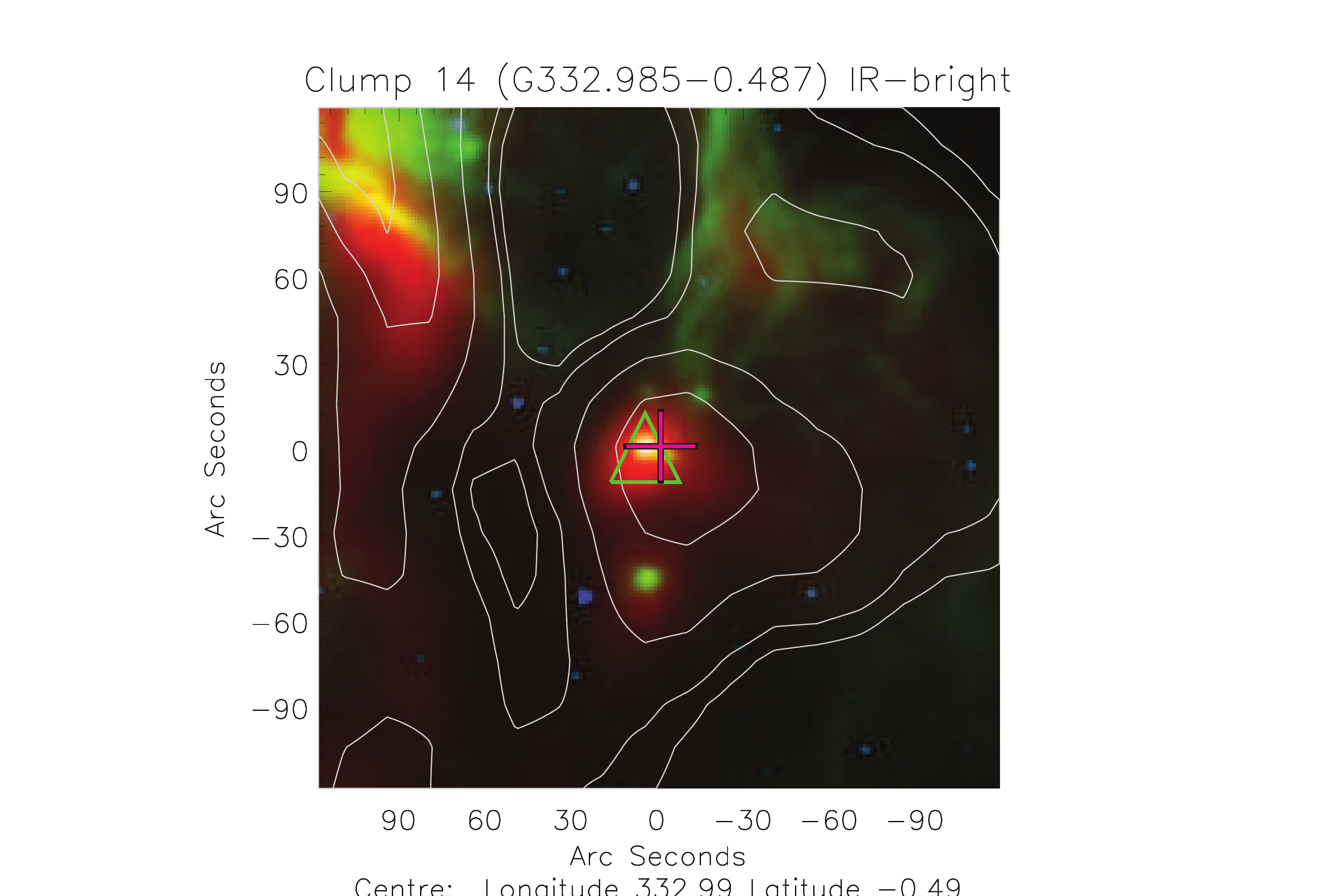}
\includegraphics[trim=100 20 190 40,clip,width=0.32\textwidth]{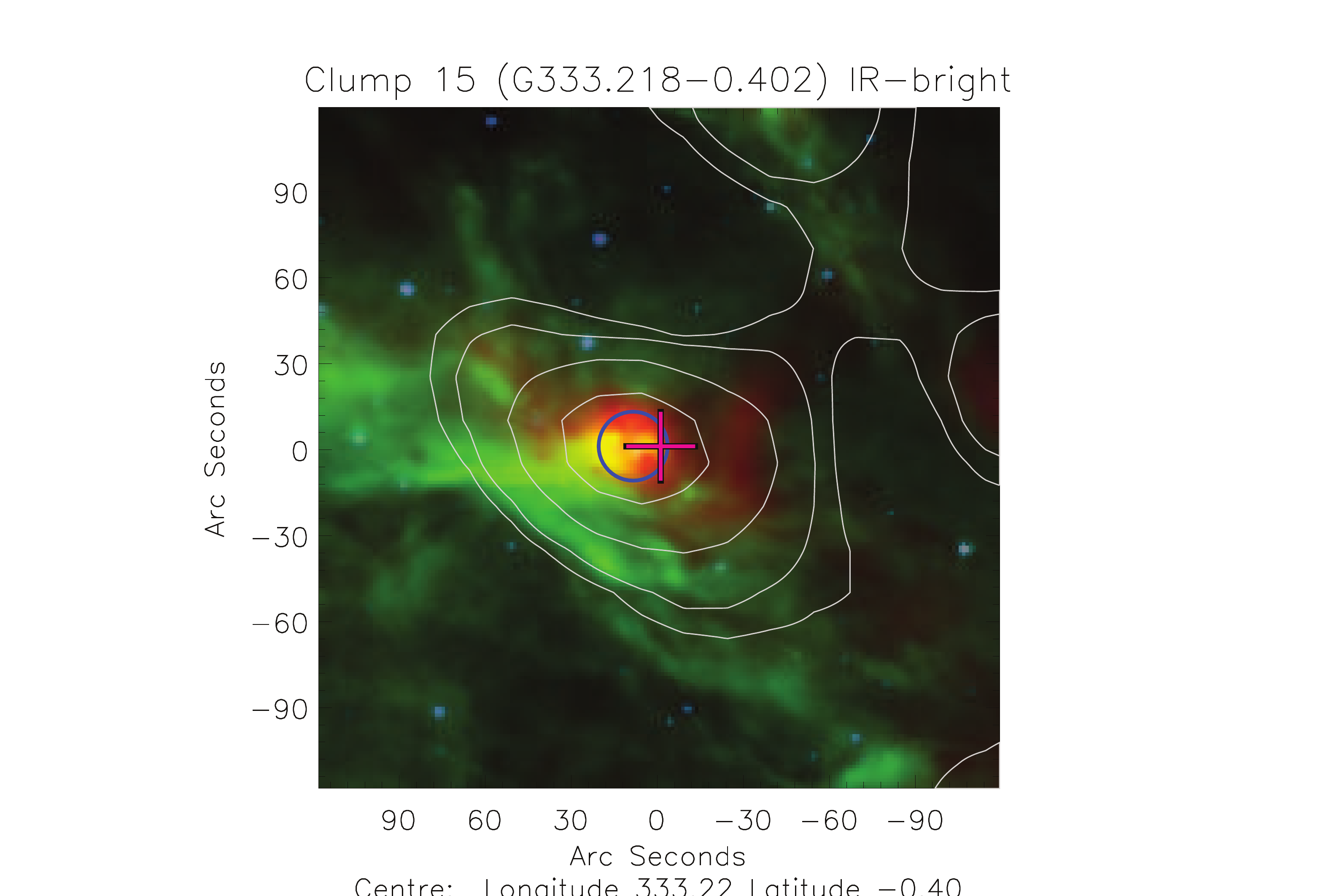}\\
\vspace*{0.5cm}
\includegraphics[trim=100 20 210 40,clip,width=0.32\textwidth]{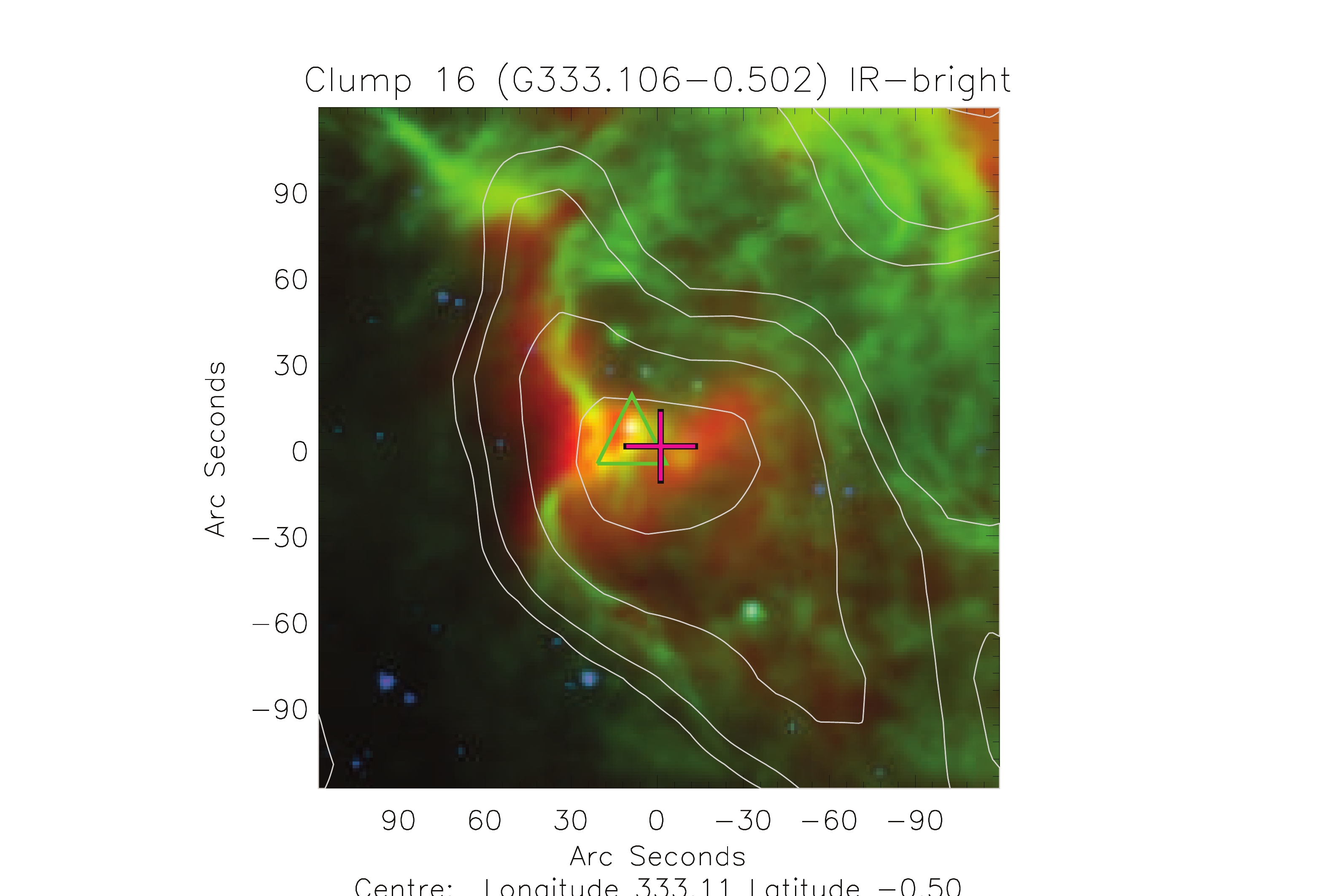}
\includegraphics[trim=100 20 210 40,clip,width=0.32\textwidth]{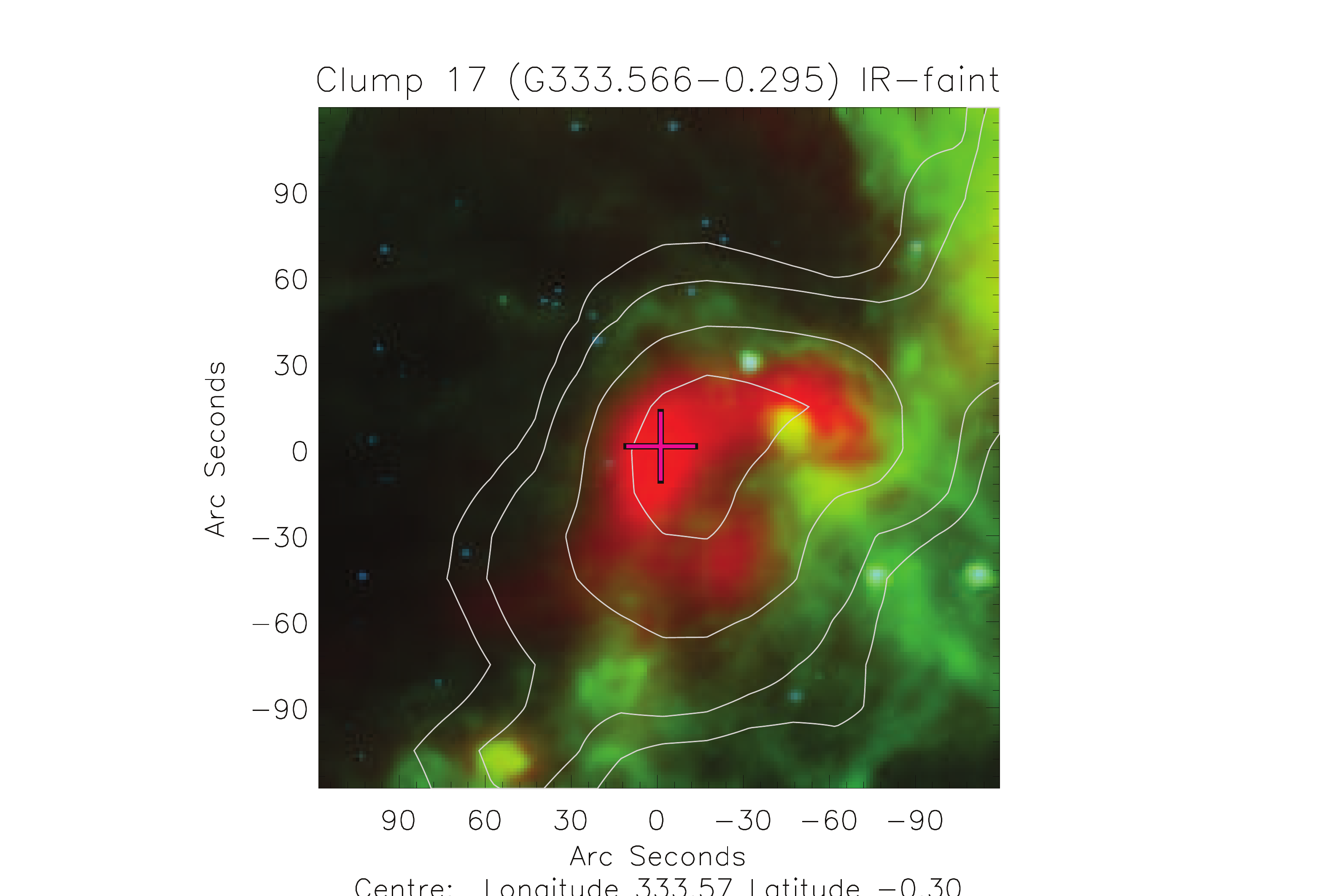}
\includegraphics[trim=100 20 210 40,clip,width=0.32\textwidth]{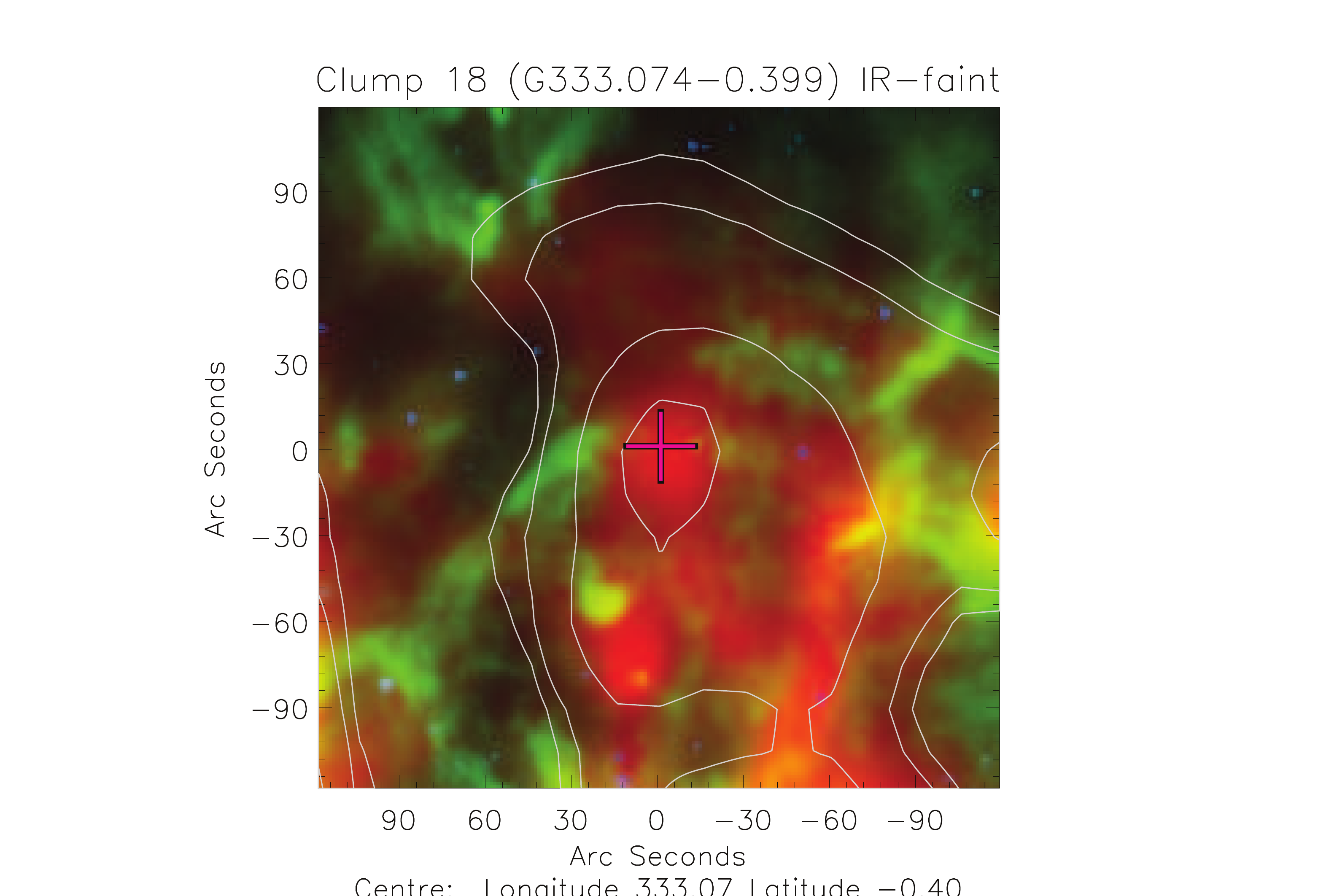}\\
\vspace*{0.5cm}
\includegraphics[trim=100 20 190 40,clip,width=0.32\textwidth]{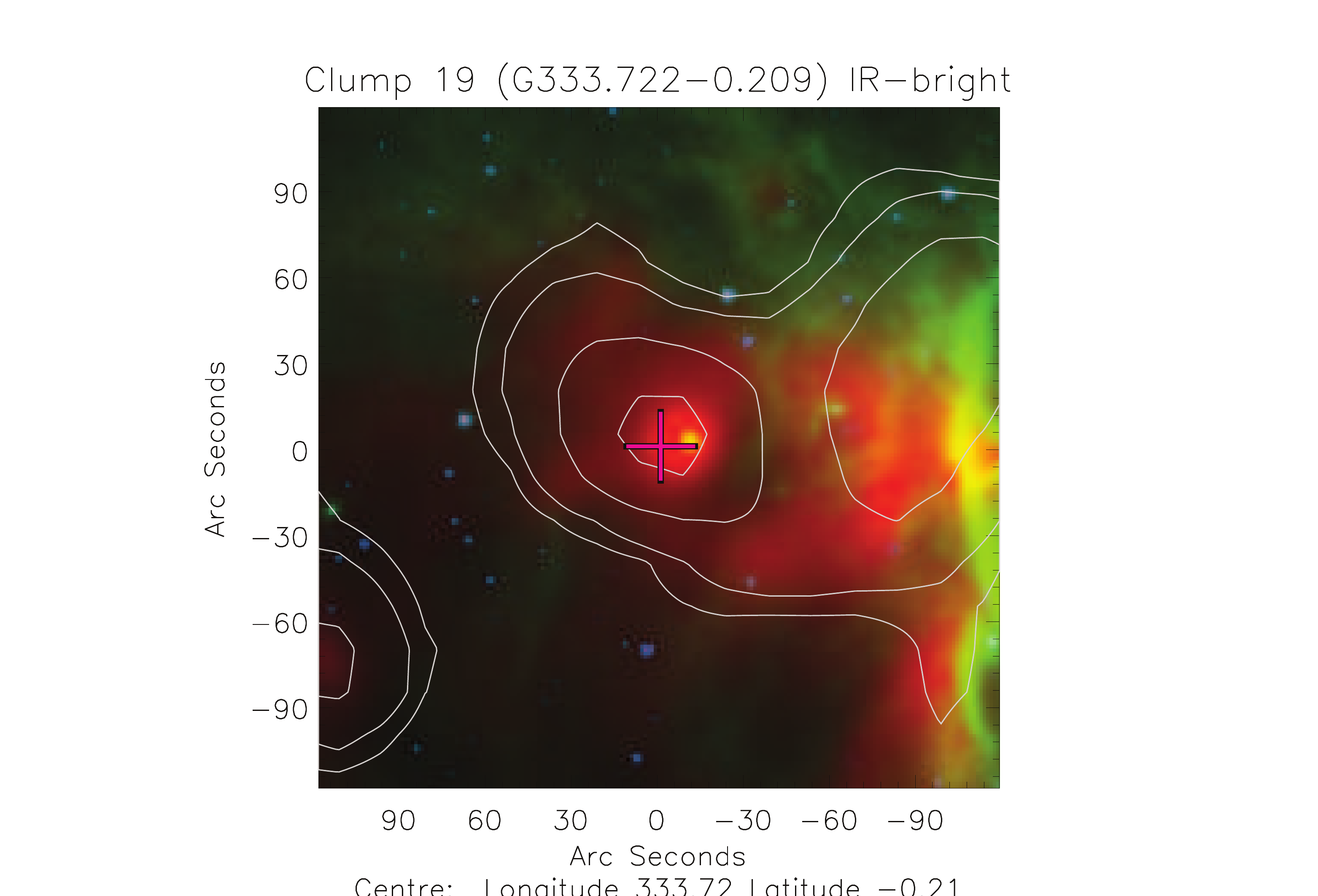}
\includegraphics[trim=100 20 190 40,clip,width=0.32\textwidth]{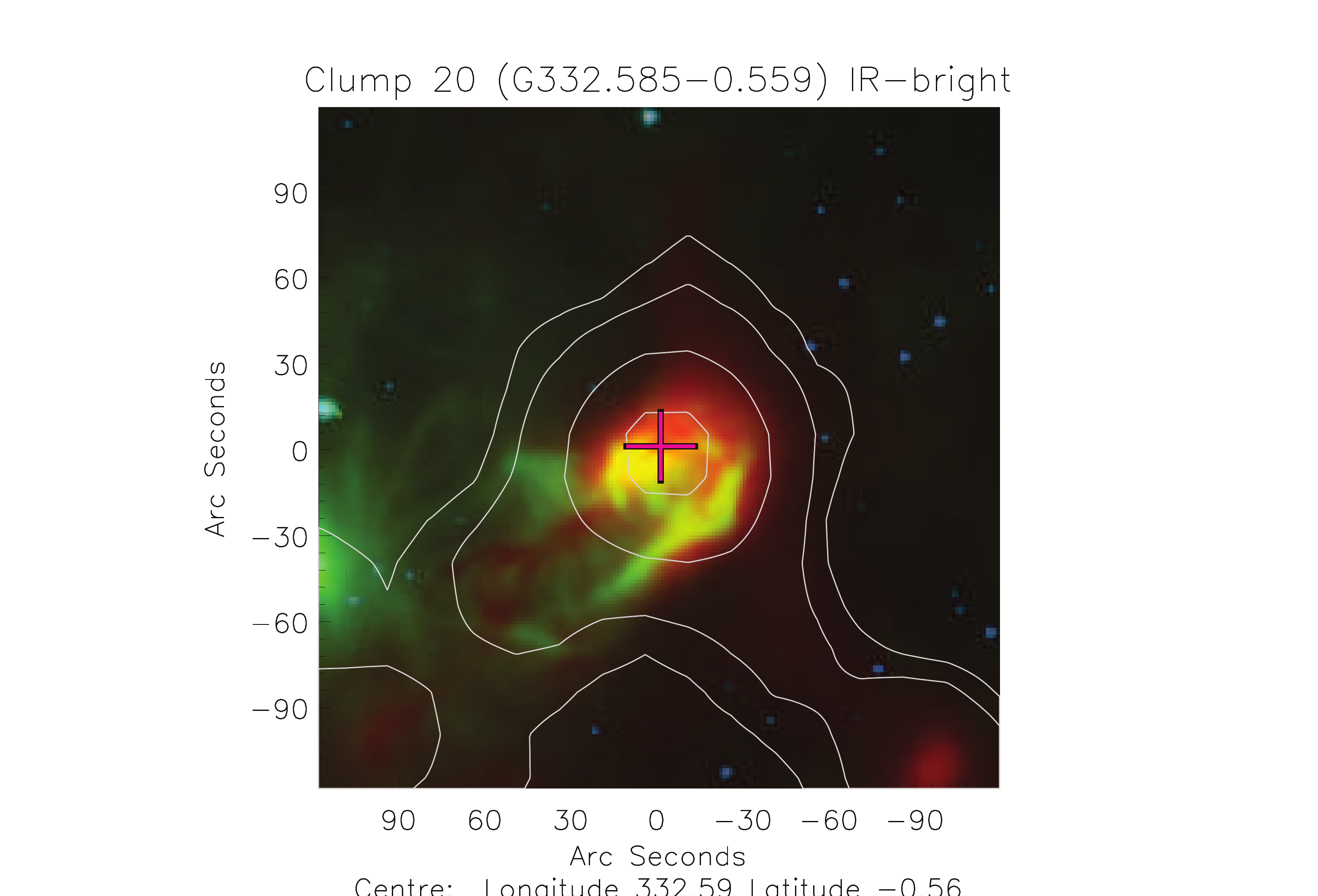}
\includegraphics[trim=100 20 190 40,clip,width=0.32\textwidth]{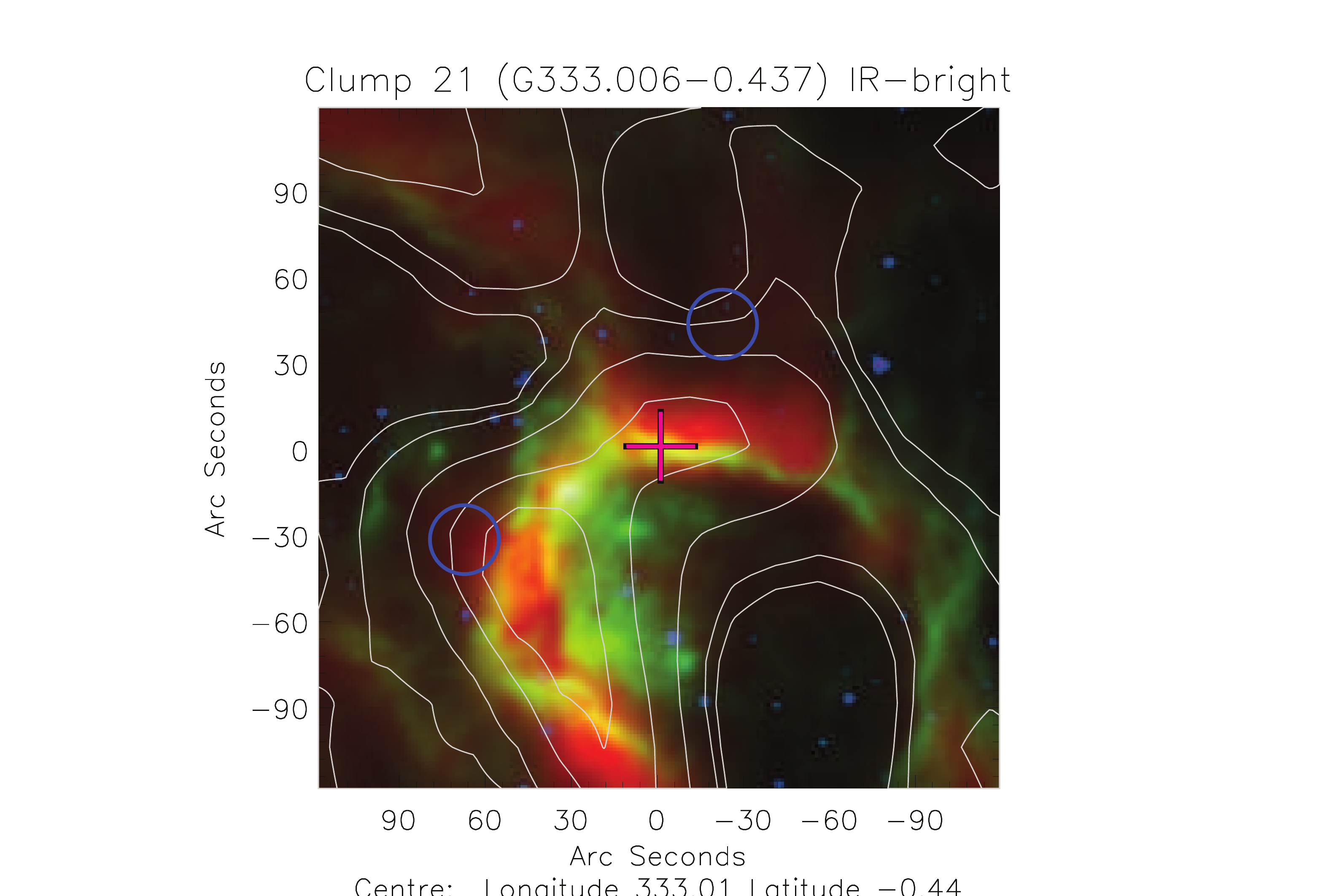}\\
\vspace*{0.5cm}
\includegraphics[trim=100 20 210 40,clip,width=0.32\textwidth]{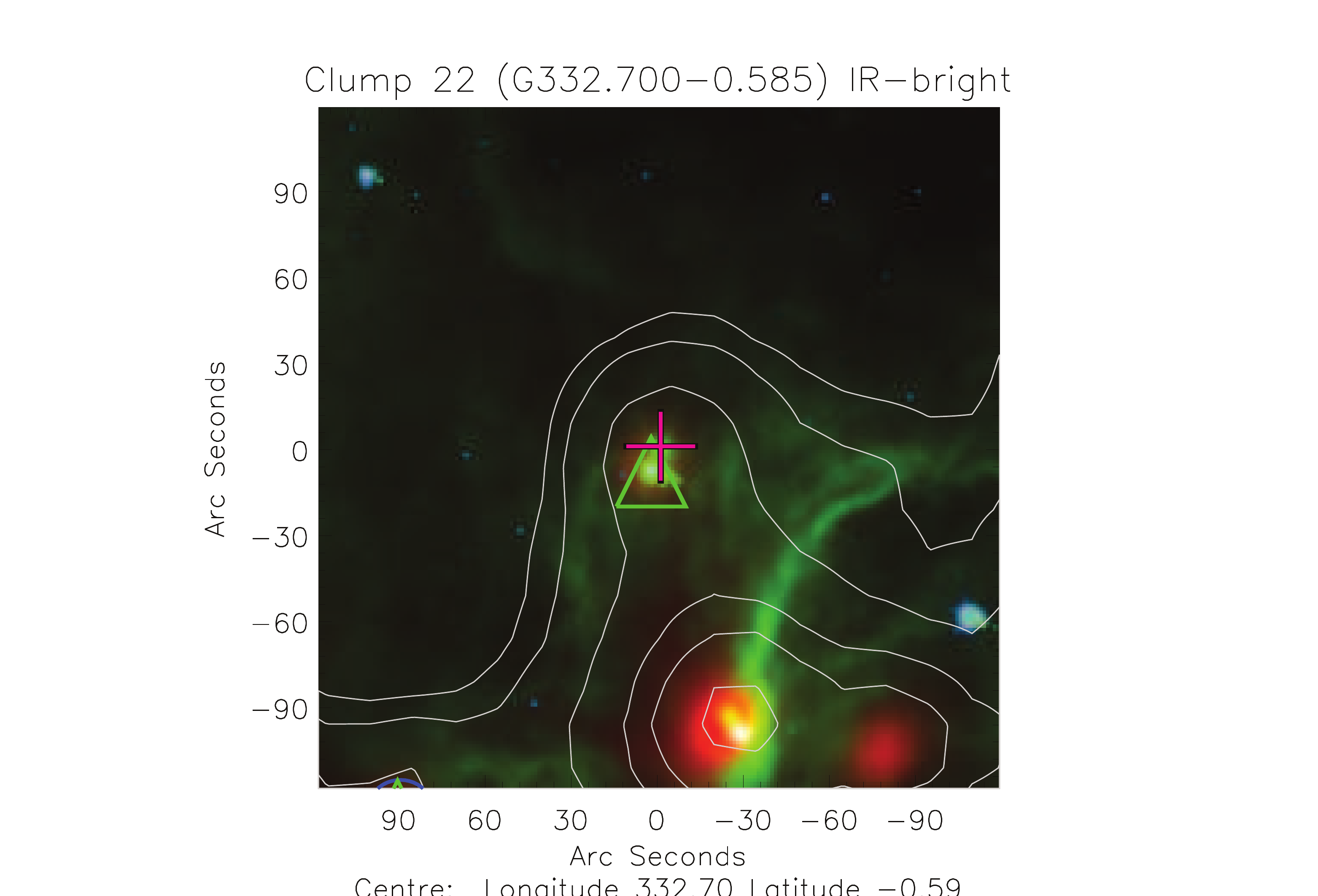}
\includegraphics[trim=100 20 210 40,clip,width=0.32\textwidth]{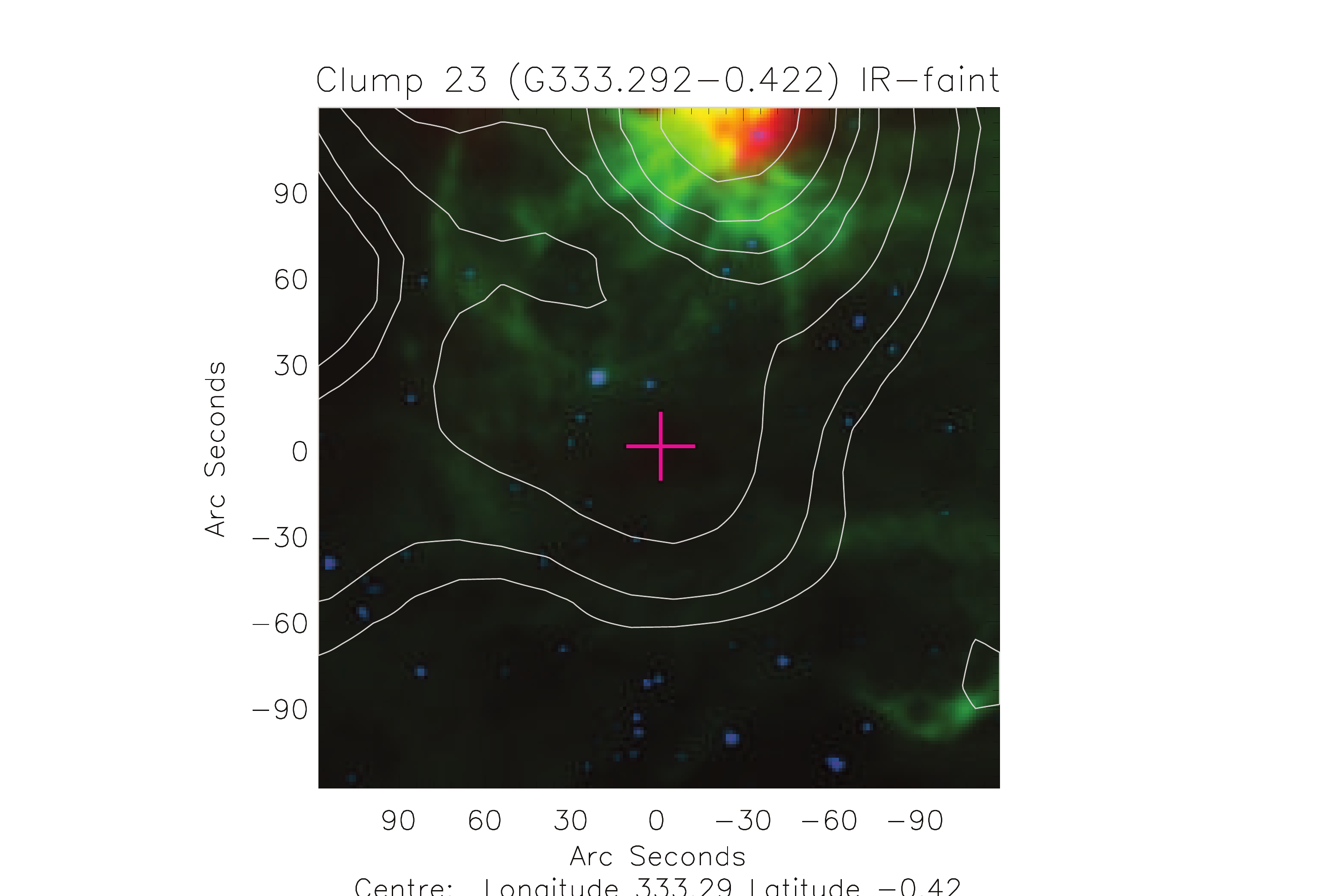}
\includegraphics[trim=100 20 210 40,clip,width=0.32\textwidth]{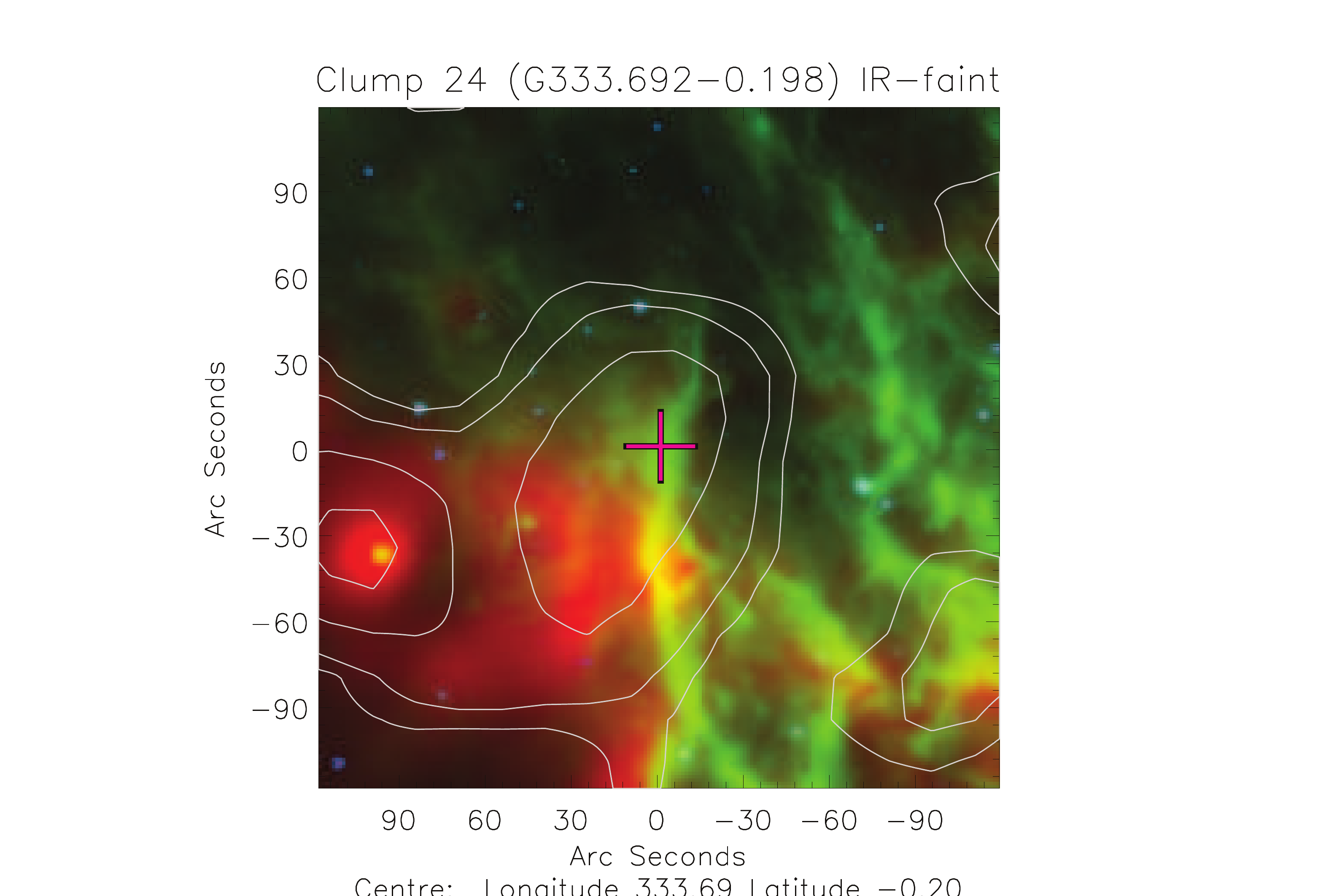}\\
\vspace*{0.3cm}\contcaption{}
\end{figure*}

\begin{figure*}
\centering
\includegraphics[trim=100 20 190 40,clip,width=0.32\textwidth]{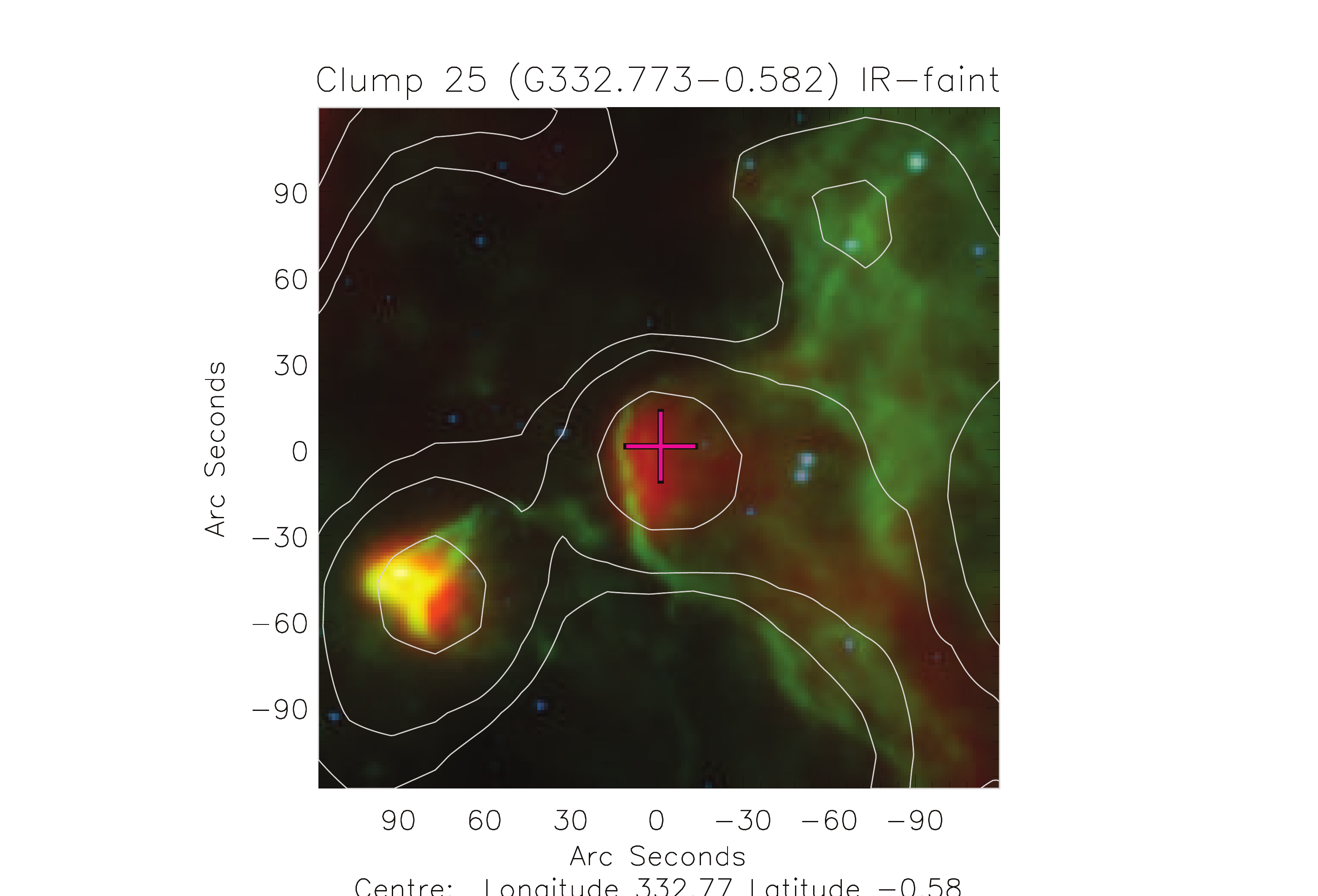}
\includegraphics[trim=100 20 190 40,clip,width=0.32\textwidth]{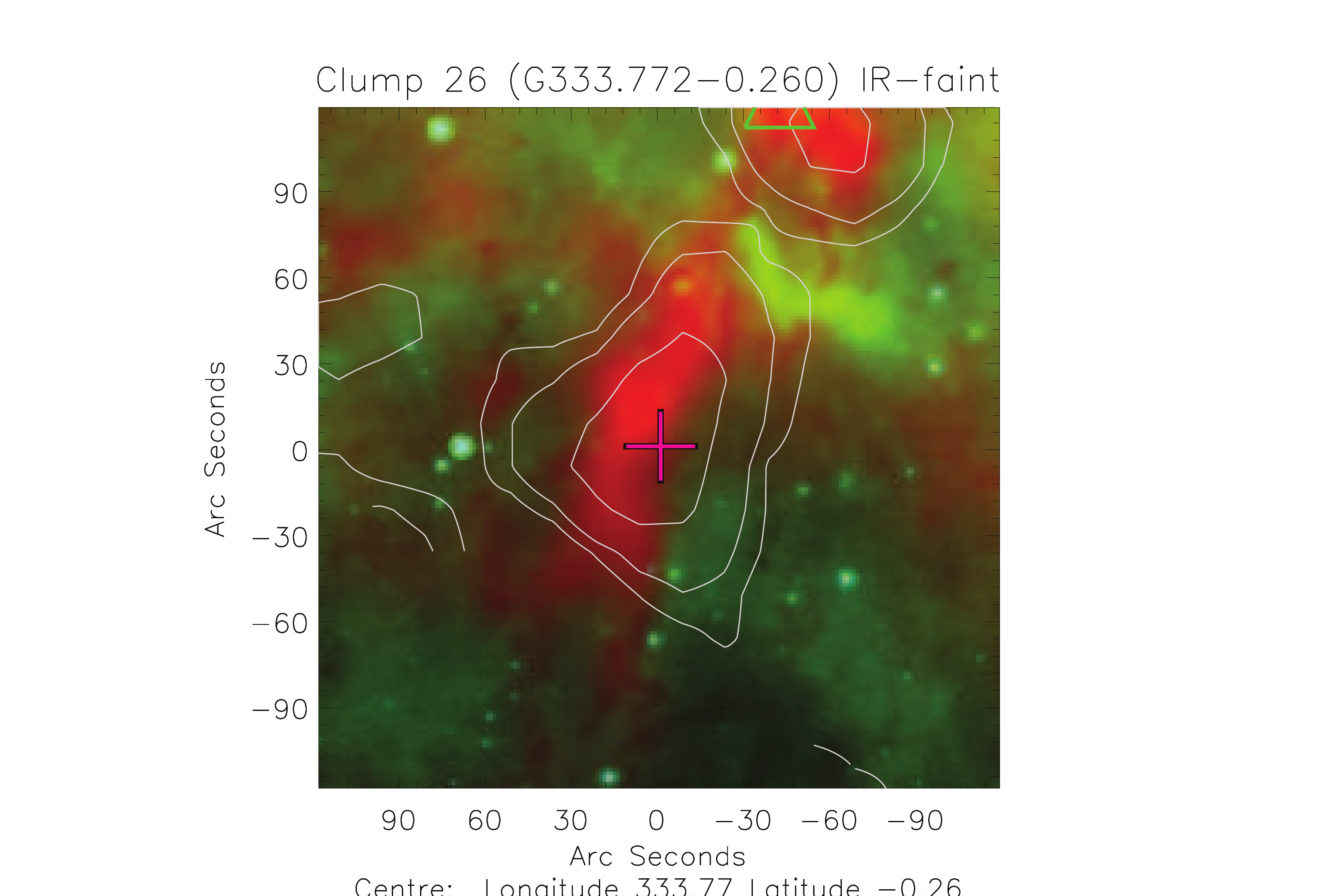}
\includegraphics[trim=100 20 190 40,clip,width=0.32\textwidth]{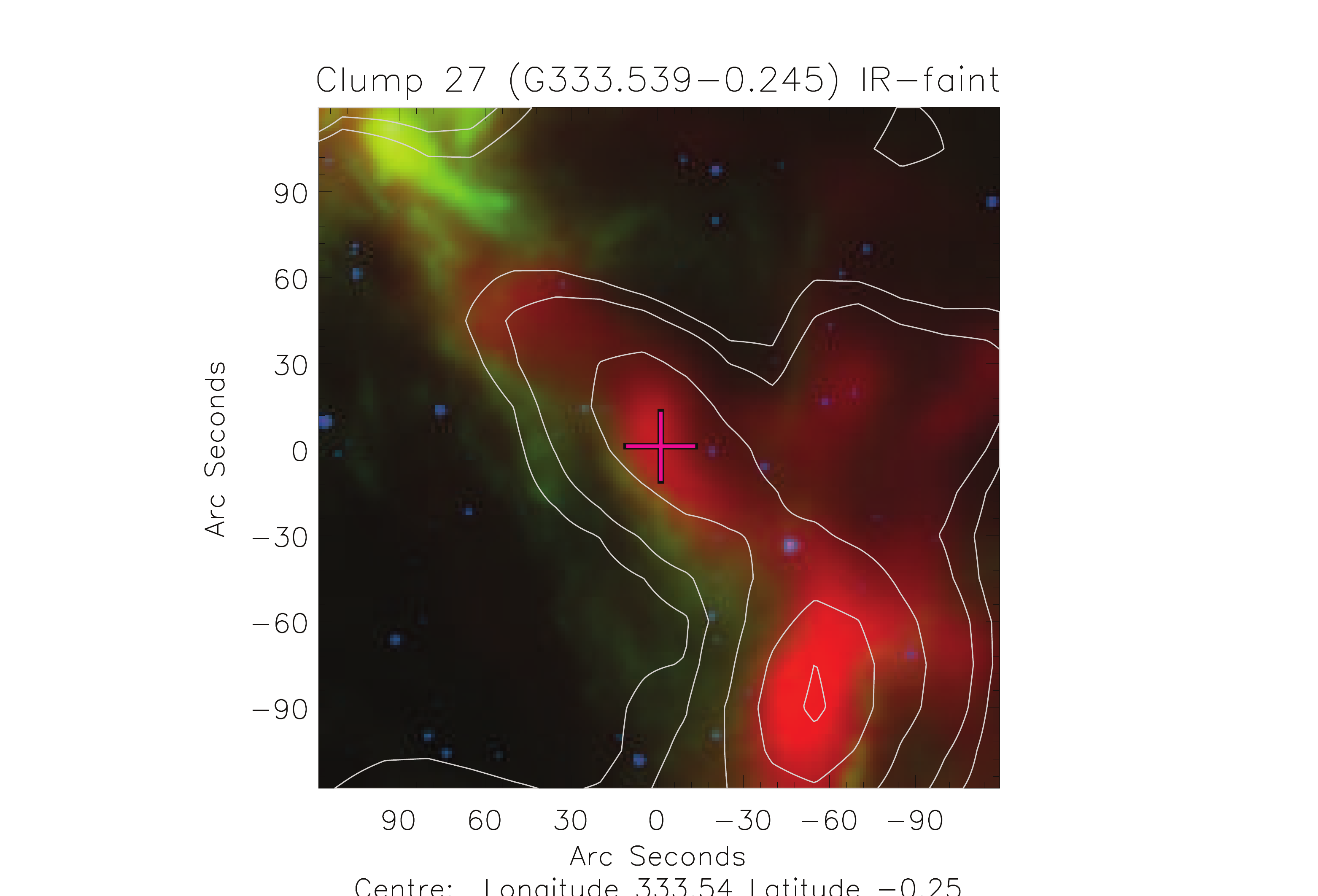}\\
\vspace*{0.5cm}
\includegraphics[trim=100 20 210 40,clip,width=0.32\textwidth]{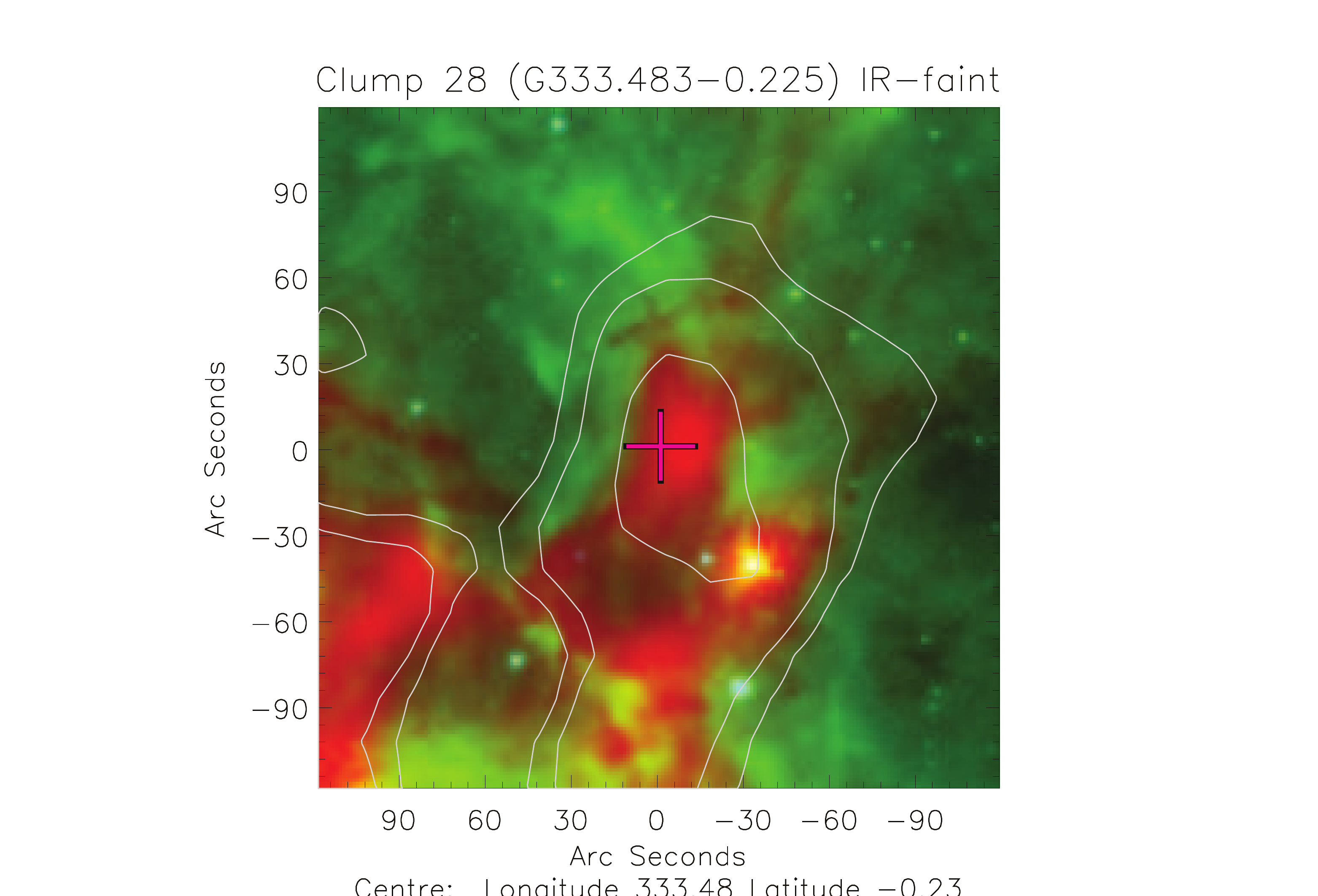}
\includegraphics[trim=100 20 210 40,clip,width=0.32\textwidth]{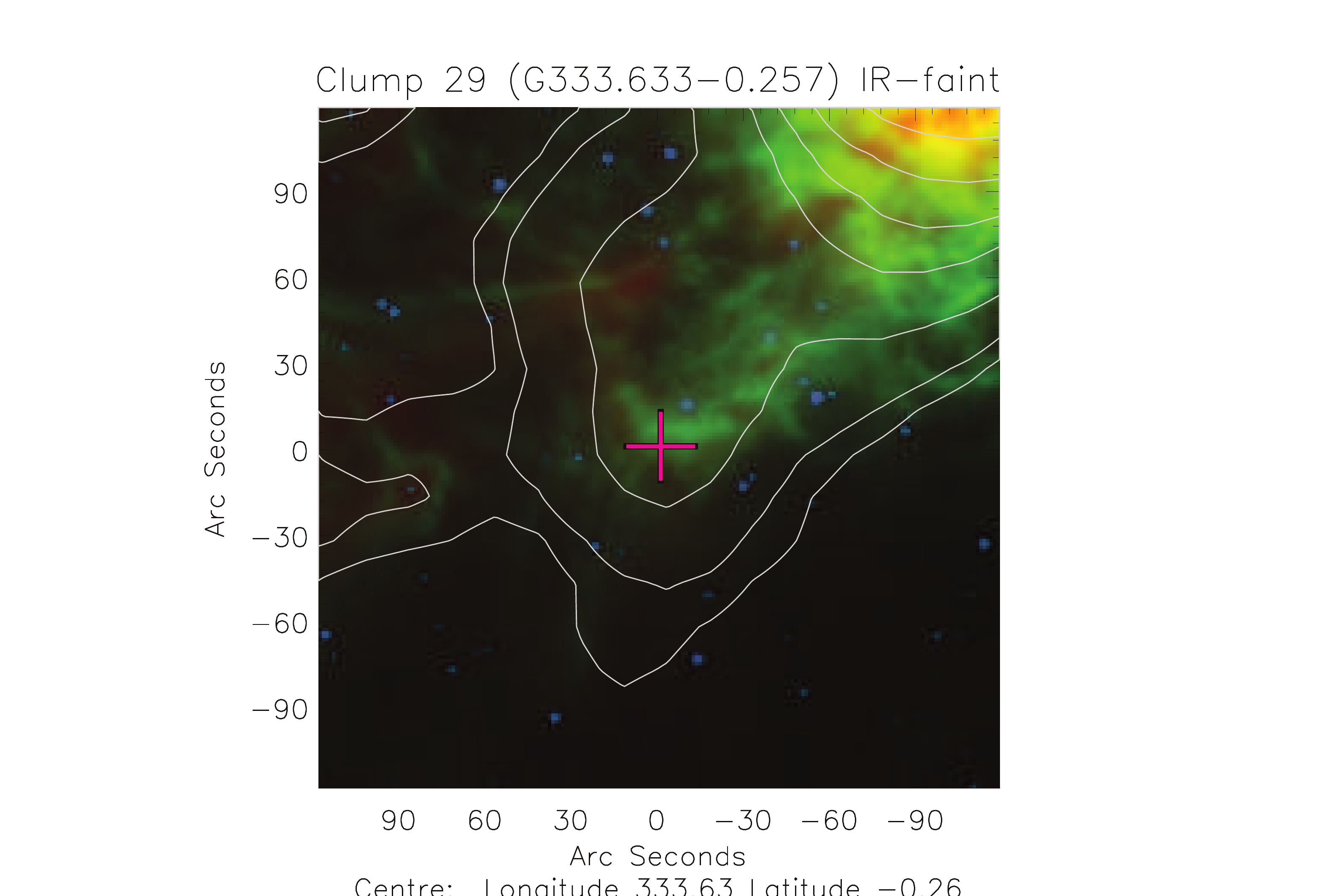}
\includegraphics[trim=100 20 210 40,clip,width=0.32\textwidth]{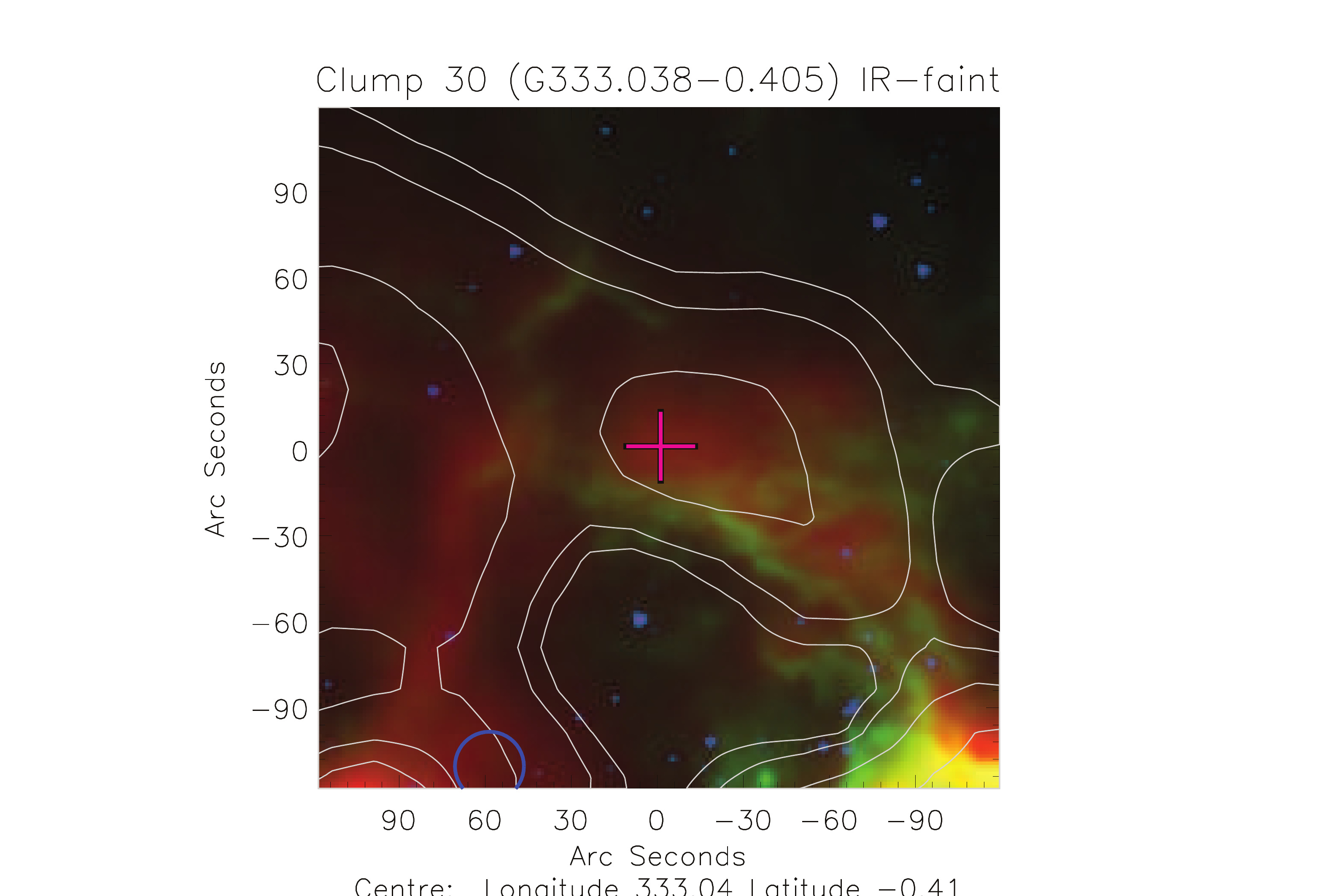}\\
\vspace*{0.5cm}
\includegraphics[trim=100 20 190 40,clip,width=0.32\textwidth]{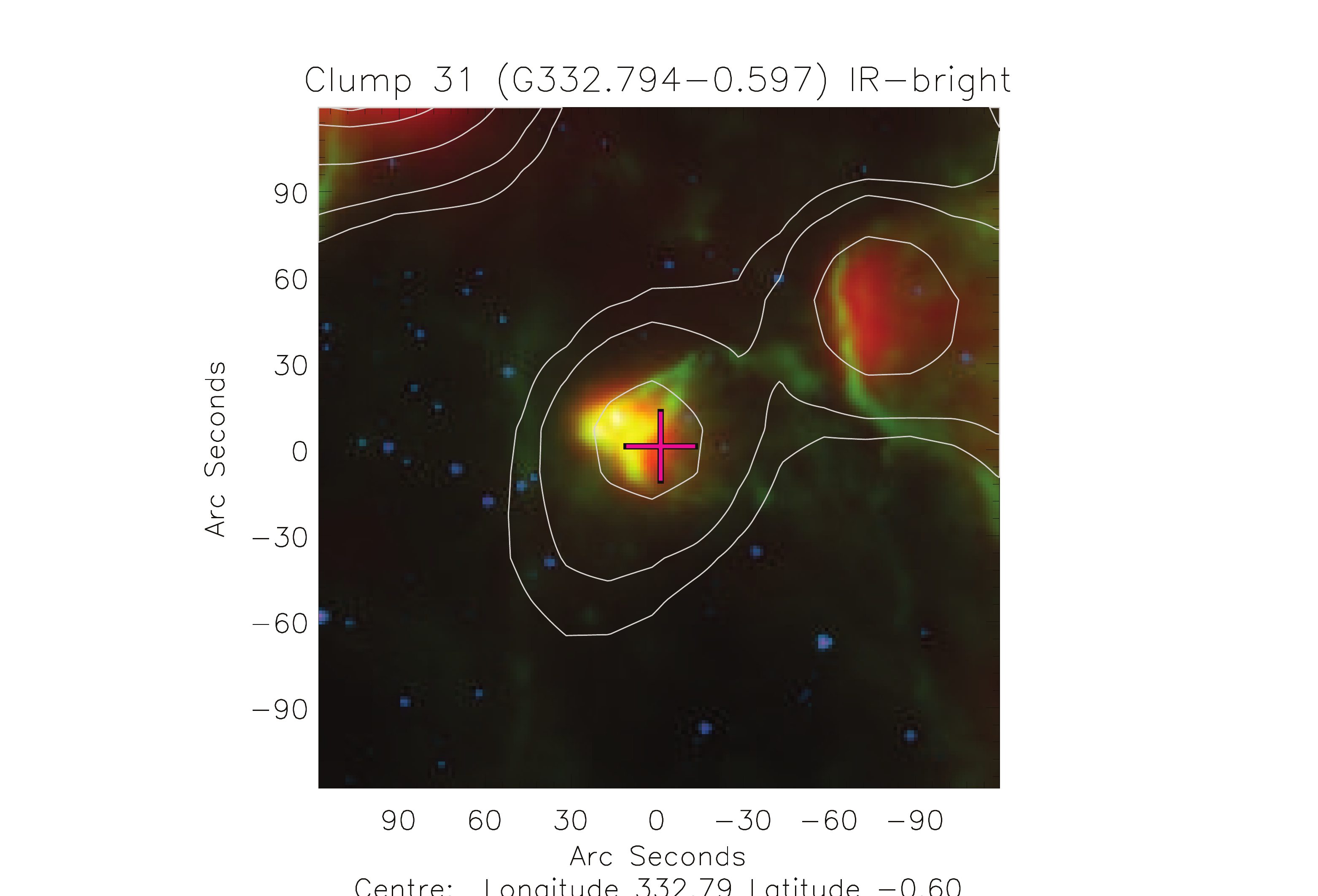}
\includegraphics[trim=100 20 190 40,clip,width=0.32\textwidth]{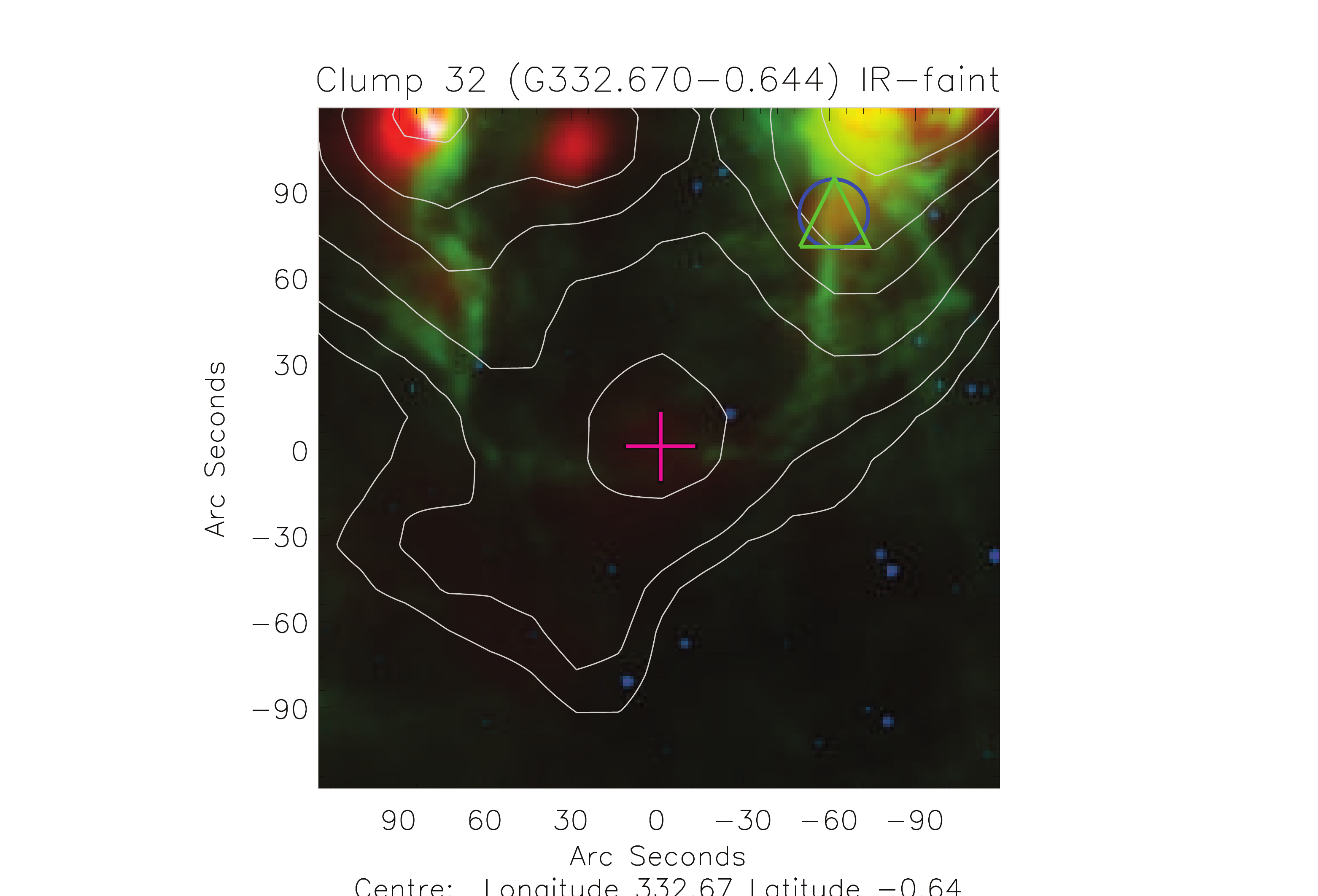}
\includegraphics[trim=100 20 190 40,clip,width=0.32\textwidth]{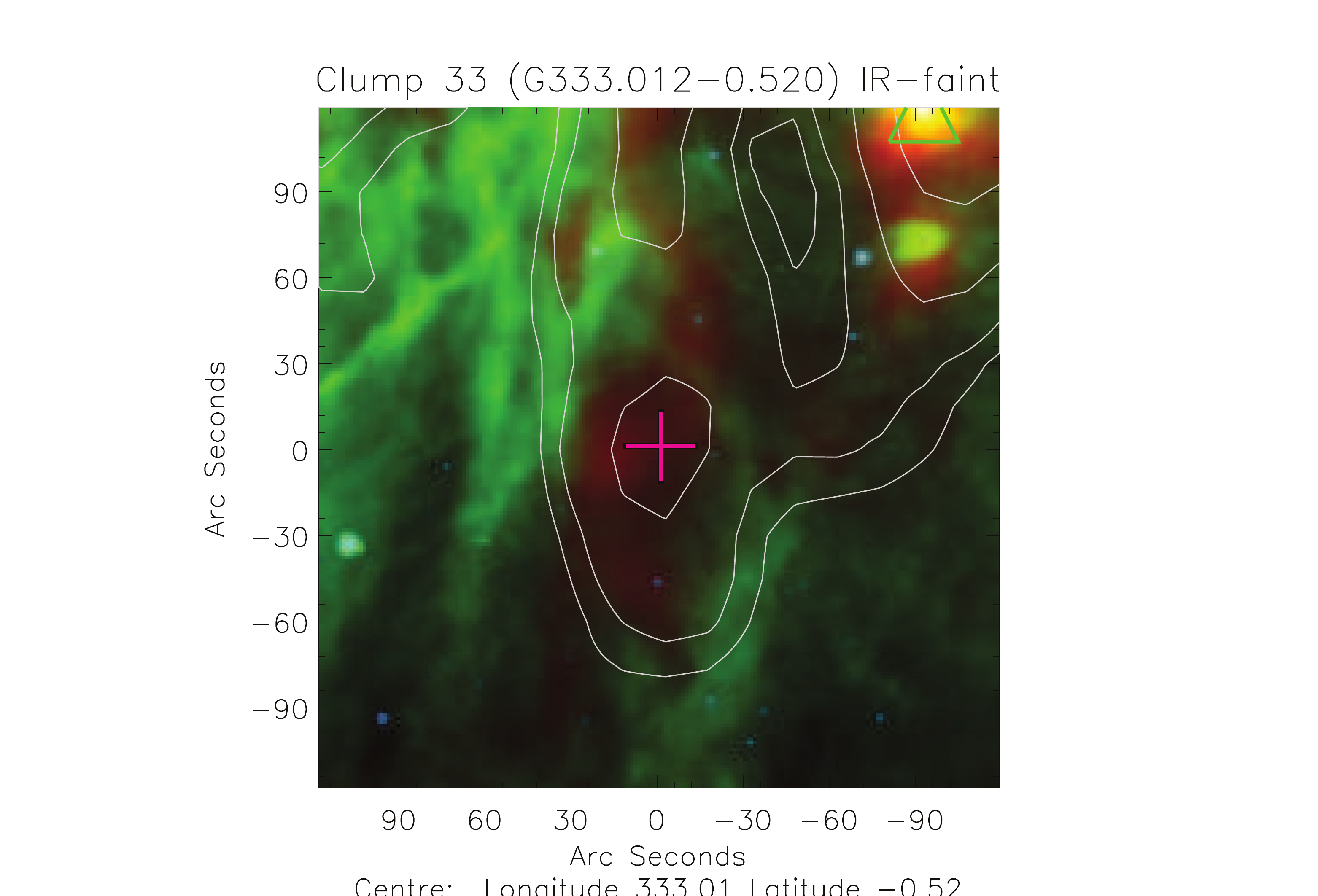}\\
\vspace*{0.5cm}
\includegraphics[trim=100 20 210 40,clip,width=0.32\textwidth]{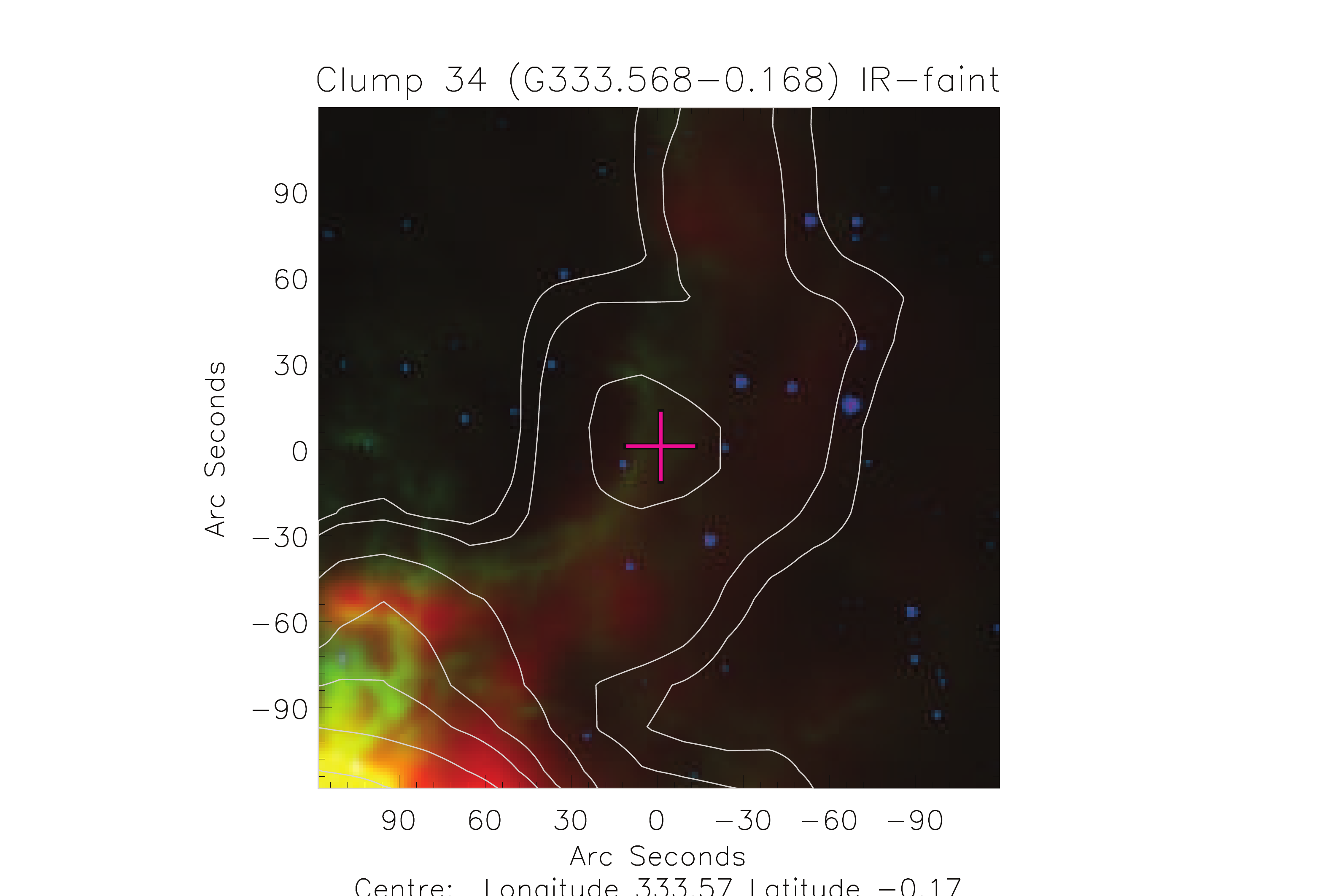}
\includegraphics[trim=100 20 210 40,clip,width=0.32\textwidth]{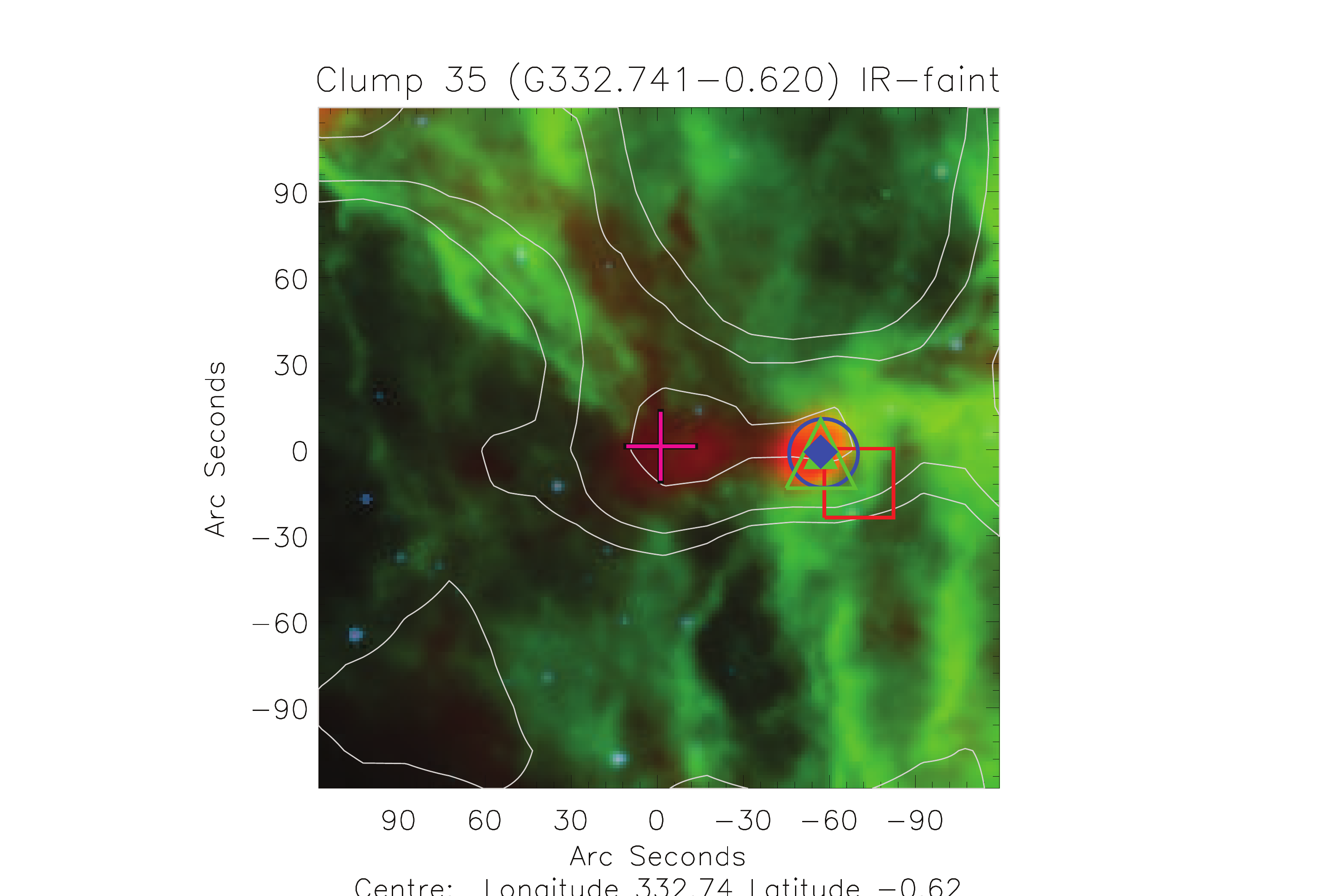}
\includegraphics[trim=100 20 210 40,clip,width=0.32\textwidth]{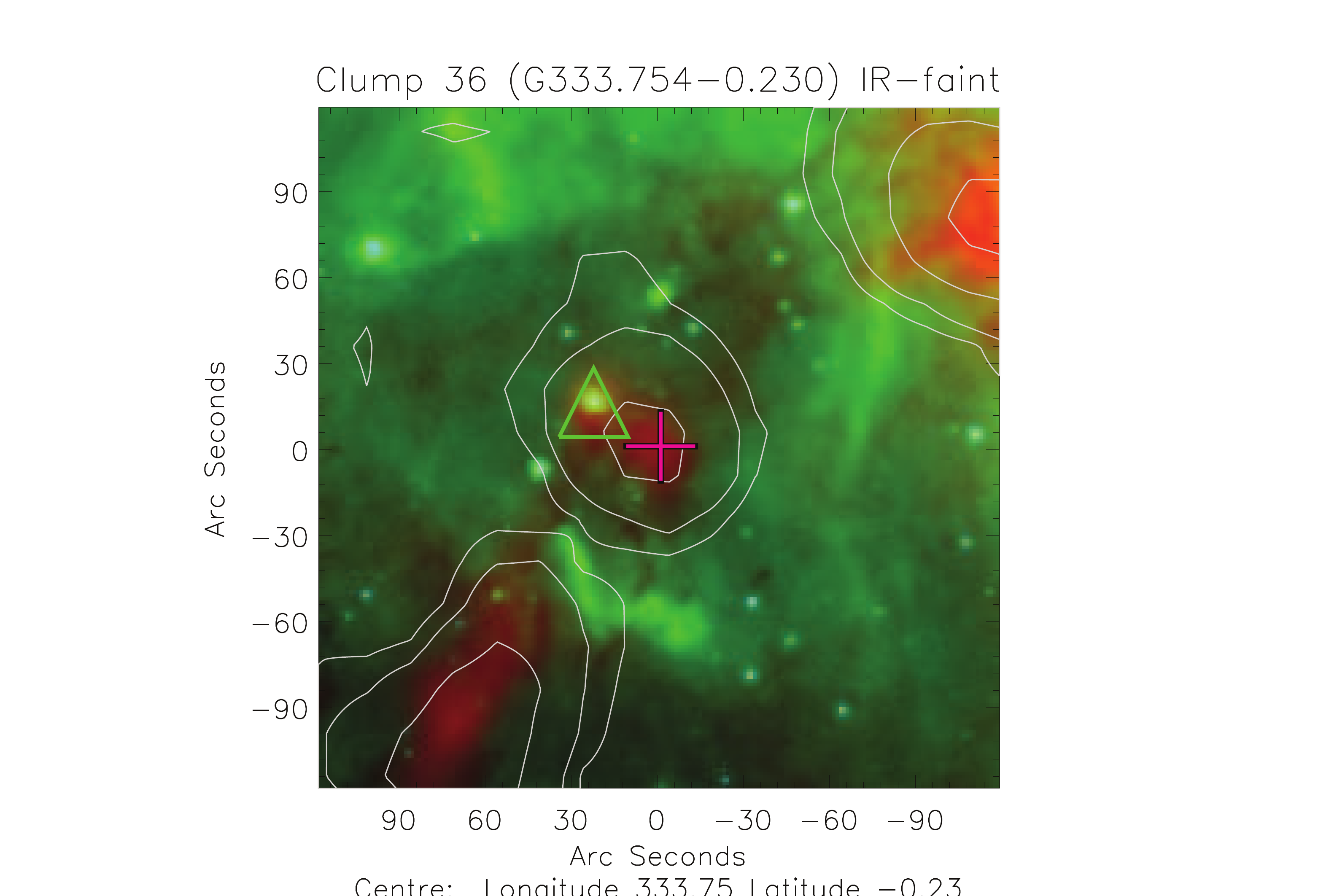}\\
\vspace*{0.3cm}\contcaption{}
\end{figure*}

\begin{figure*}
\centering
\includegraphics[trim=100 20 190 40,clip,width=0.32\textwidth]{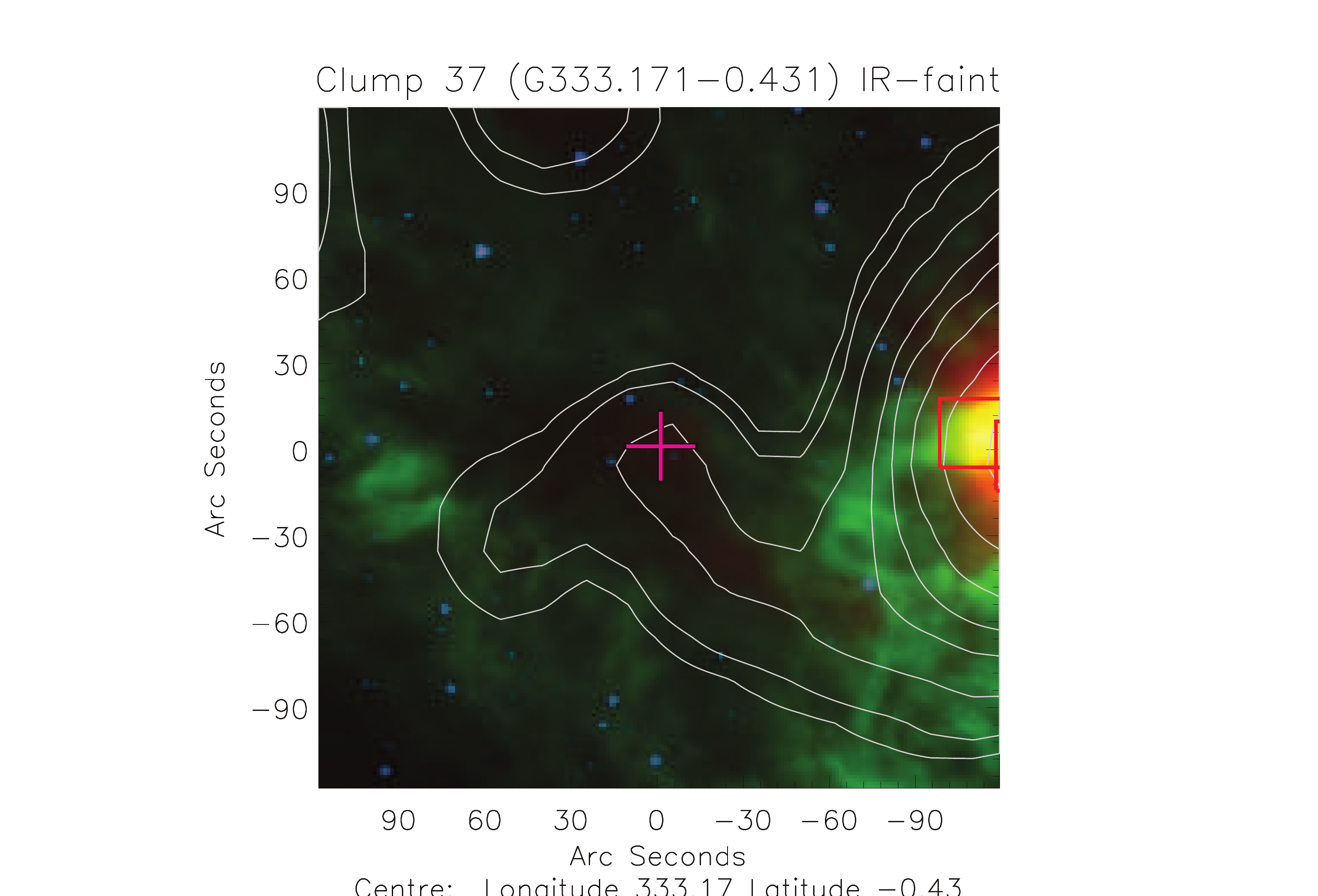}
\includegraphics[trim=100 20 190 40,clip,width=0.32\textwidth]{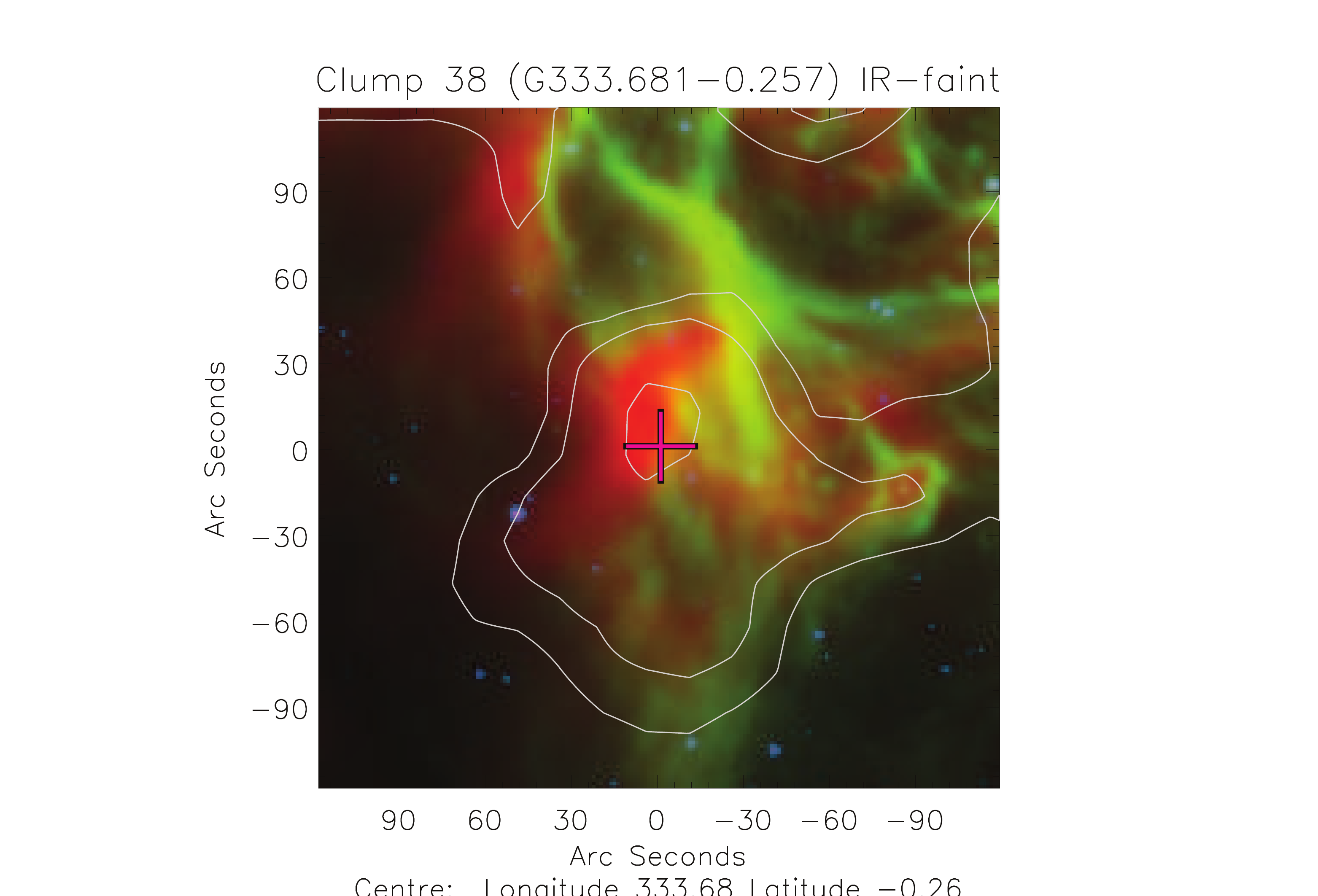}
\includegraphics[trim=100 20 190 40,clip,width=0.32\textwidth]{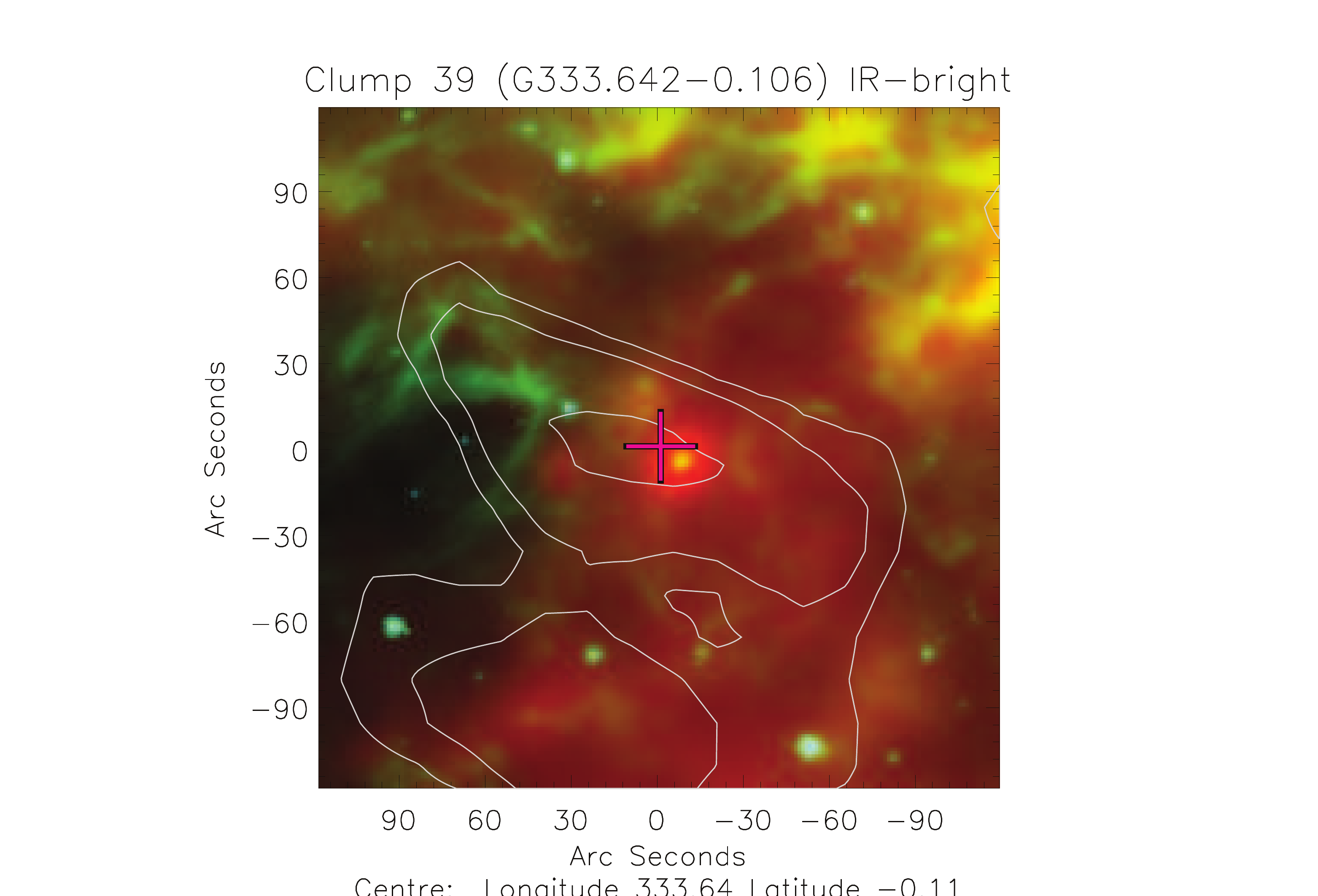}\\
\vspace*{0.5cm}
\includegraphics[trim=100 20 210 40,clip,width=0.32\textwidth]{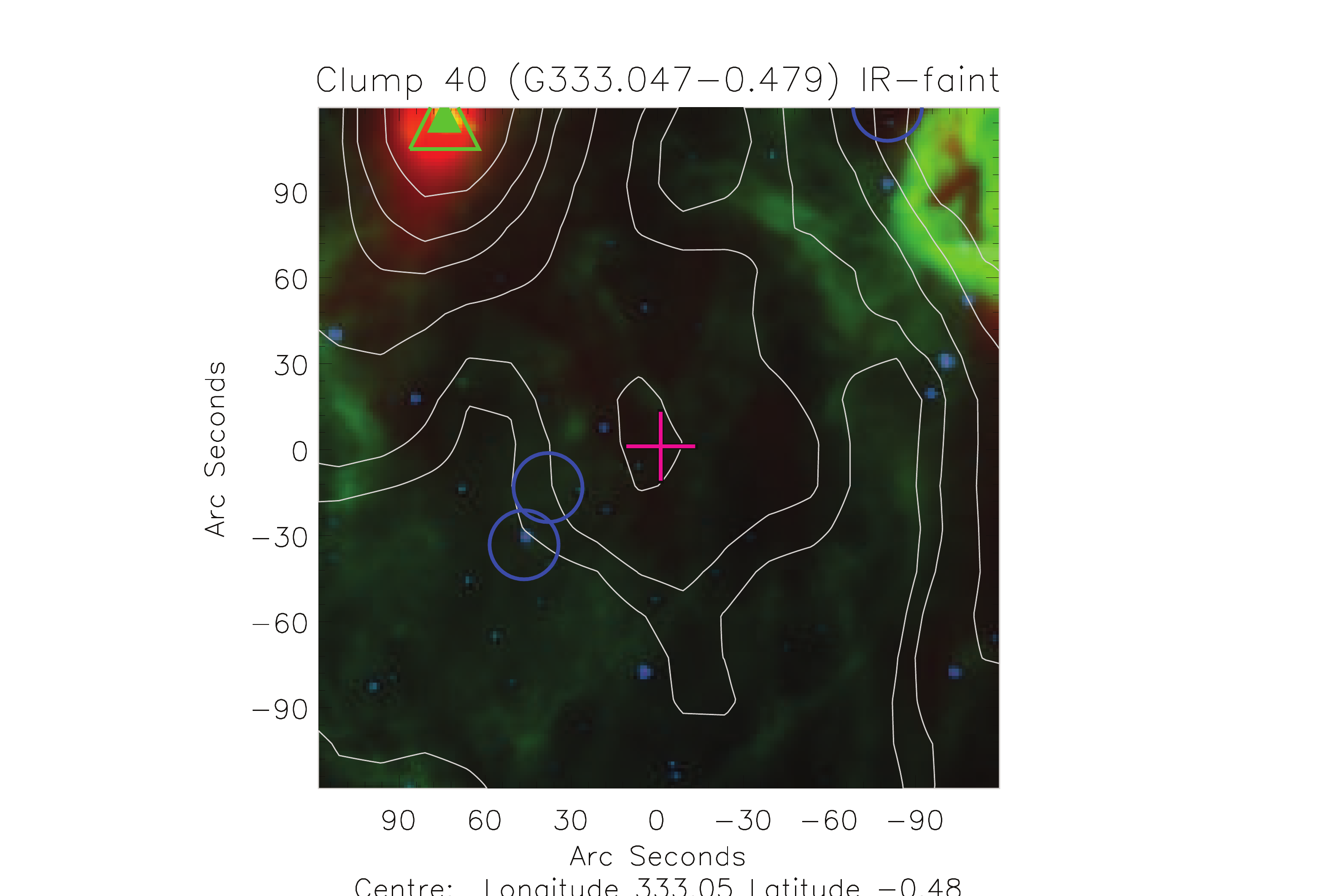}
\includegraphics[trim=100 20 210 40,clip,width=0.32\textwidth]{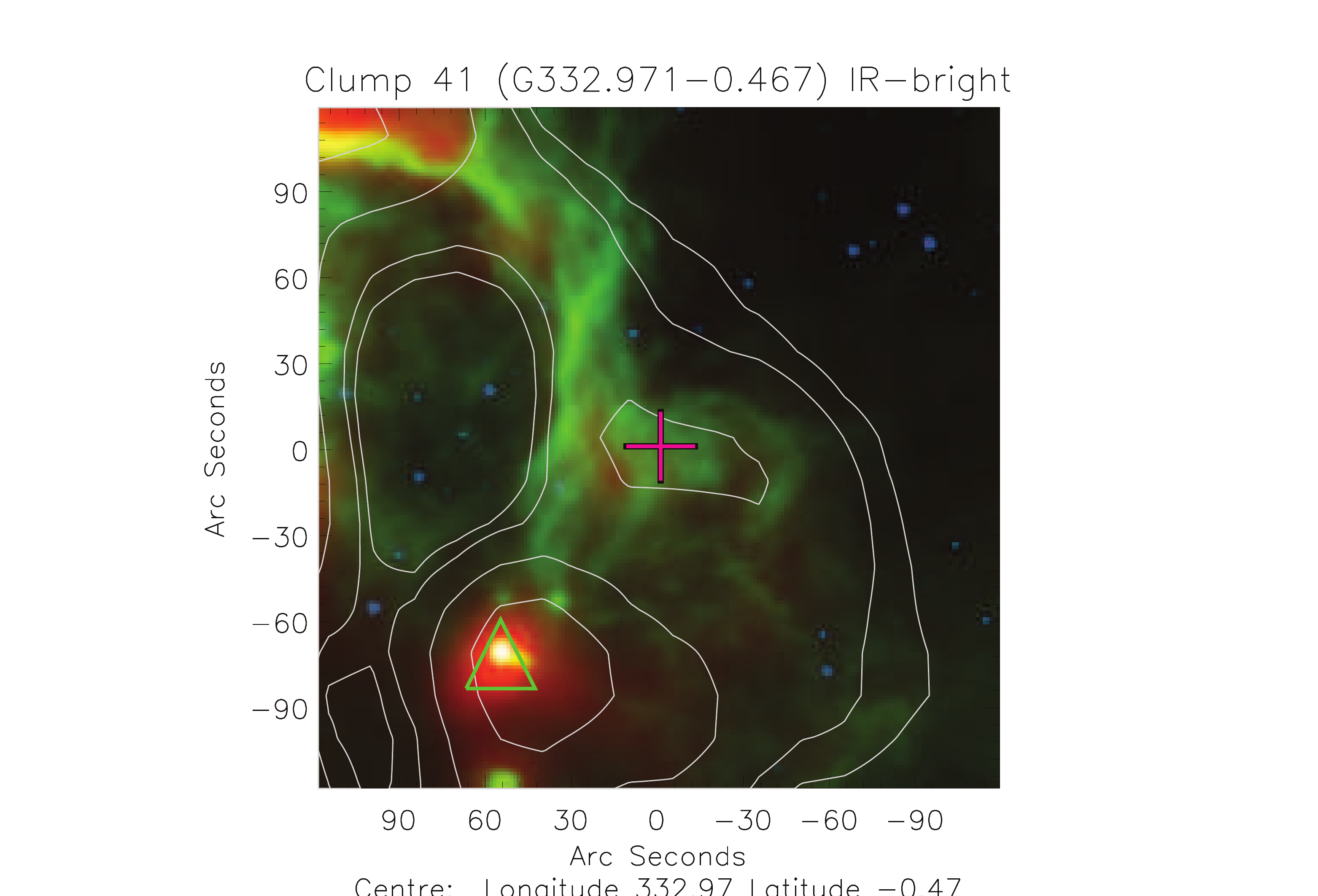}
\includegraphics[trim=100 20 210 40,clip,width=0.32\textwidth]{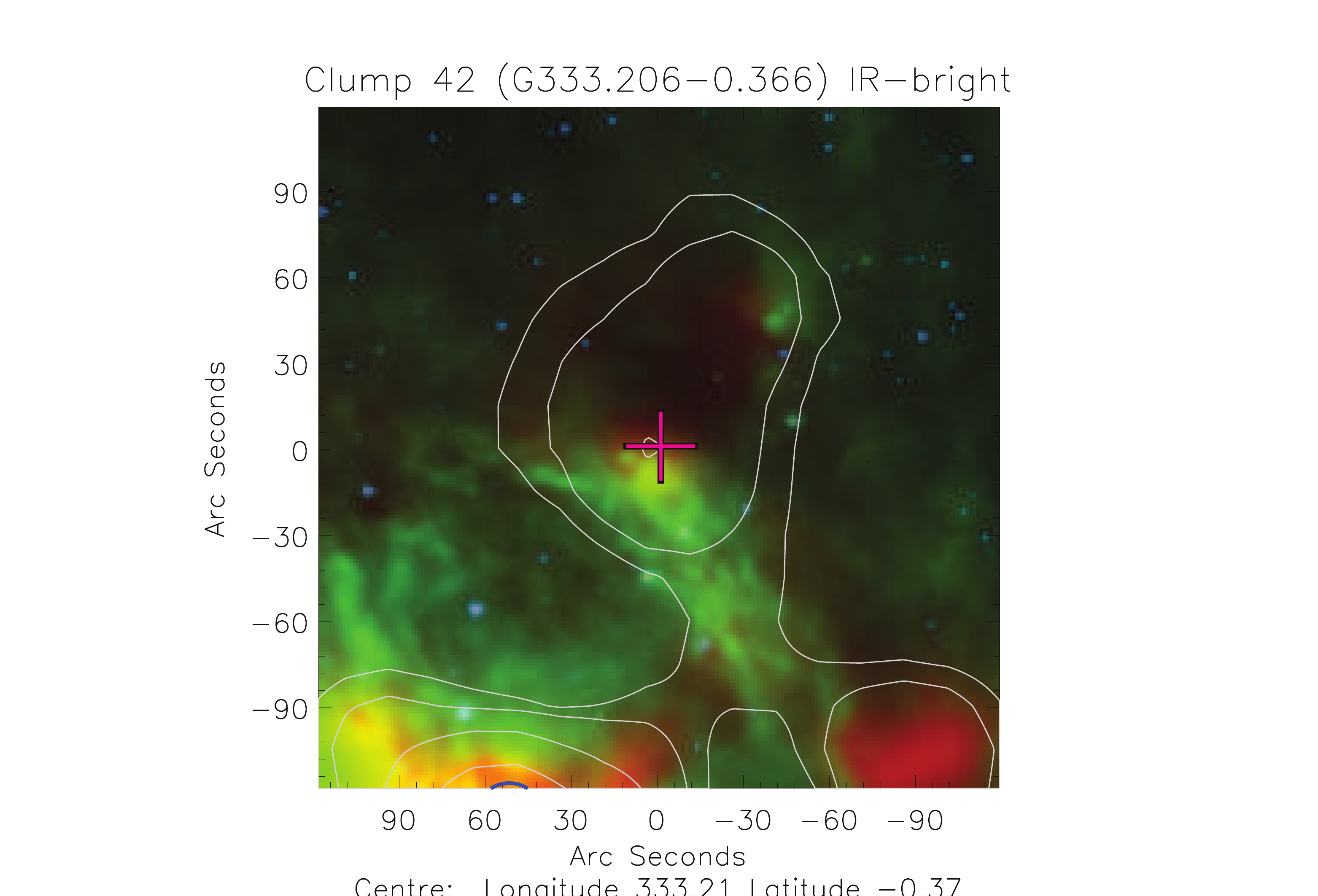}\\
\vspace*{0.5cm}
\includegraphics[trim=100 20 190 40,clip,width=0.32\textwidth]{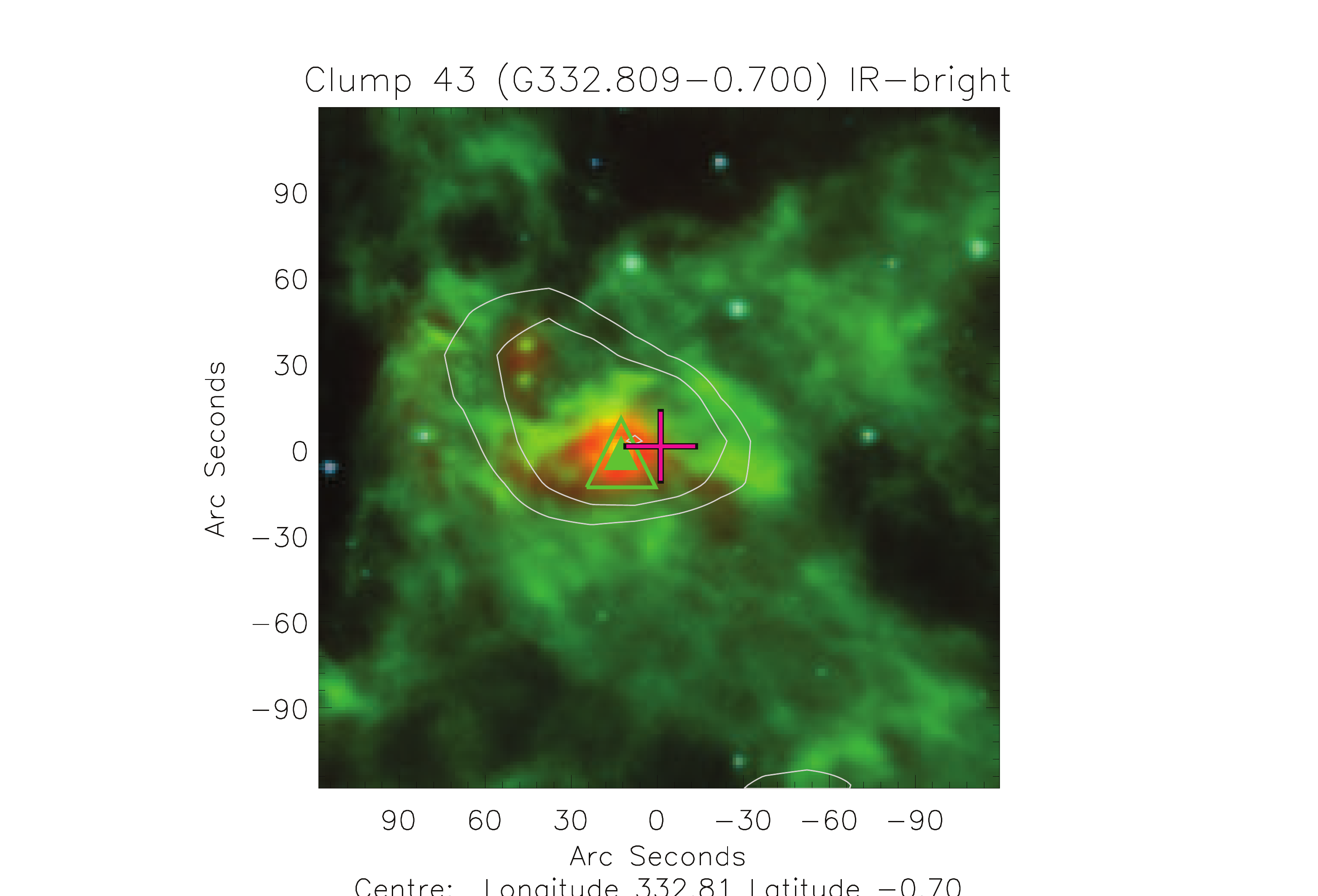}
\includegraphics[trim=100 20 190 40,clip,width=0.32\textwidth]{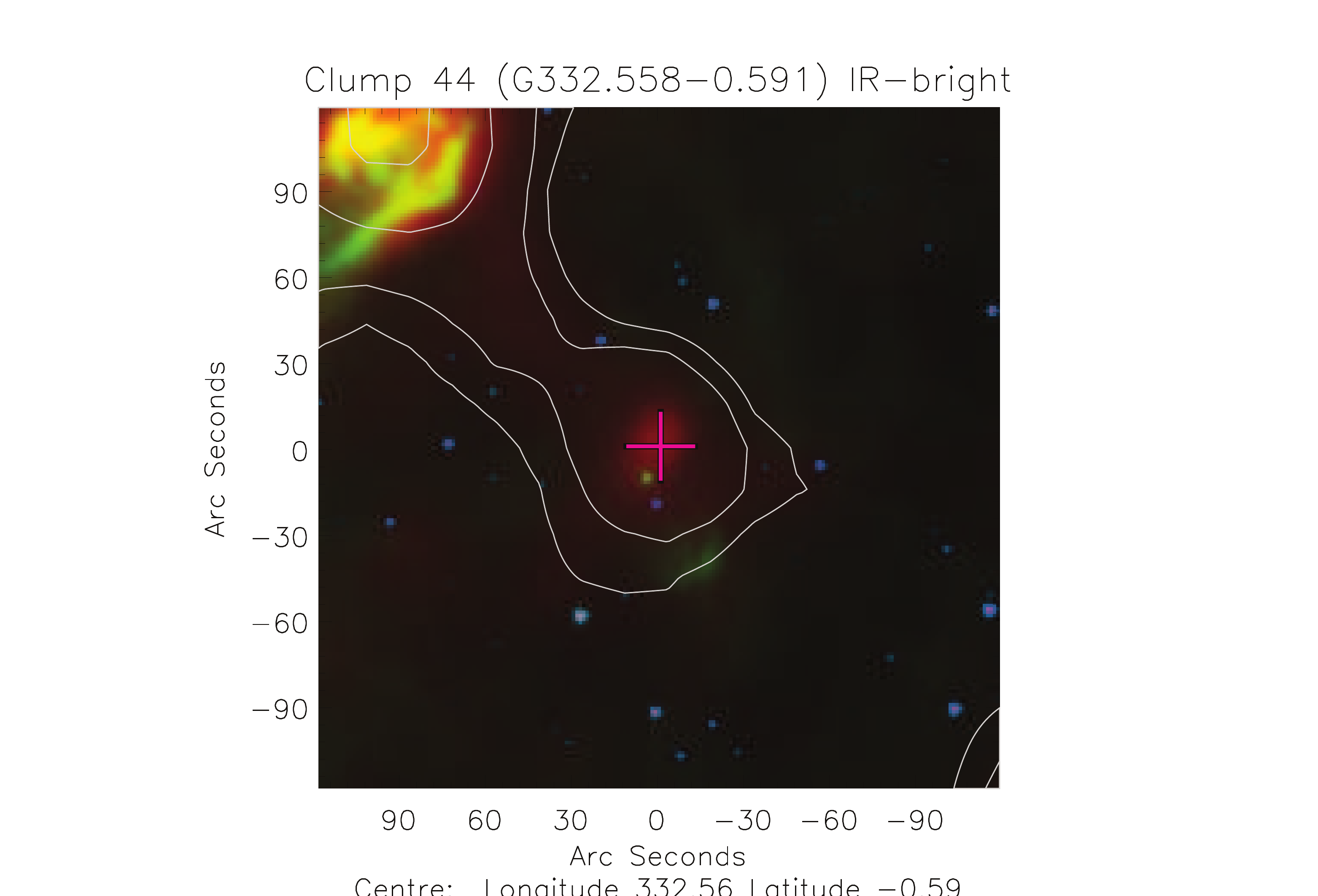}
\includegraphics[trim=100 20 190 40,clip,width=0.32\textwidth]{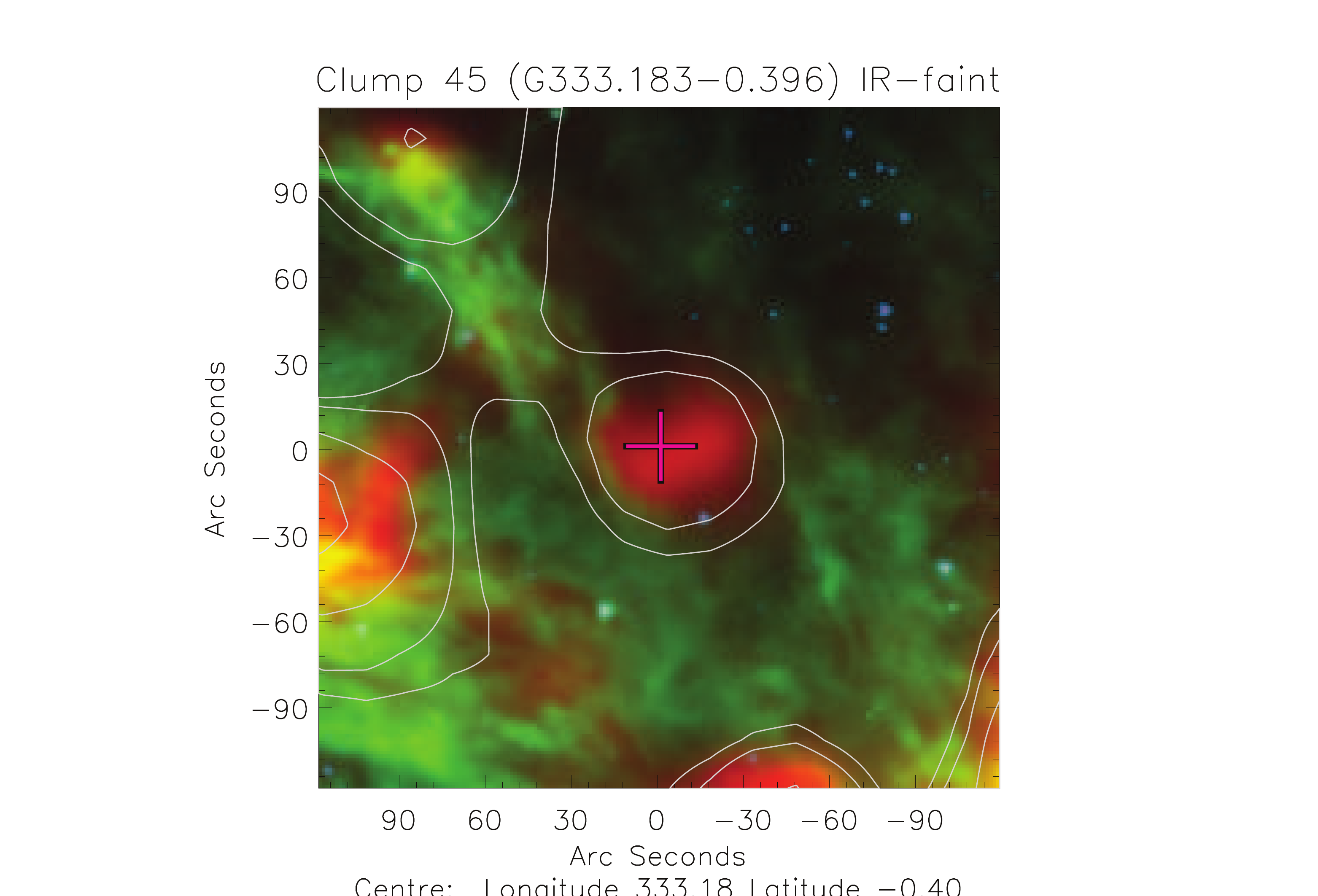}\\
\vspace*{0.5cm}
\includegraphics[trim=100 20 210 40,clip,width=0.32\textwidth]{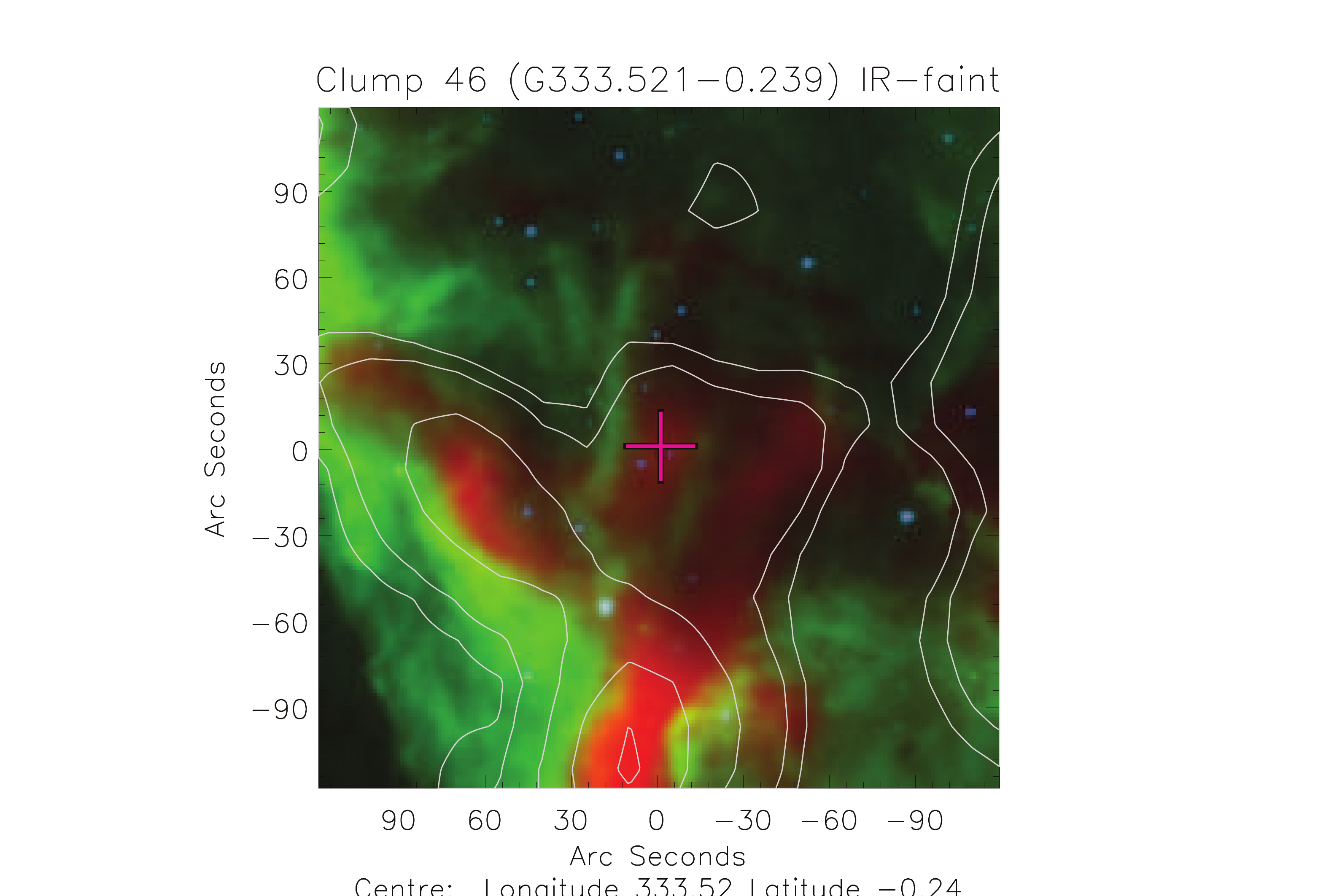}
\includegraphics[trim=100 20 210 40,clip,width=0.32\textwidth]{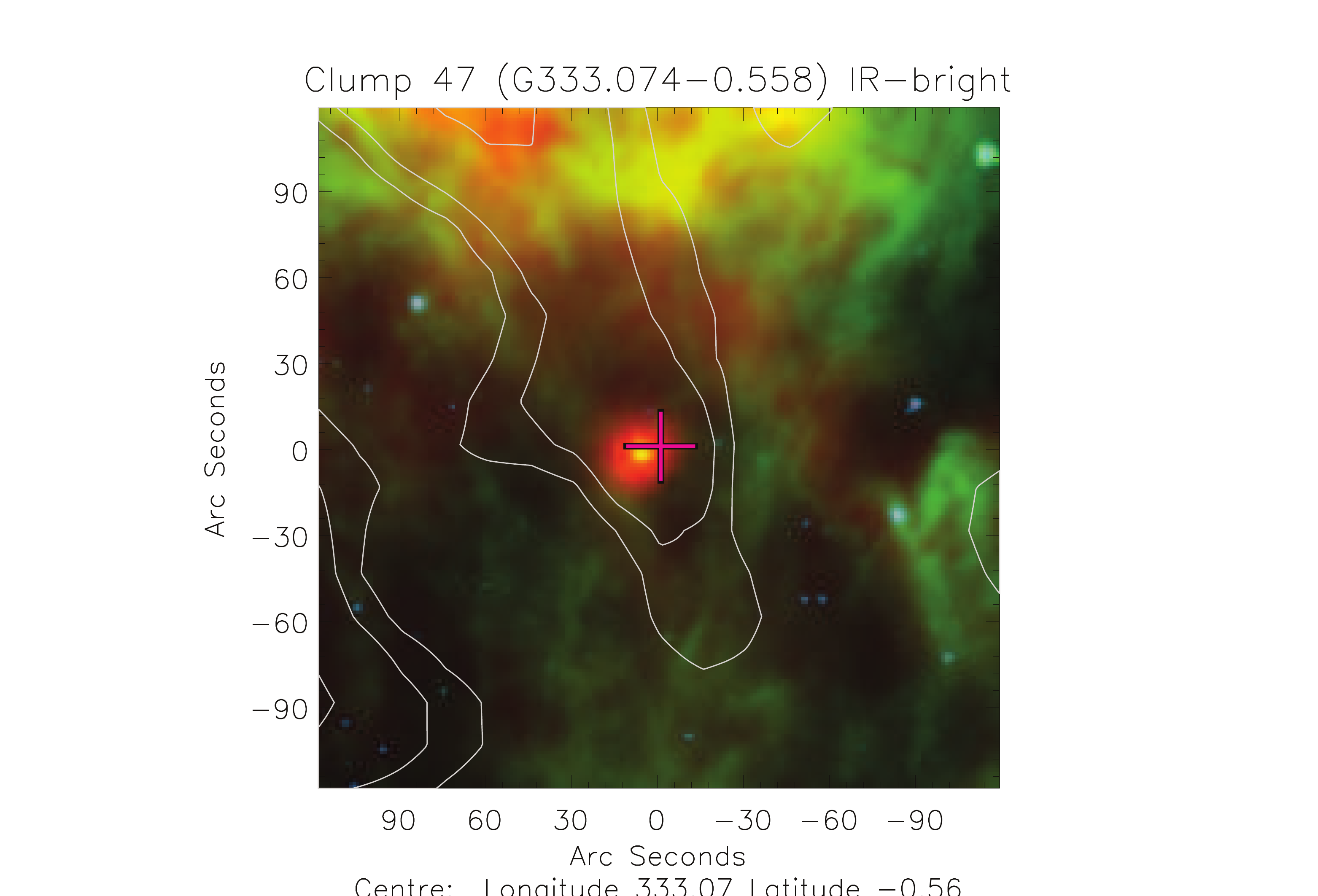}
\includegraphics[trim=100 20 210 40,clip,width=0.32\textwidth]{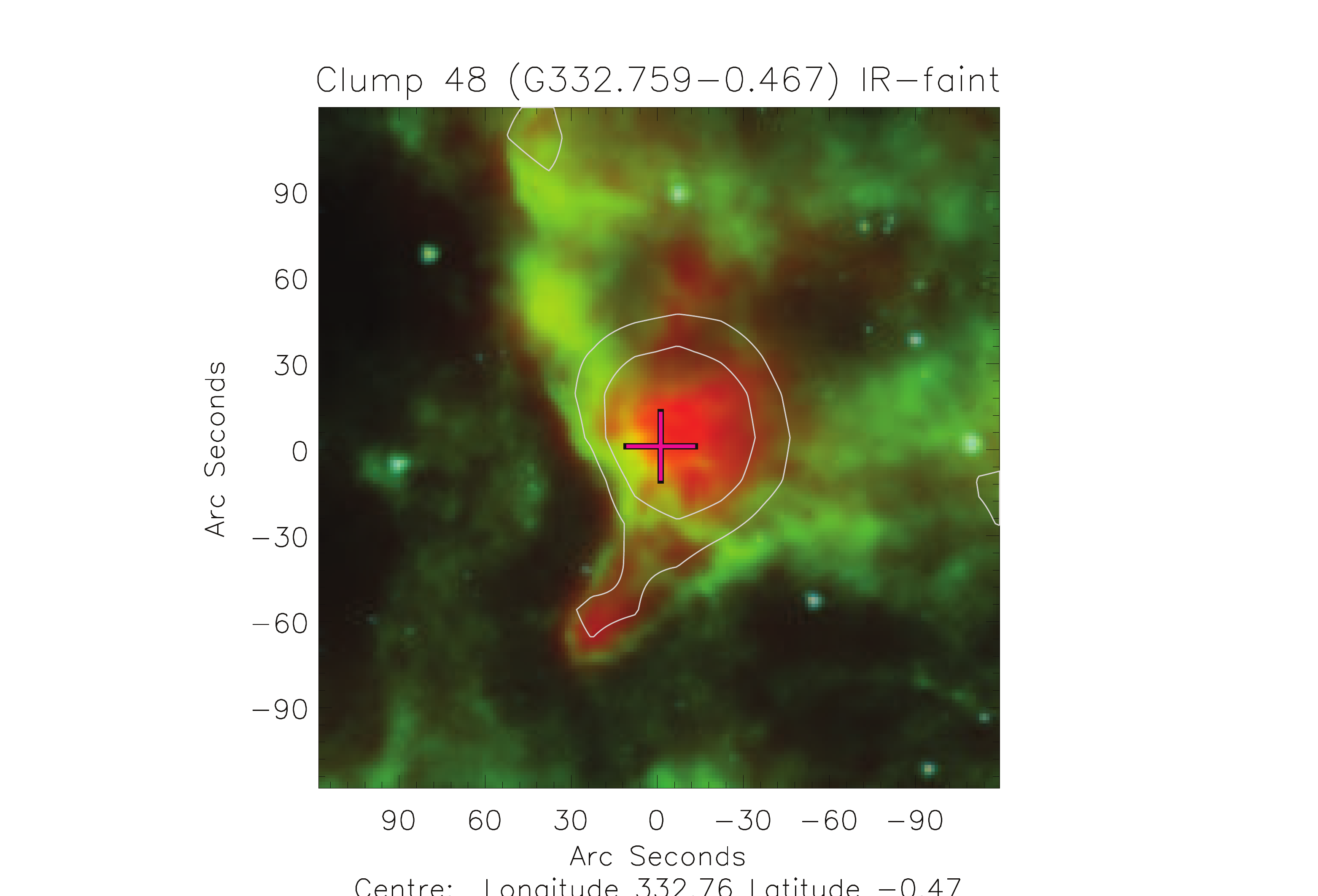}\\
\vspace*{0.3cm}\contcaption{}
\end{figure*}

\begin{figure*}
\centering
\includegraphics[trim=100 20 190 40,clip,width=0.32\textwidth]{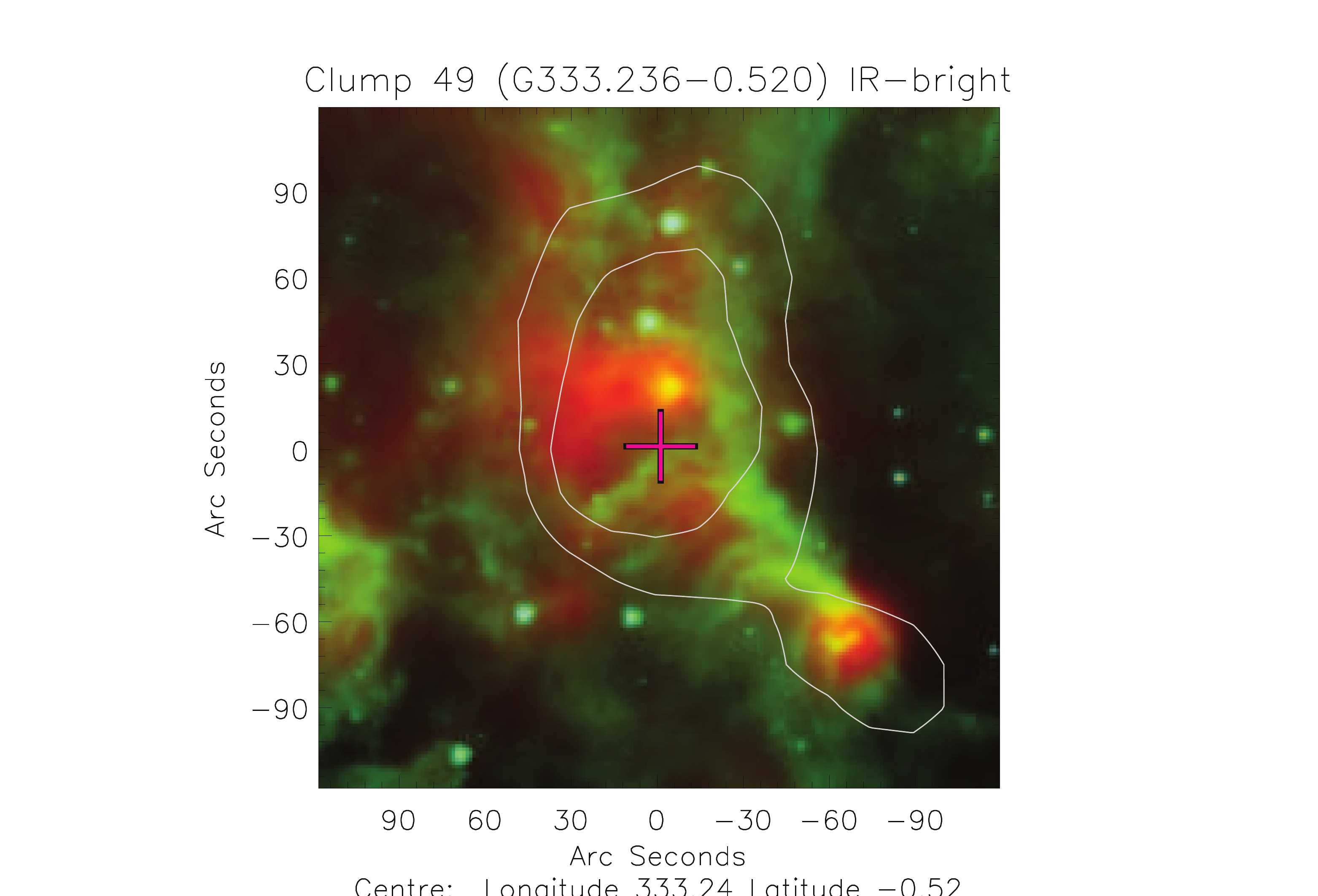}
\includegraphics[trim=100 20 190 40,clip,width=0.32\textwidth]{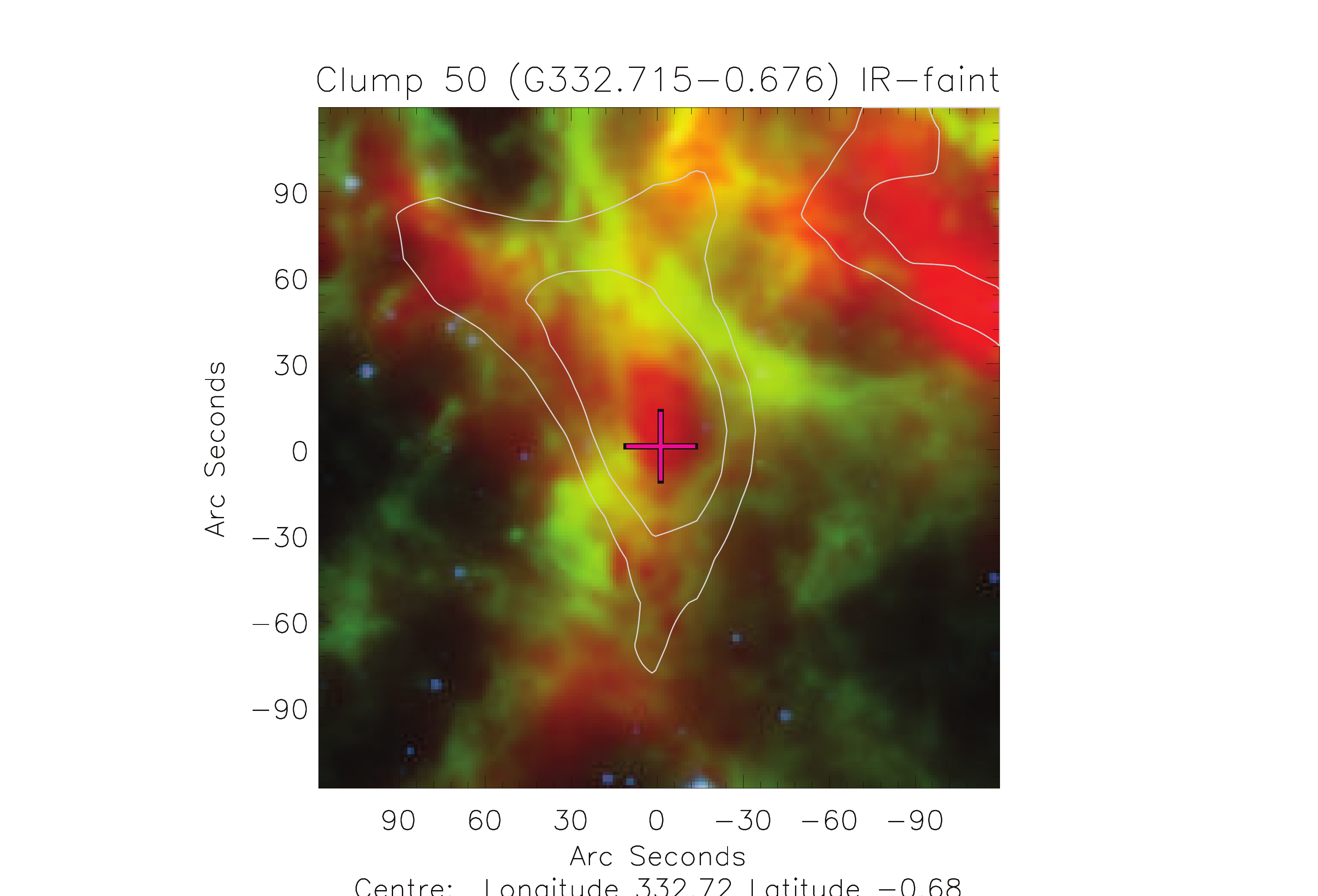}
\includegraphics[trim=100 20 190 40,clip,width=0.32\textwidth]{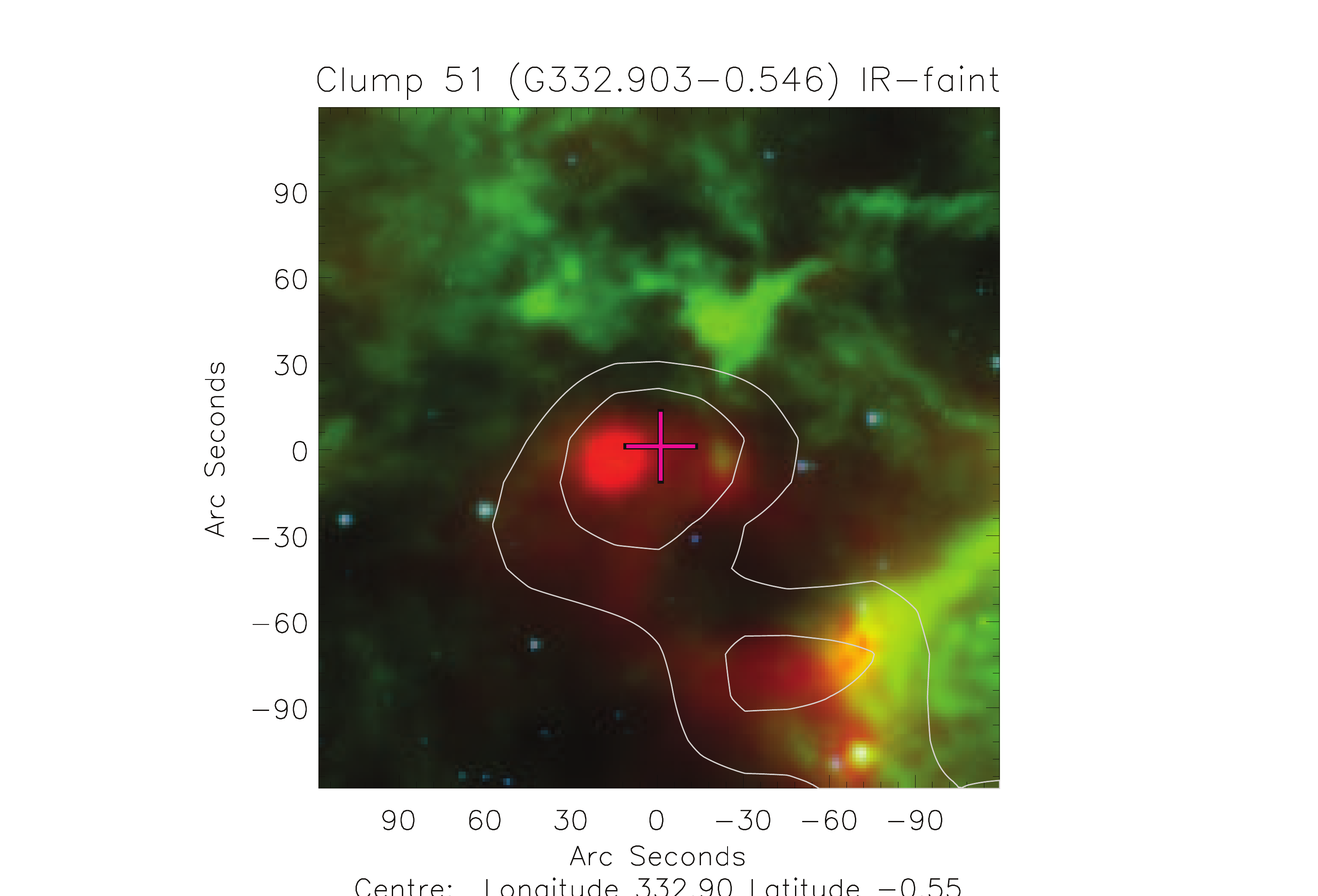}\\
\vspace*{0.5cm}
\includegraphics[trim=100 20 210 40,clip,width=0.32\textwidth]{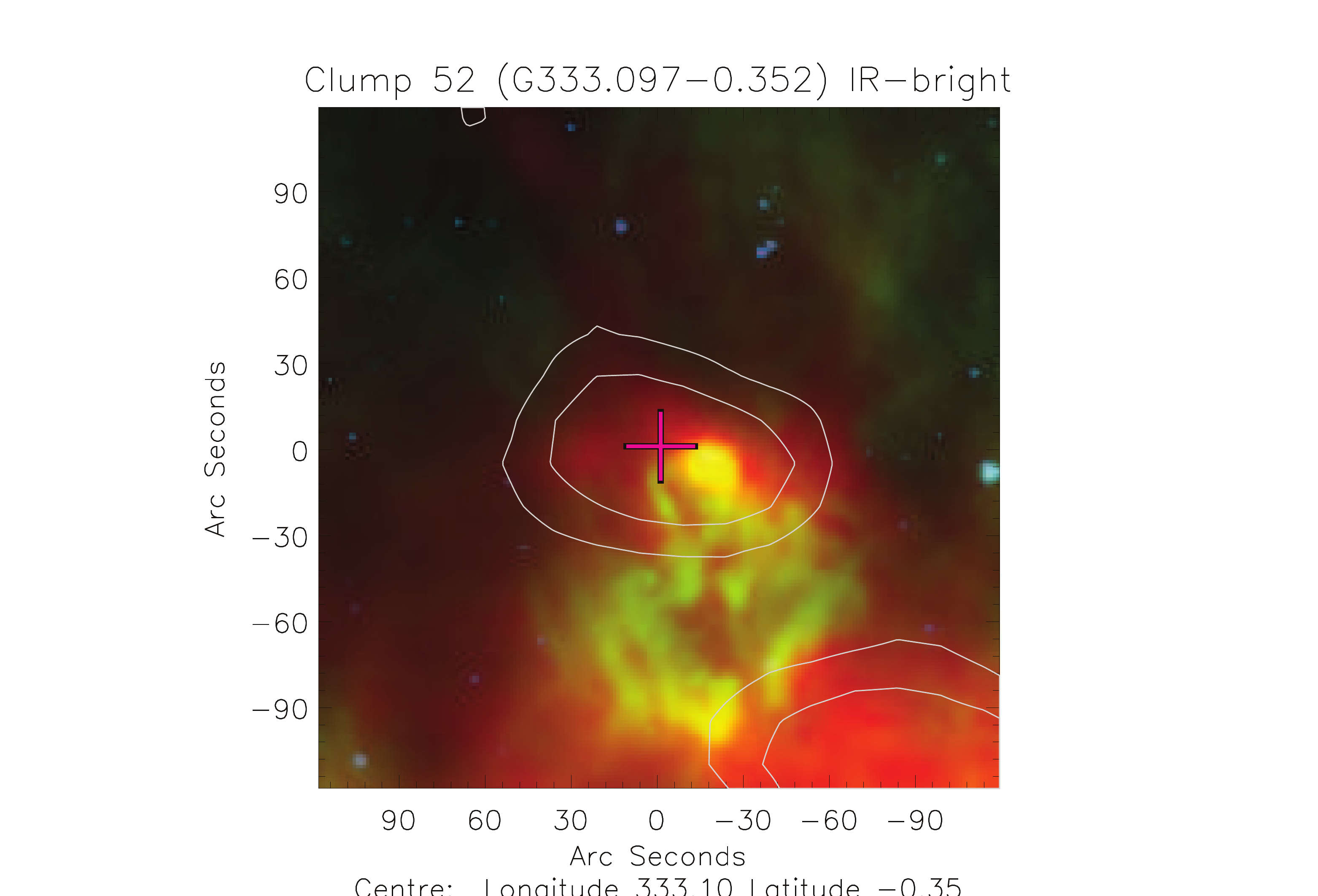}
\includegraphics[trim=100 20 210 40,clip,width=0.32\textwidth]{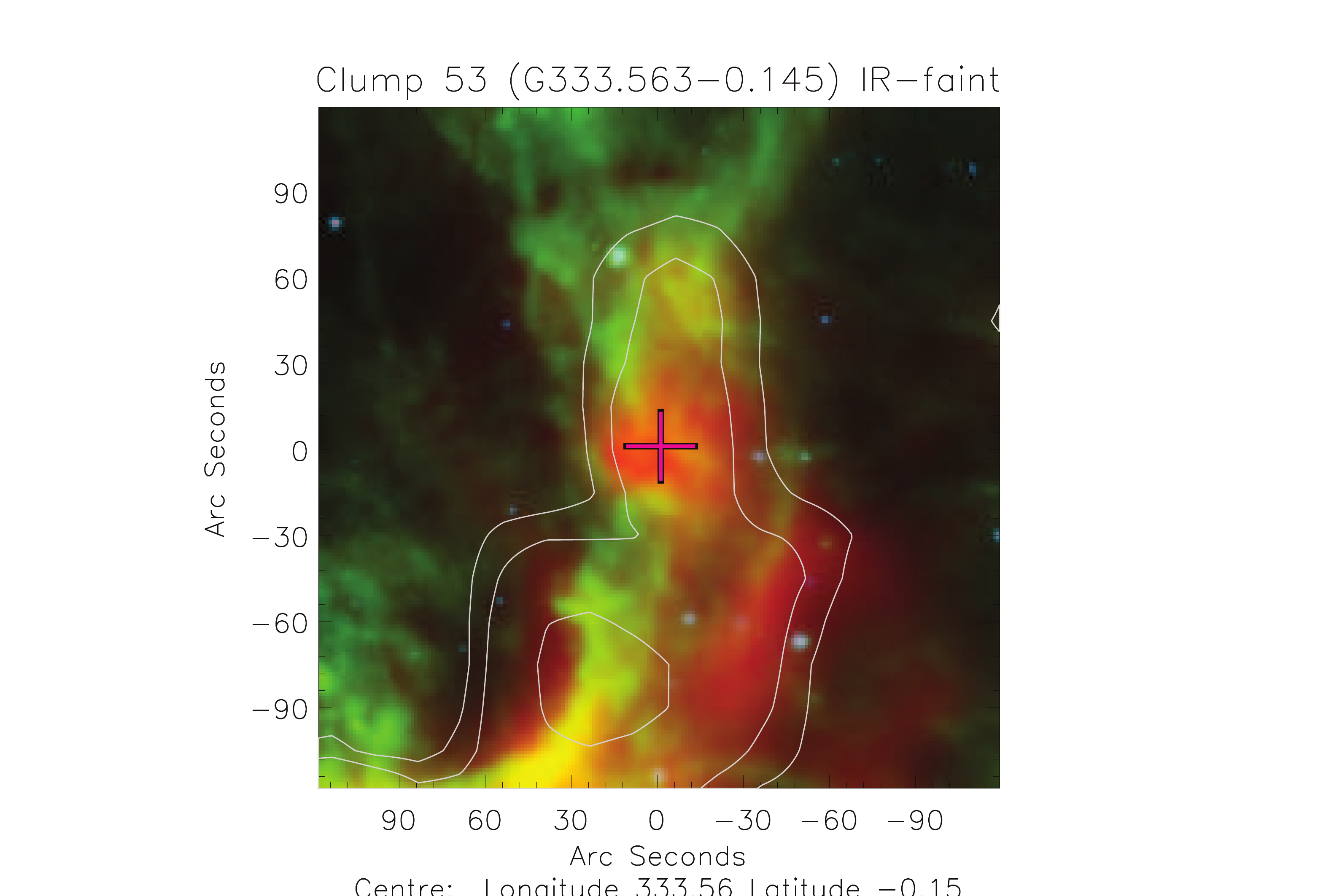}
\includegraphics[trim=100 20 210 40,clip,width=0.32\textwidth]{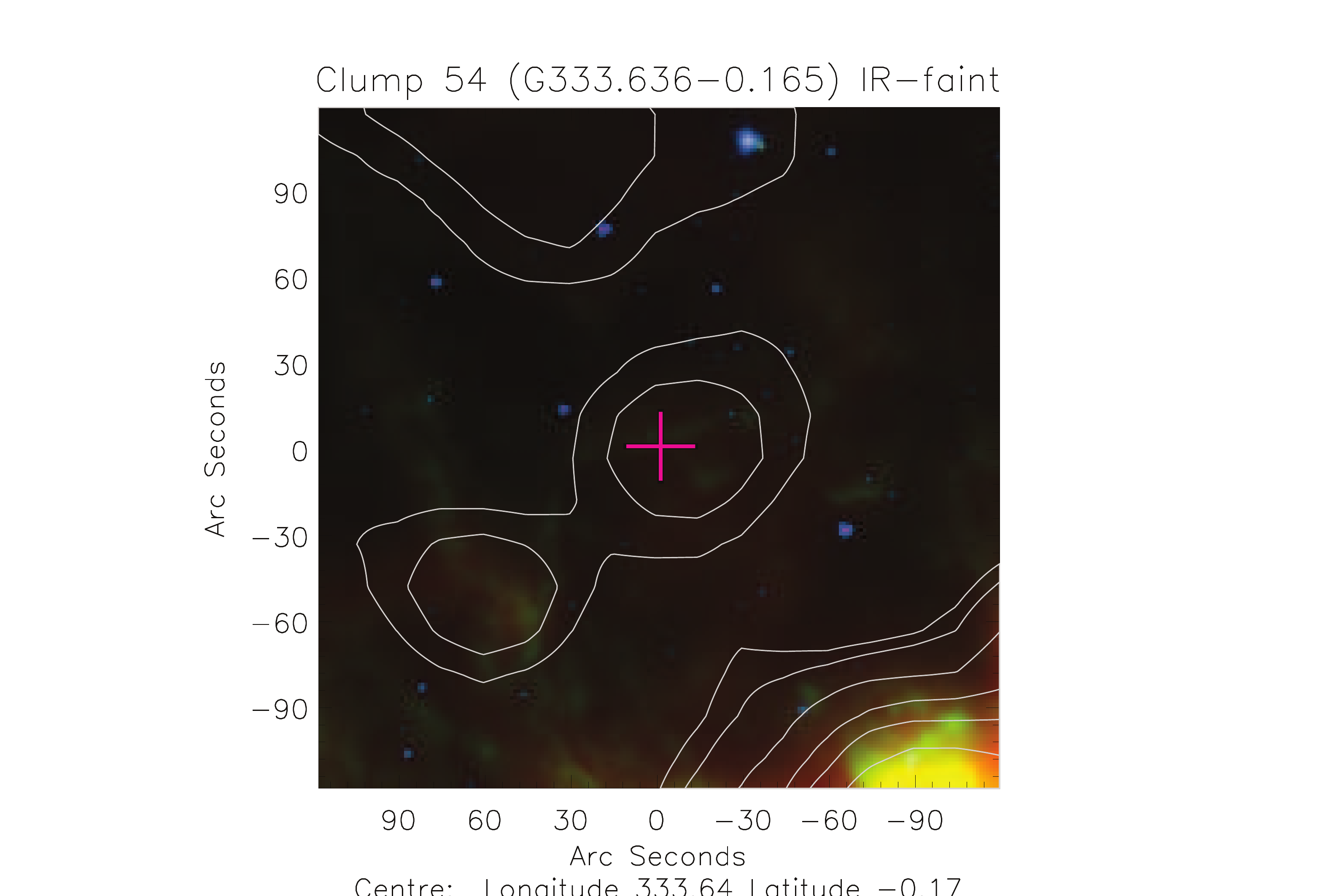}\\
\vspace*{0.5cm}
\includegraphics[trim=100 20 190 40,clip,width=0.32\textwidth]{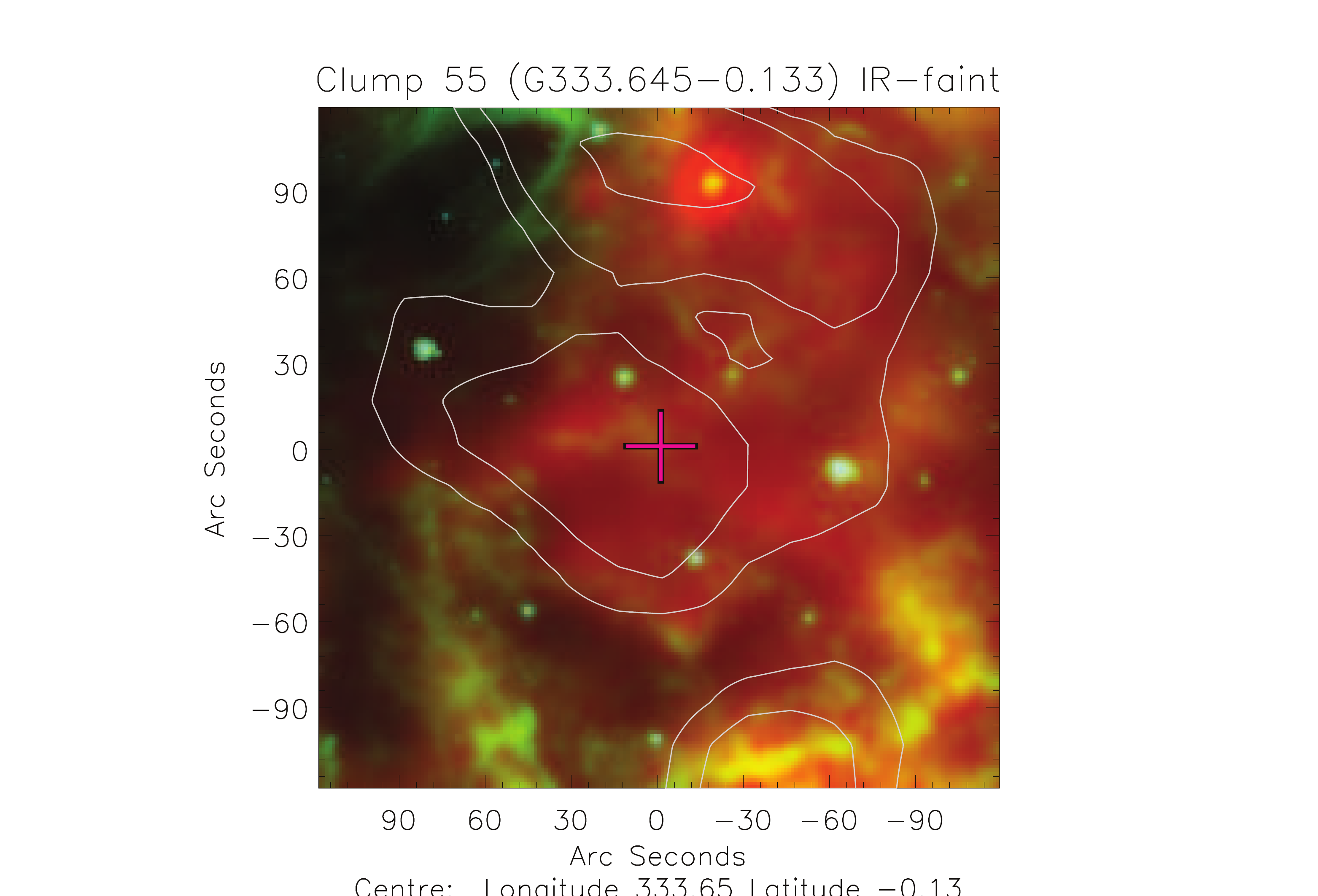}
\includegraphics[trim=100 20 190 40,clip,width=0.32\textwidth]{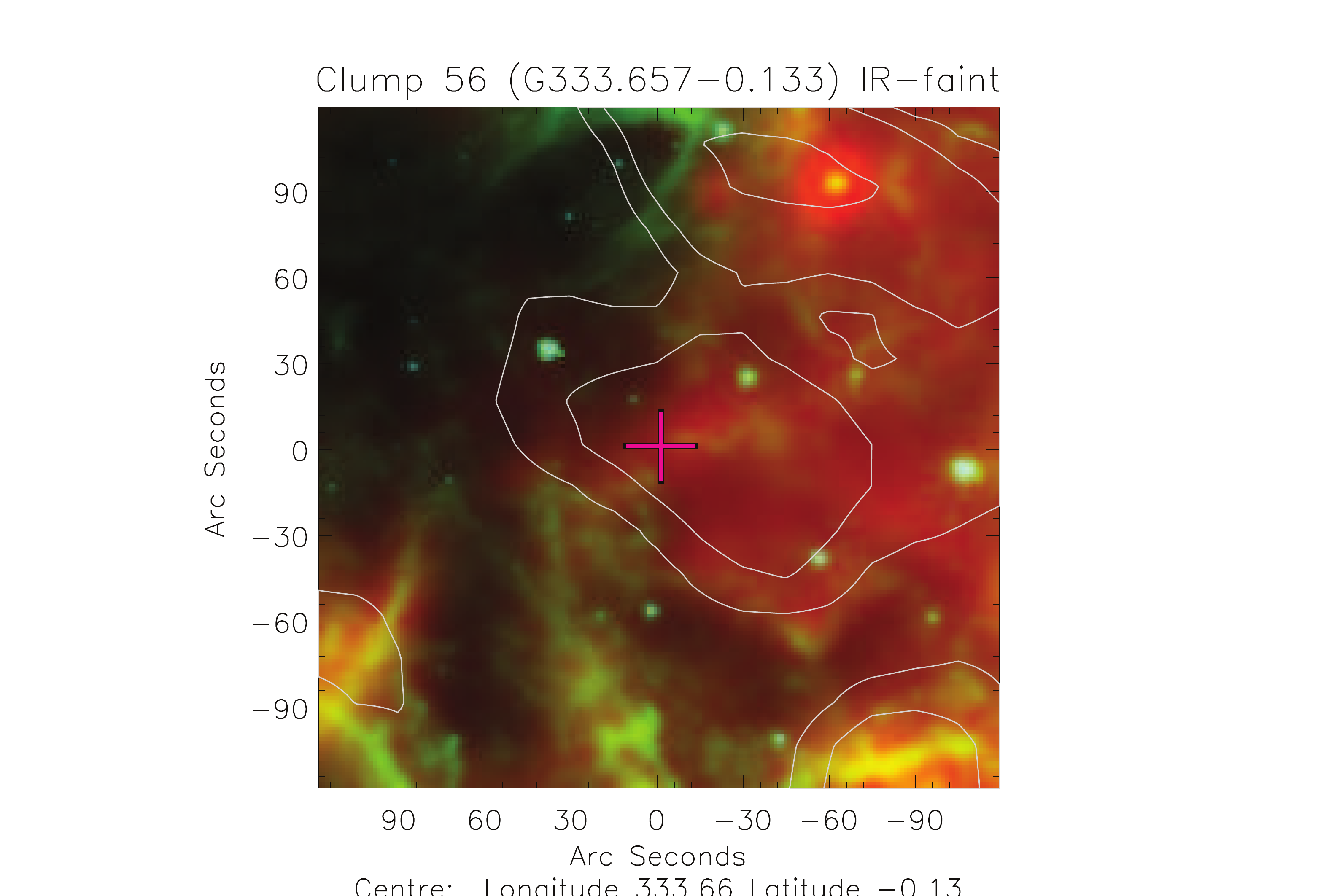}
\includegraphics[trim=100 20 190 40,clip,width=0.32\textwidth]{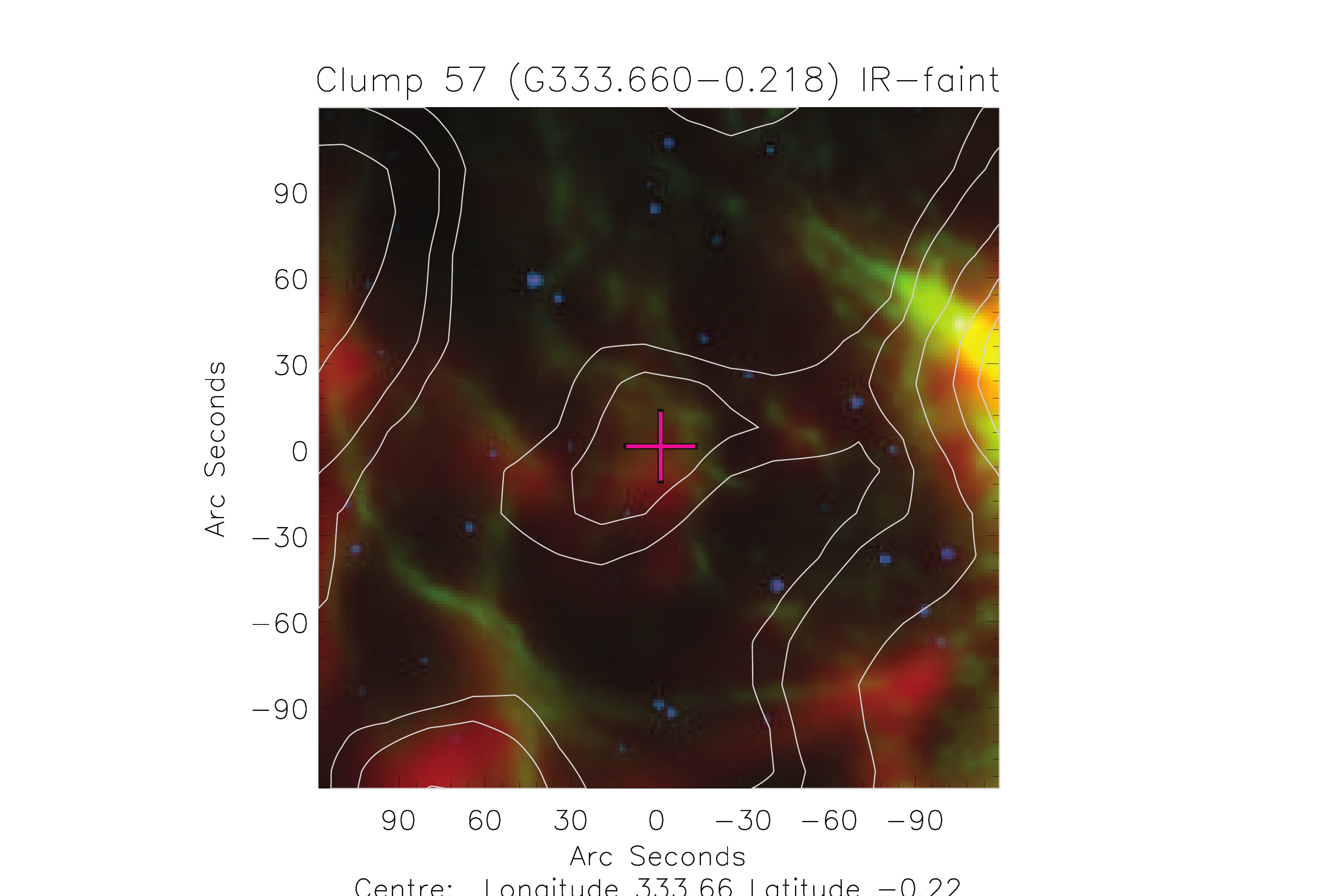}\\
\vspace*{0.5cm}
\includegraphics[trim=100 20 210 40,clip,width=0.32\textwidth]{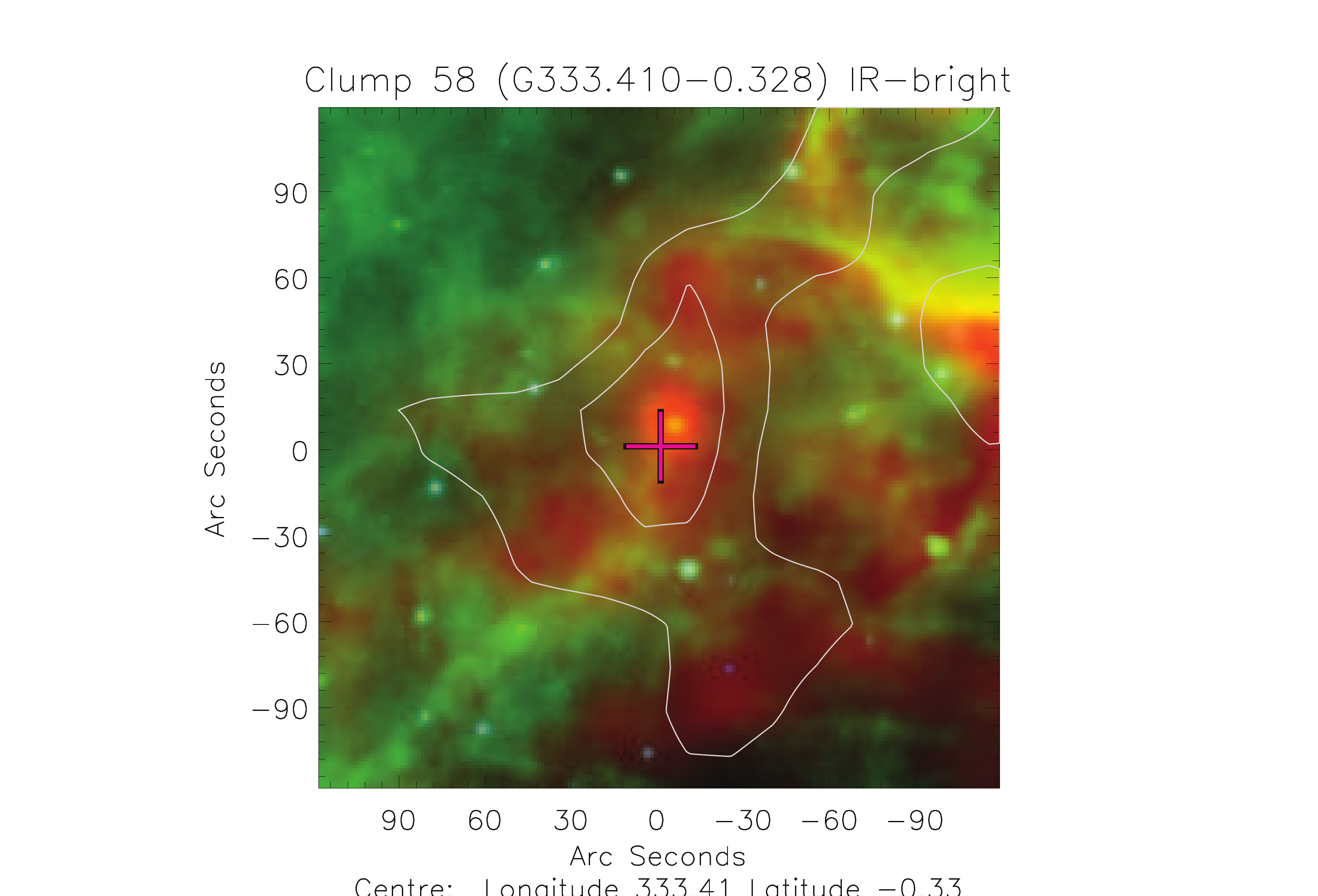}
\includegraphics[trim=100 20 210 40,clip,width=0.32\textwidth]{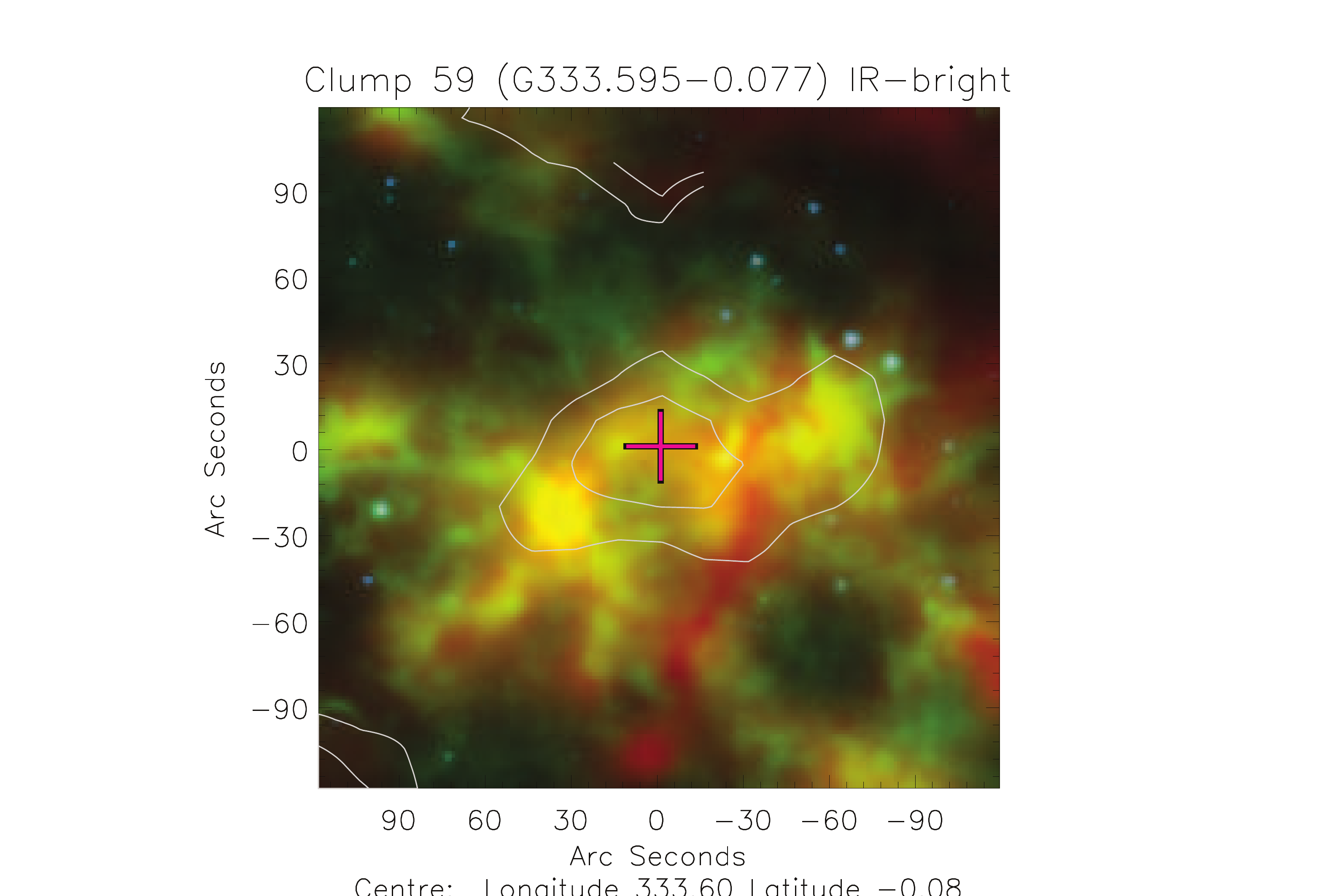}
\includegraphics[trim=100 20 210 40,clip,width=0.32\textwidth]{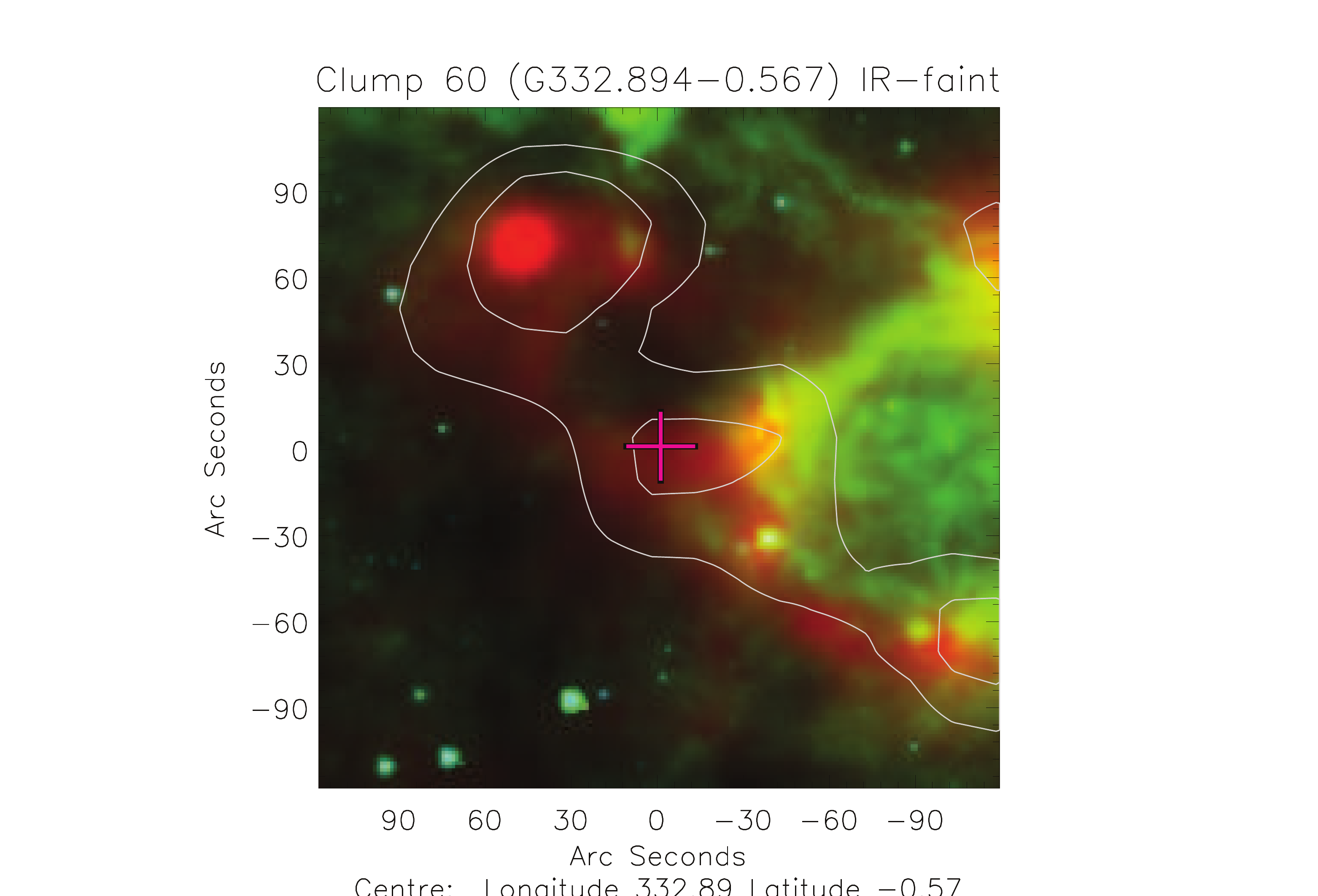}\\
\vspace*{0.3cm}\contcaption{}
\end{figure*}

\begin{figure*}
\centering
\includegraphics[trim=100 20 190 40,clip,width=0.32\textwidth]{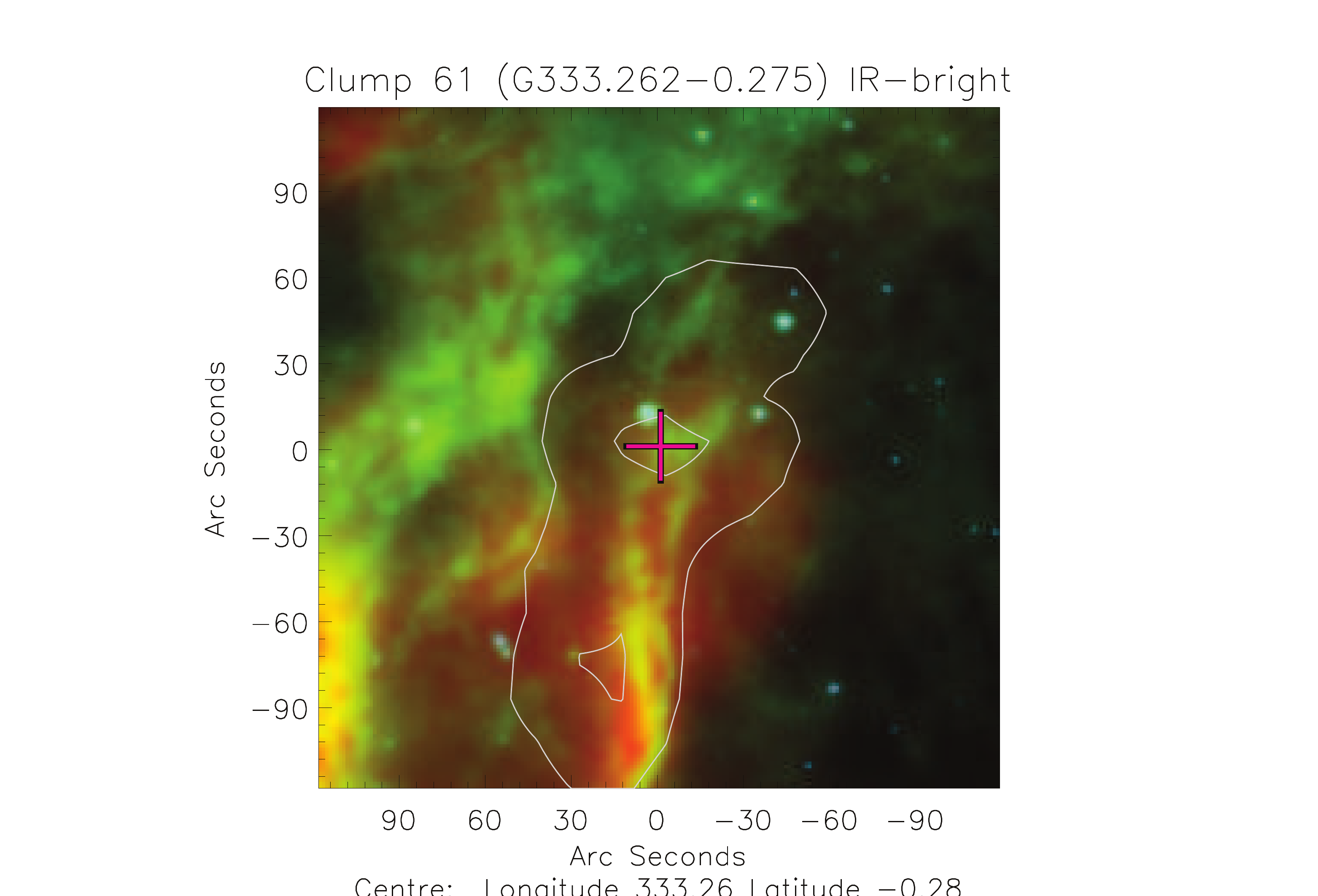}
\includegraphics[trim=100 20 190 40,clip,width=0.32\textwidth]{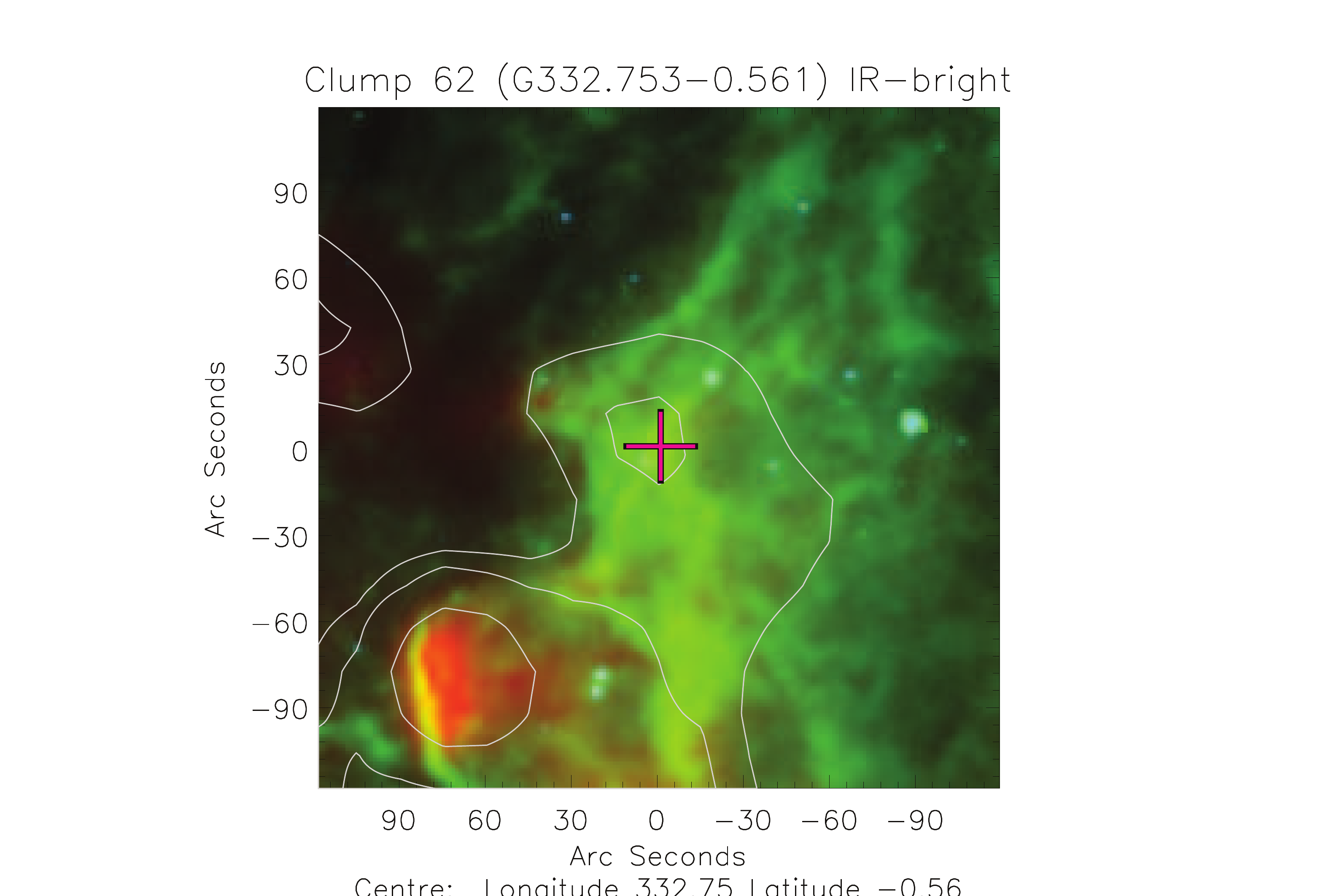}
\includegraphics[trim=100 20 190 40,clip,width=0.32\textwidth]{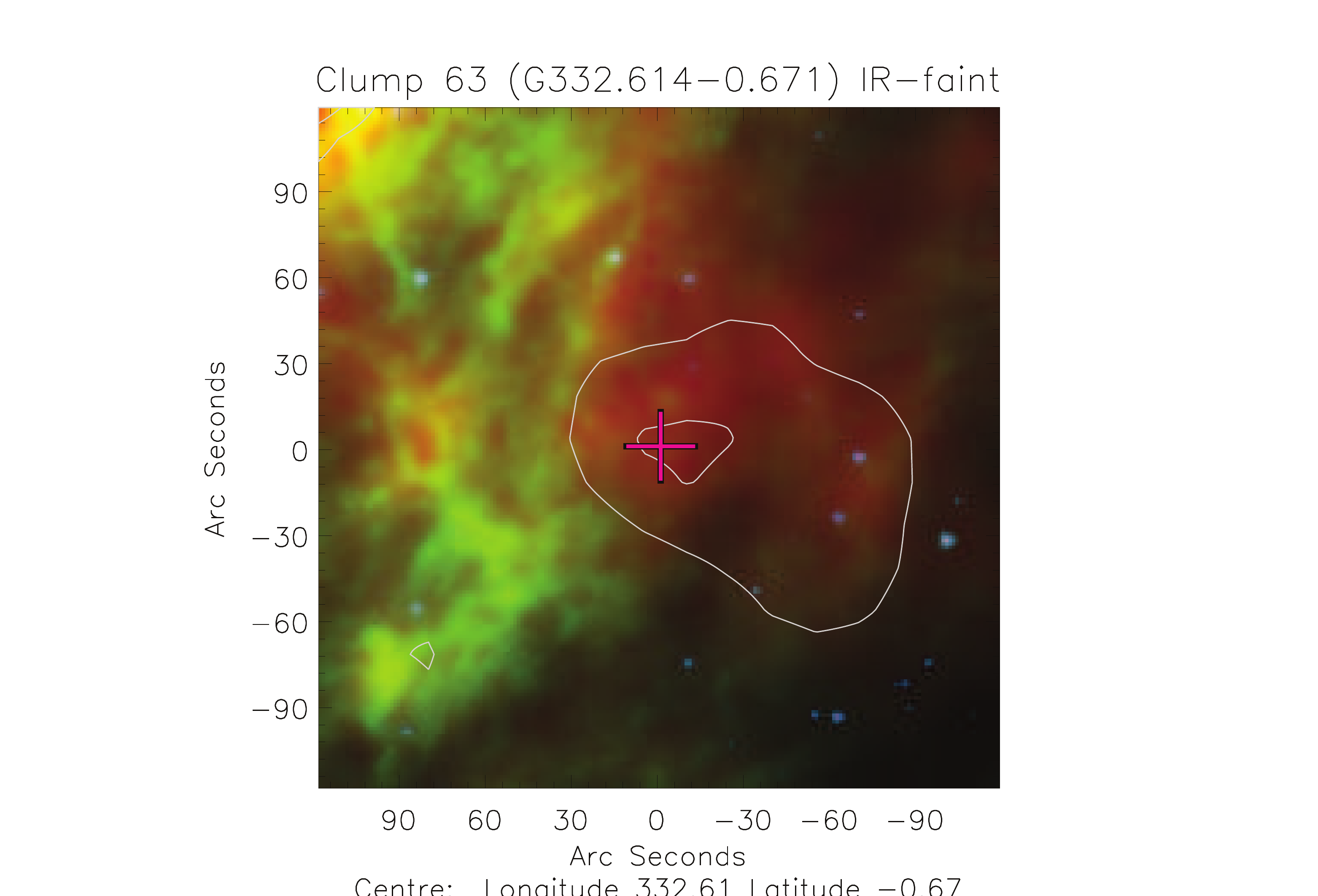}\\
\vspace*{0.3cm}\contcaption{}
\end{figure*}
\clearpage

\renewcommand\thefigure{B.\arabic{figure}}
\renewcommand\thefigure{B}
\setcounter{figure}{0}


\begin{figure*}
\centering
\includegraphics[width=0.49\textwidth, trim=-5 -30 -5 -30]{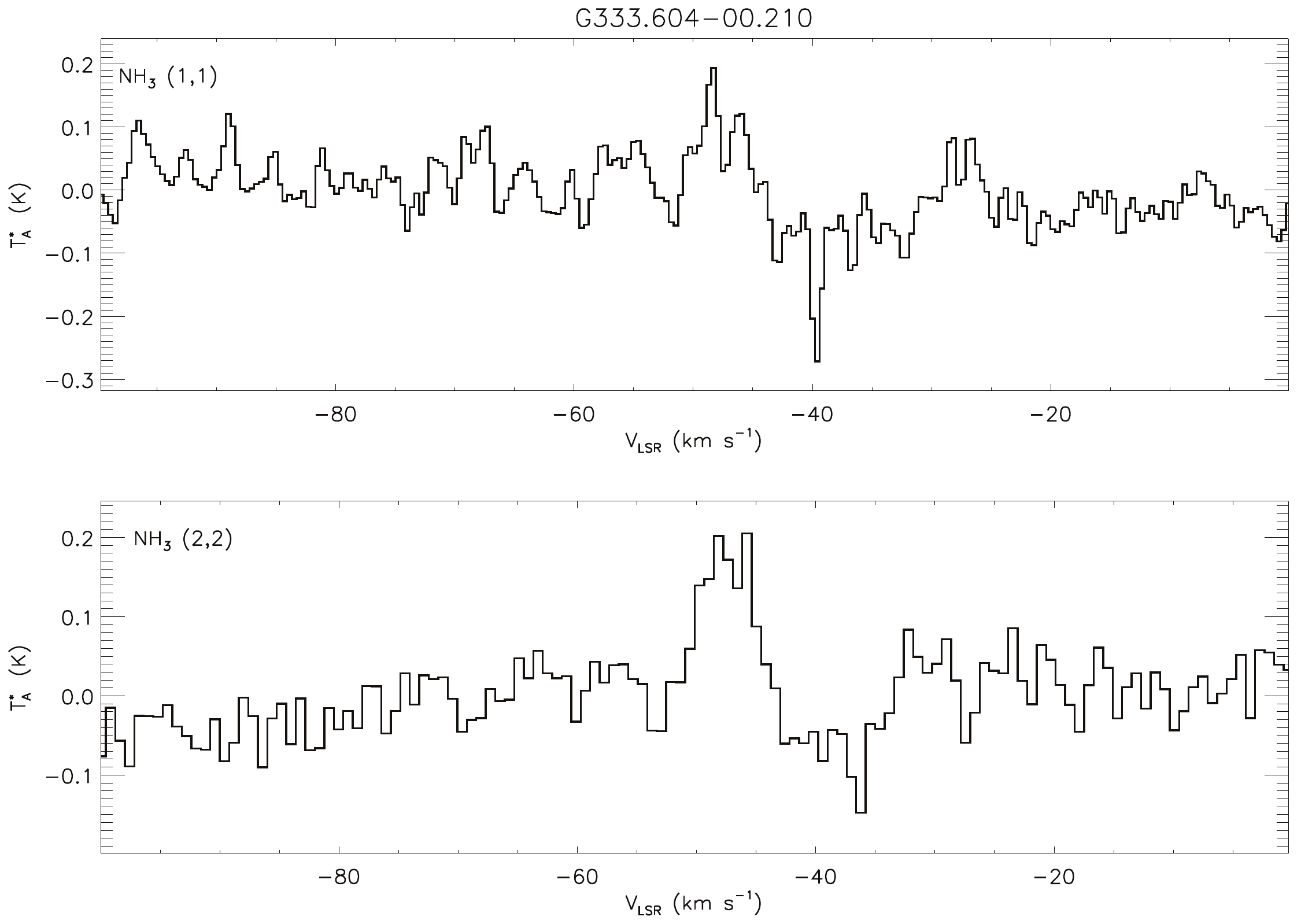}
\includegraphics[width=0.49\textwidth, trim=-5 -30 -5 -30]{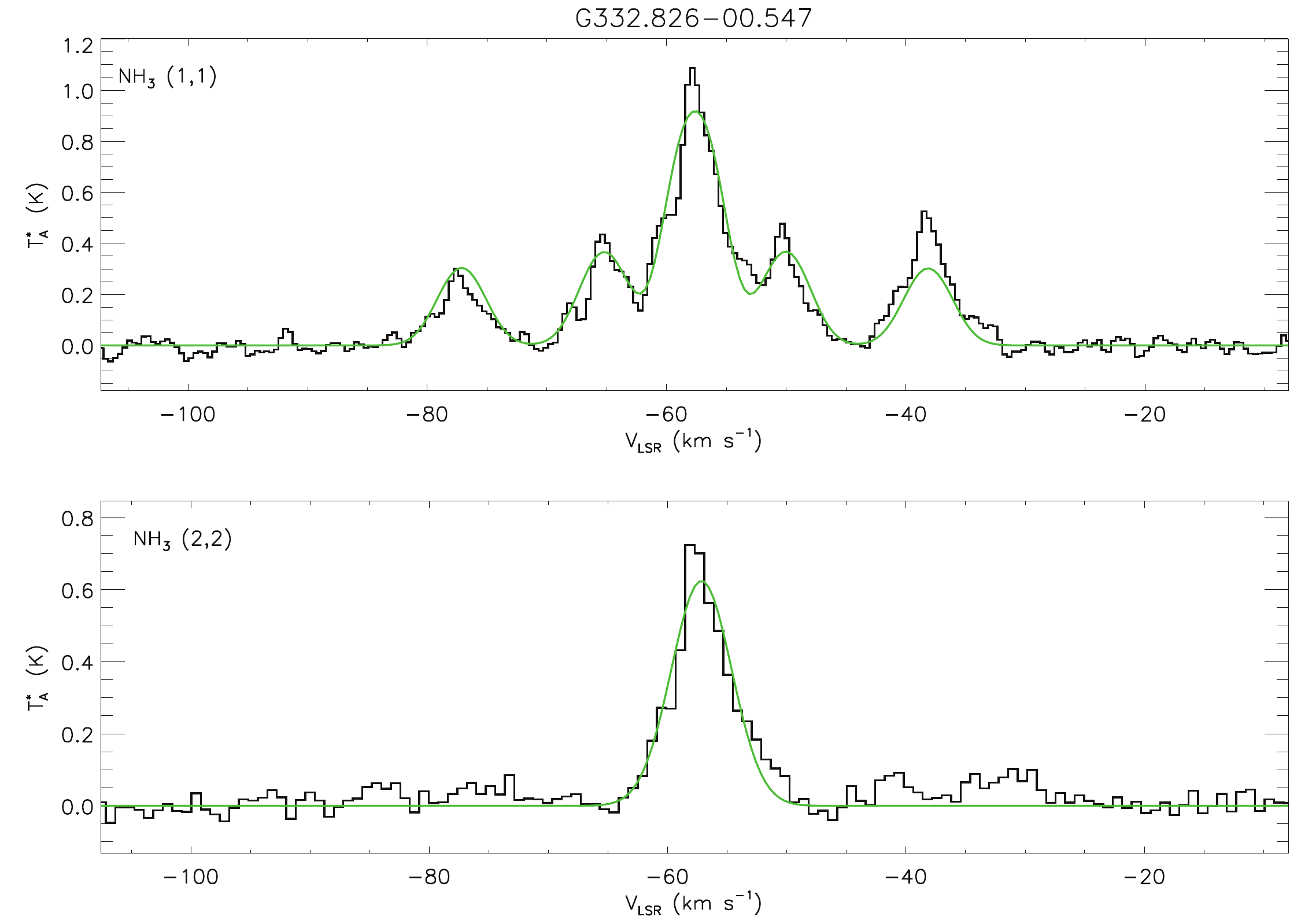}\\
\vspace*{0.5cm}
\includegraphics[width=0.49\textwidth, trim=-5 -30 -5 -30]{FIGURES/FigB/clump3_NH3}
\includegraphics[width=0.49\textwidth, trim=-5 -30 -5 -30]{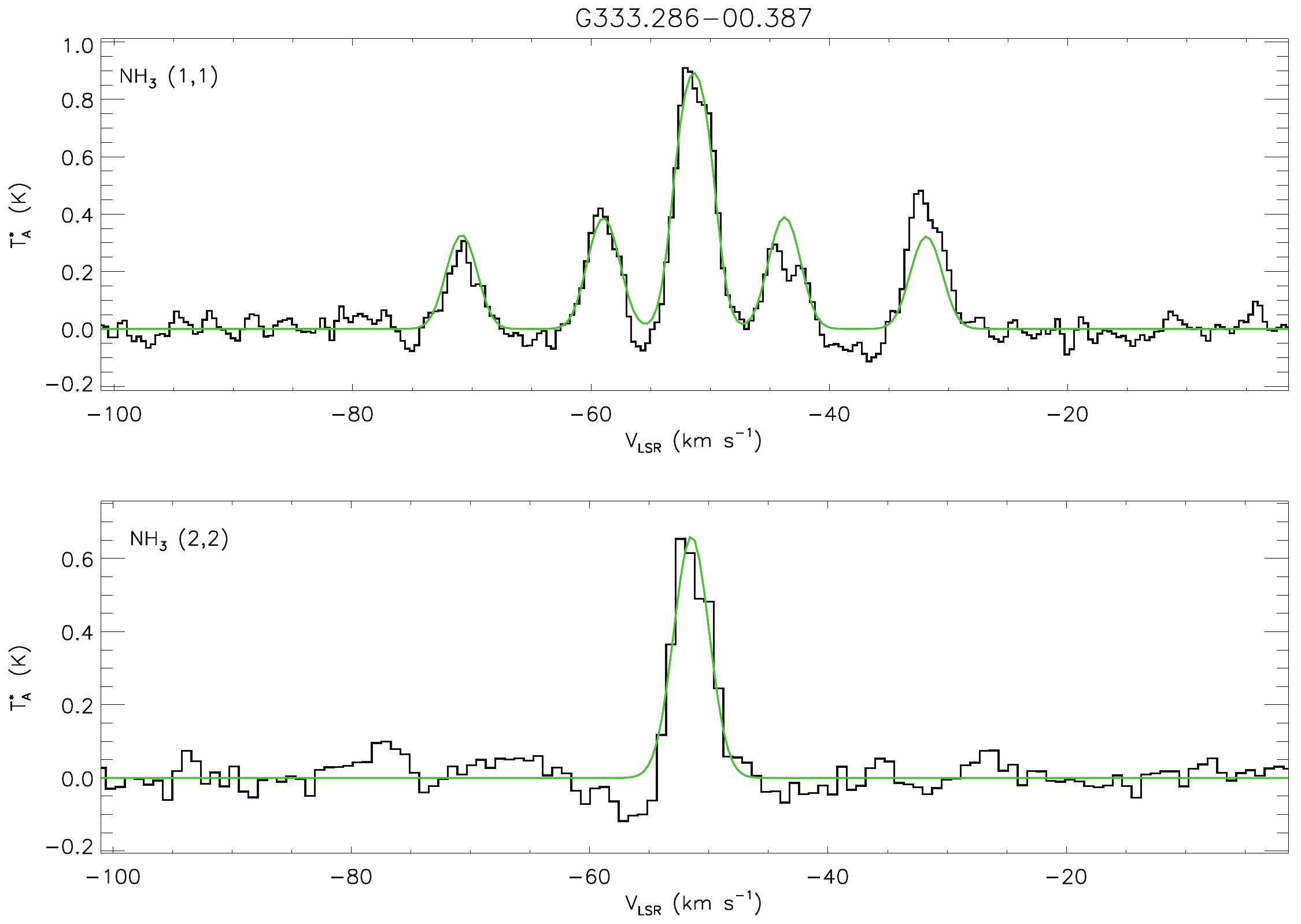}\\
\vspace*{0.5cm}
\includegraphics[width=0.49\textwidth, trim=-5 -30 -5 -30]{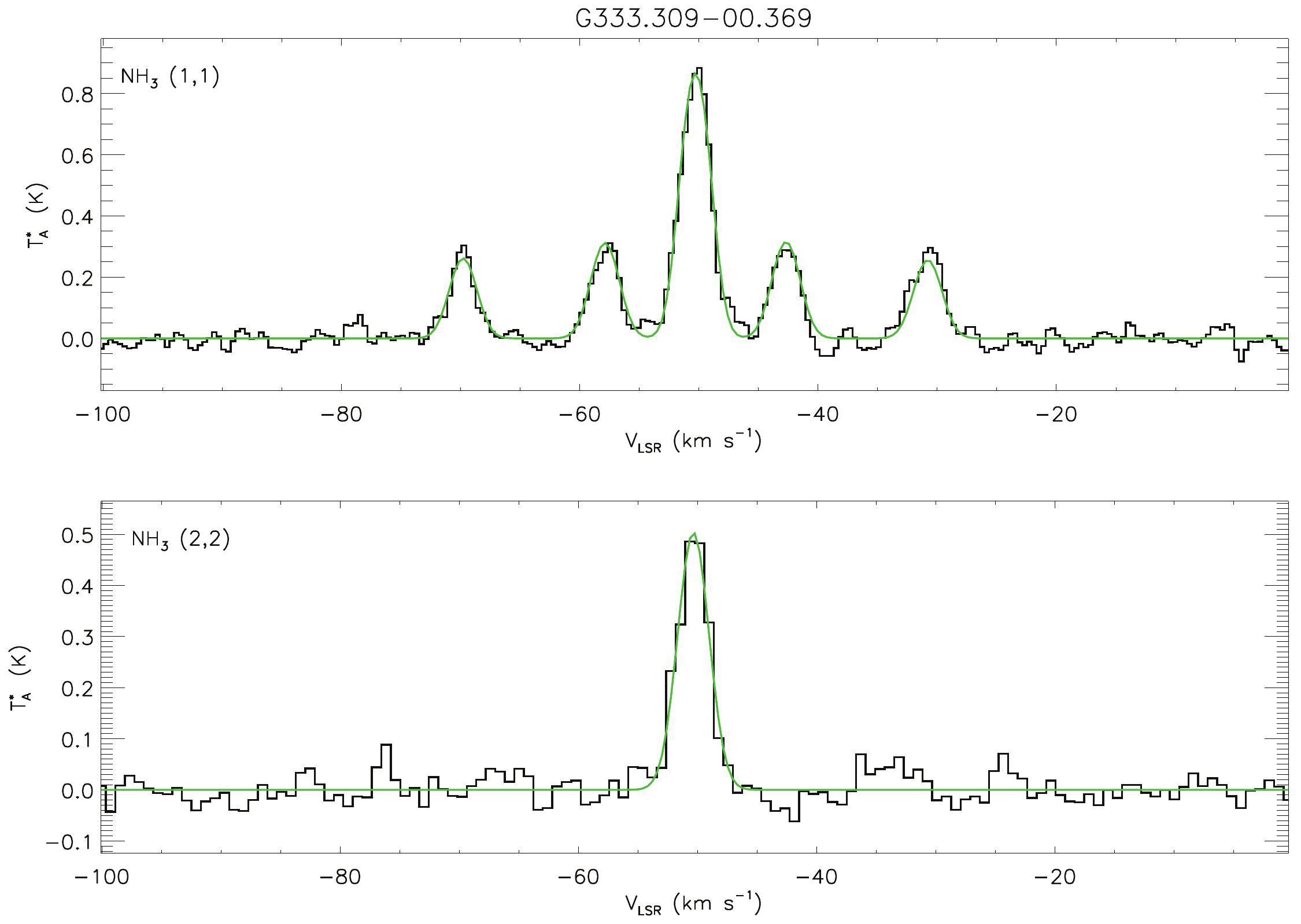}
\includegraphics[width=0.49\textwidth, trim=-5 -30 -5 -30]{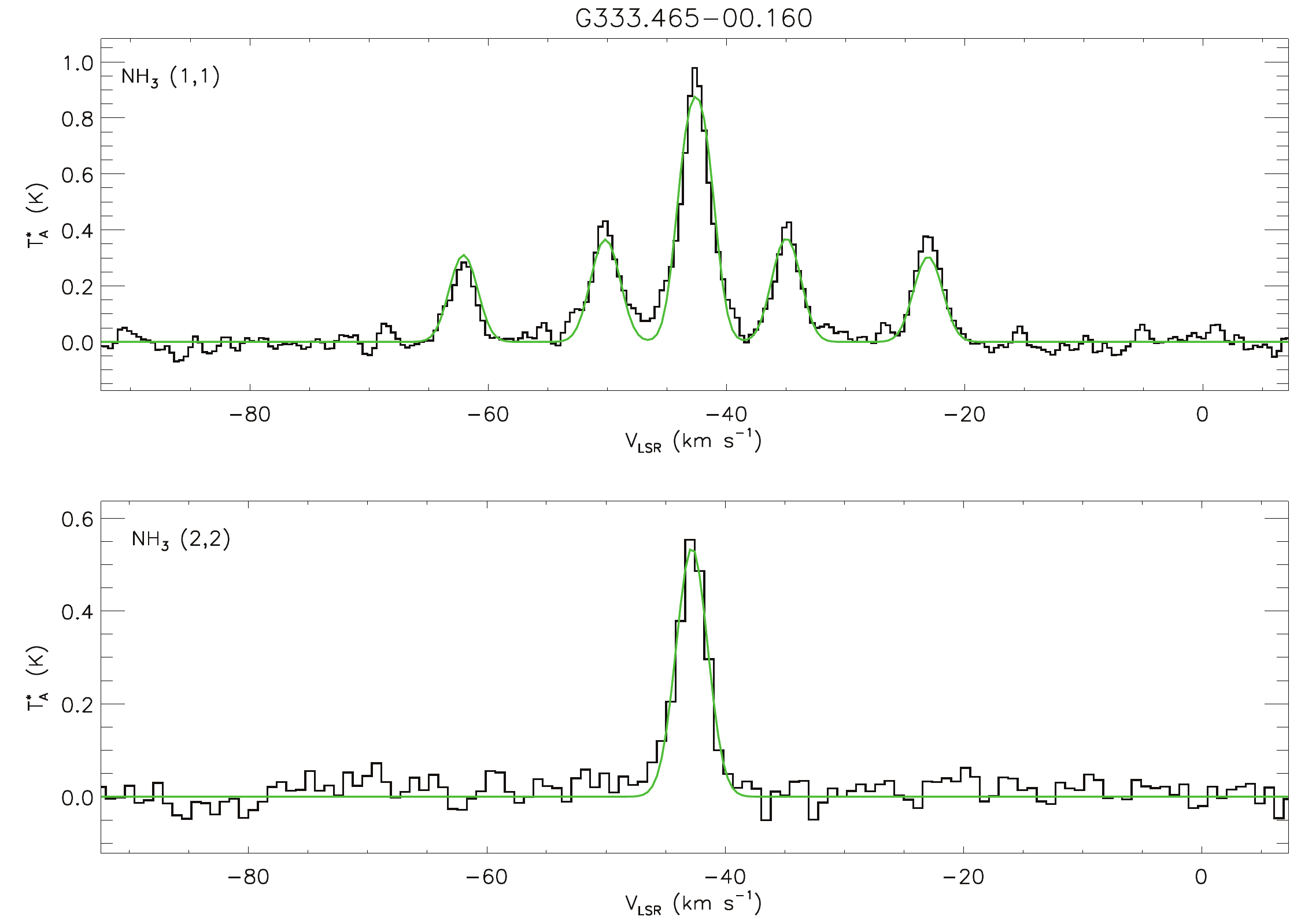}\\
\vspace*{0.5cm}
\caption{\ammonia\ emission detected towards the peak dust emission of each clump identified with {\sc clumpfind}. The \ammonia~\one\ and \two\ spectra are in the top and bottom panel, respectively. The model fits to these data are shown in green. Clumps~1-6 are shown l-r, t-b.}
\label{app:spectra}
\end{figure*}

\begin{figure*}
\centering
\includegraphics[width=0.49\textwidth, trim=-5 -30 -5 -30]{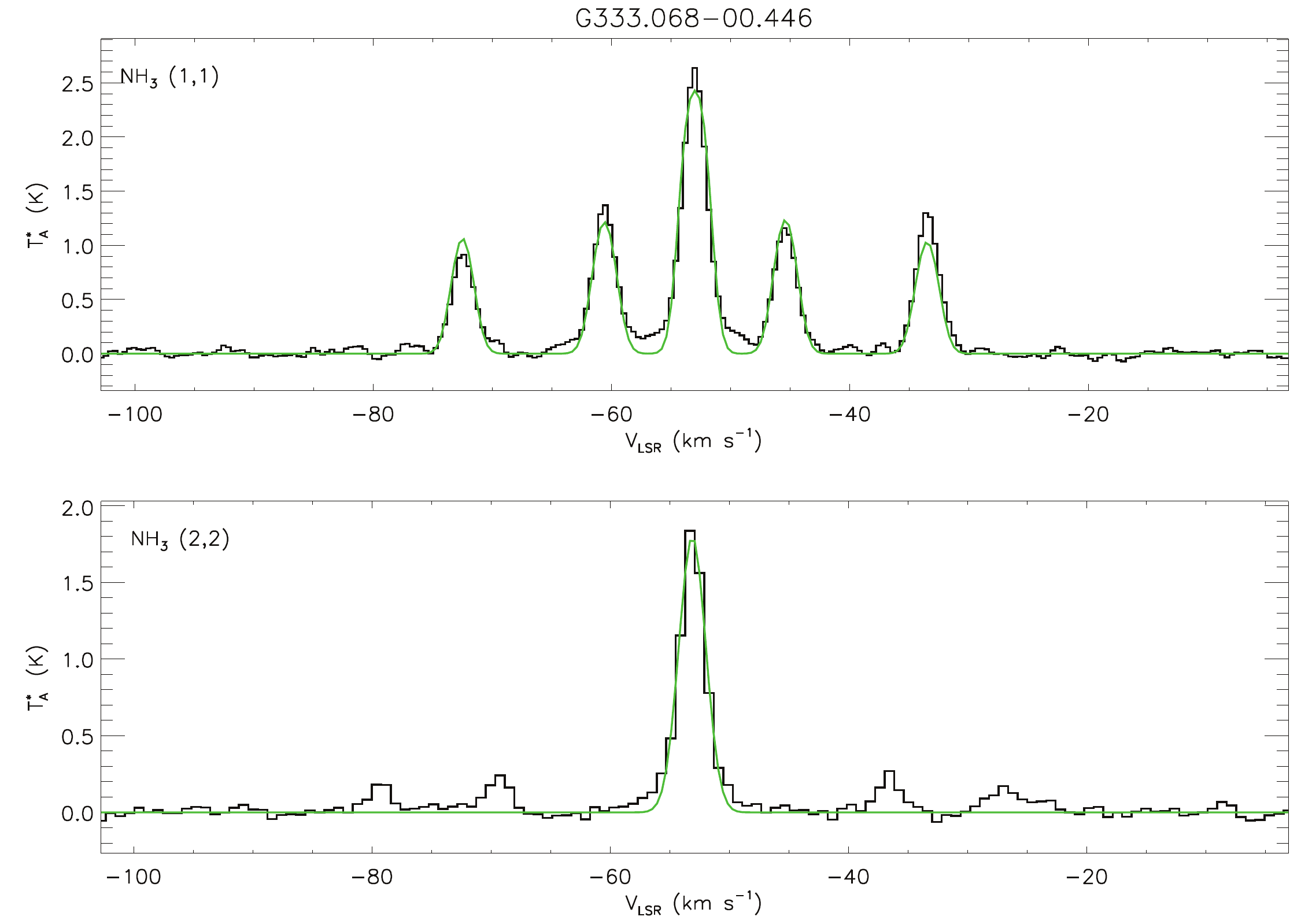}
\includegraphics[width=0.49\textwidth, trim=-5 -30 -5 -30]{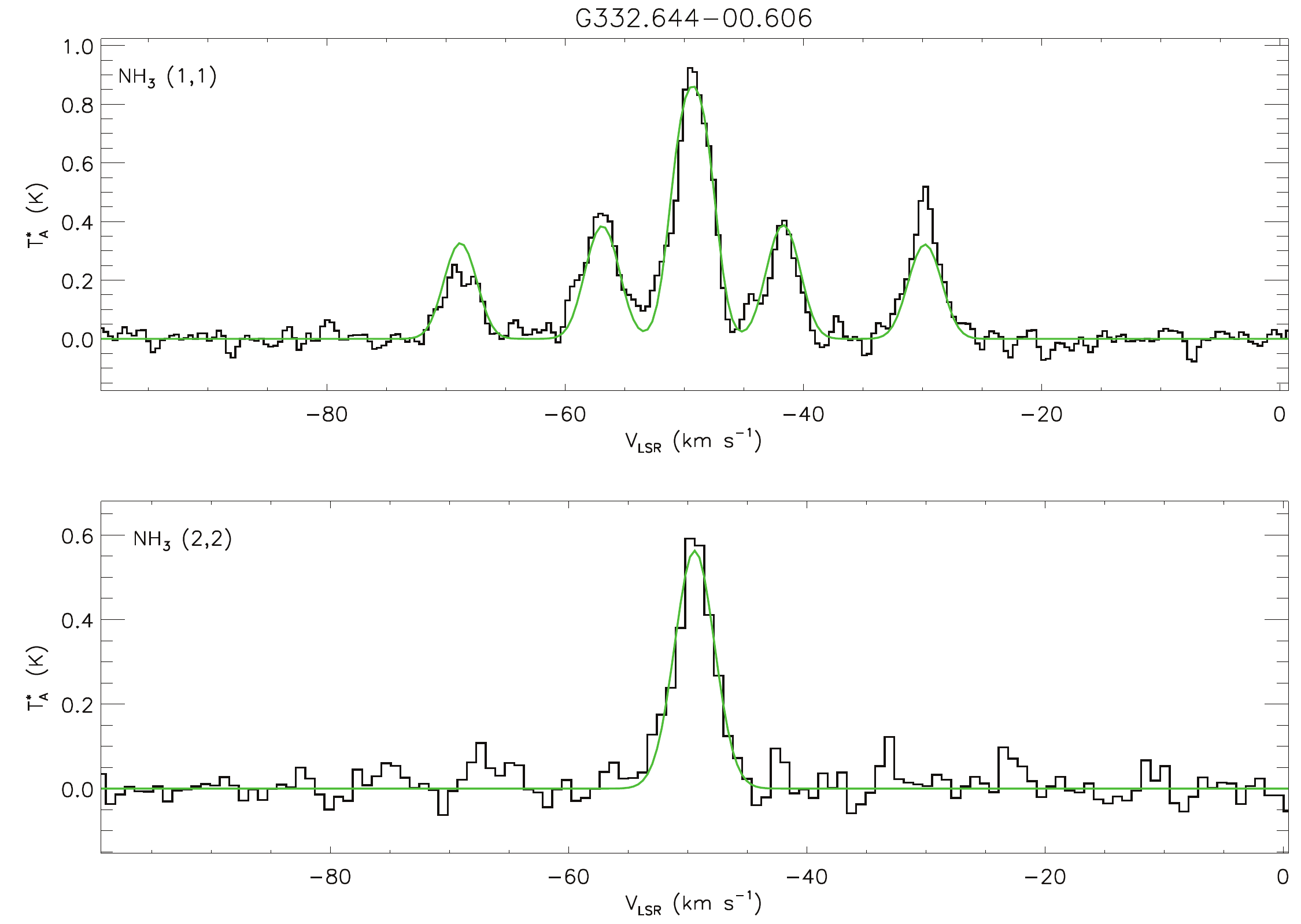}\\
\vspace*{0.5cm}
\includegraphics[width=0.49\textwidth, trim=-5 -30 -5 -30]{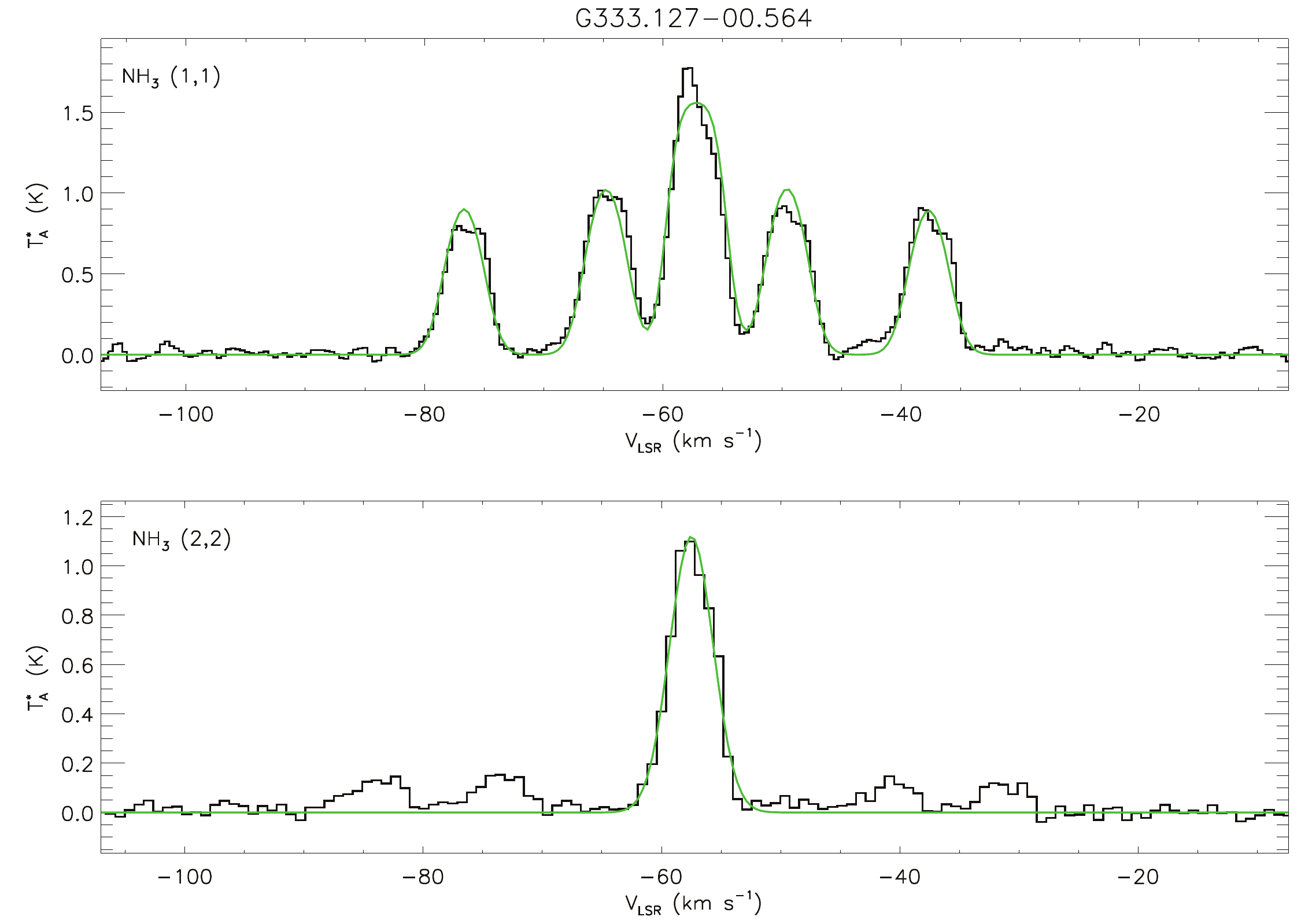}
\includegraphics[width=0.49\textwidth, trim=-5 -30 -5 -30]{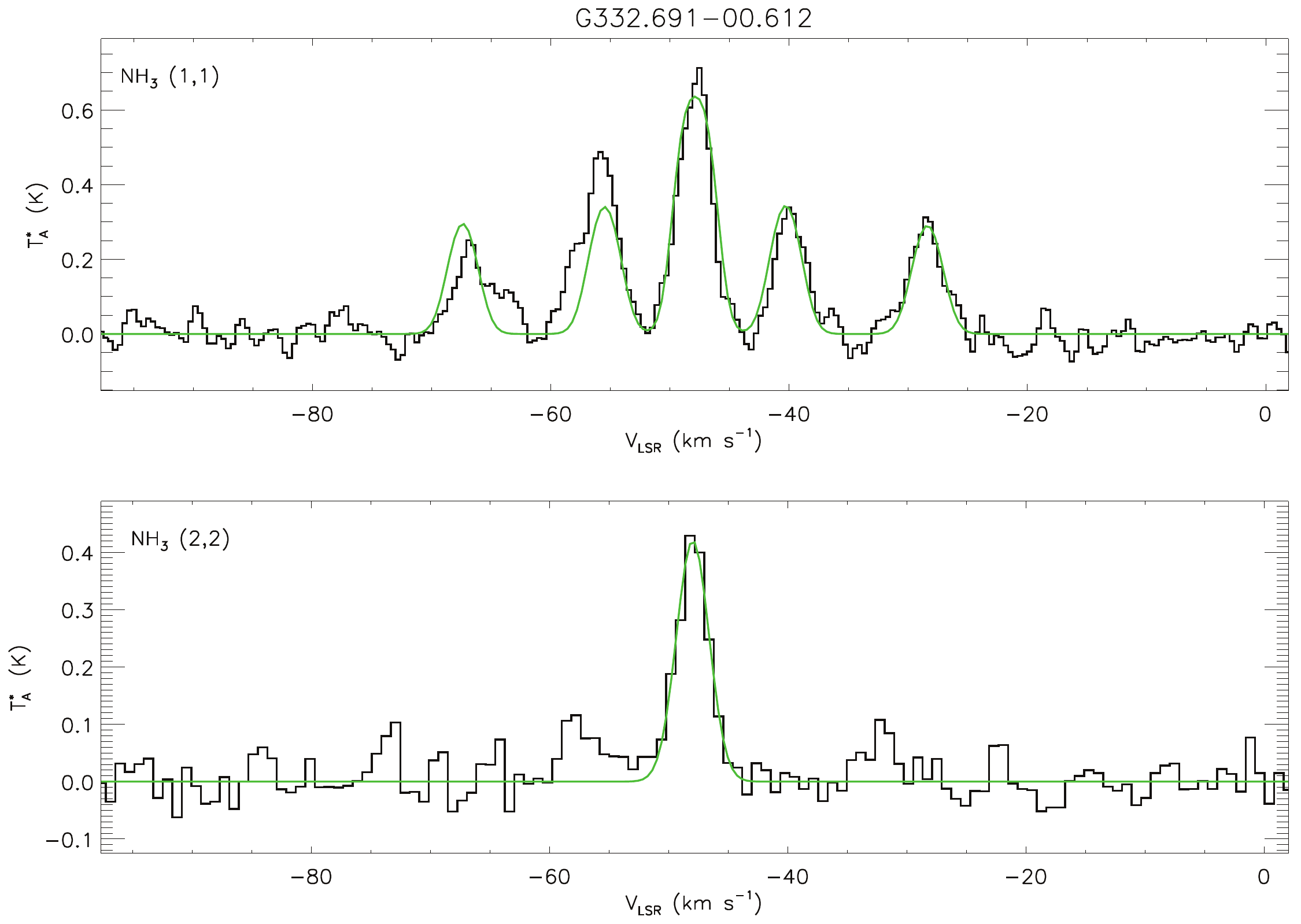}\\
\vspace*{0.5cm}
\includegraphics[width=0.49\textwidth, trim=-5 -30 -5 -30]{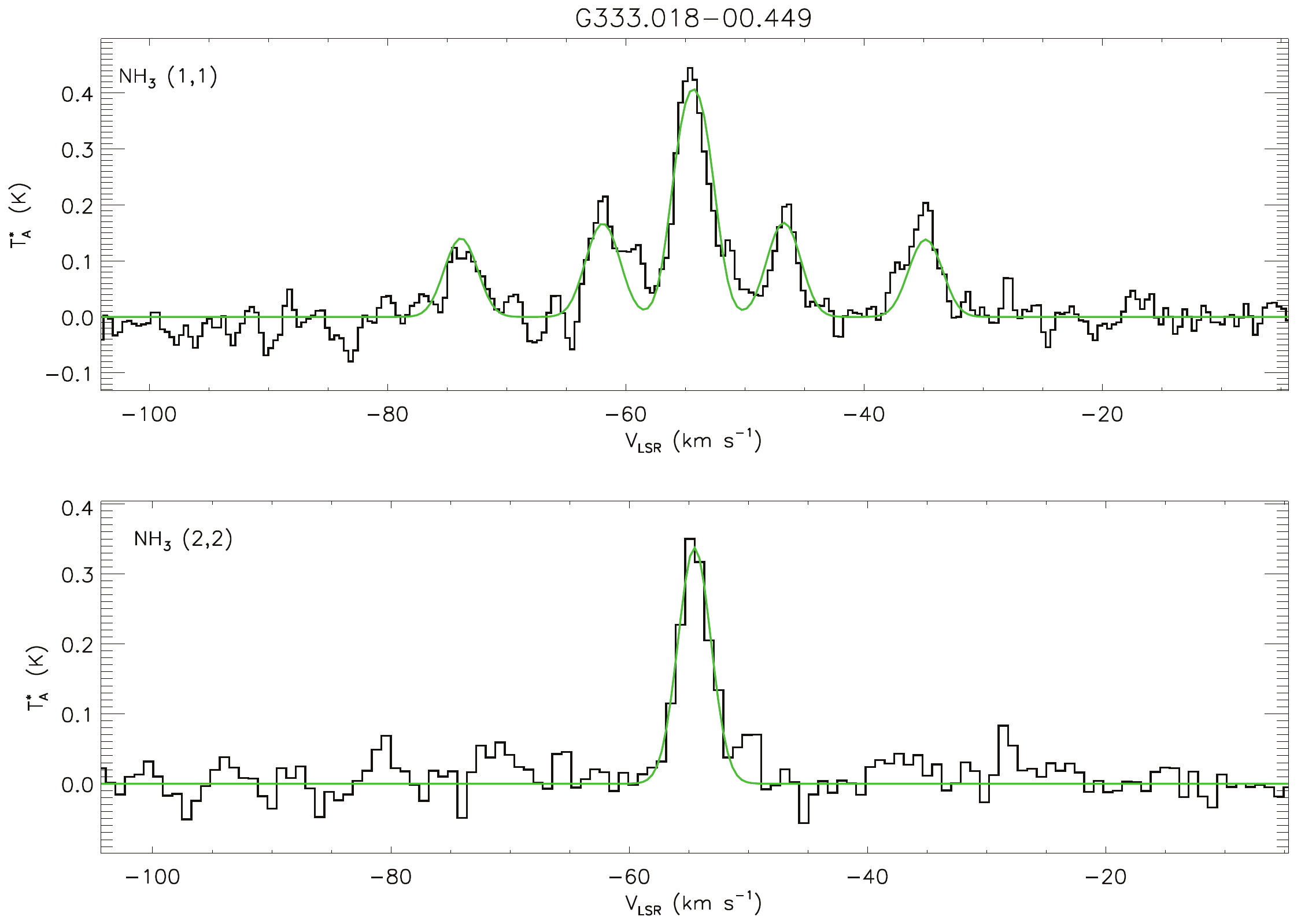}
\includegraphics[width=0.49\textwidth, trim=-5 -30 -5 -30]{FIGURES/FigB/clump12_NH3}\\
\vspace*{0.5cm}
\contcaption{Clumps~7-12  are shown l-r, t-b.}
\end{figure*}

\begin{figure*}
\centering
\includegraphics[width=0.49\textwidth, trim=-5 -30 -5 -30]{FIGURES/FigB/clump13_NH3}
\includegraphics[width=0.49\textwidth, trim=-5 -30 -5 -30]{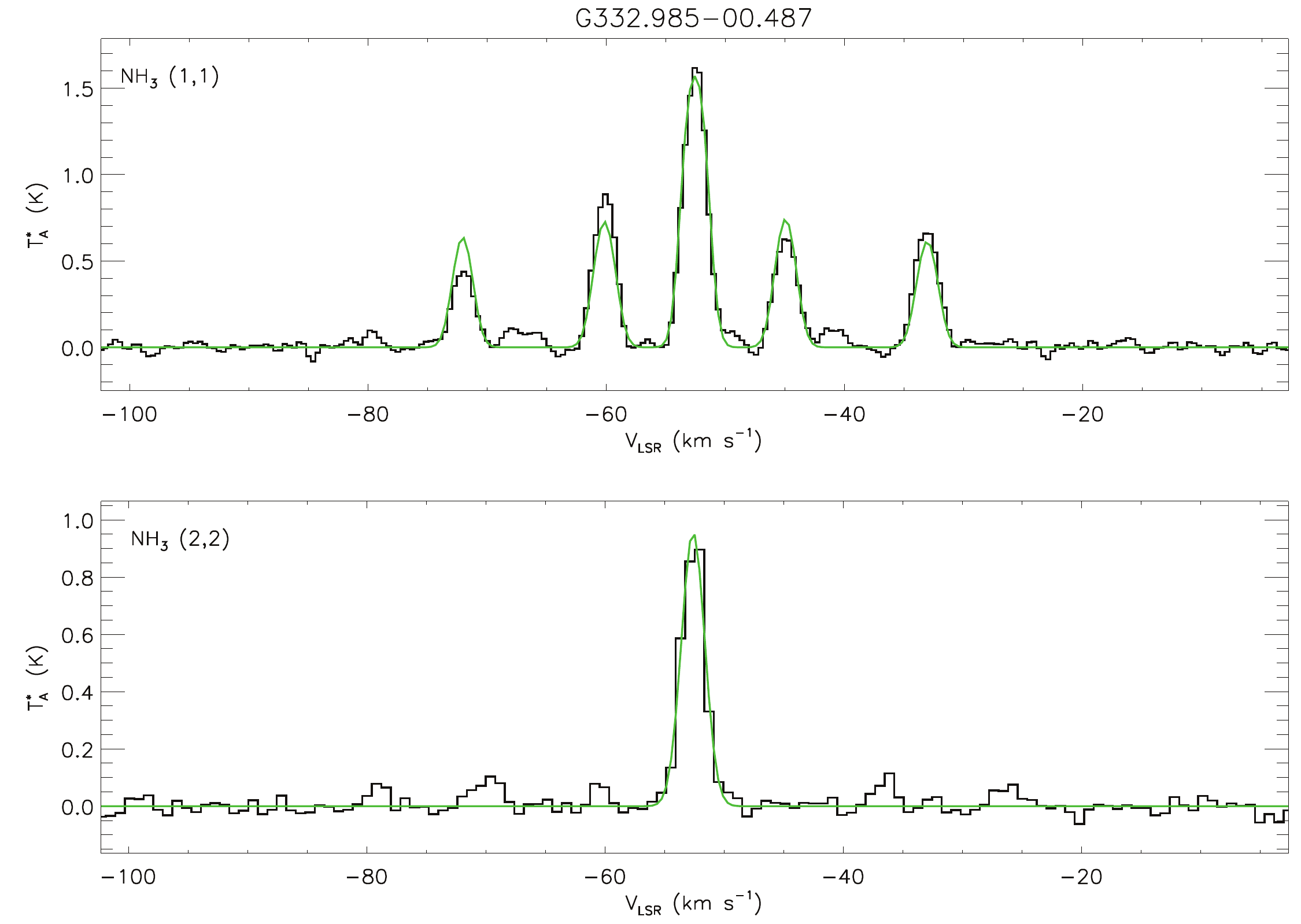}\\
\vspace*{0.5cm}
\includegraphics[width=0.49\textwidth, trim=-5 -30 -5 -30]{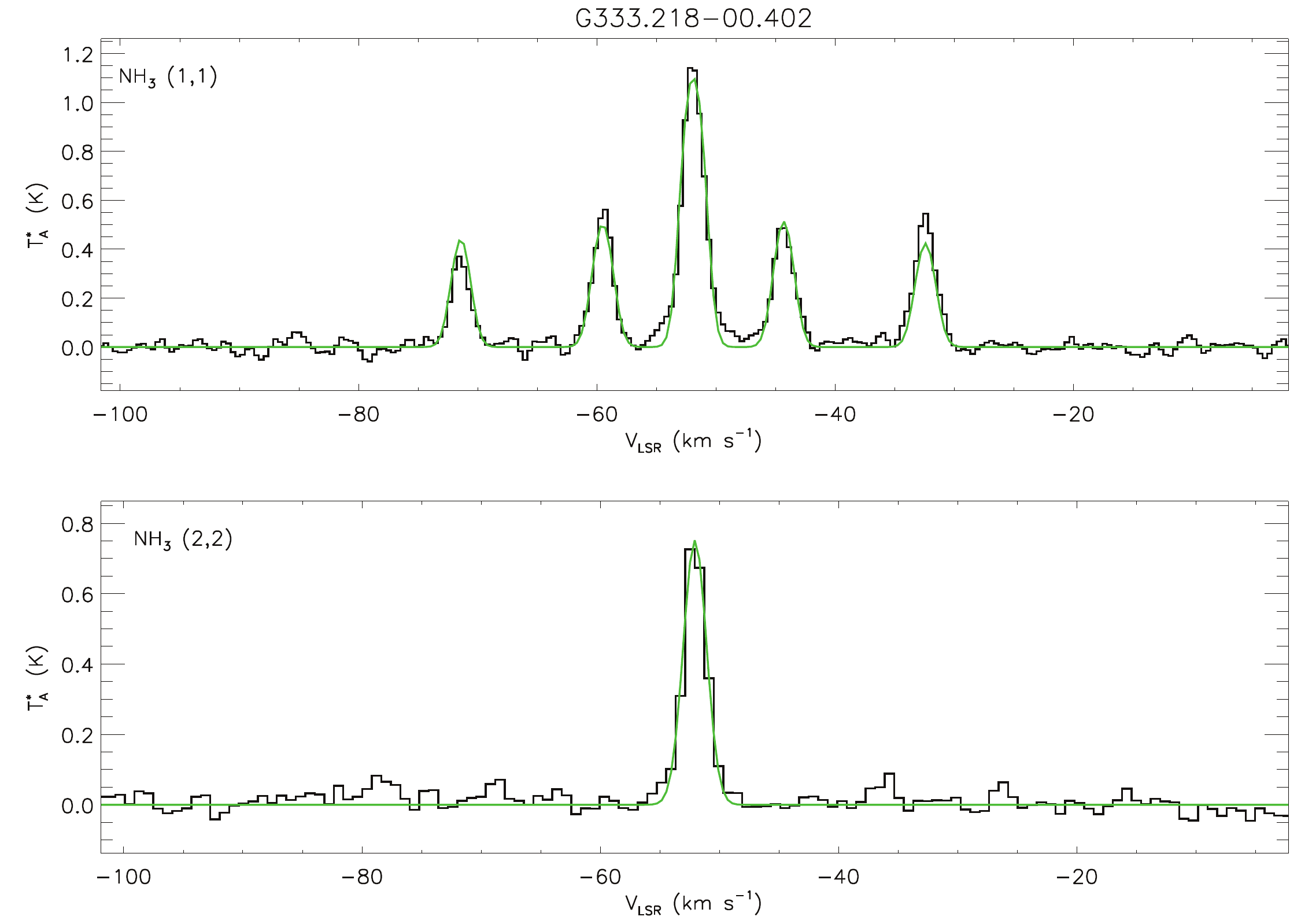}
\includegraphics[width=0.49\textwidth, trim=-5 -30 -5 -30]{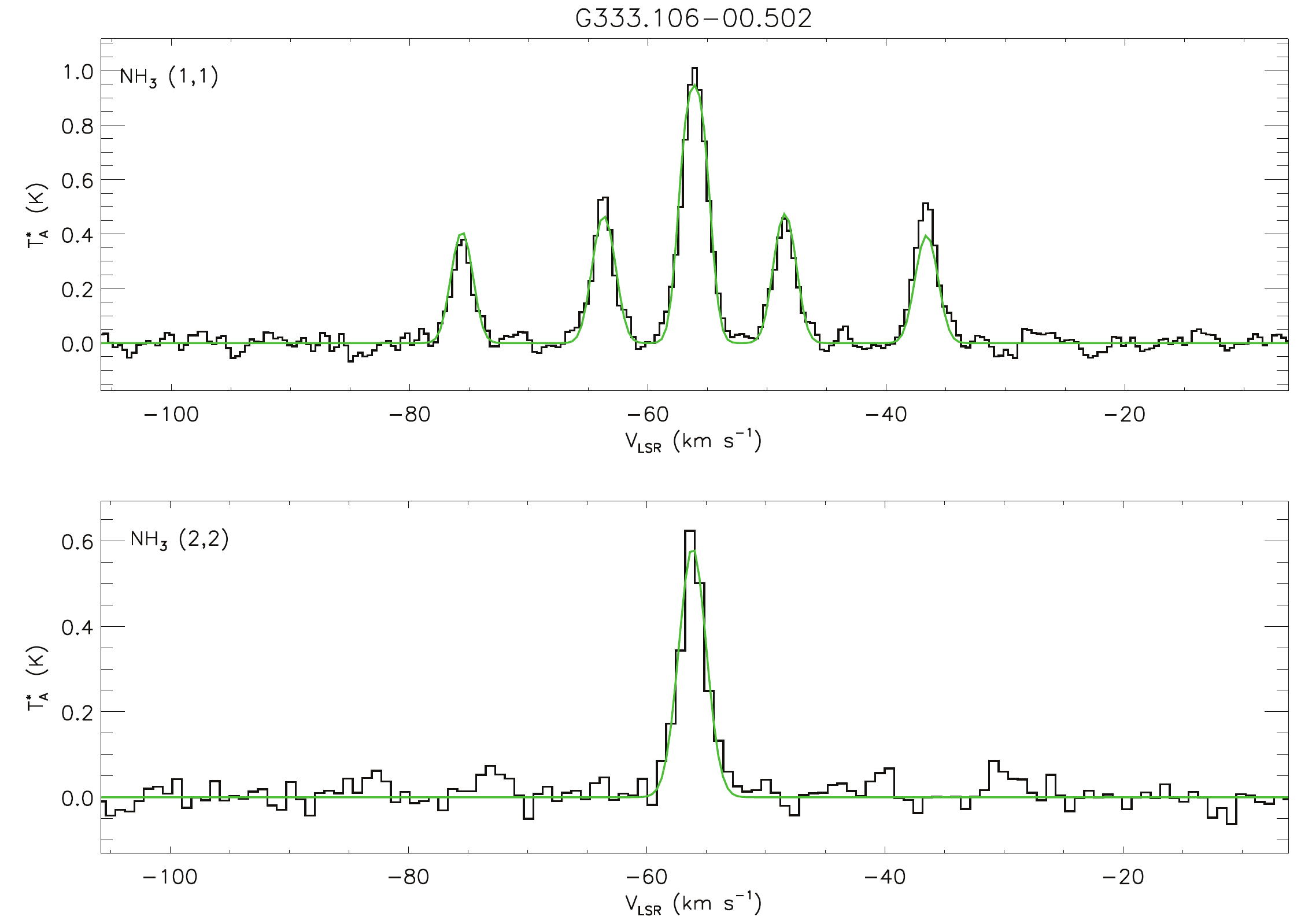}\\
\vspace*{0.5cm}
\includegraphics[width=0.49\textwidth, trim=-5 -30 -5 -30]{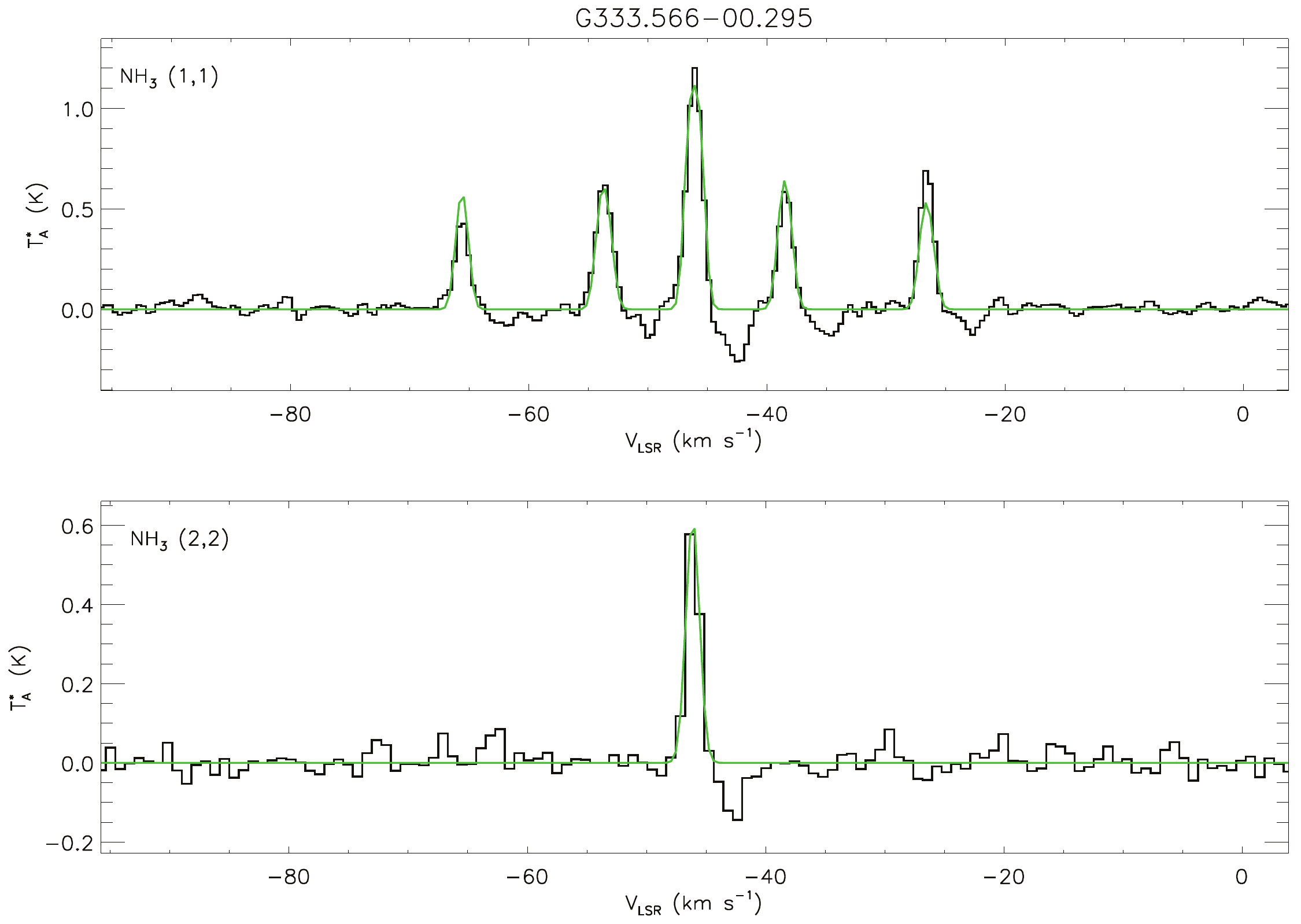}
\includegraphics[width=0.49\textwidth, trim=-5 -30 -5 -30]{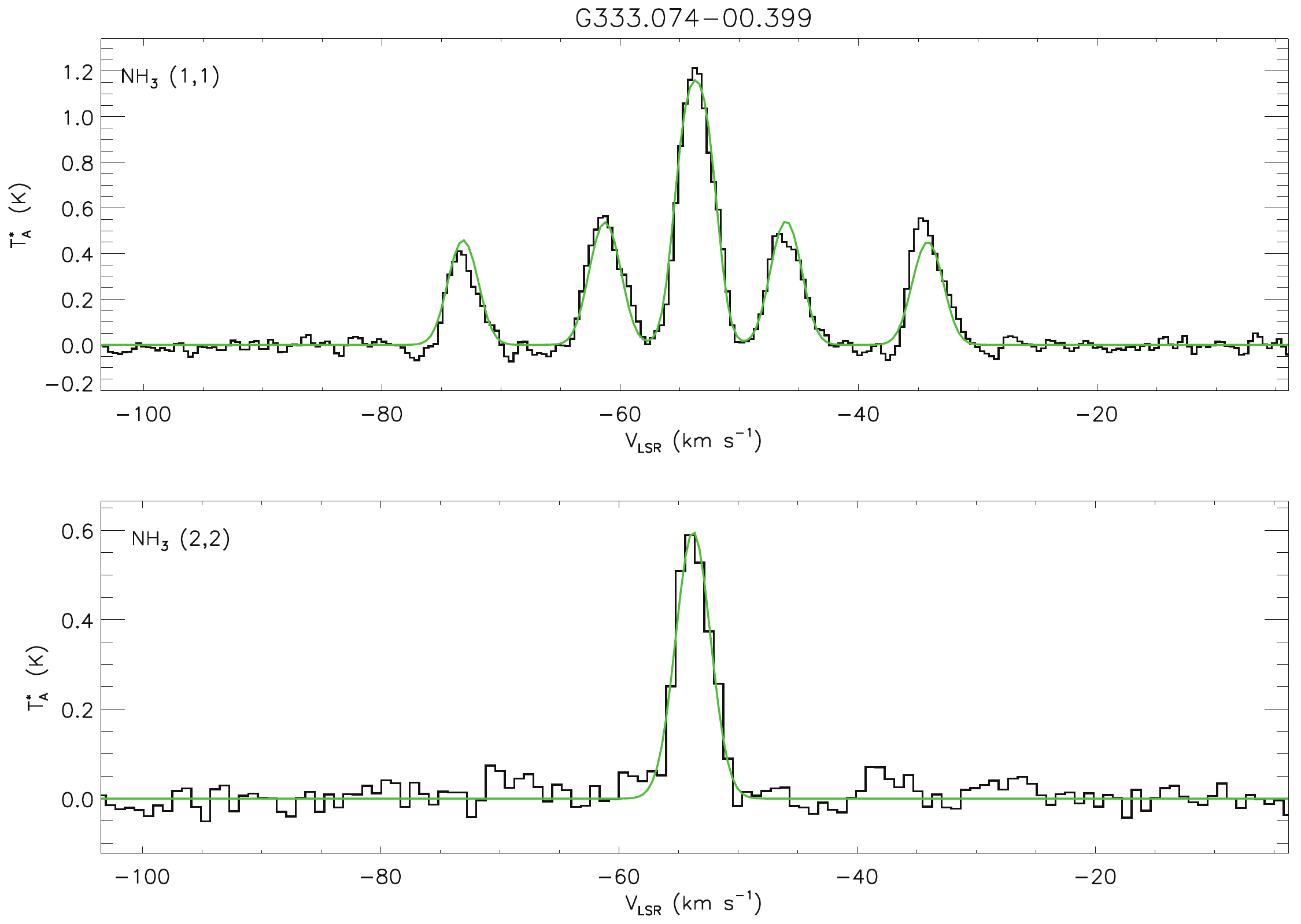}\\
\vspace*{0.5cm}
\contcaption{Clumps~13-18 are shown l-r, t-b.}
\end{figure*}

\begin{figure*}
\centering
\includegraphics[width=0.49\textwidth, trim=-5 -30 -5 -30]{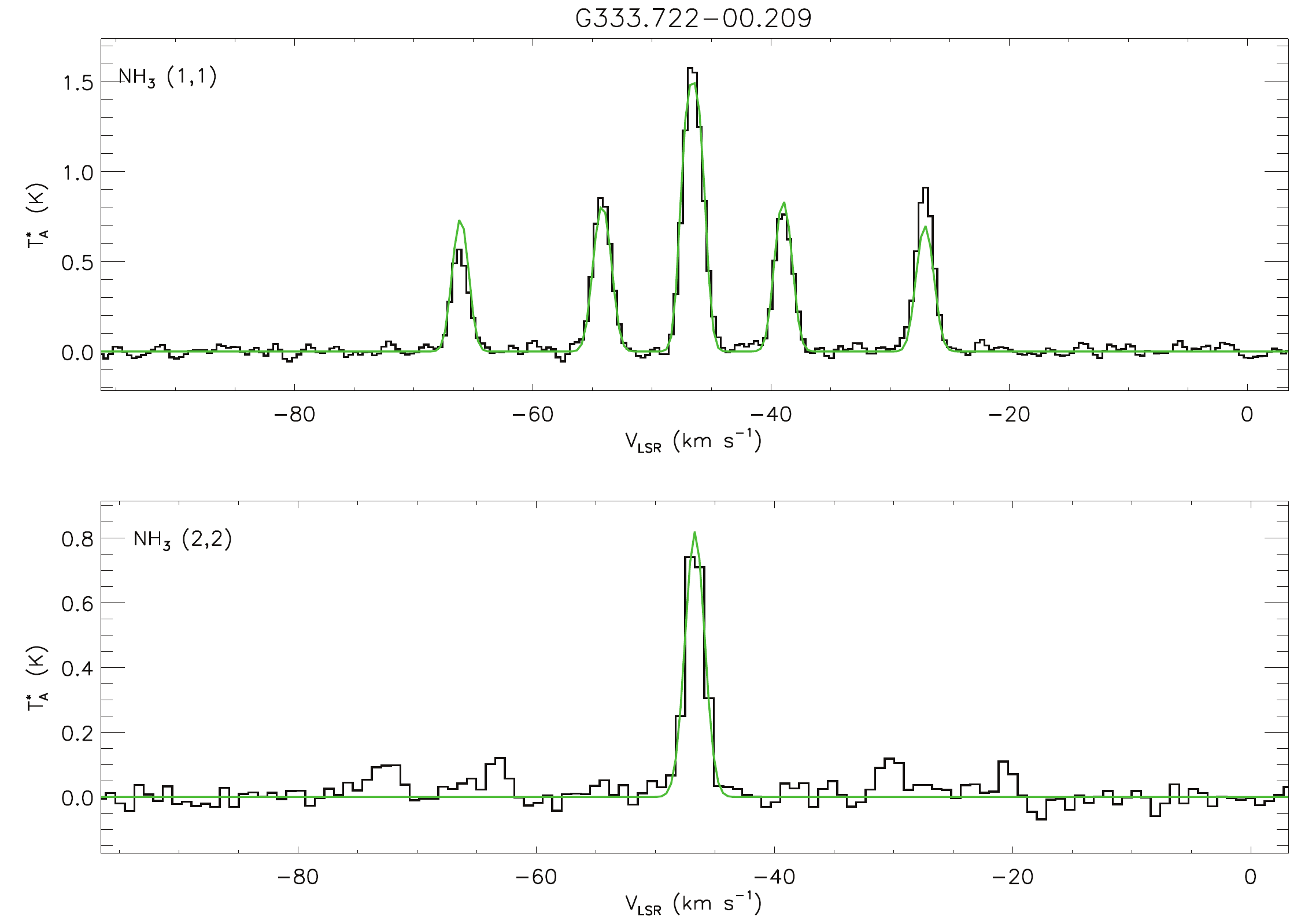}
\includegraphics[width=0.49\textwidth, trim=-5 -30 -5 -30]{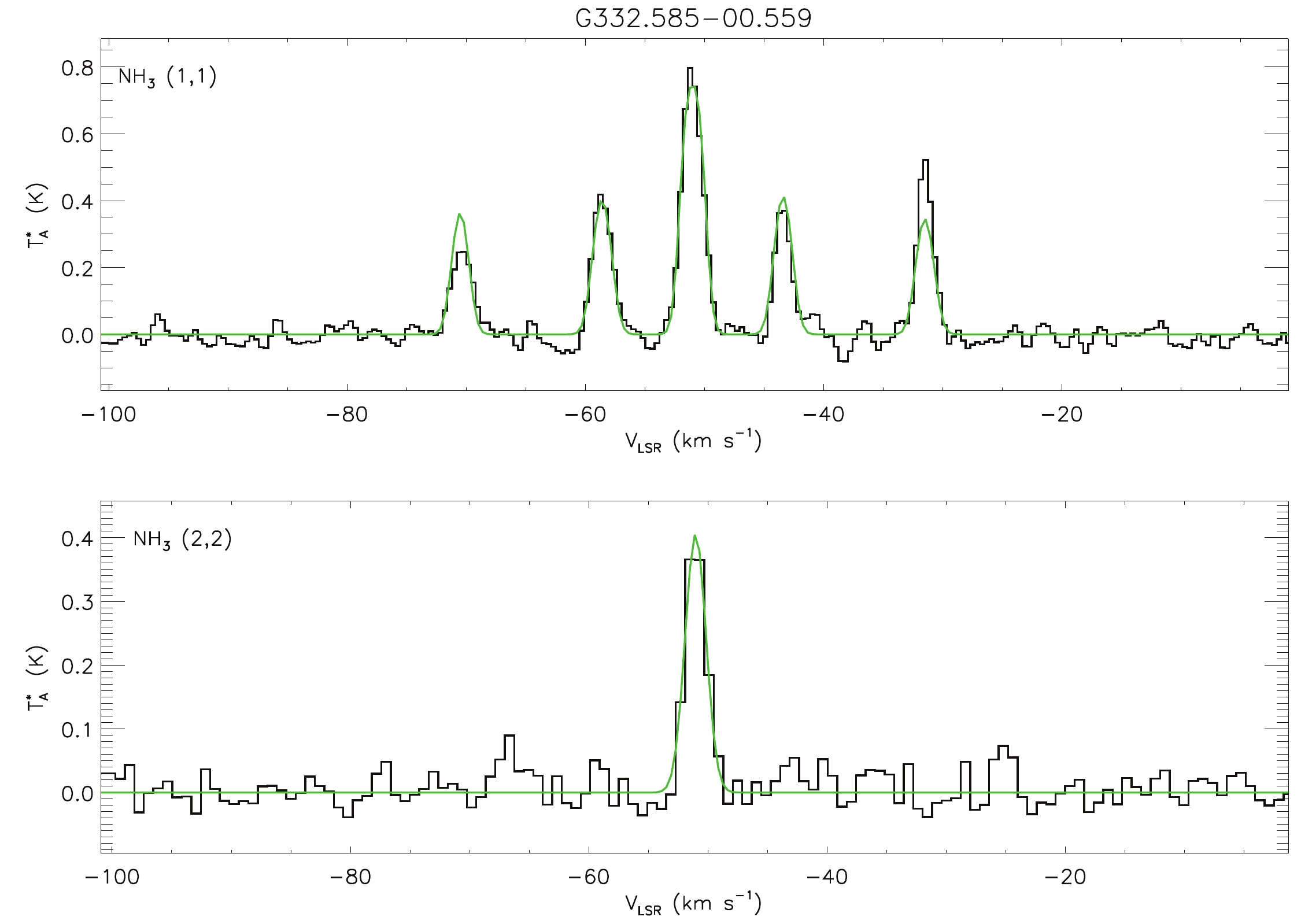}\\
\vspace*{0.5cm}
\includegraphics[width=0.49\textwidth, trim=-5 -30 -5 -30]{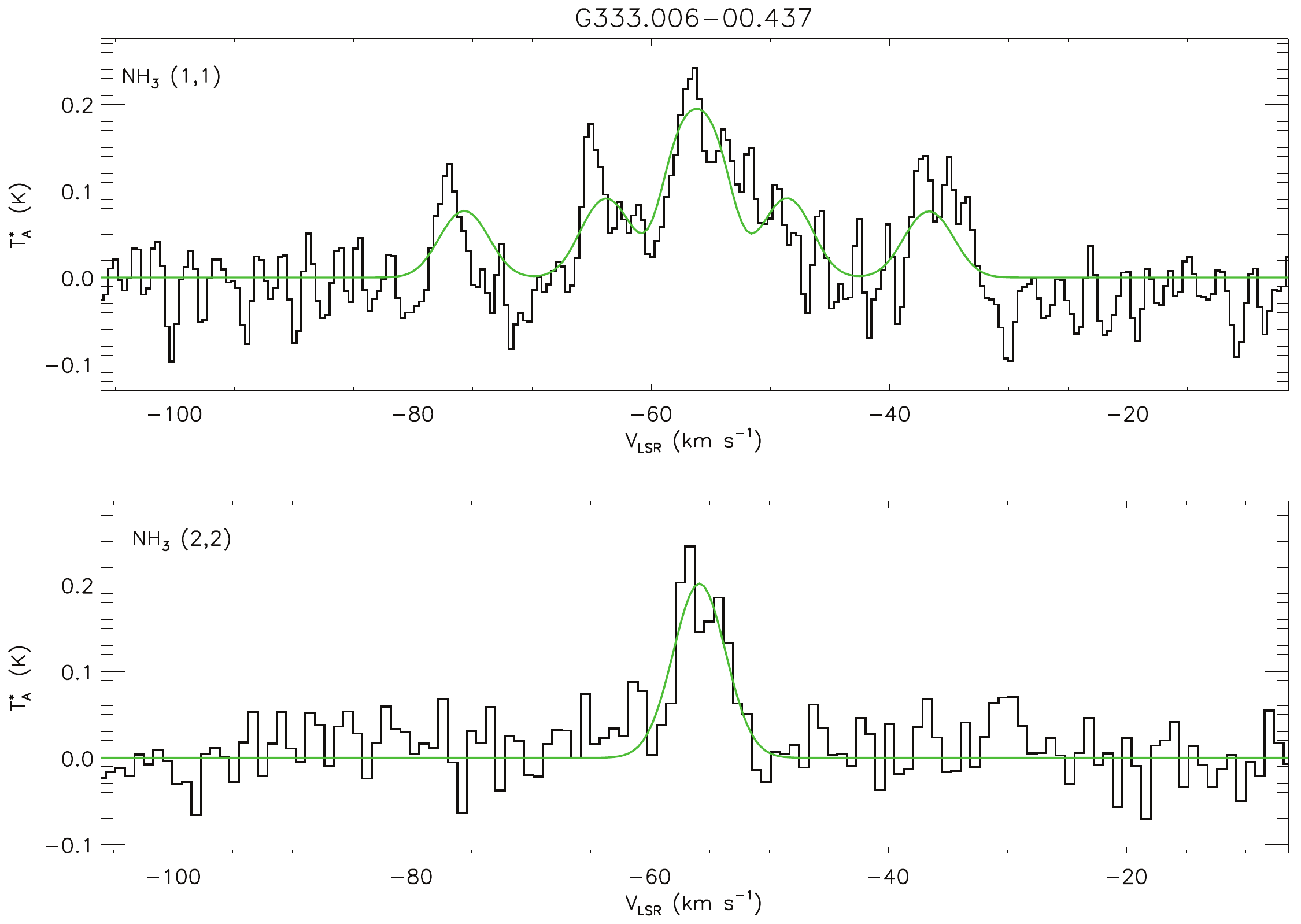}
\includegraphics[width=0.49\textwidth, trim=-5 -30 -5 -30]{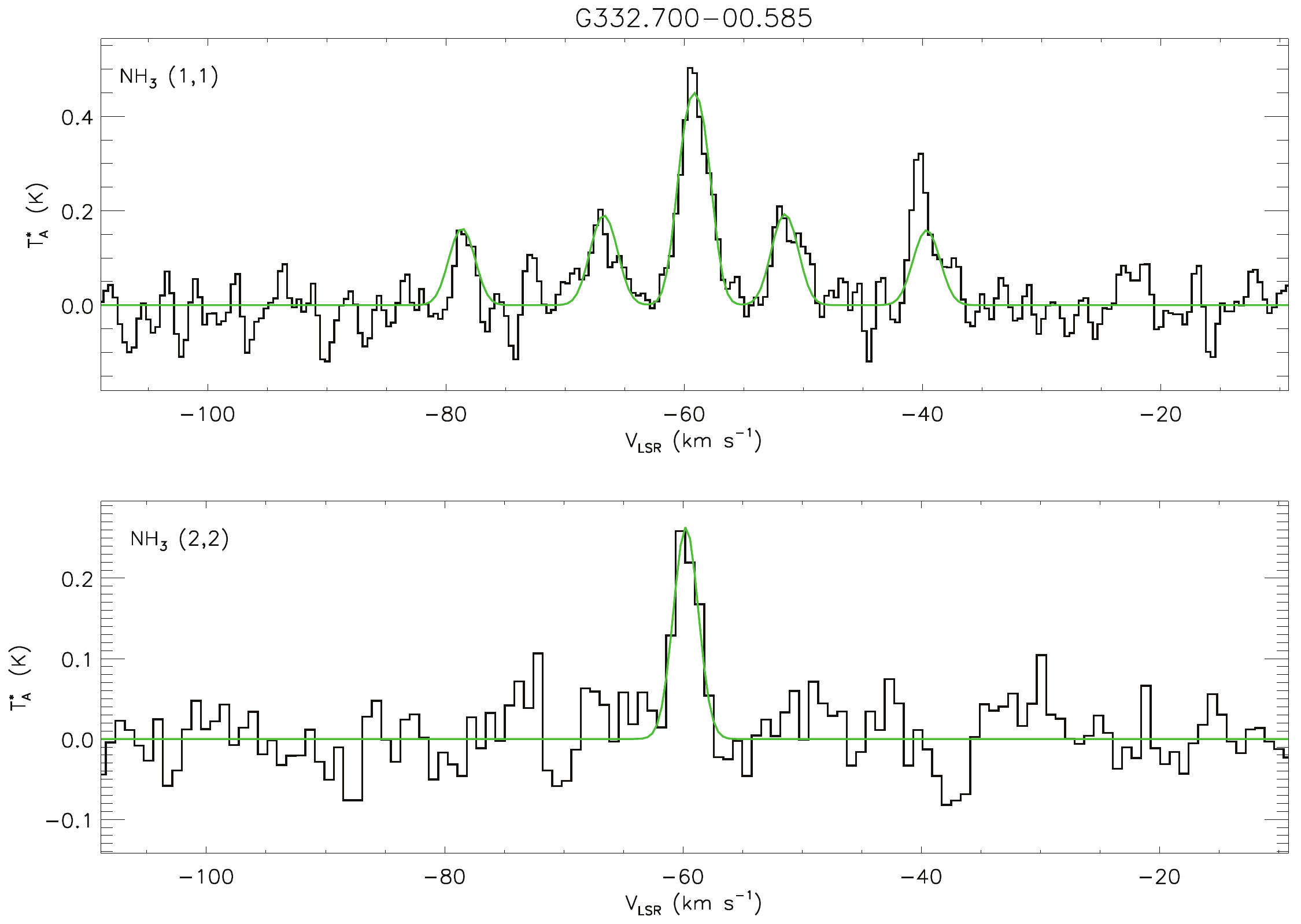}\\
\vspace*{0.5cm}
\includegraphics[width=0.49\textwidth, trim=-5 -30 -5 -30]{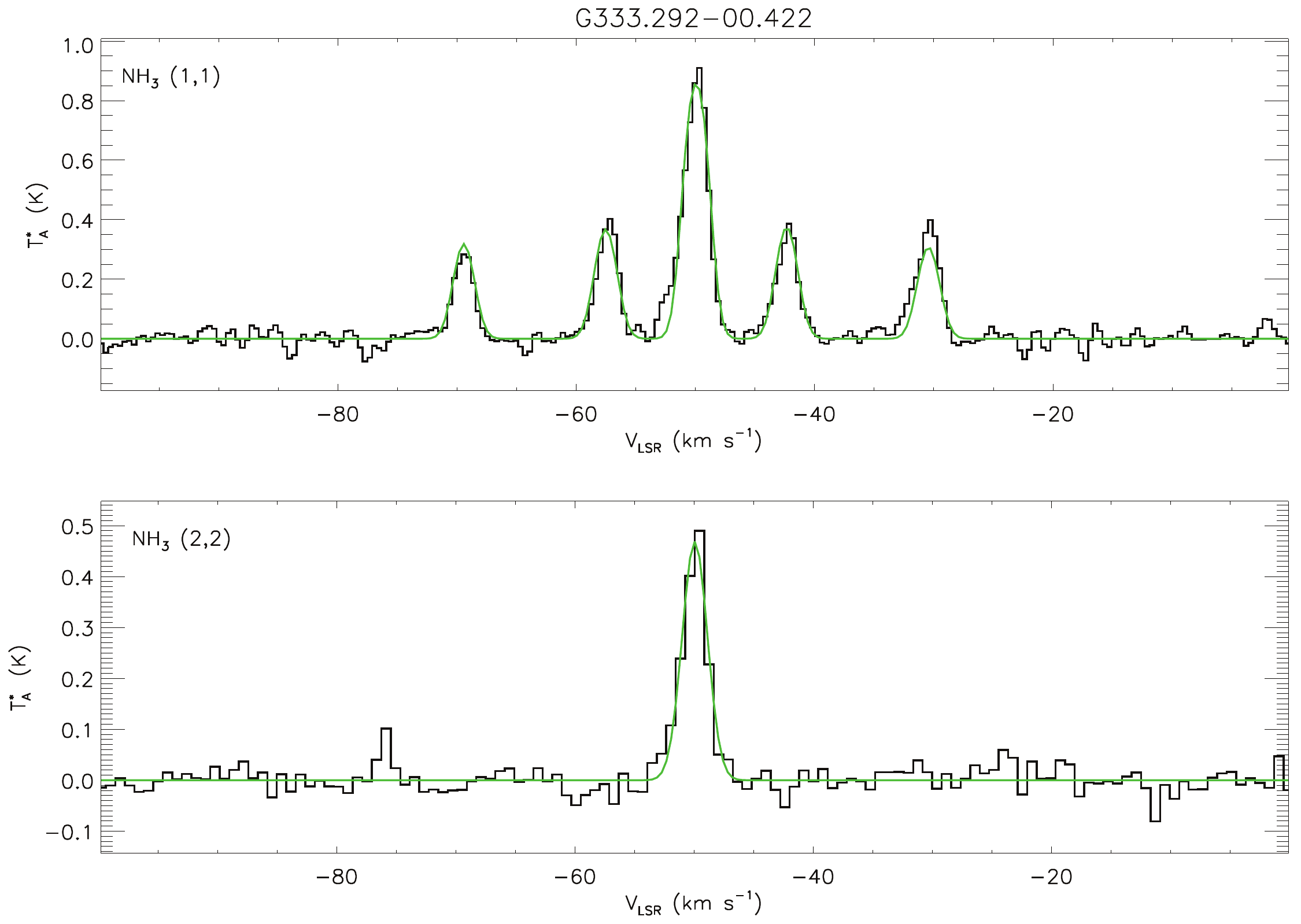}
\includegraphics[width=0.49\textwidth, trim=-5 -30 -5 -30]{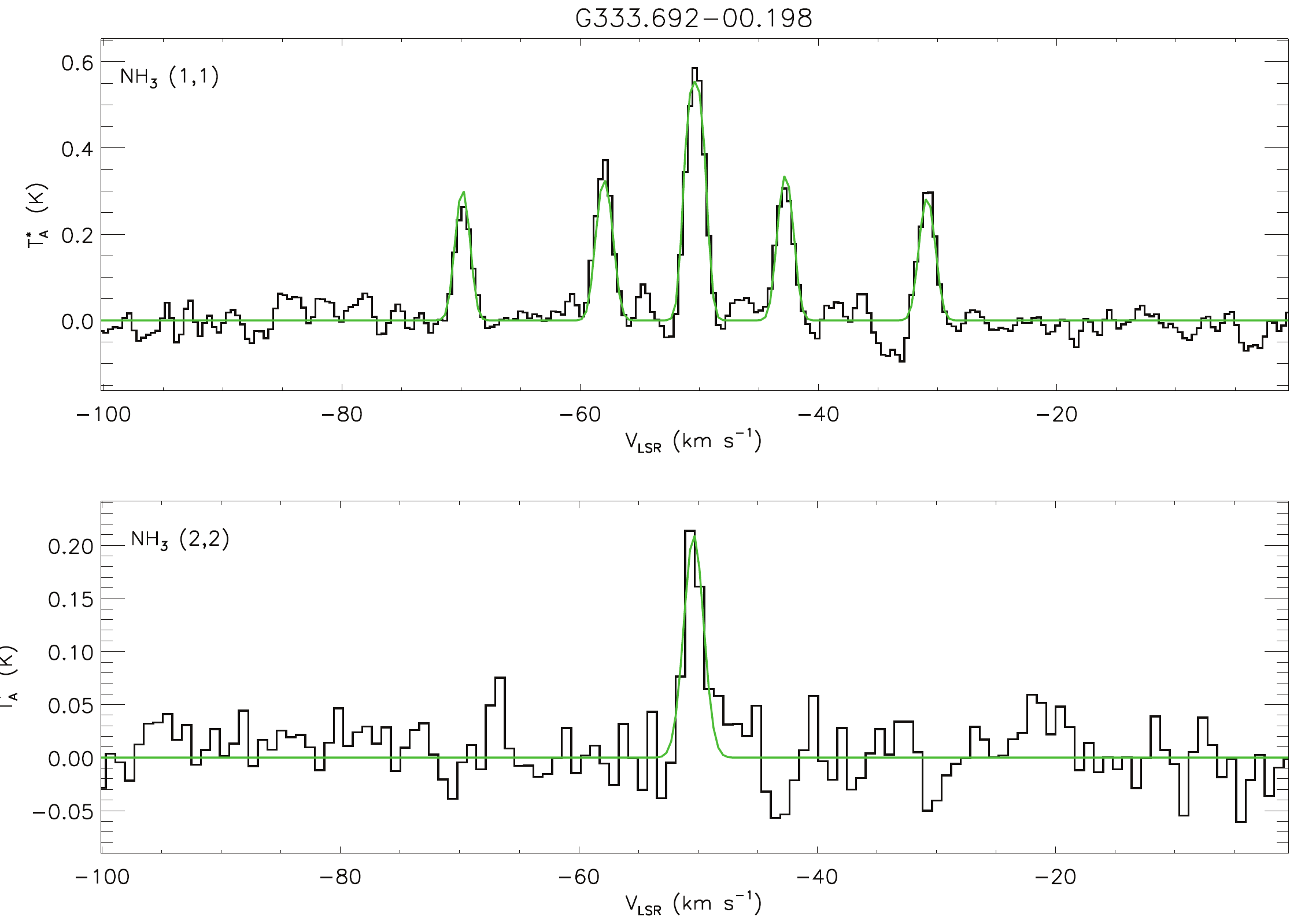}\\
\vspace*{0.5cm}
\contcaption{Clumps~19-24 are shown l-r, t-b.}
\end{figure*}

\begin{figure*}
\centering
\includegraphics[width=0.49\textwidth, trim=-5 -30 -5 -30]{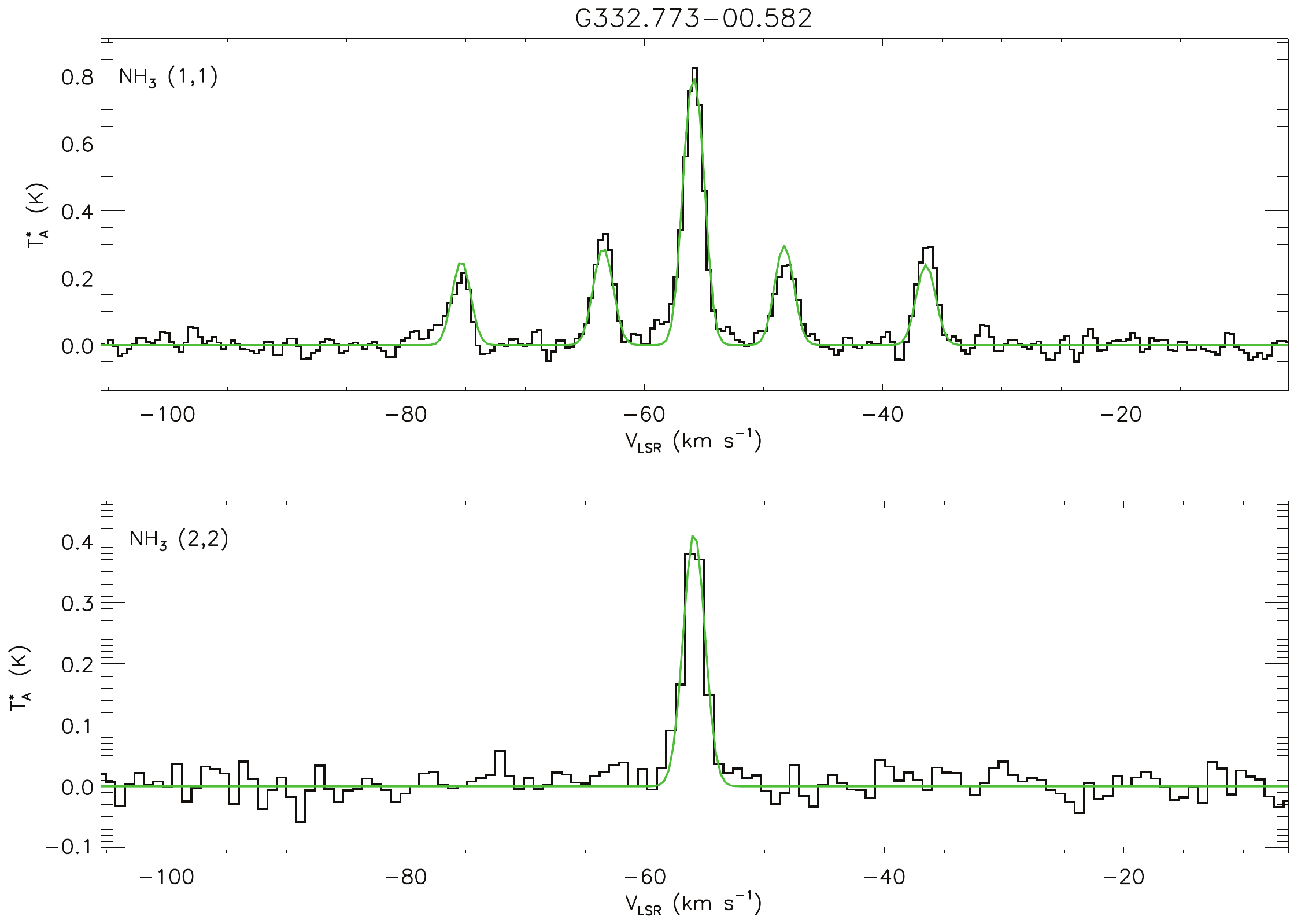}
\includegraphics[width=0.49\textwidth, trim=-5 -30 -5 -30]{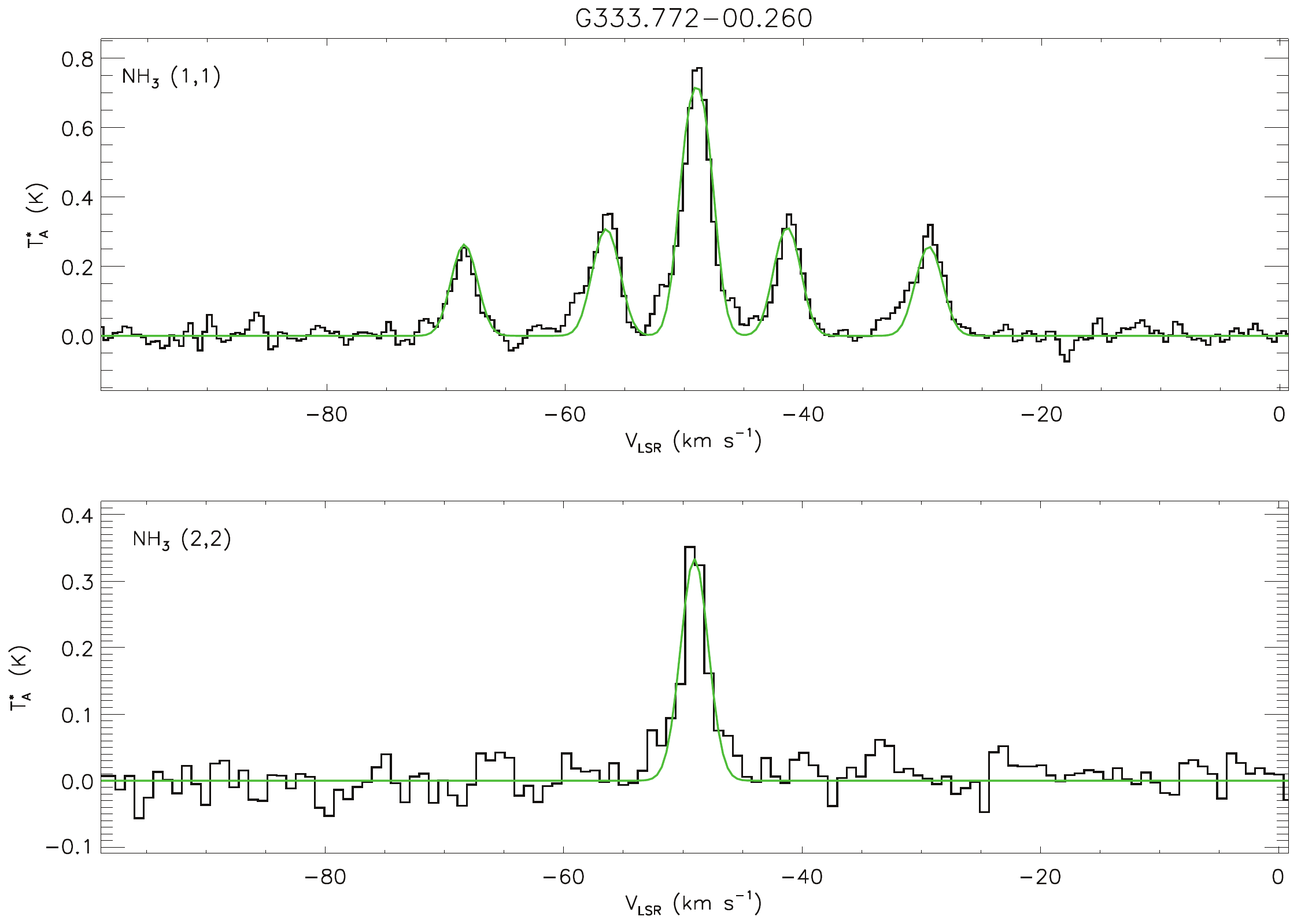}\\
\vspace*{0.5cm}
\includegraphics[width=0.49\textwidth, trim=-5 -30 -5 -30]{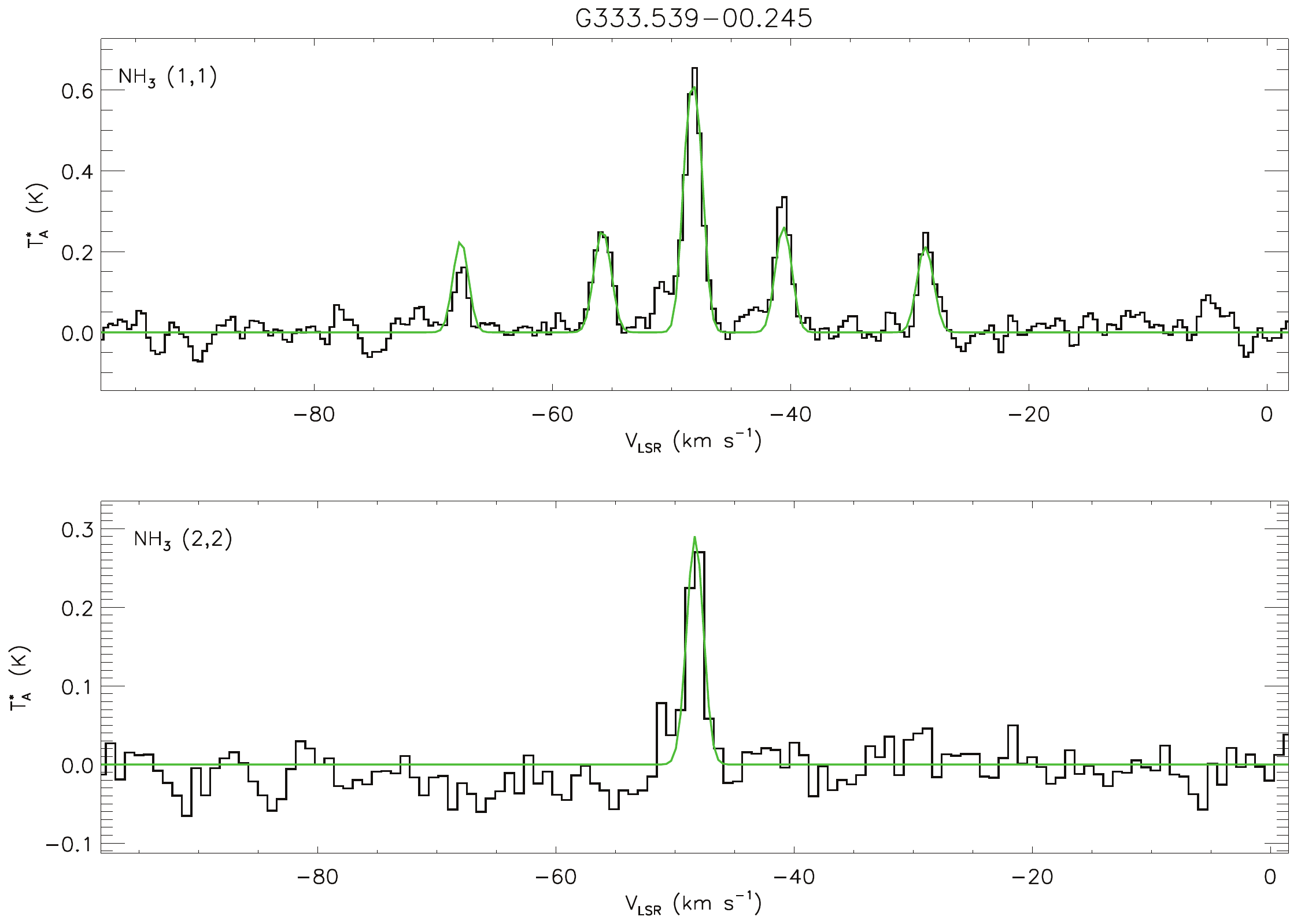}
\includegraphics[width=0.49\textwidth, trim=-5 -30 -5 -30]{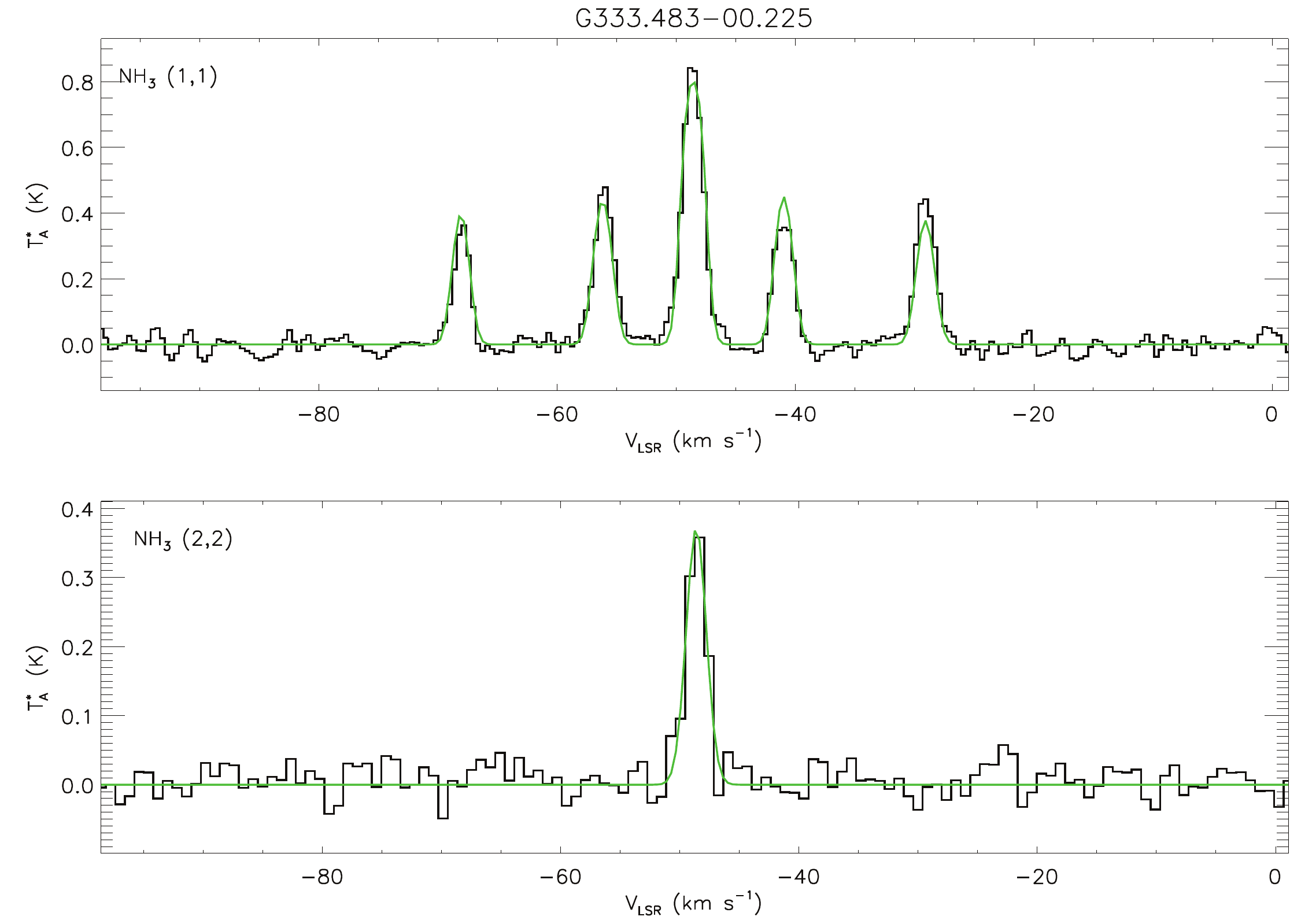}\\
\vspace*{0.5cm}
\includegraphics[width=0.49\textwidth, trim=-5 -30 -5 -30]{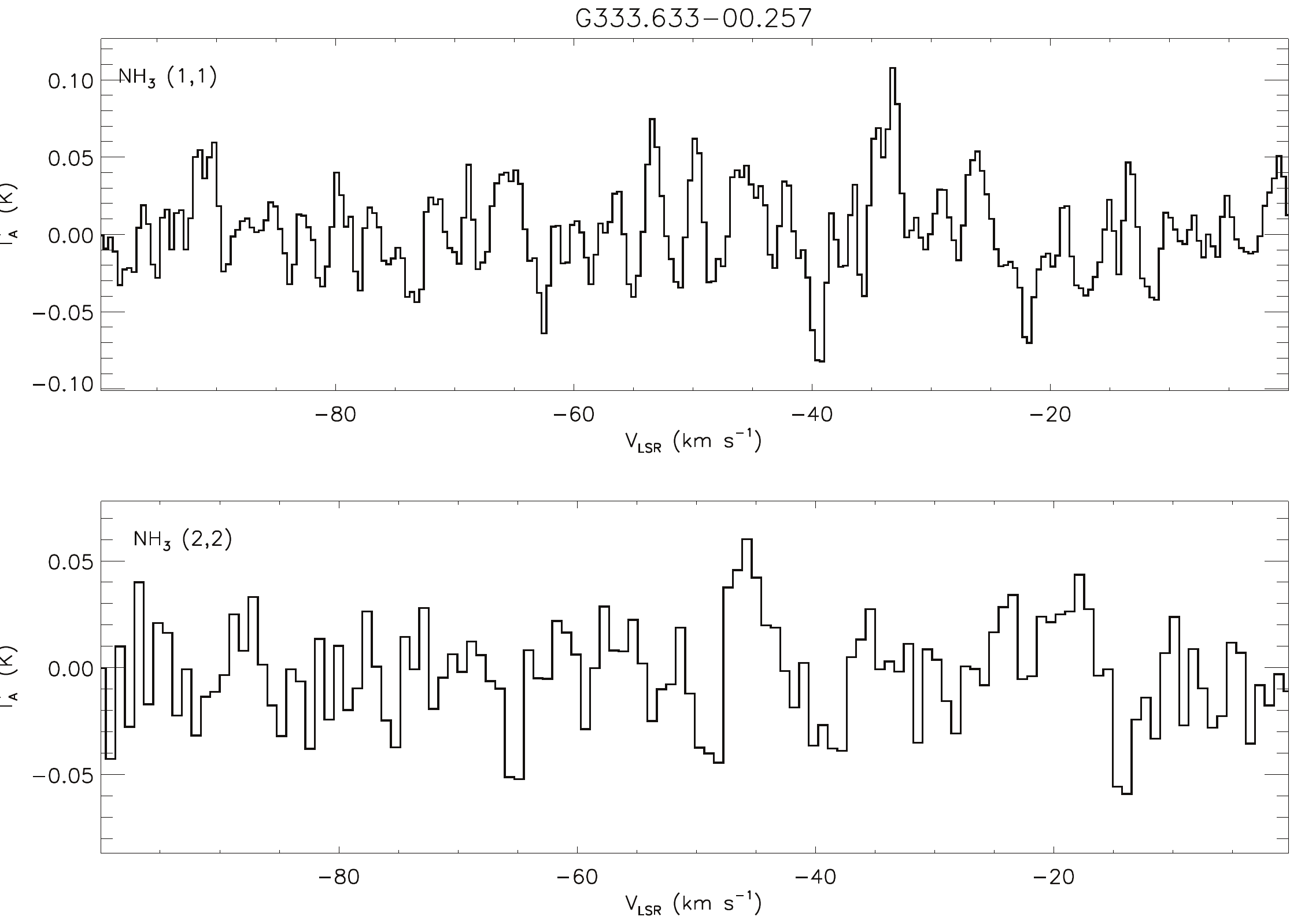}
\includegraphics[width=0.49\textwidth, trim=-5 -30 -5 -30]{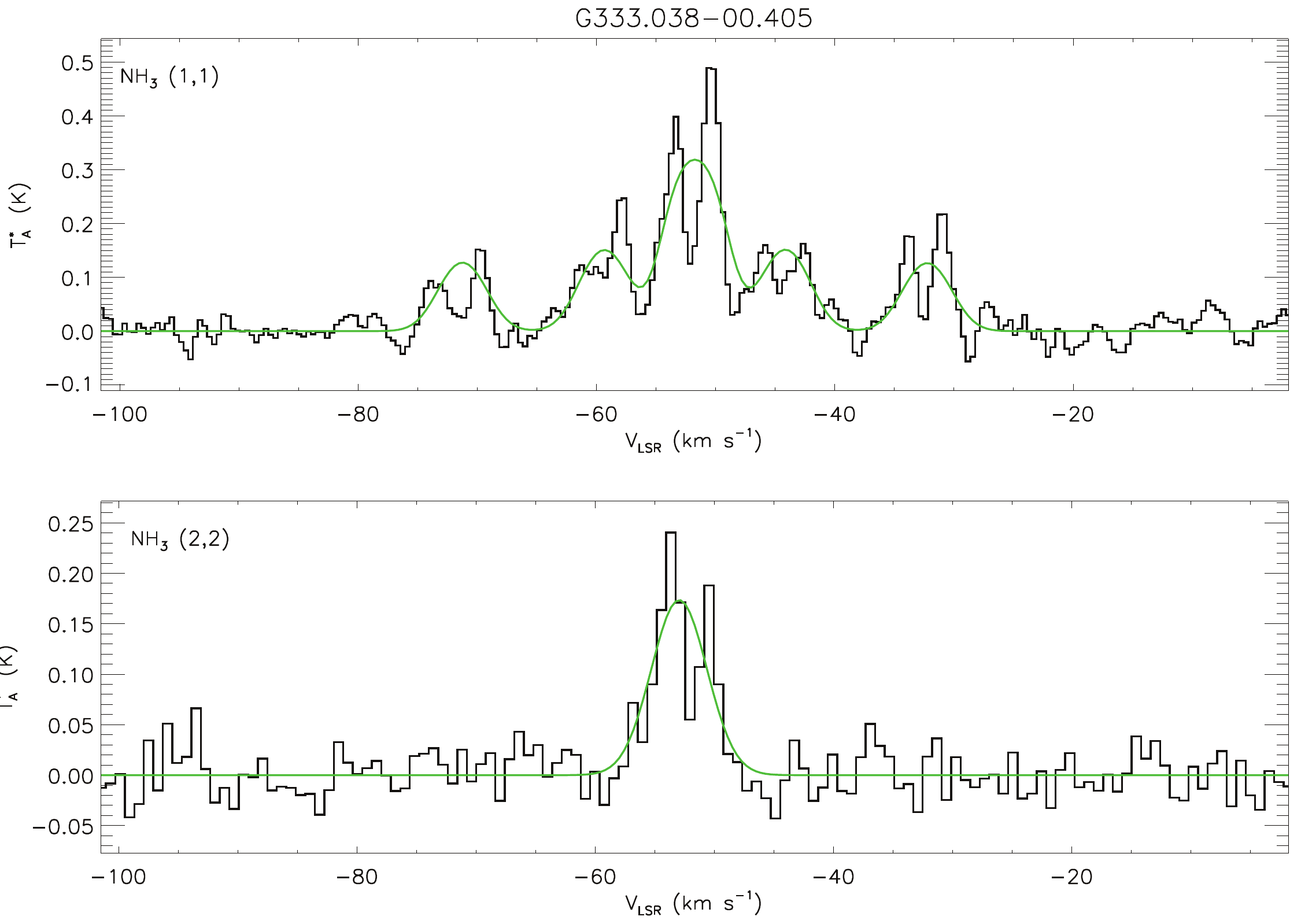}\\
\vspace*{0.5cm}
\contcaption{Clumps~25-30 are shown l-r, t-b.}
\end{figure*}

\begin{figure*}
\centering
\includegraphics[width=0.49\textwidth, trim=-5 -30 -5 -30]{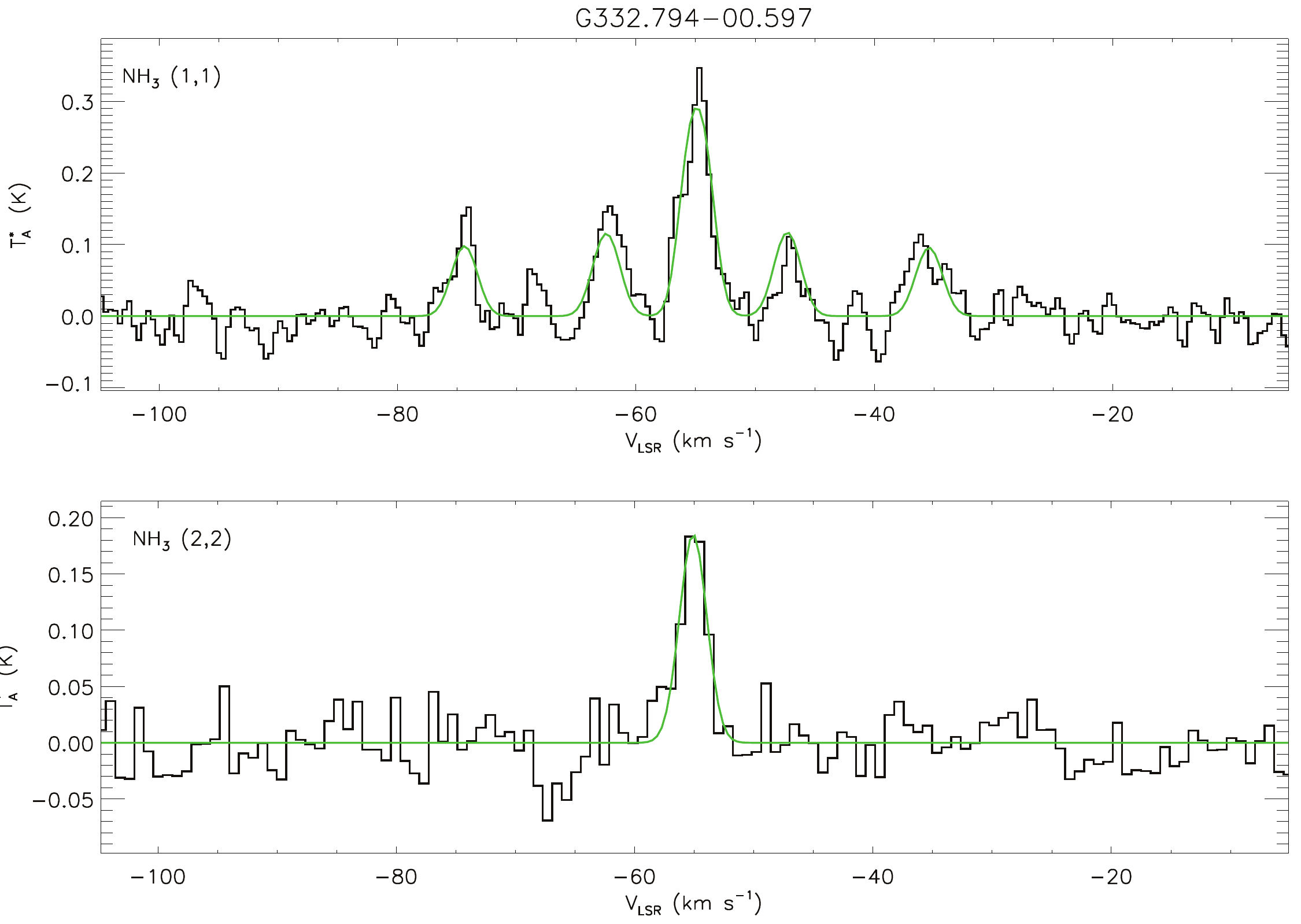}
\includegraphics[width=0.49\textwidth, trim=-5 -30 -5 -30]{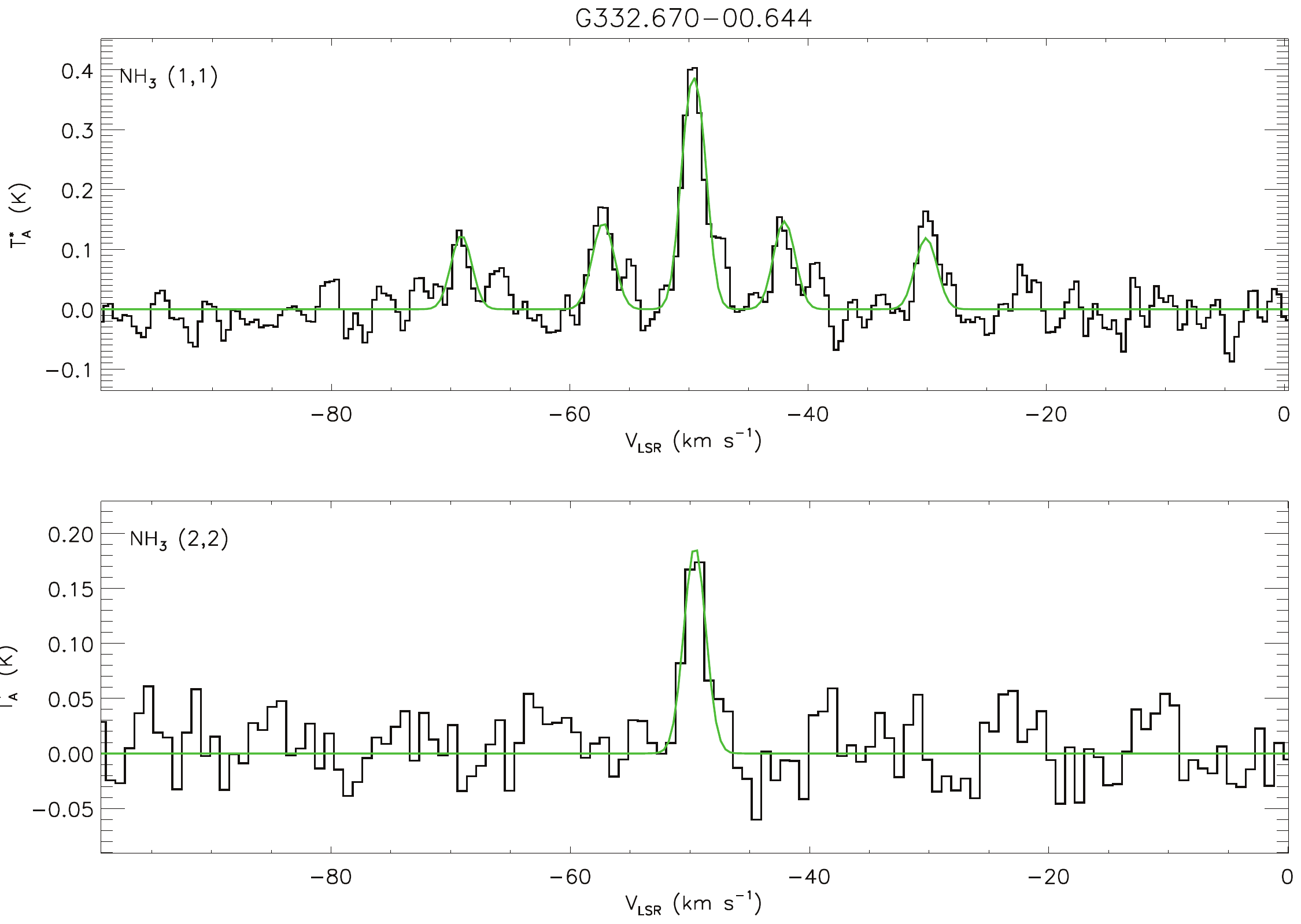}\\
\vspace*{0.5cm}
\includegraphics[width=0.49\textwidth, trim=-5 -30 -5 -30]{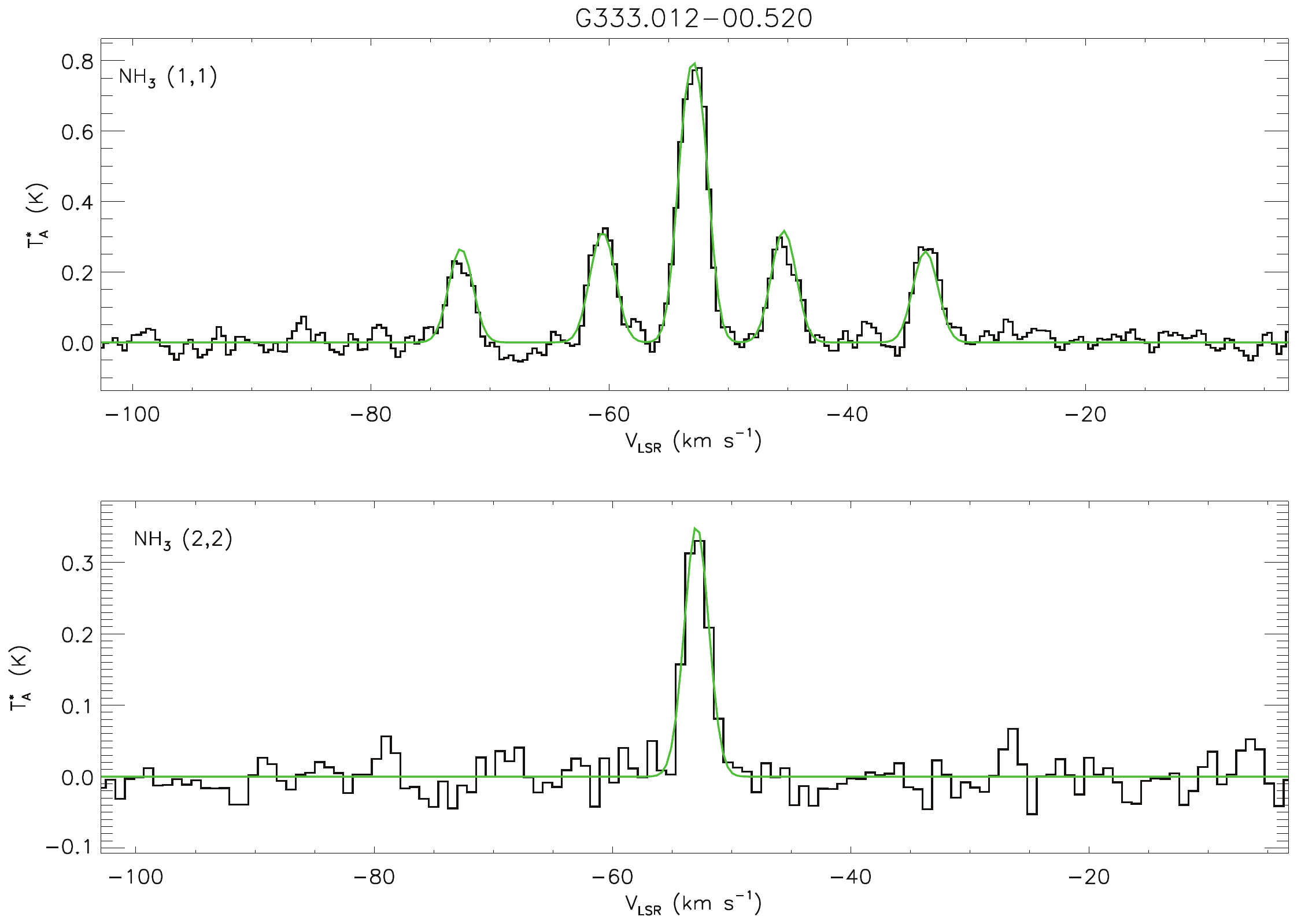}
\includegraphics[width=0.49\textwidth, trim=-5 -30 -5 -30]{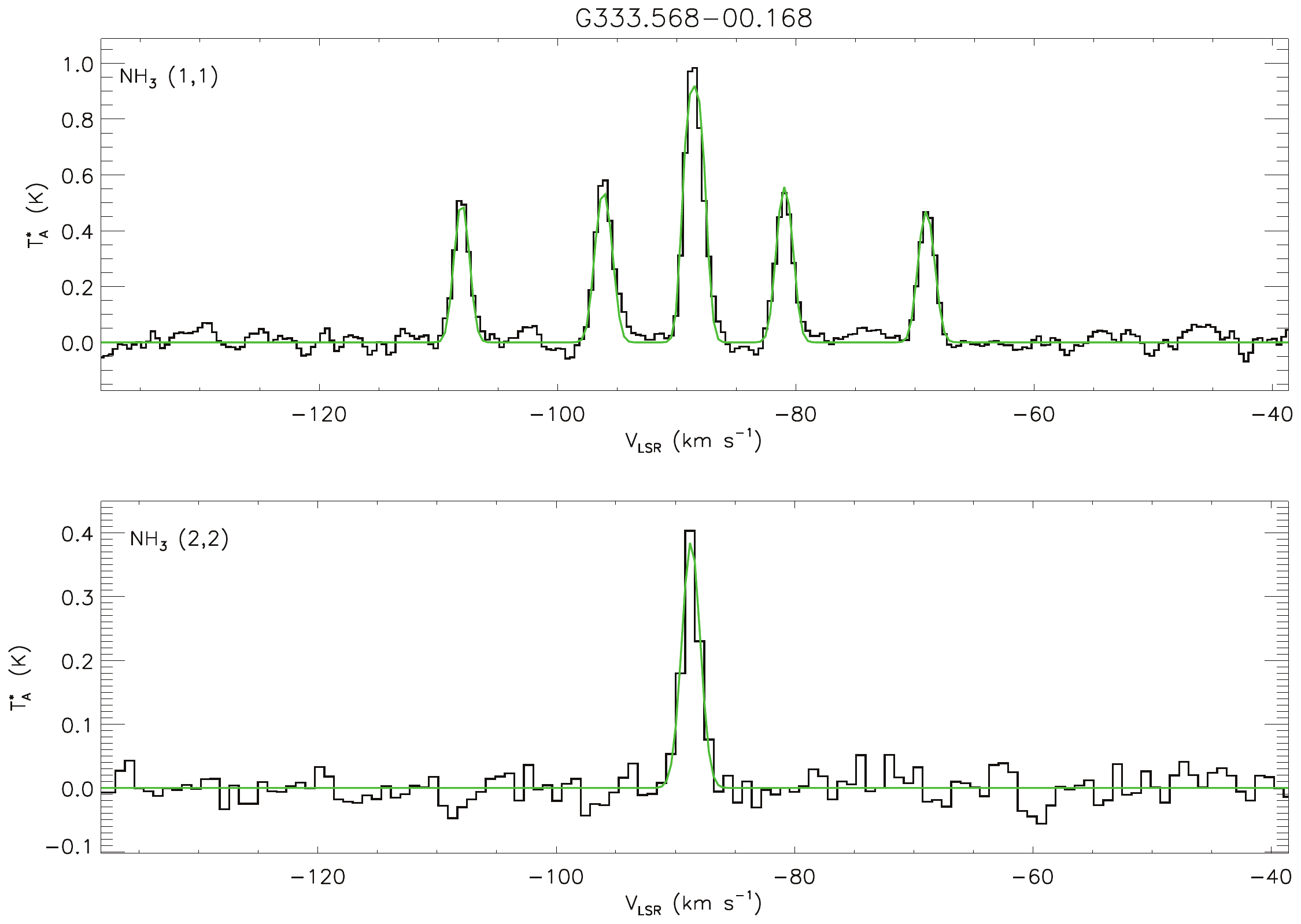}\\
\vspace*{0.5cm}
\includegraphics[width=0.49\textwidth, trim=-5 -30 -5 -30]{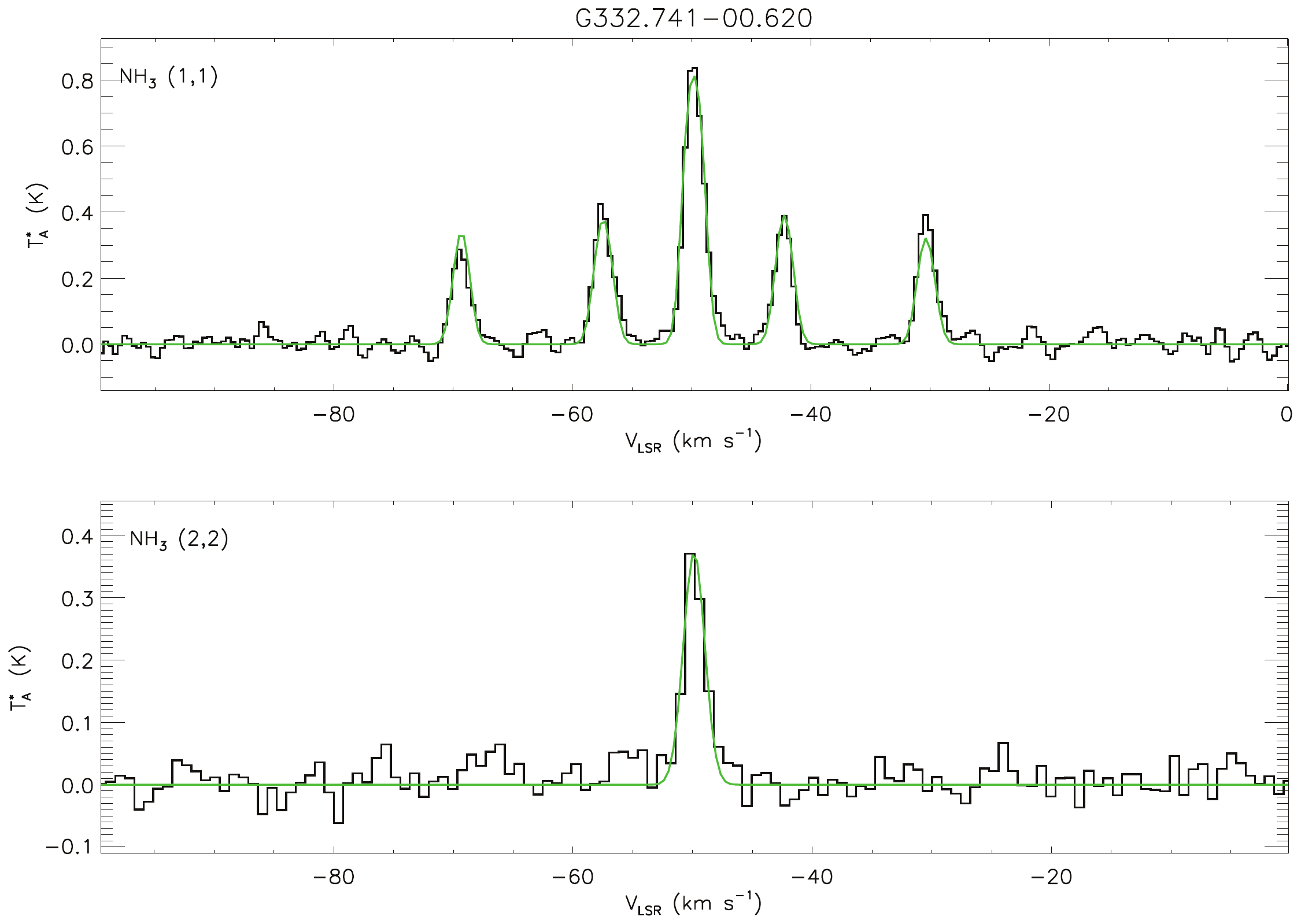}
\includegraphics[width=0.49\textwidth, trim=-5 -30 -5 -30]{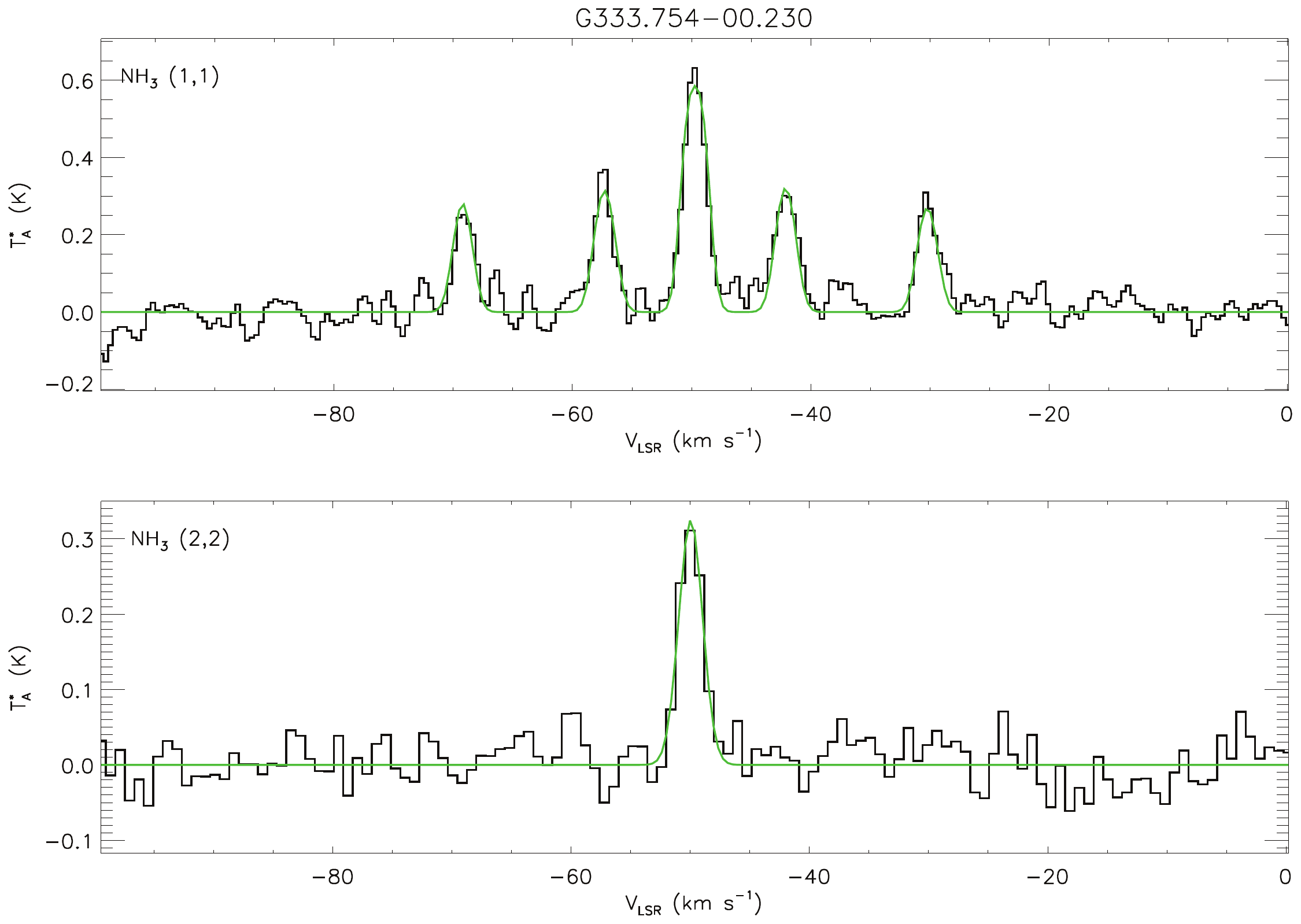}\\
\vspace*{0.5cm}
\contcaption{Clumps 31-36 are shown l-r, t-b.}
\end{figure*}

\begin{figure*}
\centering
\includegraphics[width=0.49\textwidth, trim=-5 -30 -5 -30]{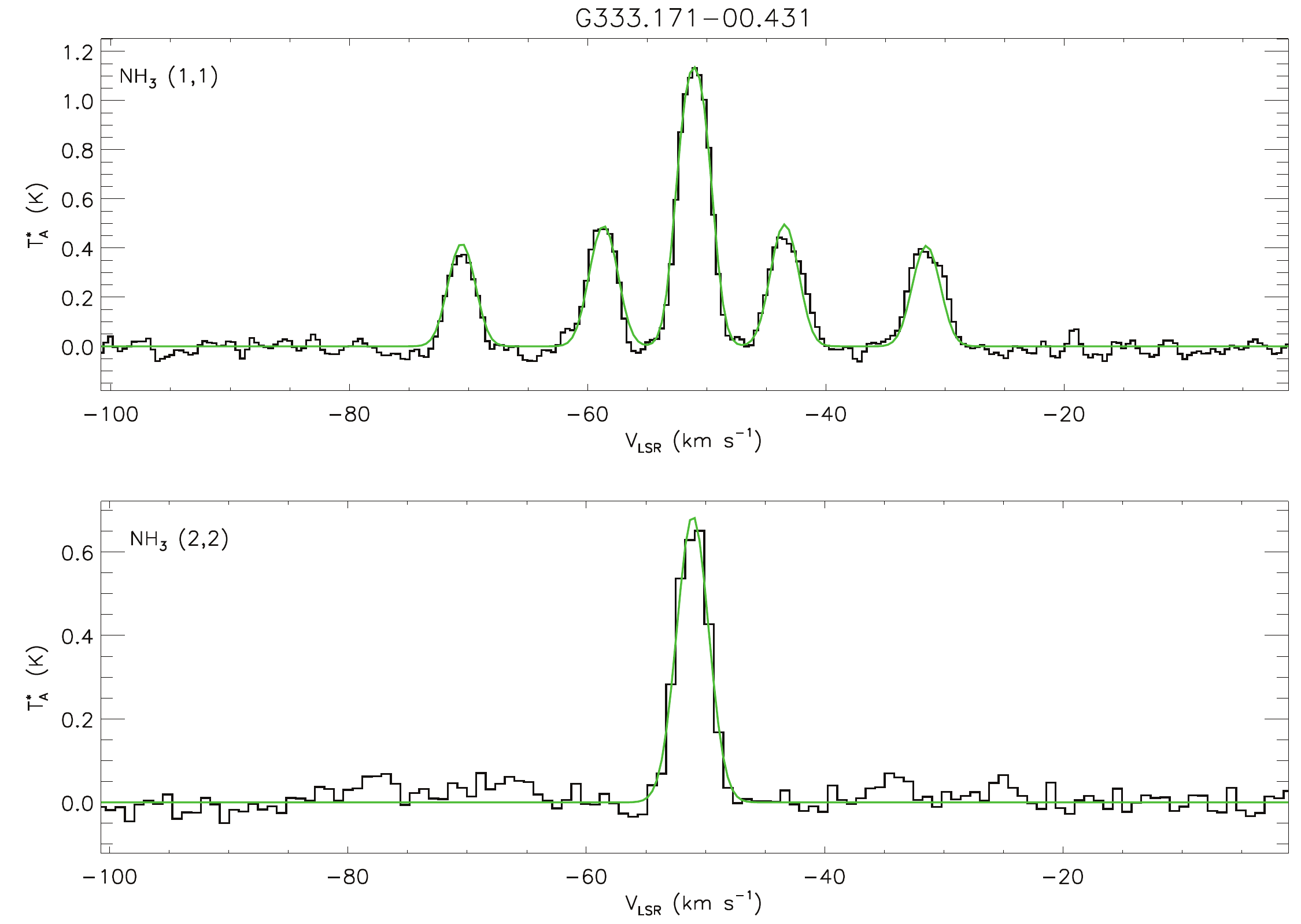}
\includegraphics[width=0.49\textwidth, trim=-5 -30 -5 -30]{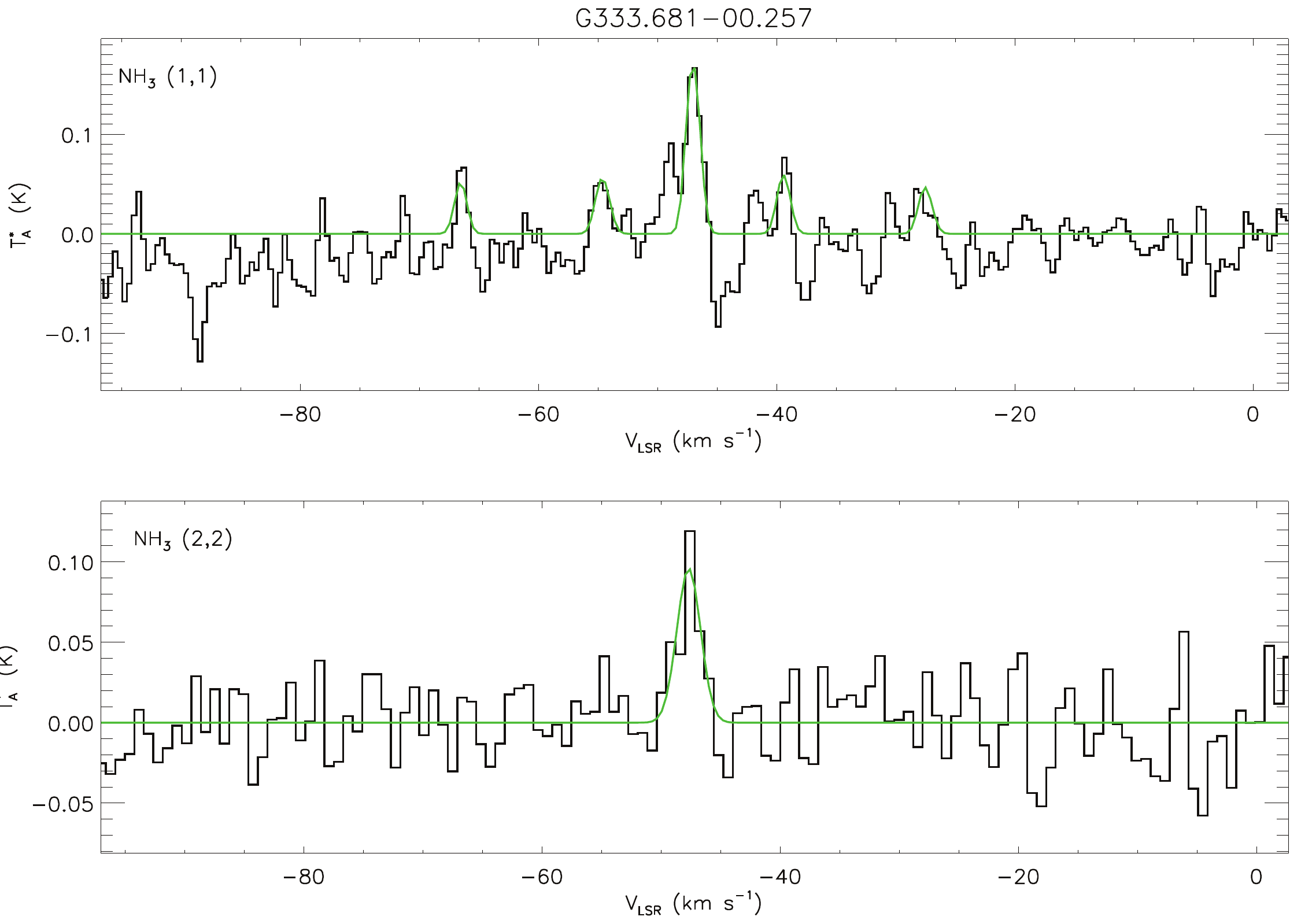}\\
\vspace*{0.5cm}
\includegraphics[width=0.49\textwidth, trim=-5 -30 -5 -30]{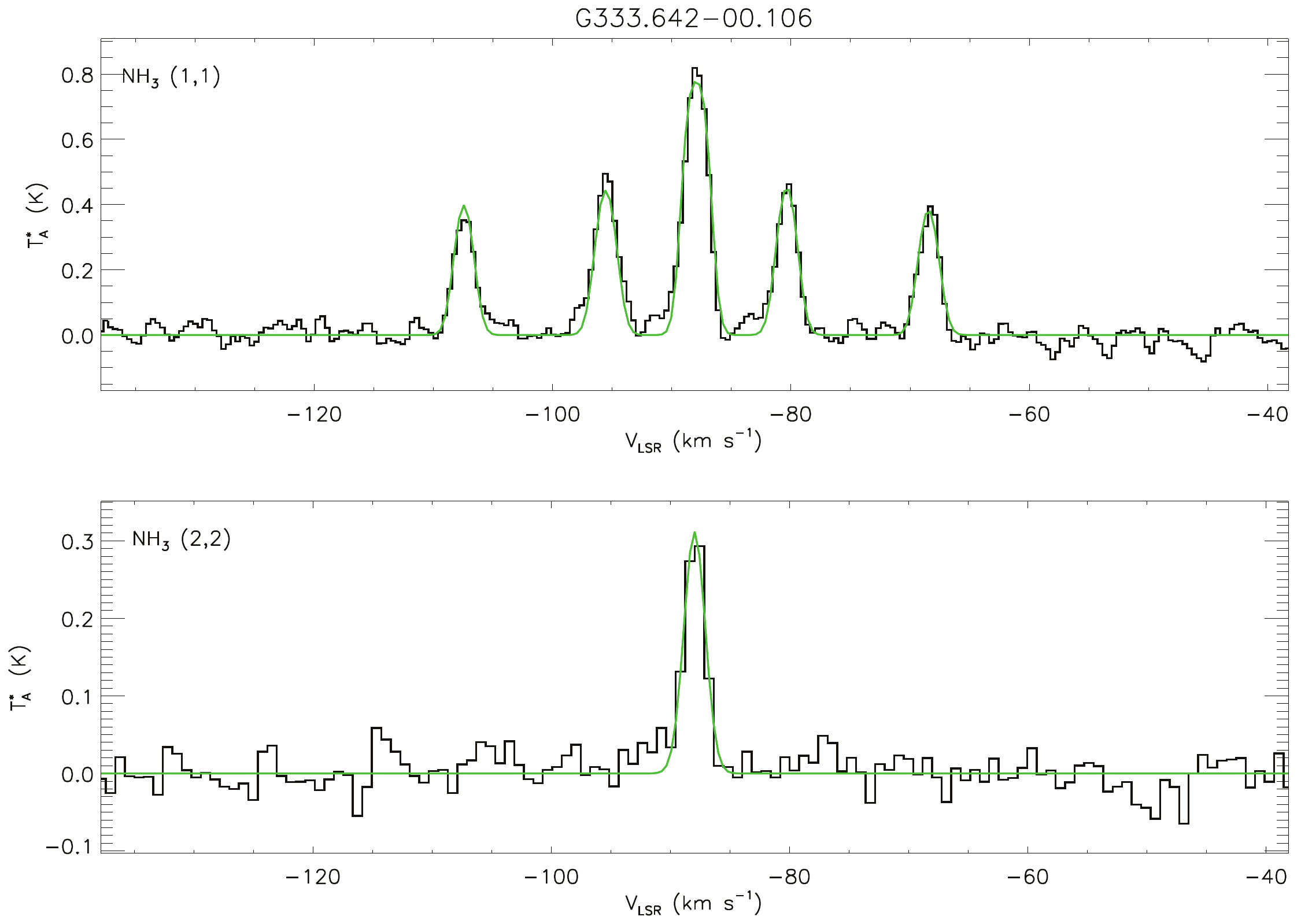}
\includegraphics[width=0.49\textwidth, trim=-5 -30 -5 -30]{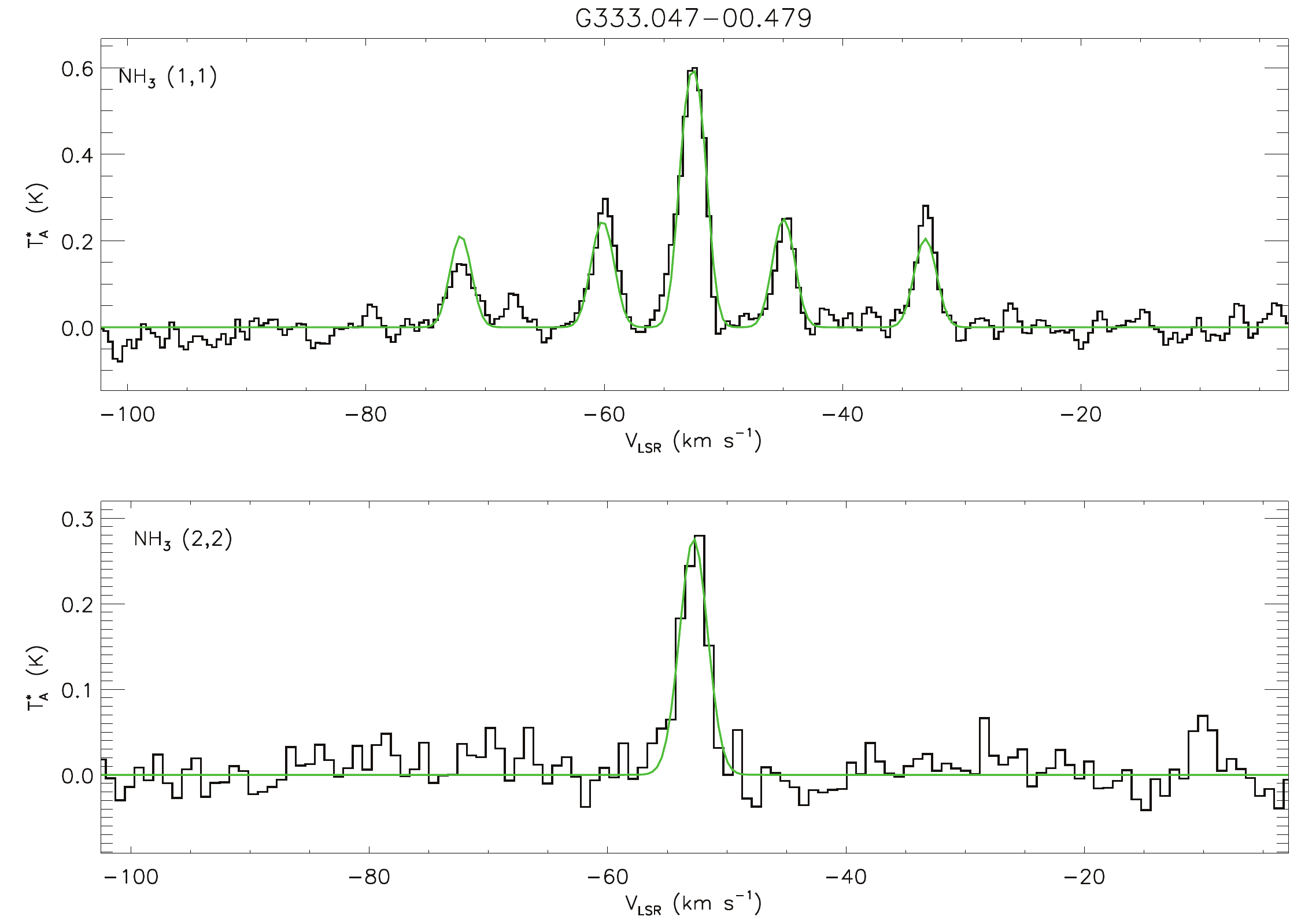}\\
\vspace*{0.5cm}
\includegraphics[width=0.49\textwidth, trim=-5 -30 -5 -30]{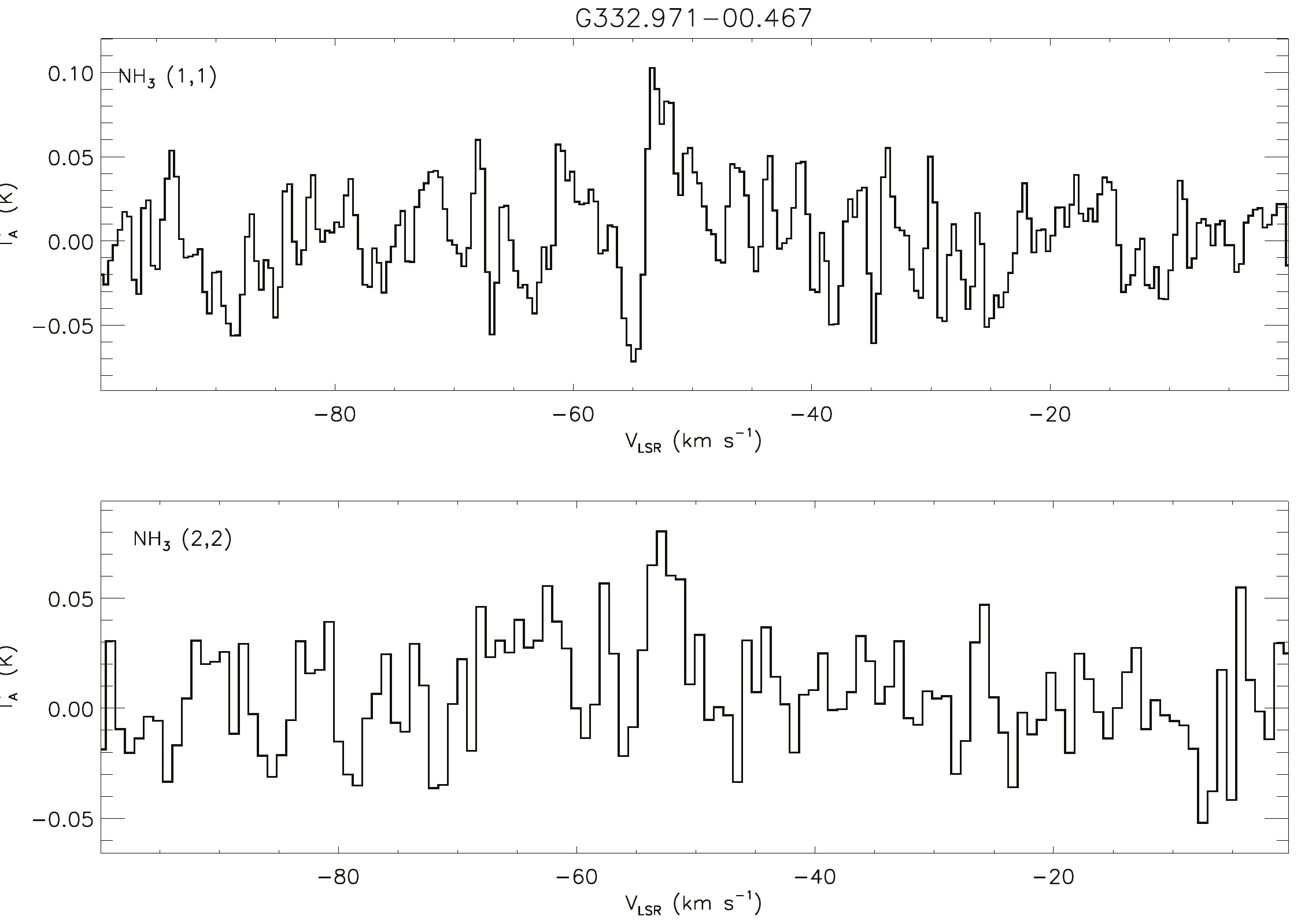}
\includegraphics[width=0.49\textwidth, trim=-5 -30 -5 -30]{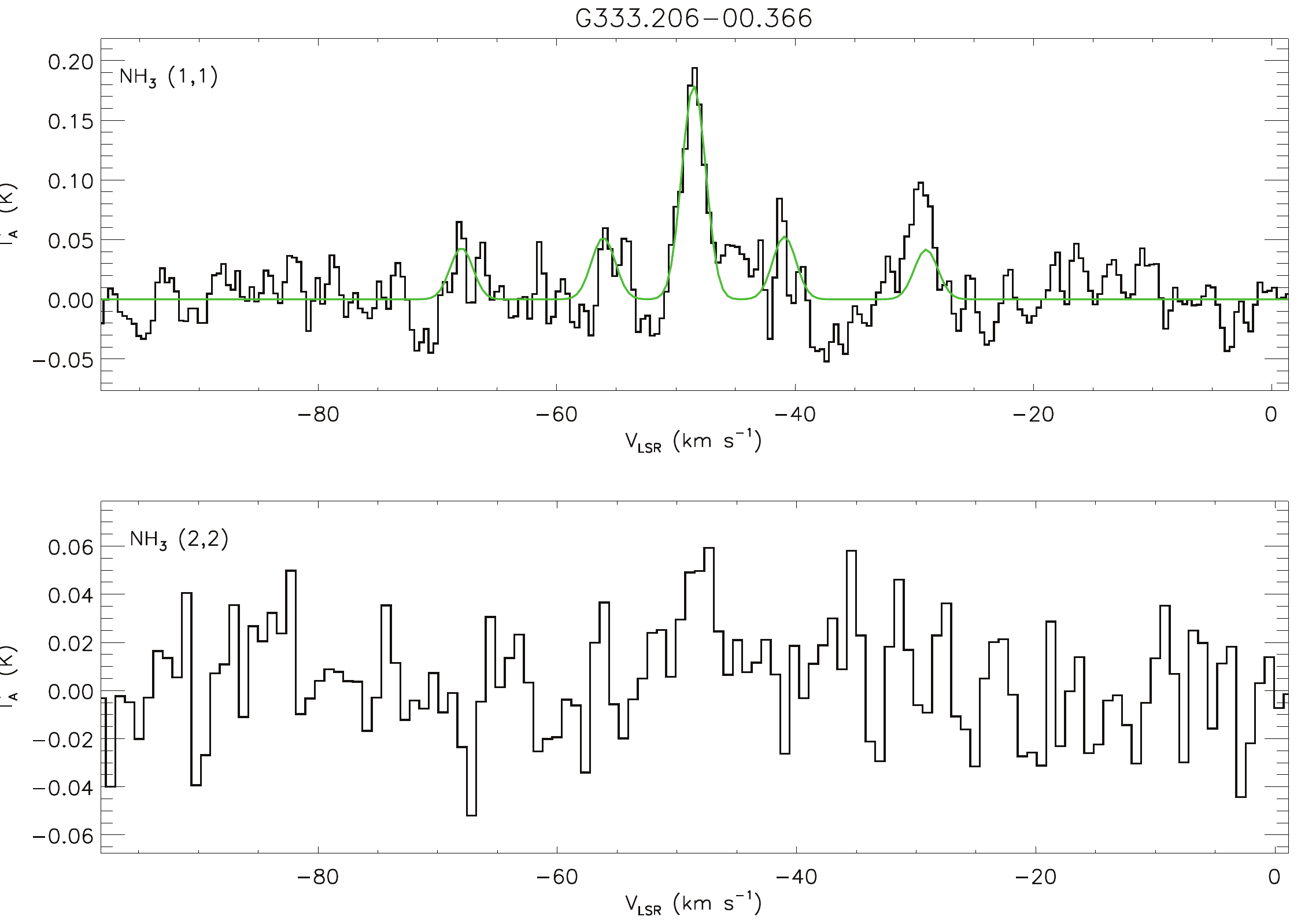}\\
\vspace*{0.5cm}
\contcaption{Clumps~37-42 are shown l-r, t-b.}
\end{figure*}

\begin{figure*}
\centering
\includegraphics[width=0.49\textwidth, trim=-5 -30 -5 -30]{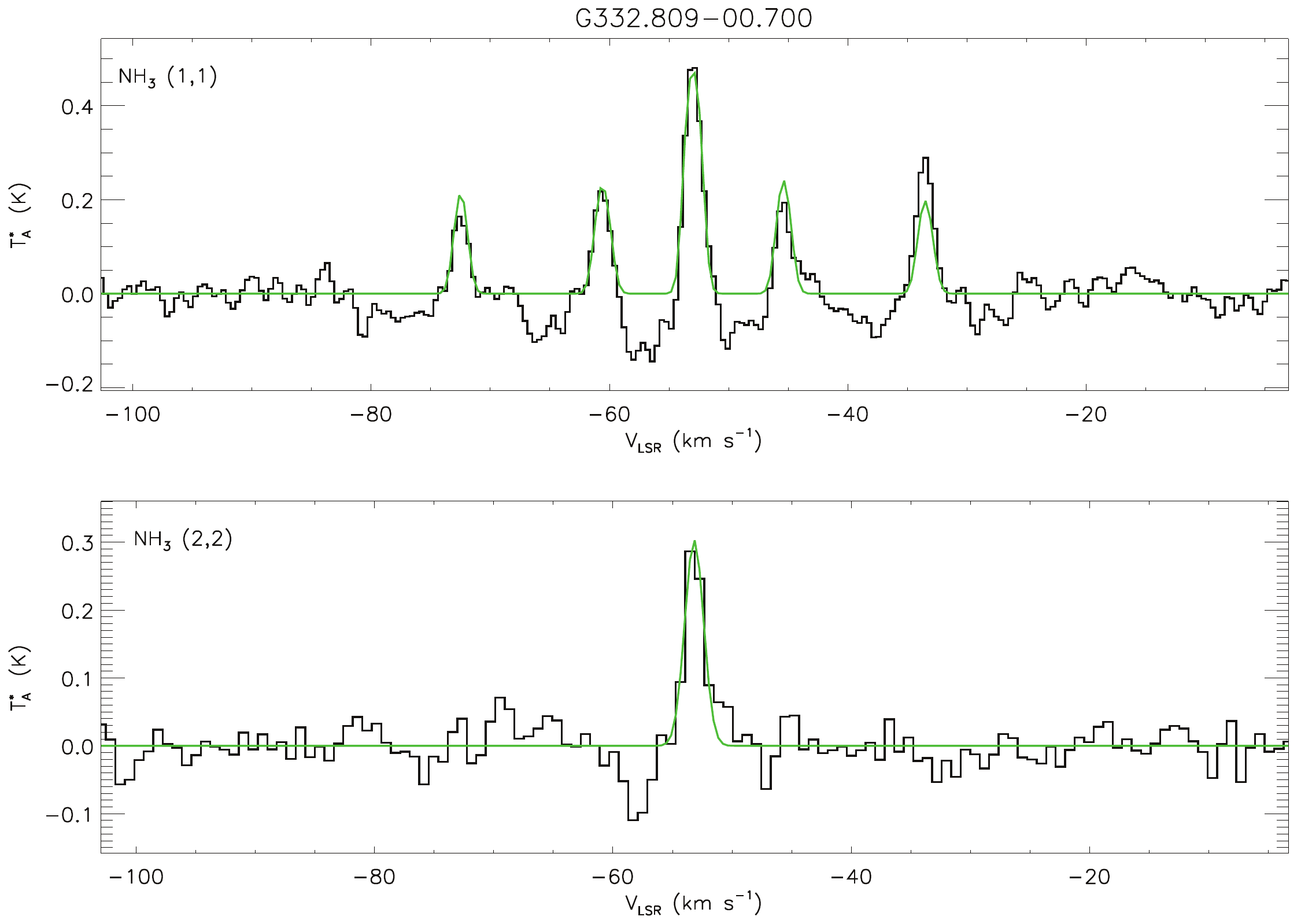}
\includegraphics[width=0.49\textwidth, trim=-5 -30 -5 -30]{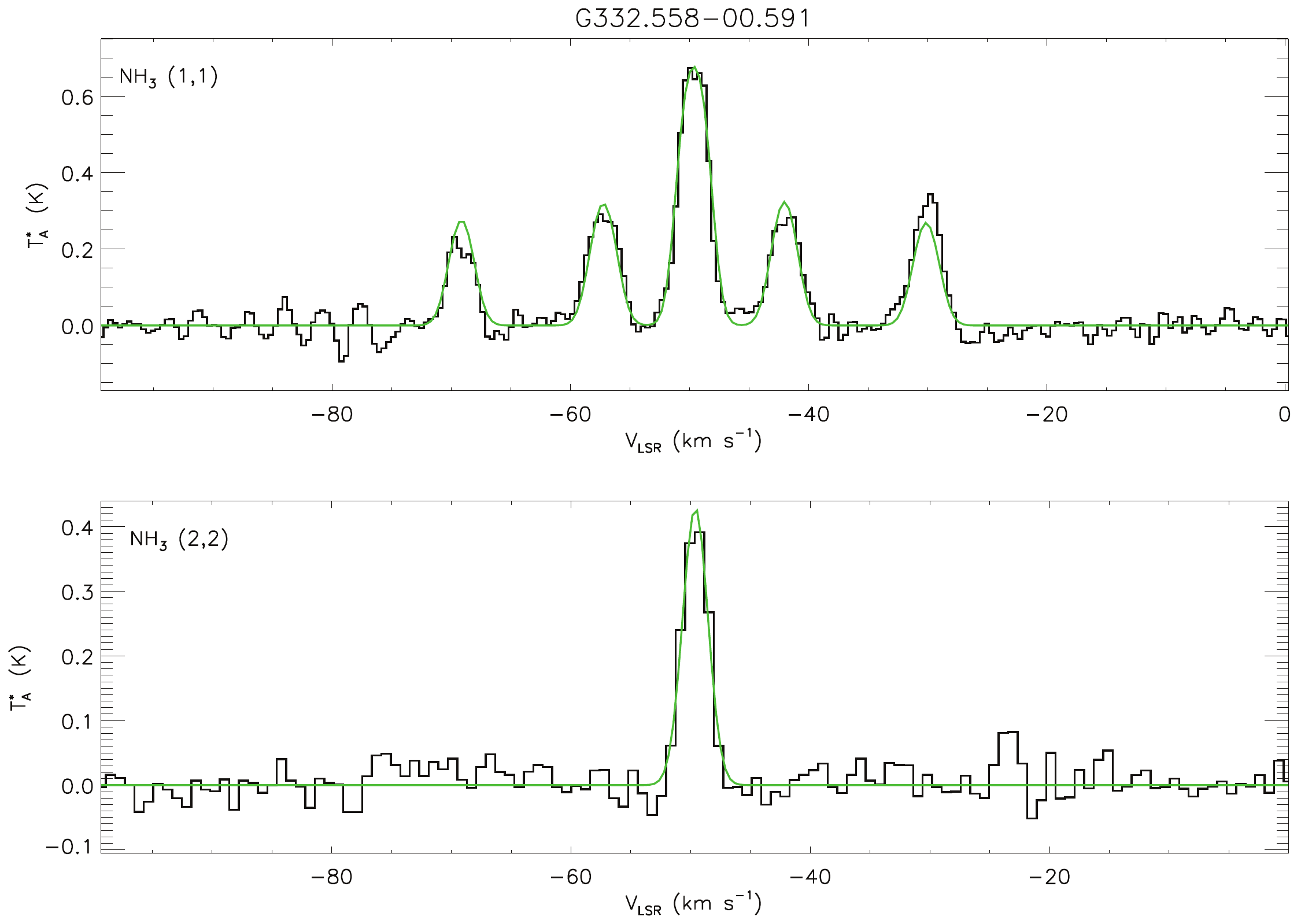}\\
\vspace*{0.5cm}
\includegraphics[width=0.49\textwidth, trim=-5 -30 -5 -30]{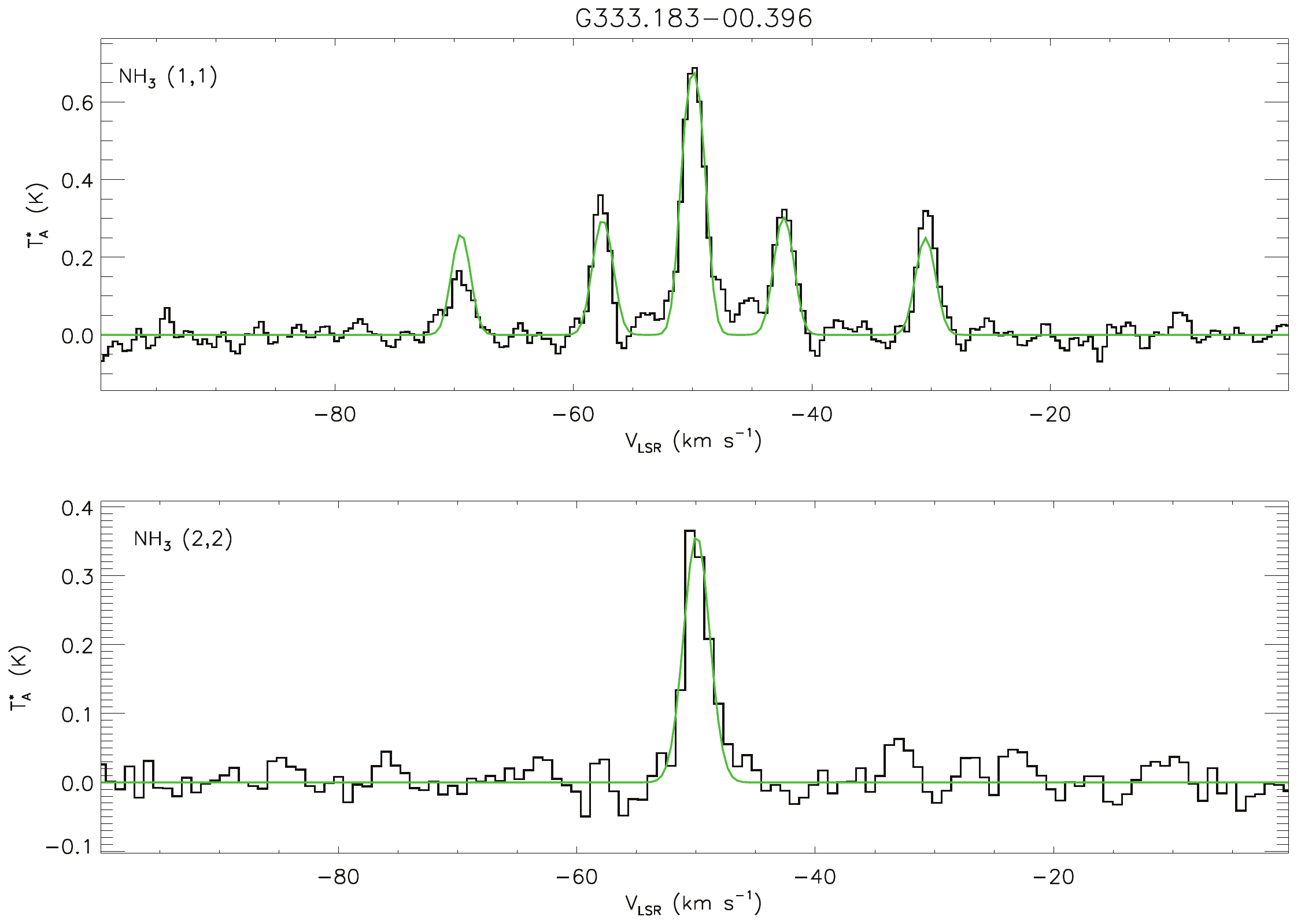}
\includegraphics[width=0.49\textwidth, trim=-5 -30 -5 -30]{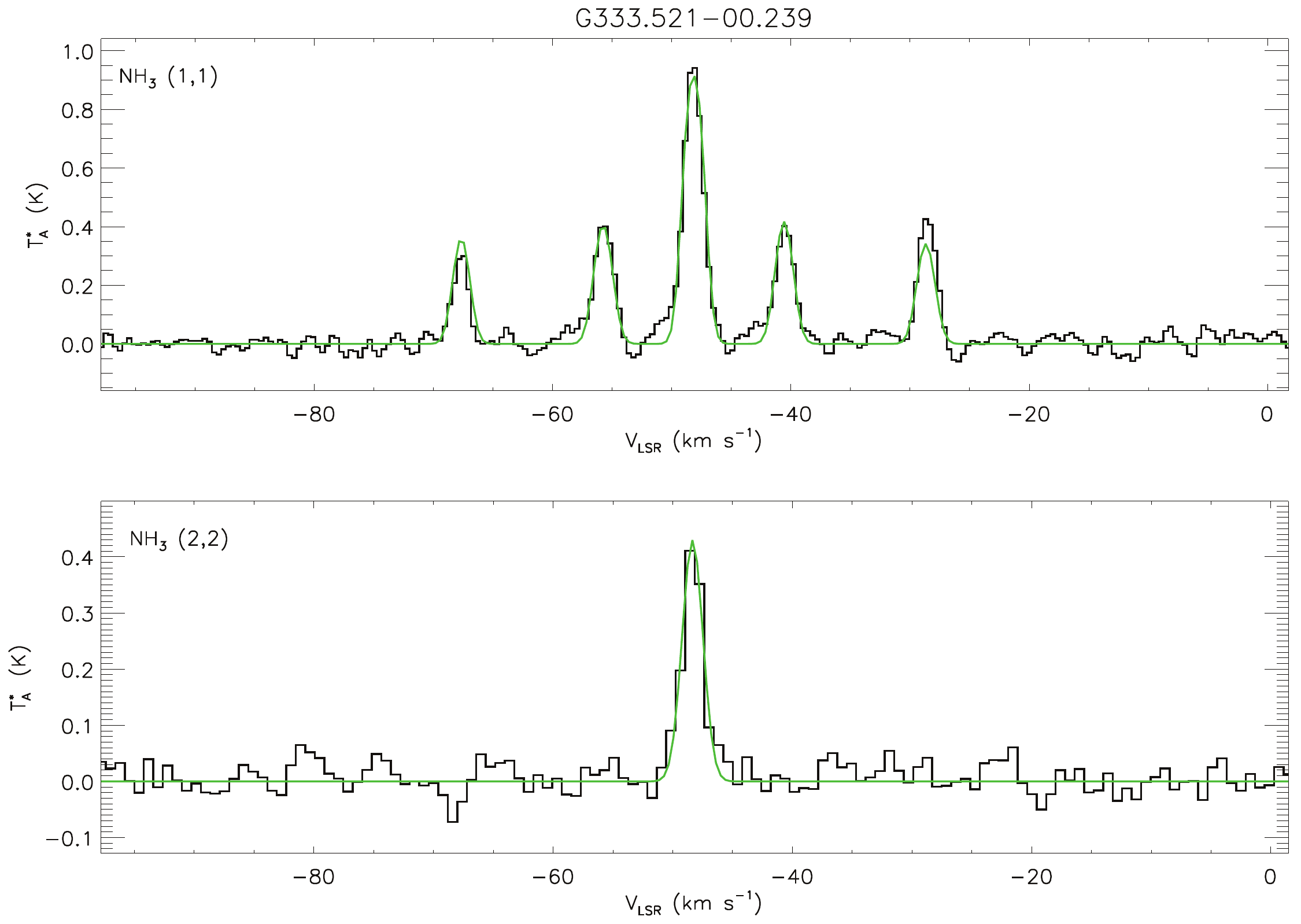}\\
\vspace*{0.5cm}
\includegraphics[width=0.49\textwidth, trim=-5 -30 -5 -30]{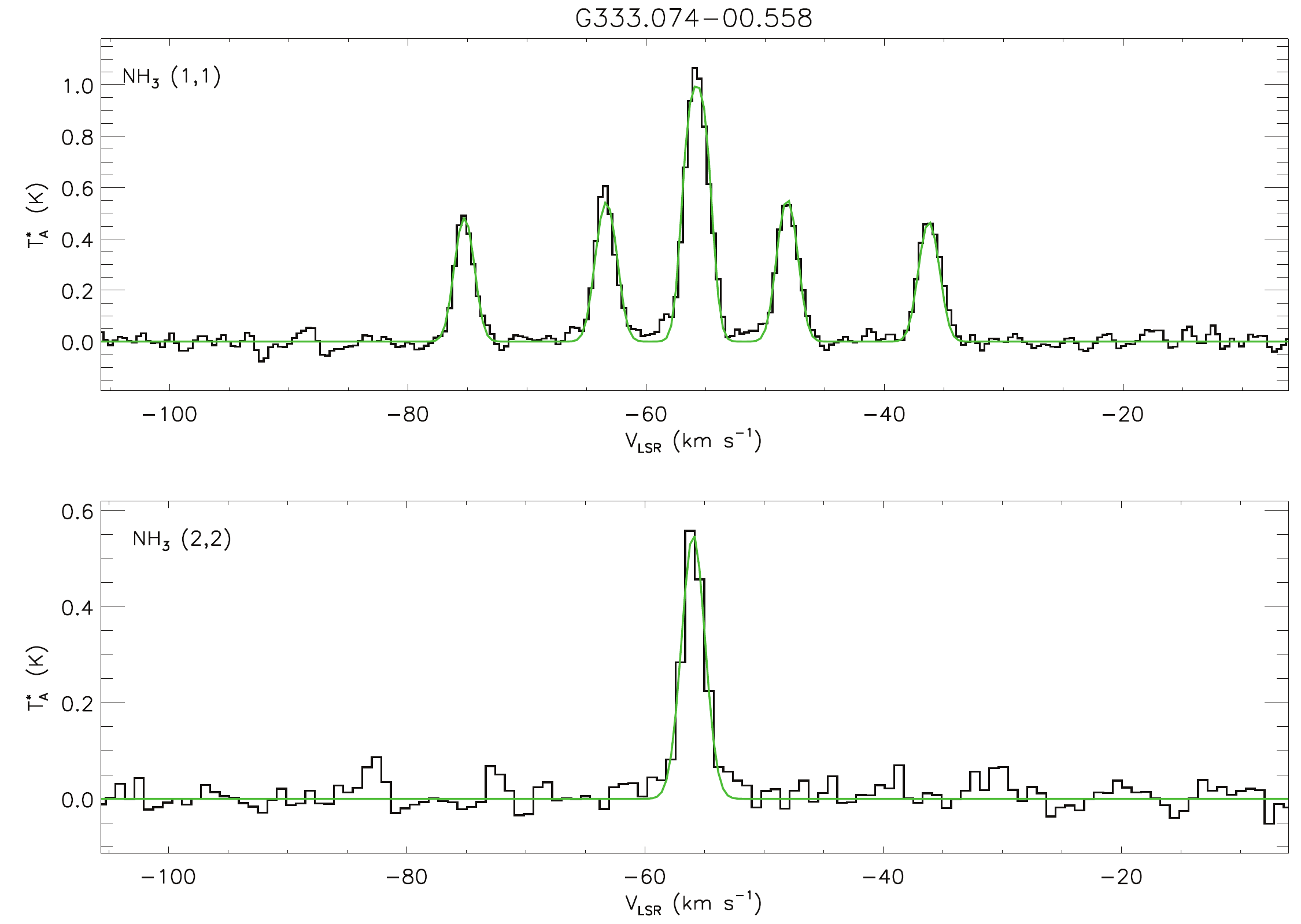}
\includegraphics[width=0.49\textwidth, trim=-5 -30 -5 -30]{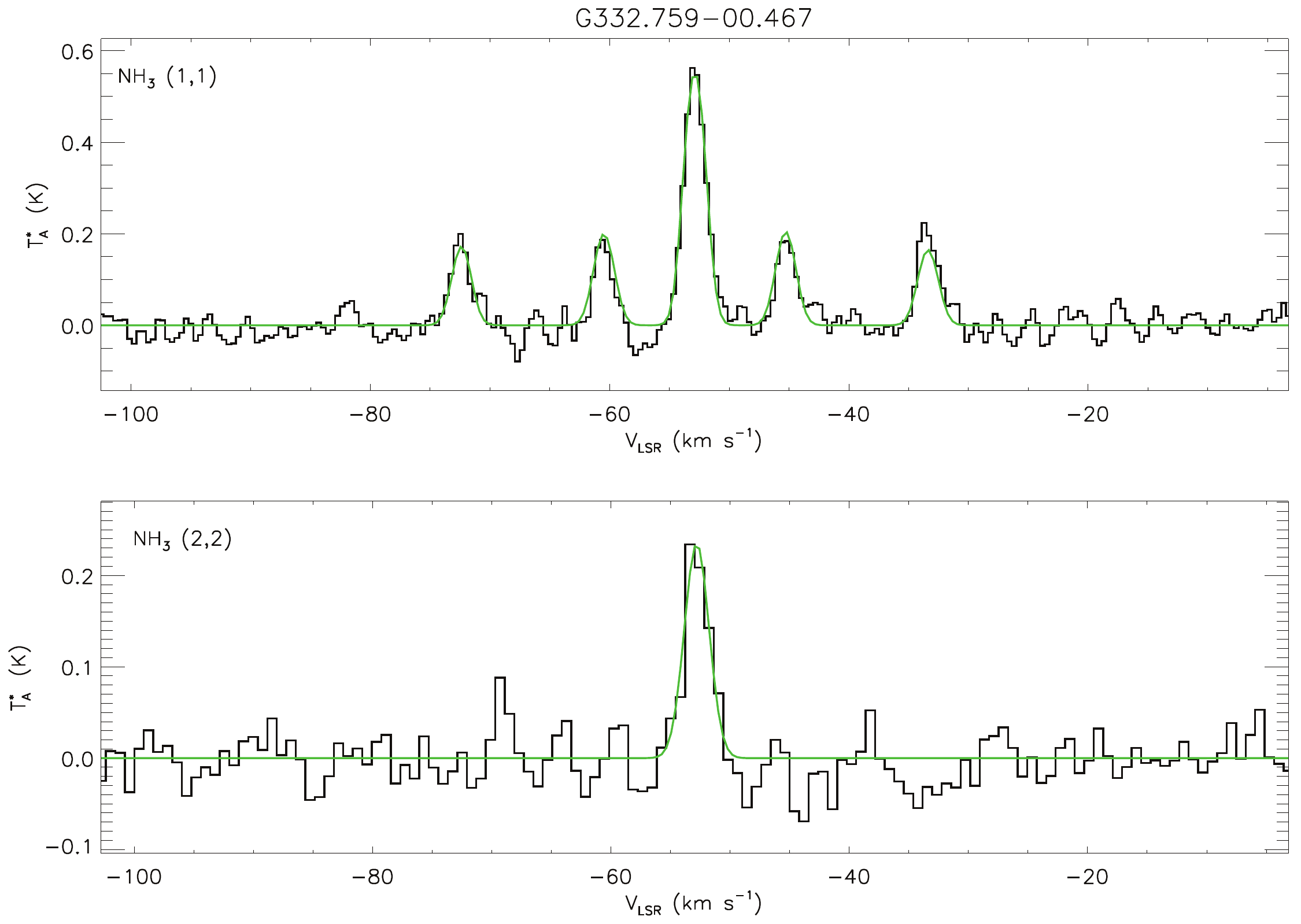}\\
\vspace*{0.5cm}
\contcaption{Clumps 43-48 are shown l-r, t-b.}
\end{figure*}

\begin{figure*}
\centering
\includegraphics[width=0.49\textwidth, trim=-5 -30 -5 -30]{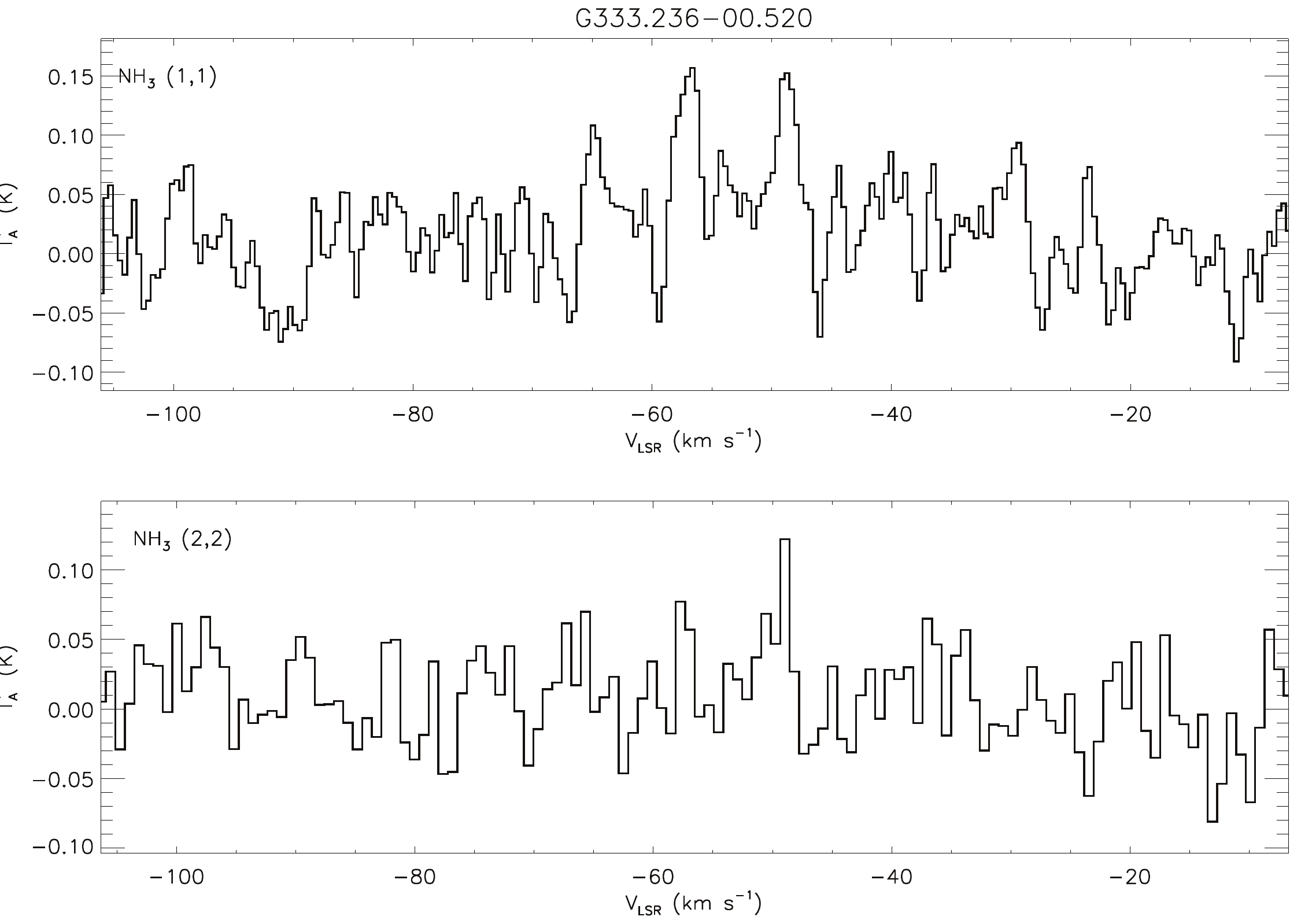}
\includegraphics[width=0.49\textwidth, trim=-5 -30 -5 -30]{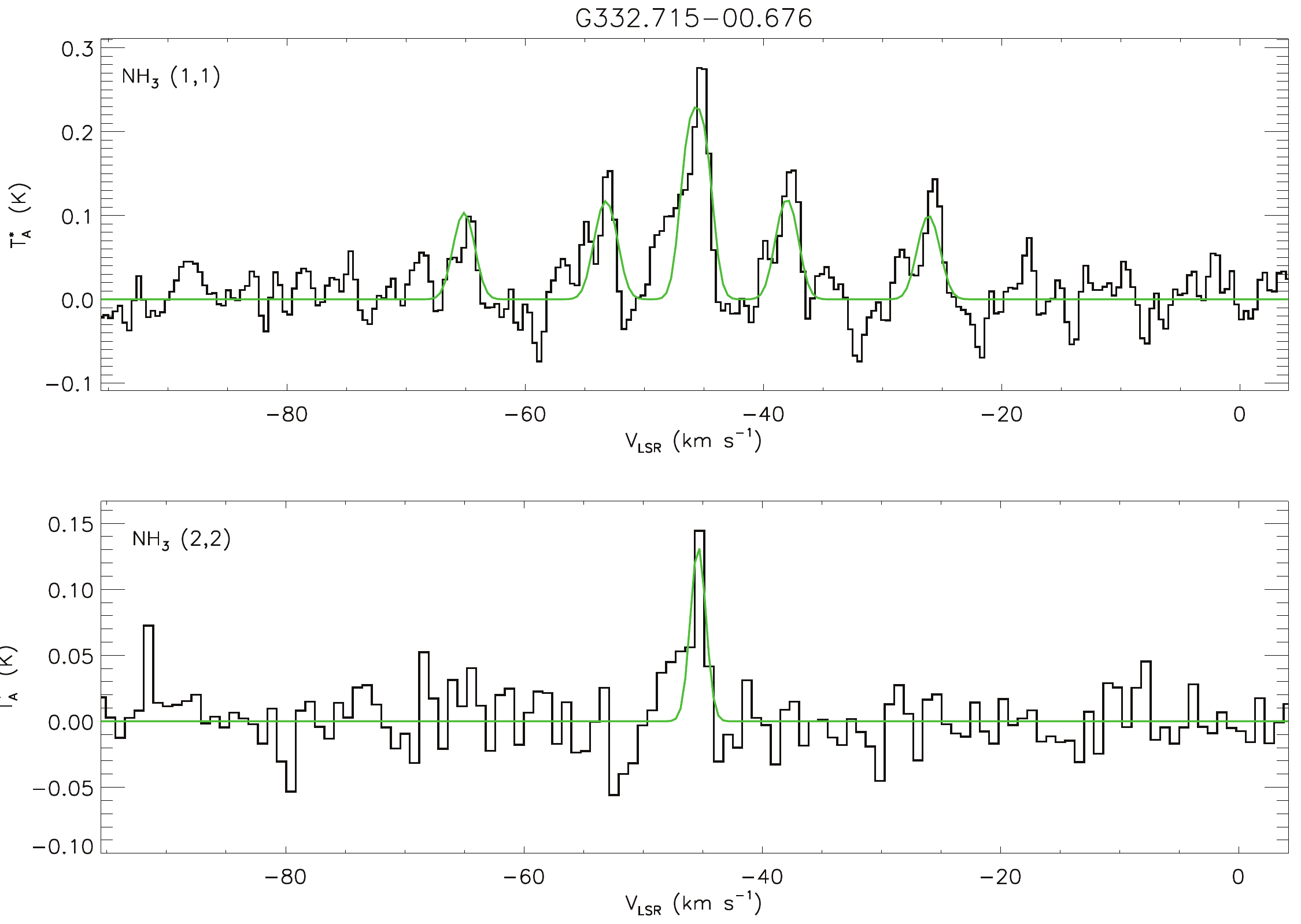}\\
\vspace*{0.5cm}
\includegraphics[width=0.49\textwidth, trim=-5 -30 -5 -30]{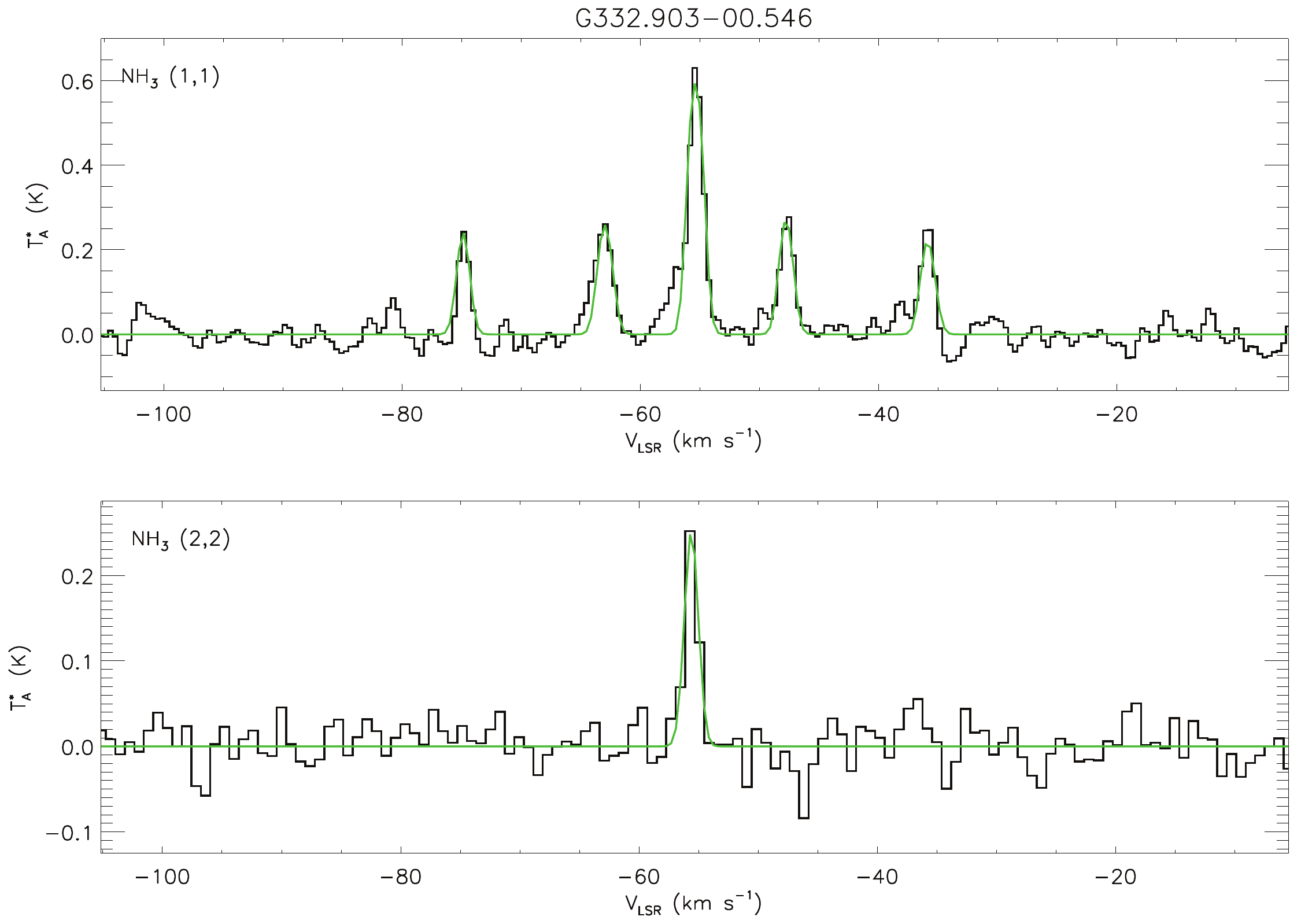}
\includegraphics[width=0.49\textwidth, trim=-5 -30 -5 -30]{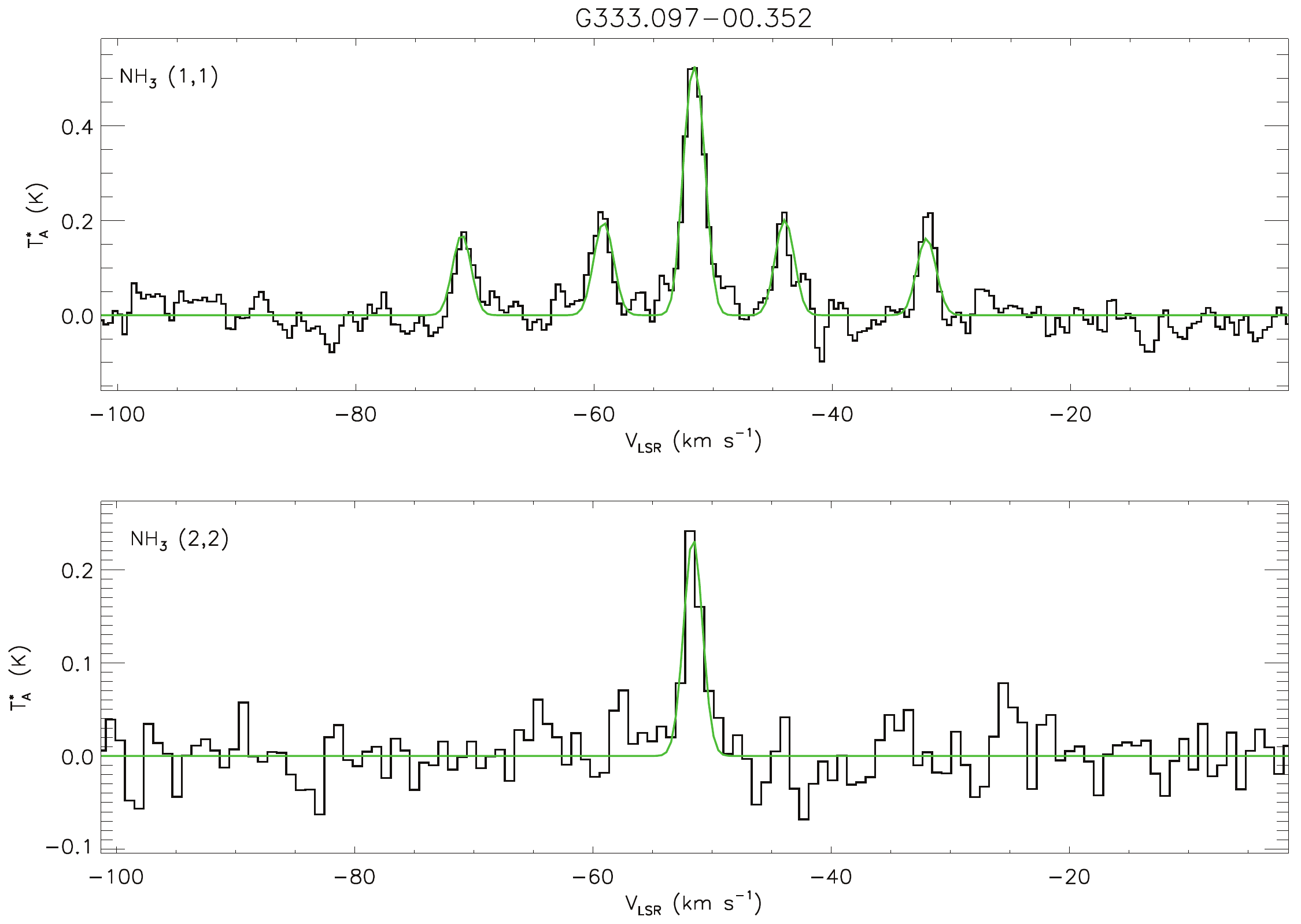}\\
\vspace*{0.5cm}
\includegraphics[width=0.49\textwidth, trim=-5 -30 -5 -30]{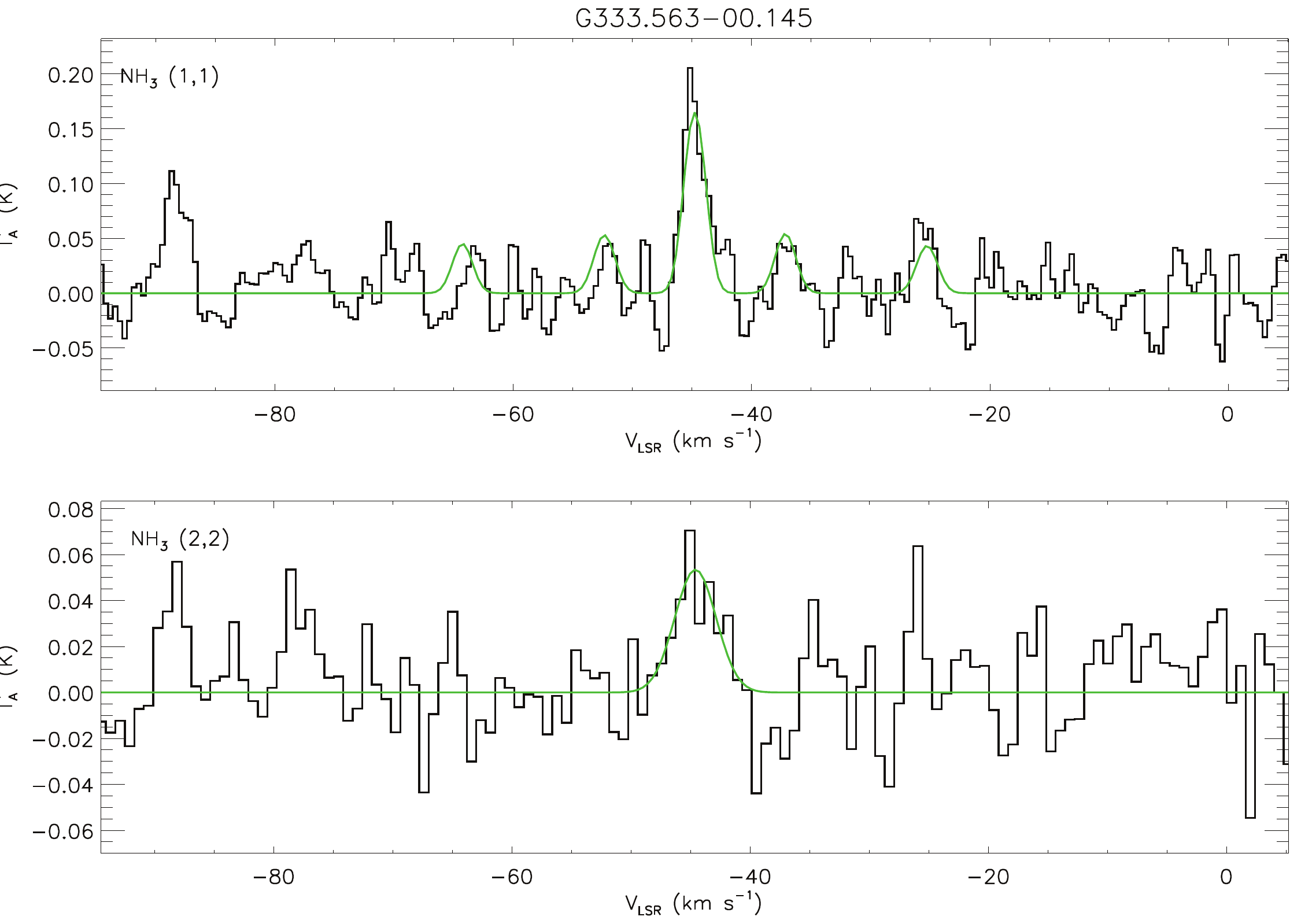}
\includegraphics[width=0.49\textwidth, trim=-5 -30 -5 -30]{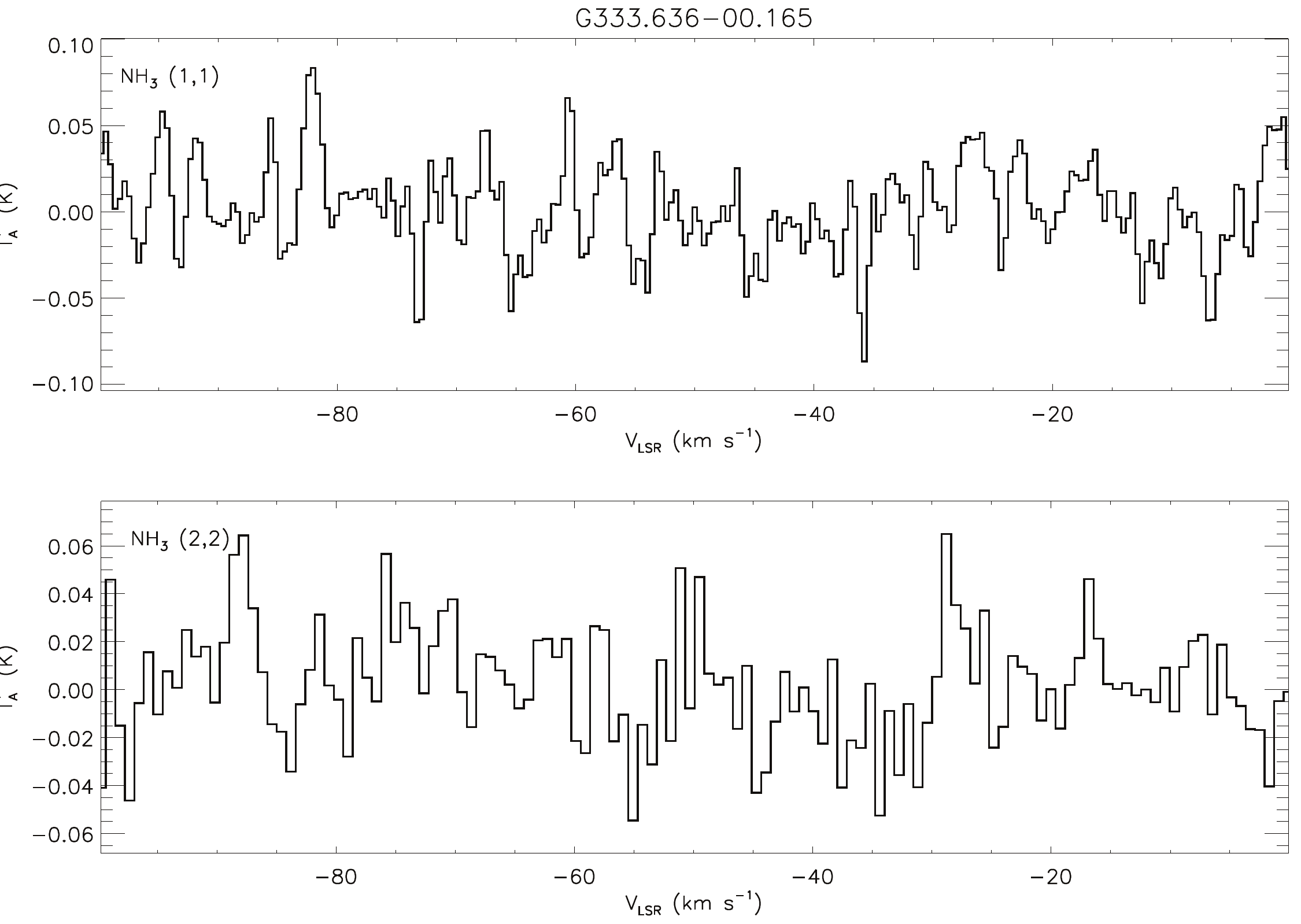}\\
\vspace*{0.5cm}
\contcaption{Clumps~49-54 are shown l-r, t-b.}
\end{figure*}

\begin{figure*}
\centering
\includegraphics[width=0.49\textwidth, trim=-5 -30 -5 -30]{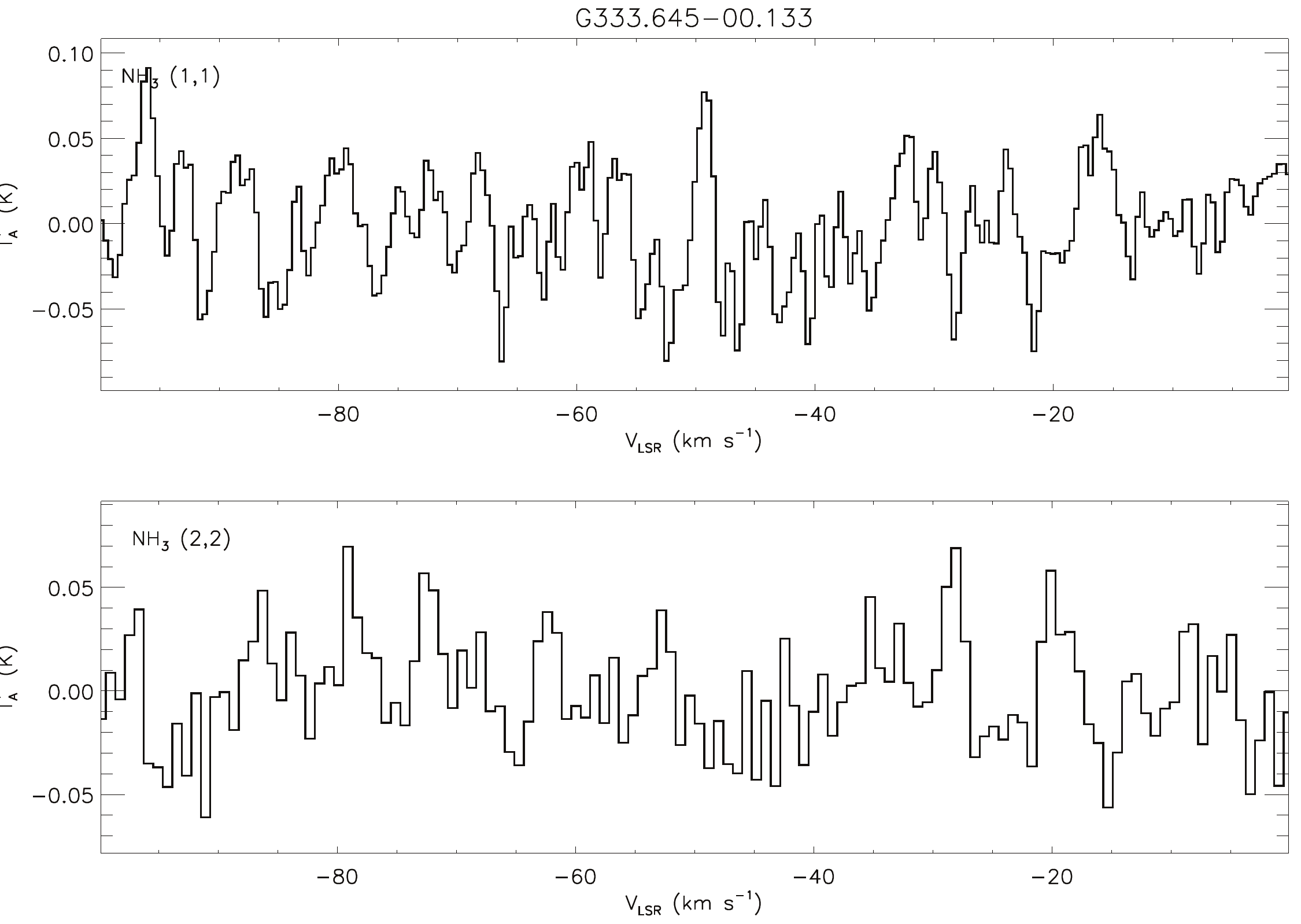}
\includegraphics[width=0.49\textwidth, trim=-5 -30 -5 -30]{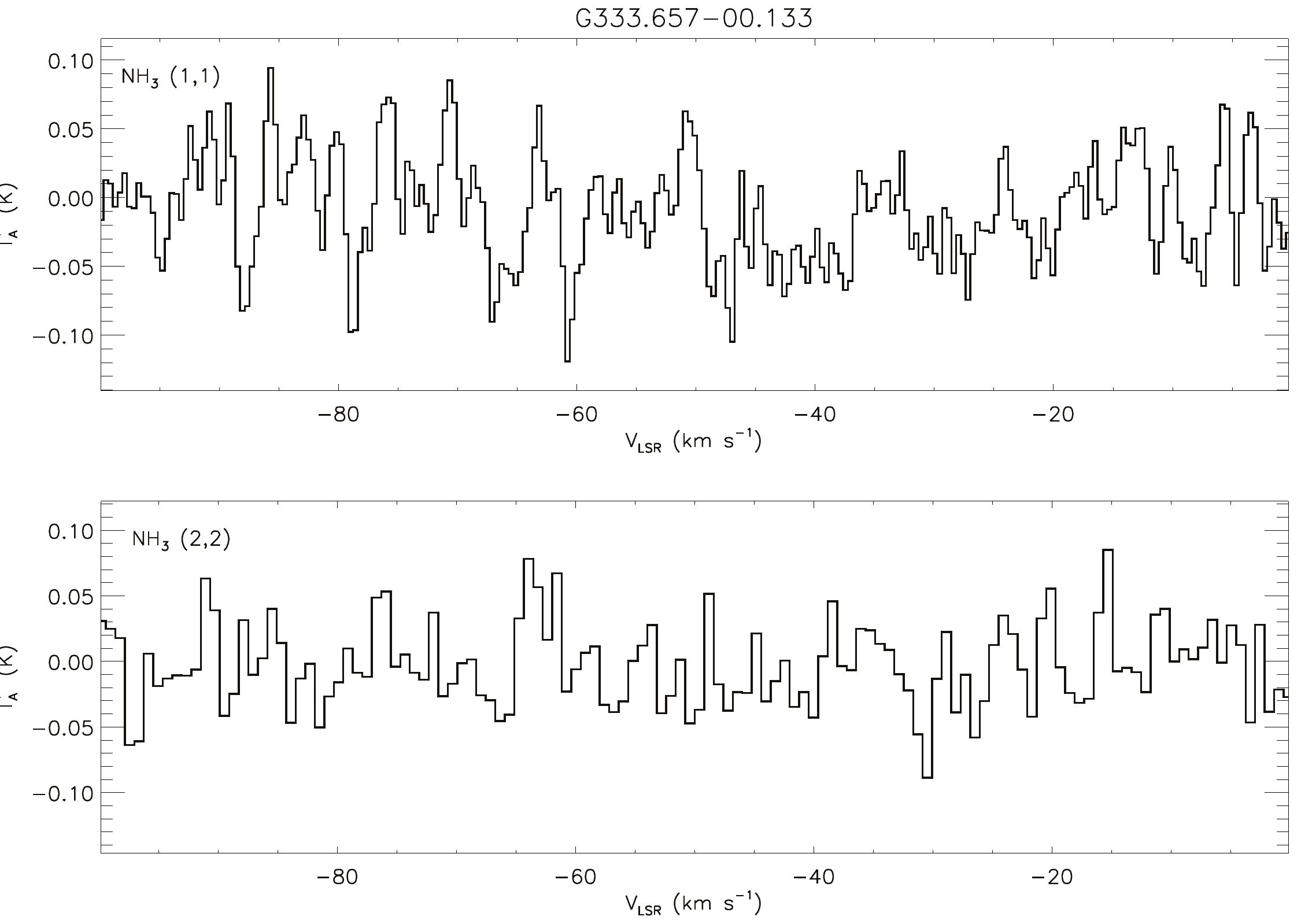}\\
\vspace*{0.5cm}
\includegraphics[width=0.49\textwidth, trim=-5 -30 -5 -30]{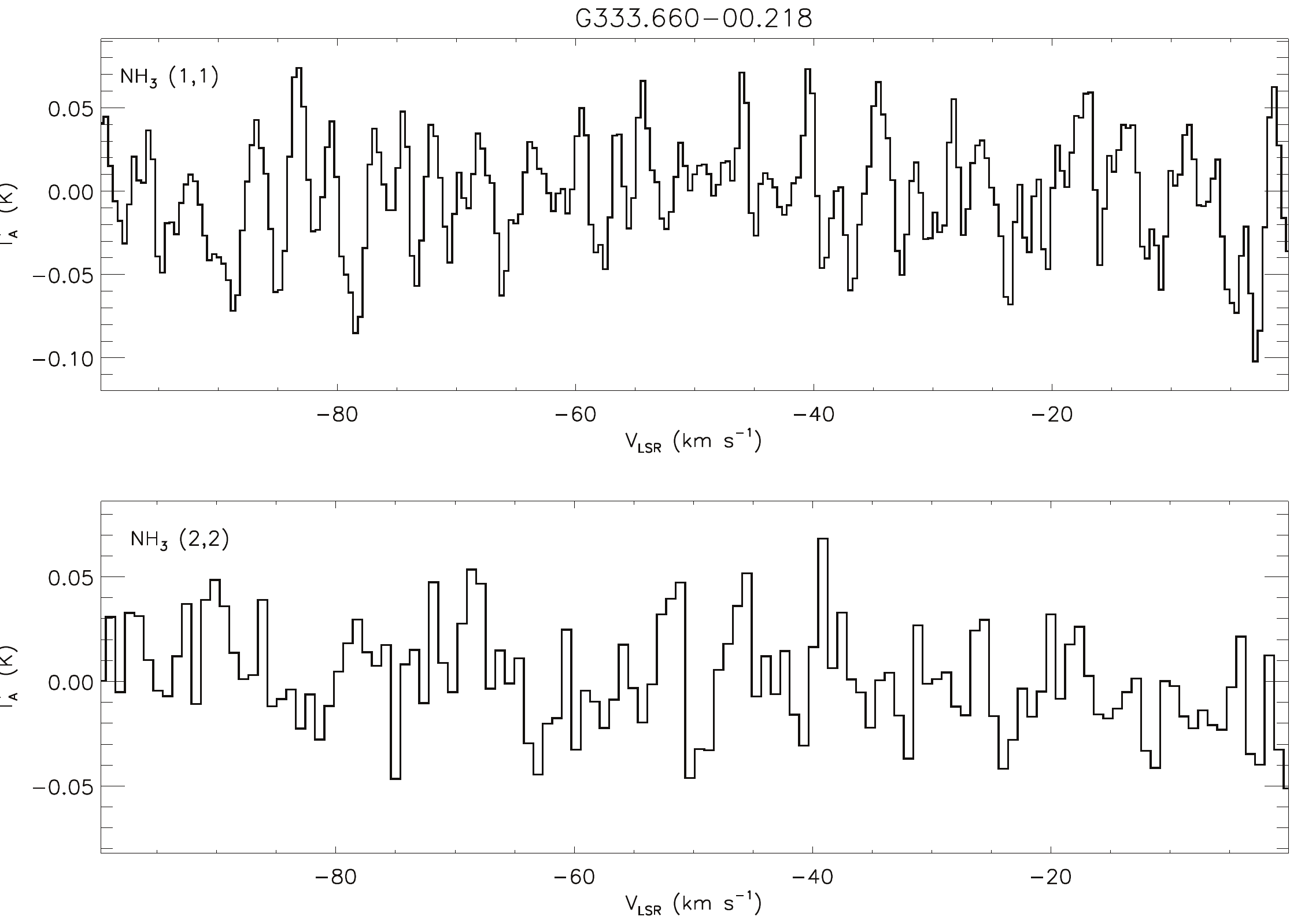}
\includegraphics[width=0.49\textwidth, trim=-5 -30 -5 -30]{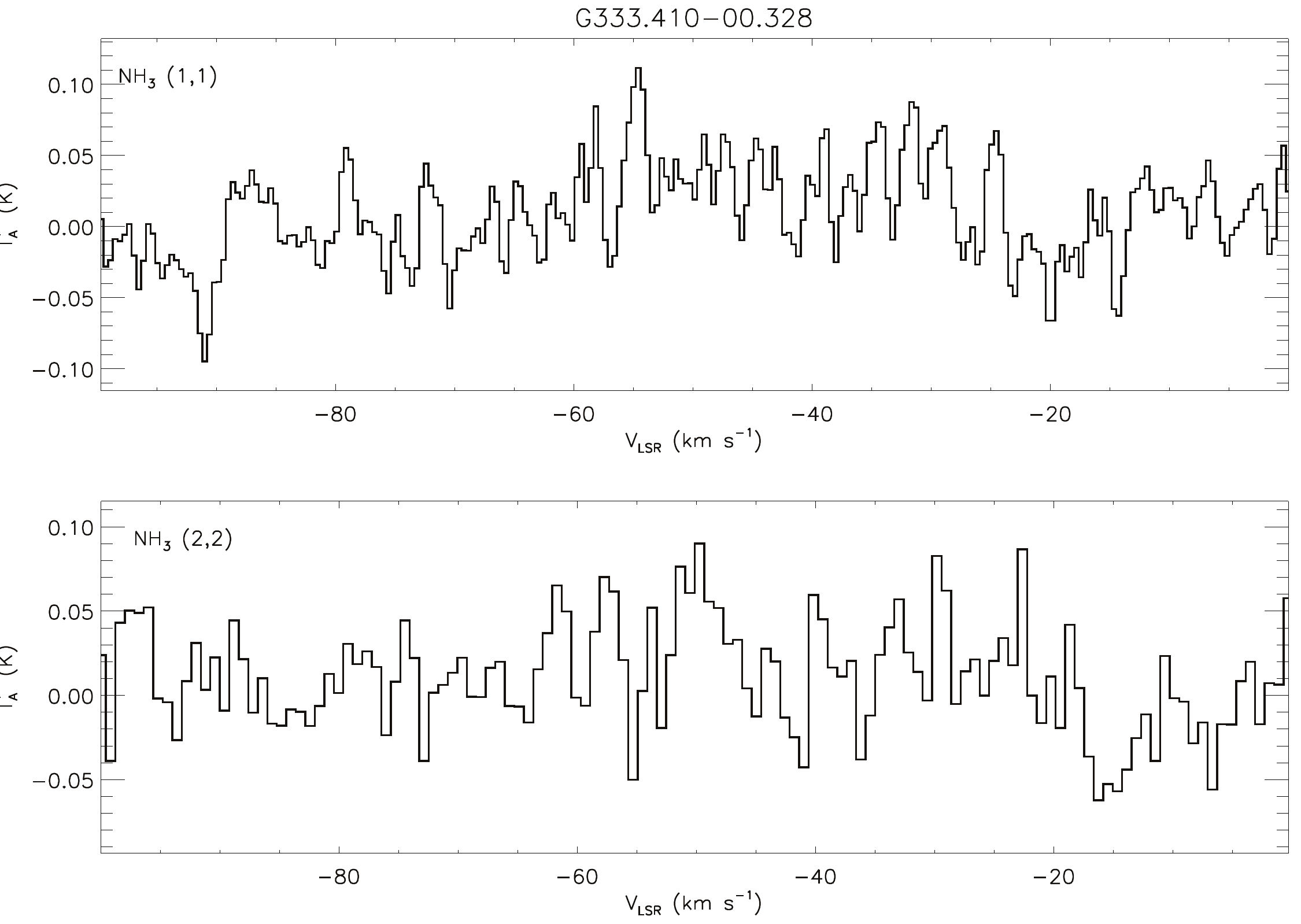}\\
\vspace*{0.5cm}
\includegraphics[width=0.49\textwidth, trim=-5 -30 -5 -30]{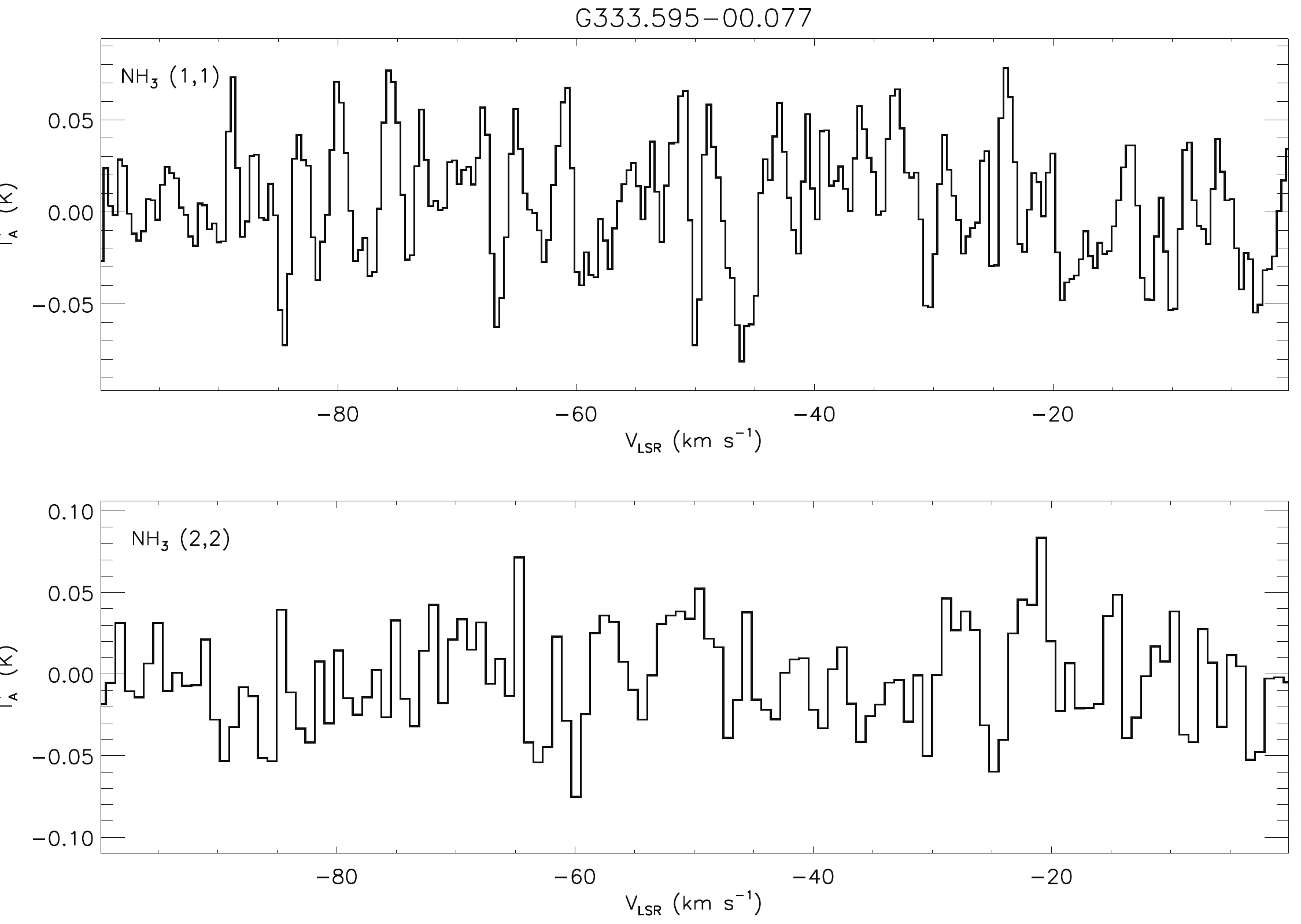}
\includegraphics[width=0.49\textwidth, trim=-5 -30 -5 -30]{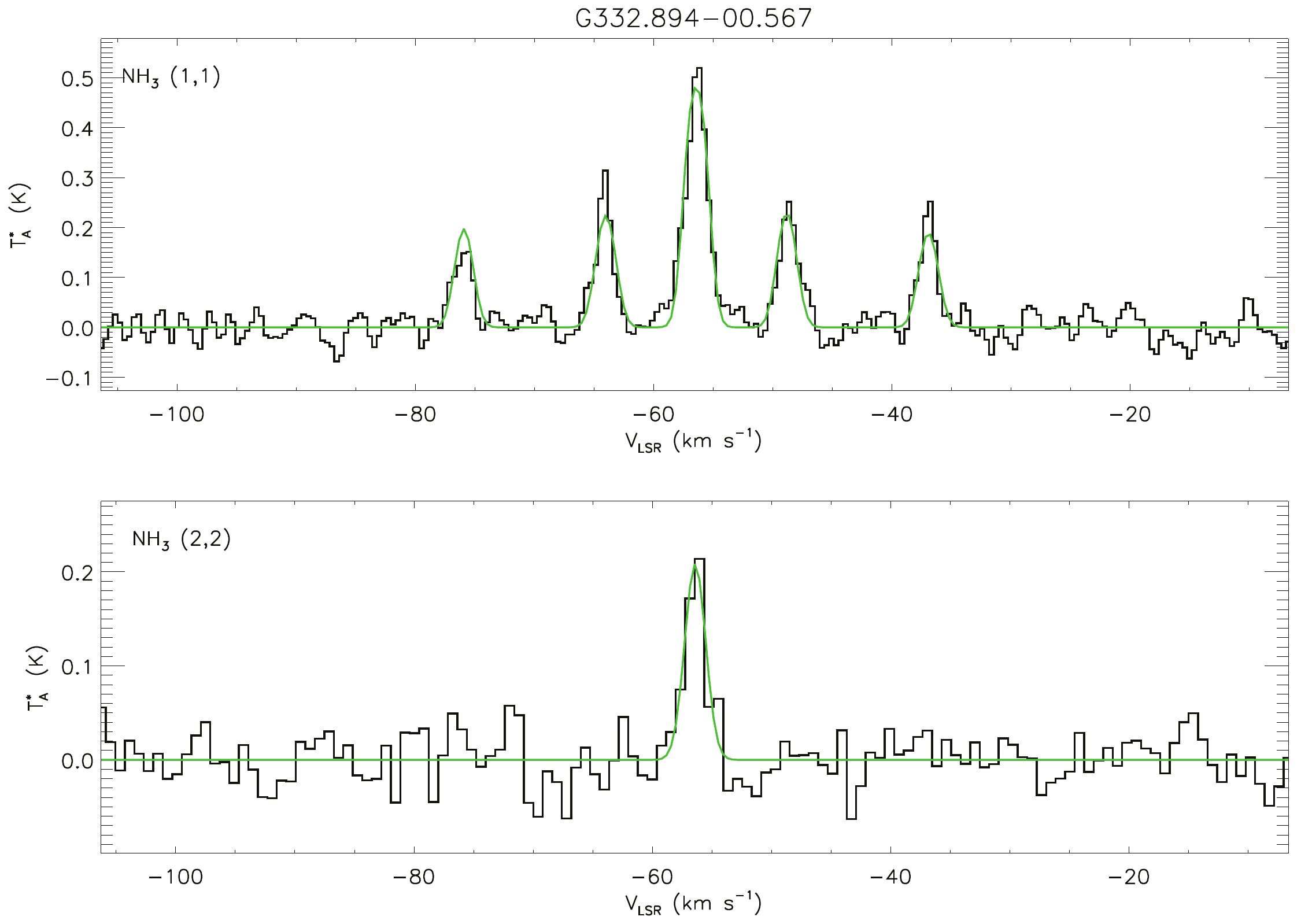}\\
\vspace*{0.5cm}
\contcaption{Clumps 55-60 are shown l-r, t-b.}
\end{figure*}

\begin{figure*}
\centering
\includegraphics[width=0.49\textwidth, trim=-5 -30 -5 -30]{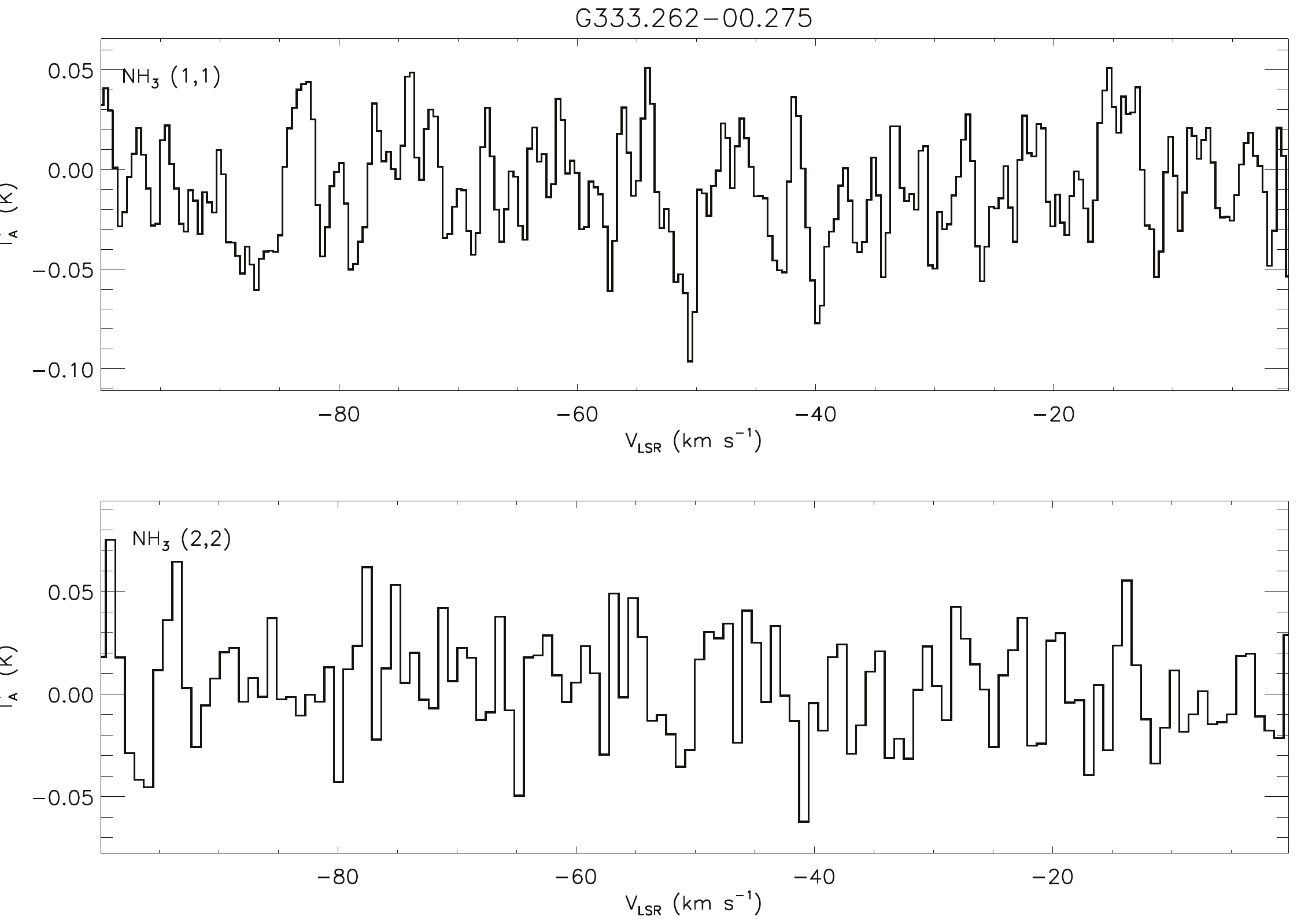}
\includegraphics[width=0.49\textwidth, trim=-5 -30 -5 -30]{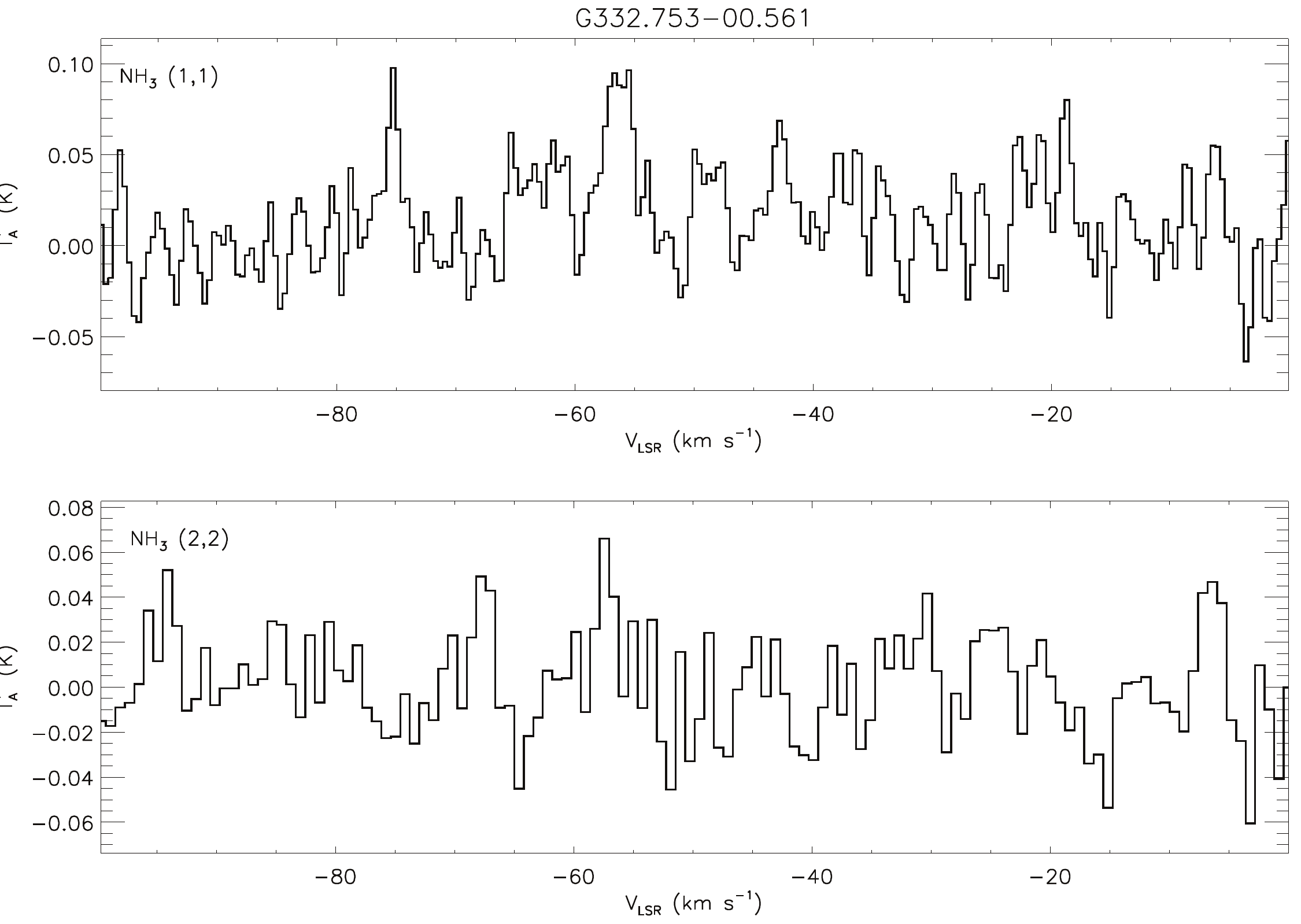}\\
\vspace*{0.5cm}
\includegraphics[width=0.49\textwidth, trim=-5 -30 -5 -30]{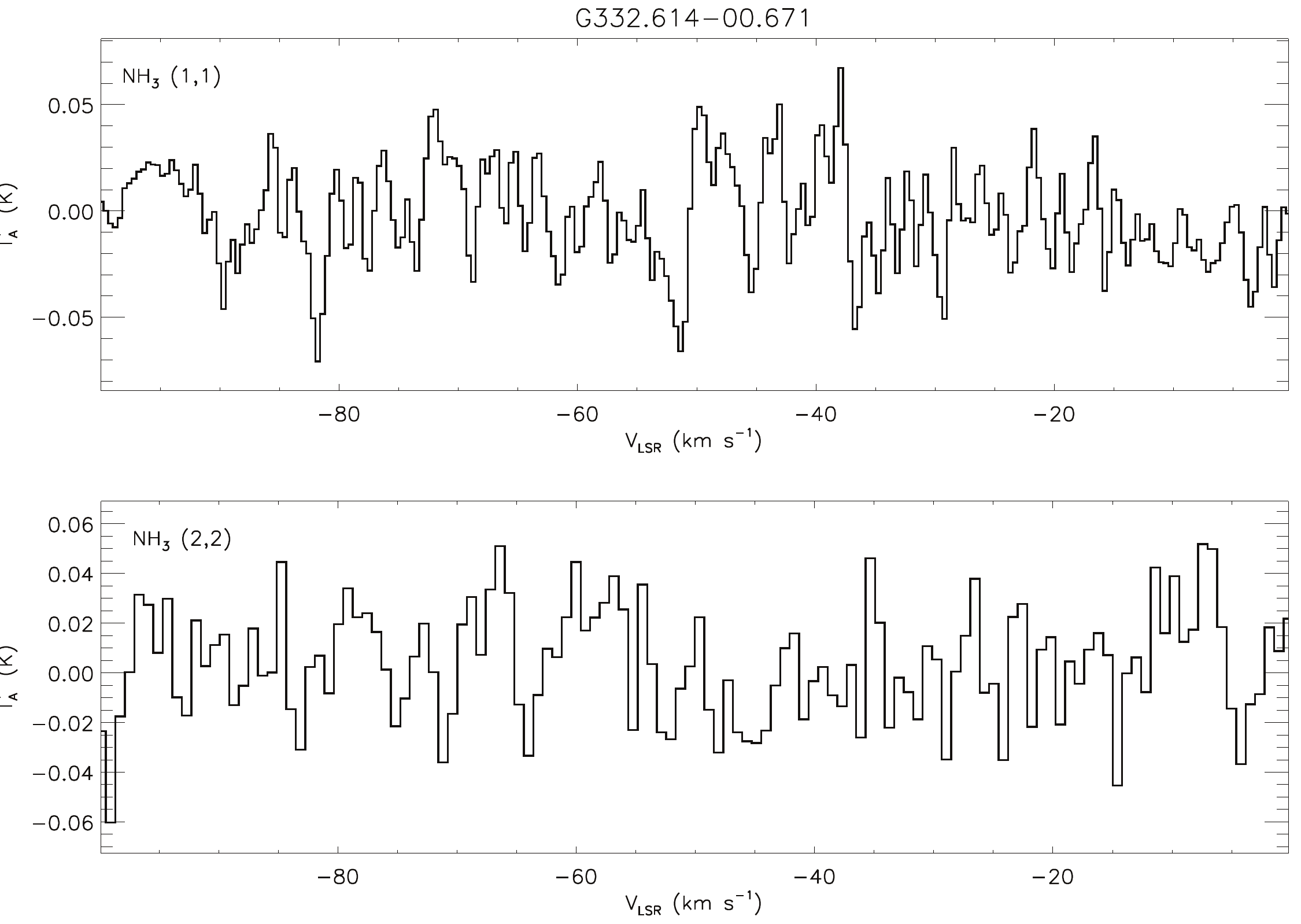}
\contcaption{Clumps 61-63 are shown l-r, t-b.}
\end{figure*}
\clearpage

\end{document}